\RequirePackage{etex}
\ifdefined\pdfsuppressptexinfo
\fi
\documentclass[11pt,letterpaper]{article}
\ifdefined\pdfobjcompresslevel
\fi

\usepackage[utf8]{inputenc}
\usepackage[T1]{fontenc}
\usepackage{lmodern}
\usepackage{etoolbox}

\usepackage{natbib}
\setcitestyle{authoryear,open={(},close={)},citesep={,}}

\usepackage{amsmath,amssymb,amsthm,mathtools}
\usepackage{mathrsfs}
\usepackage{dsfont}
\usepackage[dvipsnames]{xcolor}

\newcommand{\rev}[1]{{\normalcolor#1}}

\usepackage{graphicx}
\DeclareGraphicsExtensions{.pdf,.png,.jpg,.jpeg}
\usepackage{float}
\usepackage{placeins}
\usepackage{caption}
\usepackage{subcaption}
\usepackage{array}
\usepackage{tabularx}
\usepackage{booktabs}
\newcommand{\SphereDGPWidth}{0.96\textwidth}

\usepackage[shortlabels]{enumitem}
\setlist{
  topsep=0.35em,
  itemsep=0.15em,
  parsep=0pt,
  partopsep=0pt
}
\setlist[itemize,1]{leftmargin=*,label=\(\bullet\)}
\setlist[itemize,2]{leftmargin=*,label=\(\circ\)}
\setlist[enumerate,1]{
  leftmargin=*,
  label=\textup{(\roman*)},
  ref=\textup{(\roman*)}
}
\setlist[enumerate,2]{
  leftmargin=*,
  label=\textup{(\alph*)},
  ref=\theenumi\textup{(\alph*)}
}

\allowdisplaybreaks
\usepackage[expansion=false]{microtype}
\usepackage[letterpaper,margin=1in]{geometry}
\usepackage{setspace}

\usepackage{titling}
\usepackage{abstract}
\pretitle{%
  \begin{center}
  \begin{minipage}{0.94\textwidth}
  \centering
  \LARGE\bfseries
}
\posttitle{%
  \par
  \end{minipage}
  \end{center}
  \vspace{0.2em}
}
\preauthor{\begin{center}\large\scshape}
\postauthor{\par\end{center}\vspace{-0.6em}}
\renewcommand{\and}{\par}
\predate{\begin{center}\normalsize\itshape}
\postdate{\par\end{center}\vspace{-0.45em}}

\newcommand{\paperkeywords}[1]{%
  \par\noindent\textbf{Keywords:} #1\par
}

\usepackage[
  hypertexnames=false,
  colorlinks=true,
  linkcolor=blue,
  filecolor=blue,
  citecolor=blue,
  urlcolor=black
]{hyperref}
\IfFormatAtLeastF{2026-06-01}{\usepackage{aliascnt}}

\newtheoremstyle{plainthm}
  {6pt}{6pt}{\itshape}{0pt}{\normalfont\bfseries}{.}{0.5em}{}
\newtheoremstyle{definitionthm}
  {6pt}{6pt}{\normalfont}{0pt}{\normalfont\bfseries}{.}{0.5em}{}
\newtheoremstyle{remarkthm}
  {6pt}{6pt}{\normalfont}{0pt}{\normalfont\bfseries}{.}{0.5em}{}

\theoremstyle{plainthm}
\newtheorem{theorem}{Theorem}[section]
\IfFormatAtLeastTF{2026-06-01}{%
  \newtheorem{lemma}[theorem]{Lemma}
  \newtheorem{corollary}[theorem]{Corollary}
  \newtheorem{proposition}[theorem]{Proposition}
}{%
  \newaliascnt{lemma}{theorem}
  \newtheorem{lemma}[lemma]{Lemma}
  \aliascntresetthe{lemma}
  \newaliascnt{corollary}{theorem}
  \newtheorem{corollary}[corollary]{Corollary}
  \aliascntresetthe{corollary}
  \newaliascnt{proposition}{theorem}
  \newtheorem{proposition}[proposition]{Proposition}
  \aliascntresetthe{proposition}
}

\theoremstyle{definitionthm}
\IfFormatAtLeastTF{2026-06-01}{%
  \newtheorem{definition}[theorem]{Definition}
  \newtheorem{condition}[theorem]{Condition}
}{%
  \newaliascnt{definition}{theorem}
  
  \aliascntresetthe{definition}
  \newaliascnt{condition}{theorem}
  \newtheorem{condition}[condition]{Condition}
  \aliascntresetthe{condition}
}
\newtheorem{assumption}{Assumption}

\theoremstyle{remarkthm}
\newtheorem{remark}{Remark}[section]

\usepackage[nameinlink,noabbrev]{cleveref}
\usepackage{autonum}

\DeclareRobustCommand{\eqrefs}[1]{\labelcref{#1}}

\makeatletter
\AtBeginDocument{%
  \autonum@generatePatchedReferenceCSL{eqrefs}%
  \autonum@generatePatchedReferenceCSL{Cref}%
}
\makeatother

\DeclareRobustCommand{\crefplainname}[1]{{\normalcolor #1}}
\AtBeginDocument{%
}

\crefname{theorem}{\crefplainname{Theorem}}{\crefplainname{Theorems}}
\Crefname{theorem}{\crefplainname{Theorem}}{\crefplainname{Theorems}}
\crefname{lemma}{\crefplainname{Lemma}}{\crefplainname{Lemmas}}
\Crefname{lemma}{\crefplainname{Lemma}}{\crefplainname{Lemmas}}
\crefname{corollary}{\crefplainname{corollary}}{\crefplainname{corollaries}}
\Crefname{corollary}{\crefplainname{Corollary}}{\crefplainname{Corollaries}}
\crefname{proposition}{\crefplainname{Proposition}}{\crefplainname{Propositions}}
\Crefname{proposition}{\crefplainname{Proposition}}{\crefplainname{Propositions}}
\crefname{definition}{\crefplainname{definition}}{\crefplainname{definitions}}
\Crefname{definition}{\crefplainname{Definition}}{\crefplainname{Definitions}}
\crefname{remark}{\crefplainname{remark}}{\crefplainname{remarks}}
\Crefname{remark}{\crefplainname{Remark}}{\crefplainname{Remarks}}
\crefname{condition}{\crefplainname{condition}}{\crefplainname{conditions}}
\Crefname{condition}{\crefplainname{Condition}}{\crefplainname{Conditions}}
\crefname{assumption}{\crefplainname{Assumption}}{\crefplainname{Assumptions}}
\Crefname{assumption}{\crefplainname{Assumption}}{\crefplainname{Assumptions}}
\crefname{section}{\crefplainname{Section}}{\crefplainname{Sections}}
\Crefname{section}{\crefplainname{Section}}{\crefplainname{Sections}}
\crefname{subsection}{\crefplainname{Section}}{\crefplainname{Sections}}
\Crefname{subsection}{\crefplainname{Section}}{\crefplainname{Sections}}
\crefname{subsubsection}{\crefplainname{Section}}{\crefplainname{Sections}}
\Crefname{subsubsection}{\crefplainname{Section}}{\crefplainname{Sections}}
\crefname{figure}{\crefplainname{figure}}{\crefplainname{figures}}
\Crefname{figure}{\crefplainname{Figure}}{\crefplainname{Figures}}
\crefname{table}{\crefplainname{table}}{\crefplainname{tables}}
\Crefname{table}{\crefplainname{Table}}{\crefplainname{Tables}}
\crefname{equation}{\crefplainname{equation}}{\crefplainname{equations}}
\Crefname{equation}{\crefplainname{Equation}}{\crefplainname{Equations}}

\providecommand*{\theHremark}{}
\renewcommand*{\theHremark}{\thesection.\arabic{remark}}
\providecommand*{\theHexample}{}
\renewcommand*{\theHexample}{\thesection.\arabic{example}}

\newcommand{\E}{\mathbb{E}}
\newcommand{\R}{\mathbb{R}}

\newcommand{\cA}{\mathcal{A}}  \newcommand{\cC}{\mathcal{C}}
\newcommand{\cD}{\mathcal{D}} \newcommand{\cE}{\mathcal{E}} \newcommand{\cF}{\mathcal{F}}
\newcommand{\cG}{\mathcal{G}} \newcommand{\cH}{\mathcal{H}} \newcommand{\cI}{\mathcal{I}}
\newcommand{\cK}{\mathcal{K}} 
\newcommand{\cM}{\mathcal{M}} \newcommand{\cN}{\mathcal{N}} \newcommand{\cO}{\mathcal{O}}
 \newcommand{\cQ}{\mathcal{Q}} \newcommand{\cR}{\mathcal{R}}
\newcommand{\cS}{\mathcal{S}}  \newcommand{\cU}{\mathcal{U}}
\newcommand{\cV}{\mathcal{V}} \newcommand{\cW}{\mathcal{W}} \newcommand{\cX}{\mathcal{X}}
\newcommand{\cY}{\mathcal{Y}} \newcommand{\cZ}{\mathcal{Z}}

\DeclareMathOperator{\vol}{vol}
\DeclareMathOperator{\bias}{bias}
\DeclareMathOperator{\Hess}{Hess}
\DeclareMathOperator{\grad}{grad}

\DeclareMathOperator{\diag}{diag}
\DeclareMathOperator{\inj}{inj}
\DeclareMathOperator{\Cut}{Cut}
\DeclareMathOperator{\Log}{Log}
\DeclareMathOperator{\Exp}{Exp}
\DeclareMathOperator{\Cov}{Cov}

\DeclareMathOperator{\tr}{tr}
\DeclareMathOperator*{\argmin}{argmin}

\newcommand{\us}{\mathrm{us}}
\newcommand{\rbc}{\mathrm{rbc}}

\newcommand{\Sym}{\mathrm{Sym}}

\newcommand{\pto}{\xrightarrow{p}}
\newcommand{\dto}{\xrightarrow{d}}

\newcommand{\wh}{\widehat}

\newcommand{\PaperTitle}{%
Inference for conditional Fr\'echet means via Riemannian Robust Bias Correction%
}
\newcommand{\PaperAuthorsMain}{%
Ryunosuke Miyazaki\textsuperscript{\normalfont 1,}\thanks{Corresponding author. Email: miyazaki11051105@gmail.com}
\qquad
Chang Jun Im\textsuperscript{\normalfont 2}%
}
\newcommand{\PaperAuthorsSupp}{Ryunosuke Miyazaki and Chang Jun Im}
\newcommand{\PaperAuthorsSuppDisplay}{%
Ryunosuke Miyazaki\textsuperscript{\normalfont 1} and Chang Jun Im\textsuperscript{\normalfont 2}%
}

\newcommand{\PaperAffiliationsSupp}{%
\textsuperscript{\normalfont 1}\,Department of Social Data Science, Hitotsubashi University, Japan\\
\textsuperscript{\normalfont 2}\,Institute for Data Innovation in Science, Seoul National University, South Korea%
}

\providecommand{\noopsort}[1]{}

\usepackage{bibunits}
\defaultbibliographystyle{apalike}

\renewcommand{\SphereDGPWidth}{0.80\textwidth}

\hypersetup{
  pdftitle={\PaperTitle},
  pdfauthor={\PaperAuthorsSupp},
  colorlinks=true,
  citecolor=blue,
  linkcolor=blue,
  urlcolor=blue
}

\pretitle{\begin{center}\huge\bfseries}
\posttitle{\par\end{center}\vskip 0.5em}
\preauthor{\begin{center}\Large\scshape}

\makeatletter
\providecommand*{\theHsection}{}
\providecommand*{\theHsubsection}{}
\providecommand*{\theHsubsubsection}{}
\providecommand*{\theHequation}{}
\providecommand*{\theHfigure}{}
\providecommand*{\theHtable}{}
\providecommand*{\theHtheorem}{}
\providecommand*{\theHlemma}{}
\providecommand*{\theHproposition}{}
\providecommand*{\theHcorollary}{}
\providecommand*{\theHdefinition}{}
\providecommand*{\theHexample}{}
\providecommand*{\theHcondition}{}
\providecommand*{\theHassumption}{}
\providecommand*{\theHremark}{}

\newcommand{\switchtosupplementnumbering}{%
  \setcounter{section}{0}%
  \setcounter{subsection}{0}%
  \setcounter{subsubsection}{0}%
  \setcounter{equation}{0}%
  \setcounter{figure}{0}%
  \setcounter{table}{0}%
  \setcounter{theorem}{0}%
  \setcounter{assumption}{0}%
  \@addtoreset{equation}{section}%
  \@addtoreset{figure}{section}%
  \@addtoreset{table}{section}%
  \@addtoreset{theorem}{section}%
  \@addtoreset{assumption}{section}%
  \renewcommand{\thesection}{\Alph{section}}%
  \renewcommand{\thesubsection}{\thesection.\arabic{subsection}}%
  \renewcommand{\thesubsubsection}{\thesubsection.\arabic{subsubsection}}%
  \renewcommand{\theequation}{\thesection.\arabic{equation}}%
  \renewcommand{\thefigure}{\thesection.\arabic{figure}}%
  \renewcommand{\thetable}{\thesection.\arabic{table}}%
  \renewcommand{\thetheorem}{\thesection.\arabic{theorem}}%
  \renewcommand{\theassumption}{\thesection.\arabic{assumption}}%
  \renewcommand*{\theHsection}{supp.section.\Alph{section}}%
  \renewcommand*{\theHsubsection}{supp.section.\Alph{section}.\arabic{subsection}}%
  \renewcommand*{\theHsubsubsection}{supp.section.\Alph{section}.\arabic{subsection}.\arabic{subsubsection}}%
  \renewcommand*{\theHequation}{supp.equation.\Alph{section}.\arabic{equation}}%
  \renewcommand*{\theHfigure}{supp.figure.\Alph{section}.\arabic{figure}}%
  \renewcommand*{\theHtable}{supp.table.\Alph{section}.\arabic{table}}%
  \renewcommand*{\theHtheorem}{supp.theorem.\Alph{section}.\arabic{theorem}}%
  \renewcommand*{\theHlemma}{supp.theorem.\Alph{section}.\arabic{theorem}}%
  \renewcommand*{\theHproposition}{supp.theorem.\Alph{section}.\arabic{theorem}}%
  \renewcommand*{\theHcorollary}{supp.theorem.\Alph{section}.\arabic{theorem}}%
  \renewcommand*{\theHdefinition}{supp.theorem.\Alph{section}.\arabic{theorem}}%
  \renewcommand*{\theHexample}{supp.theorem.\Alph{section}.\arabic{theorem}}%
  \renewcommand*{\theHcondition}{supp.theorem.\Alph{section}.\arabic{theorem}}%
  \renewcommand*{\theHremark}{supp.remark.\Alph{section}.\arabic{remark}}%
  \renewcommand*{\theHassumption}{supp.assumption.\Alph{section}.\arabic{assumption}}%
}

\newenvironment{arxivsupplementformat}{%
  \small
  \setlength{\parskip}{0pt}%
  \setlength{\abovedisplayskip}{7pt plus 2pt minus 3pt}%
  \setlength{\belowdisplayskip}{7pt plus 2pt minus 3pt}%
  \setlength{\abovedisplayshortskip}{4pt plus 2pt minus 2pt}%
  \setlength{\belowdisplayshortskip}{4pt plus 2pt minus 2pt}%
  \setlist{topsep=0.25em,itemsep=0.10em,parsep=0pt,partopsep=0pt}%
  \captionsetup{font=small,labelfont=bf}%
  \renewcommand\section{%
    \@startsection{section}{1}{\z@}%
      {-2.0ex plus -0.5ex minus -0.2ex}%
      {0.9ex plus 0.2ex}%
      {\normalfont\large\bfseries}%
  }%
  \renewcommand\subsection{%
    \@startsection{subsection}{2}{\z@}%
      {-1.6ex plus -0.4ex minus -0.2ex}%
      {0.65ex plus 0.15ex}%
      {\normalfont\normalsize\bfseries}%
  }%
  \renewcommand\subsubsection{%
    \@startsection{subsubsection}{3}{\z@}%
      {-1.3ex plus -0.3ex minus -0.2ex}%
      {0.5ex plus 0.1ex}%
      {\normalfont\small\bfseries}%
  }%
}{}
\makeatother

\begin{document}

\begin{bibunit}[apalike]
\title{\PaperTitle}
\author{\PaperAuthorsMain}
\date{\PaperAffiliationsSupp}
\maketitle

\begin{abstract}
We develop an intrinsic framework for inference on conditional Fr\'echet means with predictors and responses taking values in Riemannian manifolds. \rev{We derive a second-order expansion of the smoothing displacement in the response tangent space and show that, unlike in regression with Euclidean responses, its leading term need not be determined by the conditional Fr\'echet mean function alone.} We propose Riemannian robust bias correction (RRBC), which estimates the \rev{leading smoothing displacement} using a local quadratic regression in the response tangent space and corrects the estimator through the exponential map. Our leave-one-out (LOO) covariance estimator uses intrinsic displacements between the full-sample and LOO RRBC estimates, accounting for the variability of the \rev{estimated leading smoothing displacement} and its covariance with the local linear estimator. We establish asymptotically valid pointwise confidence sets and simultaneous confidence tubes without undersmoothing. Simulations on spheres and symmetric positive-definite matrices show that RRBC can improve coverage when smoothing bias is nonnegligible. Cardiovascular and ocean-current applications illustrate simultaneous hypothesis testing and inference for \rev{smooth transformations and comparisons}.
\end{abstract}

\paperkeywords{Fr\'echet regression; local polynomial regression; manifold-valued data; simultaneous confidence tube; multiplier bootstrap.}

% !TeX root = arxiv_audited.tex

\section{Introduction}\label{sec:introduction}

Fr\'echet regression extends conditional-mean regression to responses in a general metric space by targeting the conditional Fr\'echet mean, defined by minimizing conditional expected squared distance \citep{PetersenandMuller2019}. This framework accommodates random objects such as probability distributions, covariance matrices, and spherical data without requiring linear operations on the response space or an external coordinate representation. \citet{PetersenandMuller2019} developed global and local Fr\'echet regression for Euclidean predictors, \citet{ChenandMuller2022} established uniform convergence theory for the local estimator, and \citet{ImandJeon2026a} developed local constant and local linear Fr\'echet regression for Riemannian predictors while allowing responses in general metric spaces.

Beyond estimating conditional Fr\'echet means, statistical inference is needed to assess whether they differ between populations or vary with predictor values. Pointwise inference quantifies uncertainty in conditional Fr\'echet means and their comparisons at a fixed predictor value, whereas simultaneous inference provides joint uncertainty quantification for conclusions over a predictor region.

Local smoothing creates bias that can affect inference even when the estimator is consistent. The local Fr\'echet regression estimator fluctuates around the localized Fr\'echet mean, whereas inference concerns the conditional Fr\'echet mean. At bandwidths of the mean squared error (MSE)-optimal order, smoothing bias can be of the same order as sampling variation, so uncorrected confidence sets need not attain nominal coverage even asymptotically. Undersmoothing reduces this bias, but smaller bandwidths reduce the effective local sample size and increase variability \citep{HallandHorowitz2013,Schennach2020}. For Euclidean local polynomial inference, \citet{Calonicoetal2018} studied robust bias correction and proposed a Studentization procedure that accounts for the variability of bias estimation and its covariance with the original estimator.

Central limit theorems for sample Fr\'echet means support inference on population means \citep{BhattacharyaandPatrangenaru2003,BhattacharyaandPatrangenaru2005,BhattacharyaandLin2017}. \citet{Miyazaki2026} constructs confidence sets for the entire mean set without assuming uniqueness. \citet{KurisuandOtsu2026} develop empirical-likelihood inference for manifold-valued Fr\'echet means and a localized pointwise procedure with Euclidean predictors under undersmoothing.

Geometry-specific regression methods include tests and simultaneous bands for global Wasserstein Fr\'echet regression of density responses \citep{Petersenetal2021}, regression for symmetric positive-definite matrix responses under the affine-invariant metric \citep{Yuanetal2012} and the Bures--Wasserstein metric \citep{XuandLi2025}, and simultaneous confidence tubes for tangent vector field regression on Riemannian manifolds \citep{Changetal2026}. Recent work also develops significance and specification tests for Fr\'echet regression in general metric spaces \citep{Songetal2026,Caietal2026}. To our knowledge, none combines intrinsic correction of local smoothing bias with feasible pointwise confidence sets and simultaneous confidence tubes for local Fr\'echet regression between Riemannian manifolds.

We develop an intrinsic framework for pointwise and simultaneous inference on conditional Fr\'echet means with Riemannian predictors and responses. Using logarithm maps, we construct the local polynomial design in the predictor tangent space and derive a second-order expansion of the smoothing displacement in the response tangent space. \rev{Unlike in regression with Euclidean responses, the leading smoothing displacement can depend on features of the conditional response distribution beyond the conditional Fr\'echet mean function.}

To obtain inference without undersmoothing, we develop Riemannian robust bias correction (RRBC). We estimate the \rev{leading smoothing displacement} using a local quadratic regression in the response tangent space and the inverse Riemannian Hessian of the empirical local linear objective, and correct the local linear estimator through the exponential map. We construct a leave-one-out (LOO) covariance estimator based on intrinsic displacements between the full-sample and LOO RRBC estimates, thereby accounting for the variability of the \rev{estimated leading smoothing displacement} and its covariance with the local linear estimator. We establish pointwise and uniform consistency of this covariance estimator.

We establish asymptotically valid pointwise Wald sets and simultaneous confidence tubes without undersmoothing. A Gaussian multiplier bootstrap calibrates the simultaneous confidence tubes over growing grids and compact predictor regions. The theory covers compactly supported responses away from the cut locus on general Riemannian manifolds, as well as responses with potentially unbounded support on Hadamard manifolds under moment and regularity conditions. Inference for smooth transformations \rev{and comparisons} of conditional Fr\'echet means can remain valid even when the full tangent-space covariance is singular, provided that the covariance of the transformed estimates or their comparisons is nondegenerate.

The simulation studies compare uncorrected inference, undersmoothing, and RRBC on spheres and symmetric positive-definite matrices. Cardiovascular and ocean-current applications use RRBC to assess differences in covariance trajectories across age strata and in heading fields between ENSO phases.

The rest of the paper is organized as follows. \Cref{sec:setup} introduces the problem setting and the local linear Fr\'echet estimator. \Cref{sec:method} develops RRBC and the proposed inference procedures. \Cref{sec:theory} presents the asymptotic theory for pointwise and simultaneous inference. \Cref{sec:simulation,sec:empirical-analyses} report the simulation studies and real-data analyses, respectively. \Cref{sec:discussion} concludes with a discussion. The Supplementary Material contains all proofs and additional details.

\section{Problem Setting and Local Linear Fr\'echet Estimator}\label{sec:setup}

\subsection{Problem setting}\label{subsec:problem-setting}

Let \((\cX,g_{\cX})\) and \((\cY,g_{\cY})\) be connected, complete smooth Riemannian manifolds of dimensions \(p\) and \(q\), with geodesic distances \(d_{\cX}\) and \(d_{\cY}\). Let \((X_i,Y_i)\), \(i=1,\ldots,n\), be independent copies of \((X,Y)\in\cX\times\cY\). Write \(\mathbb P_{Y\mid X=z}\) for the conditional distribution of \(Y\) given \(X=z\). Define the conditional Fr\'echet function
\begin{equation}
F(\omega;z) \coloneqq \frac12 \int_{\cY} d_{\cY}^2(y,\omega)\, \mathbb P_{Y\mid X=z}(dy), \quad \omega\in\cY,\quad z\in\cX.
\end{equation}
Our inferential target is the conditional Fr\'echet mean, defined by
\[
m(z) \coloneqq \operatorname*{argmin}_{\omega\in\cY}F(\omega;z),\quad z\in\cX,
\]
whenever the minimizer is unique.

\subsection{Geometric notation}\label{subsec:geometric-notation}

Throughout this subsection, \(\mathcal M\) denotes either \(\cX\) or \(\cY\). For \(z\in\mathcal M\), let \(T_z\mathcal M\) and \(0_z\) denote its tangent space and zero vector, respectively. We write \(\langle\cdot,\cdot\rangle_z\) for the inner product on \(T_z\mathcal M\) induced by its Riemannian metric and \(\|\cdot\|_z\) for the corresponding norm.

For a \(C^1\) map \(\Phi:\mathcal M\to\mathcal N\), where \(\mathcal N\) is a smooth manifold, let \((d\Phi)_z:T_z\mathcal M\to T_{\Phi(z)}\mathcal N\) denote its differential at \(z\). For a finite-dimensional real vector space \(V\) and \(a\in V\), identify \(T_aV\) canonically with \(V\). We use \(D\) for ordinary derivatives between fixed vector spaces and \(\nabla\) for covariant derivatives.

For a smooth map \(\Phi:U\to V\), where \(U\subseteq\mathcal M\) is open and \(V\) is a fixed finite-dimensional real inner-product space, let \(\nabla_{\mathcal M}^2\Phi(z):T_z\mathcal M\times T_z\mathcal M\to V\) denote its second covariant derivative at \(z\in U\), given by \((\nabla(d\Phi))_z\) using the Levi-Civita connection on \(\mathcal M\) and the flat connection on \(V\).

For a smooth function \(Q:\mathcal M\to\R\), let \(\grad_z^{\mathcal M}Q\in T_z\mathcal M\) and \(\Hess_z^{\mathcal M}Q:T_z\mathcal M\to T_z\mathcal M\) denote its Riemannian gradient and Hessian operator at \(z\in\mathcal M\), respectively. These representations are related by
\[
(dQ)_z(u)=\langle\grad_z^{\mathcal M}Q,u\rangle_z,\quad \nabla_{\mathcal M}^2Q(z)(u,v)=\langle(\Hess_z^{\mathcal M}Q)u,v\rangle_z,\quad u,v\in T_z\mathcal M.
\]

For \(z\in\mathcal M\), let \(\Exp_z^{\mathcal M}:T_z\mathcal M\to\mathcal M\) denote the exponential map, \(\Cut_{\mathcal M}(z)\subseteq\mathcal M\) the cut locus of \(z\), and \(\inj_{\mathcal M}(z)\) the injectivity radius at \(z\). Let \(\Log_z^{\mathcal M}:\mathcal M\setminus\Cut_{\mathcal M}(z)\to T_z\mathcal M\) denote the logarithm map. For \(z'\in\mathcal M\setminus\Cut_{\mathcal M}(z)\), let \(\Pi_{z\to z'}^{\mathcal M}:T_z\mathcal M\to T_{z'}\mathcal M\) denote parallel transport along the unique minimizing geodesic from \(z\) to \(z'\). For convenience, for \(x\in\cX\), we extend \(\Log_x^{\cX}\) to all of \(\cX\) by setting \(\Log_x^{\cX}z=0_x\) for \(z\in\Cut_{\cX}(x)\).

For finite-dimensional real inner-product spaces \(V\) and \(W\), and vectors \(a\in V\) and \(b\in W\), define \(a\otimes b:W\to V\) by \((a\otimes b)v\coloneqq\langle b,v\rangle_W a\) for \(v\in W\). When \(V=W\), write \(a^{\otimes2}\coloneqq a\otimes a\).

\subsection{Local linear Fr\'echet estimator}\label{subsec:target-estimator}

Following \citet{PetersenandMuller2019} and \citet{ImandJeon2026a}, we construct local linear weights for Riemannian predictors at each evaluation point \(x\in\cX\) using logarithm vectors in \(T_x\cX\). To include an intercept, let \(\cR_{1,x}\coloneqq\R\oplus T_x\cX\) be equipped with the product inner product \(\langle(s,u),(s',u')\rangle_{\cR_{1,x}}\coloneqq ss'+\langle u,u'\rangle_x\) for \(s,s'\in\R\) and \(u,u'\in T_x\cX\). For \(u\in T_x\cX\), define \(r_{1,x}(u)\coloneqq(1,u)\in\cR_{1,x}\), and let \(e_{1,x}\coloneqq(1,0_x)\in\cR_{1,x}\). Let \(K_1:[0,\infty)\to[0,\infty)\) be a kernel supported on \([0,R_1]\). For \(h>0\), define
\begin{equation}\label{eq:main-predictor-kernel}
K_{1,h}(z,x) \coloneqq h^{-p} K_1\left(d_{\cX}(x,z)/h\right) \mathds1\{d_{\cX}(x,z)<\inj_{\cX}(x)\}.
\end{equation}
Define the population and empirical local linear design operators by
\begin{align}
S_{1,h}(x) \coloneqq \E\left[ K_{1,h}(X,x) r_{1,x}(\Log_x^{\cX}X/h)^{\otimes2} \right],\quad \widehat S_{1,h}(x) \coloneqq \frac1n \sum_{i=1}^n K_{1,h}(X_i,x) r_{1,x}(\Log_x^{\cX}X_i/h)^{\otimes2}.
\end{align}
When \(S_{1,h}(x)\) and \(\widehat S_{1,h}(x)\) are invertible, define the population and empirical local linear weights by
\begin{equation+} \label{eq:main-population-local-linear-weight}
\begin{aligned}
L_{1,h}(X;x) &\coloneqq K_{1,h}(X,x) \left\langle S_{1,h}(x)^{-1} r_{1,x}(\Log_x^{\cX}X/h), e_{1,x} \right\rangle_{\cR_{1,x}},\\
\widehat L_{1,h,i}(x) &\coloneqq K_{1,h}(X_i,x) \left\langle \widehat S_{1,h}(x)^{-1} r_{1,x}(\Log_x^{\cX}X_i/h), e_{1,x} \right\rangle_{\cR_{1,x}}, \quad i=1,\ldots,n.
\end{aligned}
\end{equation+}
These weights satisfy
\[
\E\left[ L_{1,h}(X;x)\, r_{1,x}(\Log_x^{\cX}X/h) \right] = \frac1n \sum_{i=1}^n \widehat L_{1,h,i}(x)\, r_{1,x}(\Log_x^{\cX}X_i/h) = e_{1,x}.
\]
Define the population and empirical objective functions by
\[
Q_{h,x}(\omega) \coloneqq \frac12 \E\left[ L_{1,h}(X;x)\,d_{\cY}^2(Y,\omega) \right], \quad \widehat Q_{h,x}(\omega) \coloneqq \frac1{2n} \sum_{i=1}^n \widehat L_{1,h,i}(x)\,d_{\cY}^2(Y_i,\omega),\quad \omega\in\cY.
\]
Whenever the corresponding minimizers exist, choose
\[
m_h(x) \in \operatorname*{argmin}_{\omega\in\cY}Q_{h,x}(\omega), \qquad \widehat m_h(x) \in \operatorname*{argmin}_{\omega\in\cY}\widehat Q_{h,x}(\omega).
\]
\rev{We refer to} \(m_h(x)\) \rev{as the localized Fr\'echet mean.} Existence of these minimizers is established in \Cref{sec:supp_consistency_localization} of the Supplementary Material.

\section{Riemannian Robust Bias Correction and Intrinsic Inference}\label{sec:method}

\subsection{Smoothing displacement and the undersmoothing estimator}\label{subsec:undersmoothing-confidence-set}

\rev{The local linear estimator targets the localized Fr\'echet mean} \(m_h(x)\), \rev{whereas inference concerns the conditional Fr\'echet mean} \(m(x)\). Write \(B_h(x)\coloneqq\Log_{m(x)}^{\cY}m_h(x)\) for the smoothing displacement. For Euclidean responses, \(B_h(x)=m_h(x)-m(x)\). By \Cref{thm:local_linear_expansion_bias},
\[
\bigl\|\Log_{m(x)}^{\cY}\widehat m_h(x)-B_h(x)\bigr\|_{m(x)}=O_p((nh^p)^{-1/2}).
\]
Define the conditional score at \(m(x)\) by
\[
s_x(z)\coloneqq\grad_{m(x)}^{\cY}F(\cdot;z)=-\E[\Log_{m(x)}^{\cY}Y\mid X=z]\in T_{m(x)}\cY,\quad z\in\cV_x,
\]
where \(\cV_x\) is the neighborhood of \(x\) satisfying \Cref{ass:local_support_log_geometry}. Since \(m(x)\) minimizes \(F(\cdot;x)\), \(s_x(x)=0_{m(x)}\).  Let \(\nabla_{\cX}^2s_x(x):T_x\cX\times T_x\cX\to T_{m(x)}\cY\) denote \(\left.\nabla_{\cX}^2s_x(z)\right|_{z=x}\), with \(x\) and the response center \(m(x)\) held fixed under differentiation in \(z\). Let \(A(x):T_{m(x)}\cY\to T_{m(x)}\cY\) denote the Riemannian Hessian of the conditional Fr\'echet function \(F(\cdot;x)\) at \(m(x)\), defined by
\begin{equation}
A(x)\coloneqq\Hess_{m(x)}^{\cY}F(\cdot;x).
\end{equation}
Define the local linear second-moment operator by
\begin{equation+}\label{eq:main-local-linear-second-moment}
M_{2,h}(x) \coloneqq \E\left[ L_{1,h}(X;x) \bigl(\Log_x^{\cX}X/h\bigr)^{\otimes2} \right].
\end{equation+}
Although \(s_x(x)=0_{m(x)}\), smoothing can leave a nonzero weighted score at \(m(x)\). The local linear weights cancel the linear variation across predictor locations, so the leading discrepancy is determined by \(\nabla_{\cX}^2s_x(x)\) and \(M_{2,h}(x)\). The factor \(-A(x)^{-1}\) converts this score discrepancy into the leading displacement. Combining these effects gives the bias coefficient
\begin{equation}\label{eq:explicit-bias-expansion}
\beta_h(x)\coloneqq-\frac12 A(x)^{-1}\E\left[L_{1,h}(X;x)\nabla_{\cX}^2s_x(x)\left(\Log_x^{\cX}X/h,\Log_x^{\cX}X/h\right)\right].
\end{equation}
By \Cref{thm:local_linear_expansion_bias},
\[
B_h(x) = h^2\beta_h(x)+o(h^2), \qquad \beta_h(x)=O(1).
\]

\begin{remark}
Suppose \(\cX=\R^p\) and \(\cY=\R\), both with their Euclidean metrics. Then \(s_x(z)=m(x)-m(z)\) and \(A(x)=1\), so
\[
\beta_h(x)=\frac12\operatorname{tr}\left(M_{2,h}(x)\Hess_x^{\cX}m\right),
\]
which coincides with the usual bias coefficient for local linear regression \citep{RuppertandWand1994,Calonicoetal2018}.
\end{remark}

Since \(B_h(x)=O(h^2)\), undersmoothing (US) makes this displacement negligible relative to sampling variation by imposing \(\sqrt{nh^p}h^2\to0\). The US estimator is defined by
\[
\widehat m_{\mathrm{us}}(x)\coloneqq\widehat m_h(x).
\]

\subsection{Riemannian robust bias correction estimator}\label{subsec:rrbc}

To construct RRBC, we estimate the leading smoothing displacement \(h^2\beta_h(x)\) in \eqref{eq:explicit-bias-expansion}. For a response center \(\omega\in\cY\) at which the required logarithms and Hessians are well defined, let \(\widehat A_{h,x}(\omega):T_\omega\cY\to T_\omega\cY\) denote the Riemannian Hessian of the empirical local linear objective \(\widehat Q_{h,x}\) at \(\omega\), given by
\[
\widehat A_{h,x}(\omega) \coloneqq \Hess_\omega^{\cY}\widehat Q_{h,x}(\cdot) = \frac1n \sum_{i=1}^n \widehat L_{1,h,i}(x)\cH_\omega(Y_i),
\]
where \(\cH_\omega(y)\coloneqq\tfrac12\Hess_\omega^{\cY}d_{\cY}^2(y,\cdot)\) is the Hessian of the loss \(d_{\cY}^2(y,\cdot)/2\), viewed as a self-adjoint operator on \(T_\omega\cY\), defined for \(y\notin\Cut_{\cY}(\omega)\). The empirical local linear second-moment operator is
\begin{equation+}\label{eq:main-empirical-local-linear-second-moment}
\widehat M_{2,h}(x) \coloneqq \frac1n \sum_{i=1}^n \widehat L_{1,h,i}(x) (\Log_x^{\cX}X_i/h)^{\otimes2}.
\end{equation+}

\begin{figure}[t]
    \centering
    \includegraphics[width=0.98\textwidth]{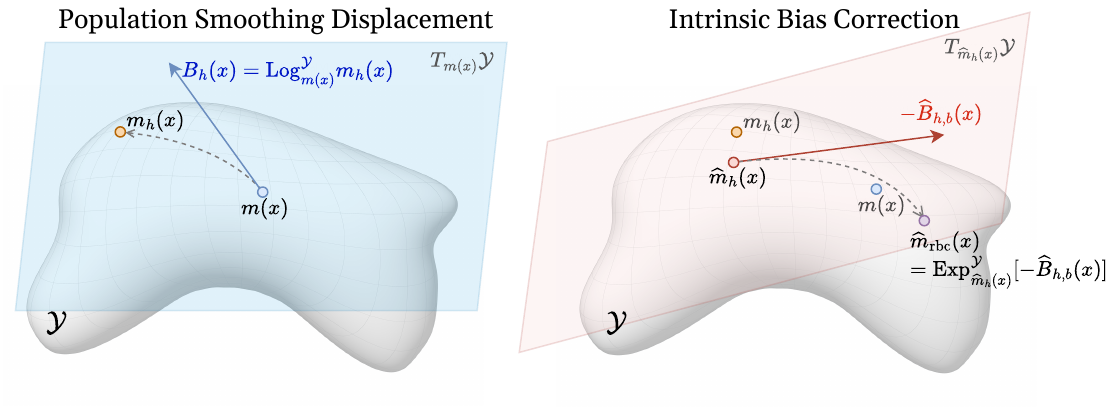}
\caption{Population smoothing displacement and intrinsic bias correction.}
    \label{fig:intrinsic-bias-correction}
\end{figure}

At a response center \(\omega\), we fit a local quadratic regression of \(\Log_\omega^{\cY}Y_i\) on \(\Log_x^{\cX}X_i/b\). Let \(\cS(T_x\cX)\) denote the space of self-adjoint operators on \(T_x\cX\), equipped with the Hilbert--Schmidt inner product. Define \(\cR_{2,x}\coloneqq\R\oplus T_x\cX\oplus\cS(T_x\cX)\), equipped with the product inner product, and, for \(u\in T_x\cX\), let \(r_{2,x}(u)\coloneqq\left(1,u,\frac12u^{\otimes2}\right)\in\cR_{2,x}\). Let \(K_2\) be the kernel for the local quadratic fit, and let \(K_{2,b}\) denote \eqref{eq:main-predictor-kernel} with \((K_1,h)\) replaced by \((K_2,b)\). Define the design operator and the coefficient operator by
\begin{align}
\widehat S_{2,b}(x) &\coloneqq \frac1n \sum_{i=1}^n K_{2,b}(X_i,x) r_{2,x}(\Log_x^{\cX}X_i/b)^{\otimes2},\\
\widehat\Theta_{2,b,x}(\omega) &\coloneqq \left[ \frac1n \sum_{i=1}^n K_{2,b}(X_i,x) \left\{ \Log_\omega^{\cY}Y_i \otimes r_{2,x}(\Log_x^{\cX}X_i/b) \right\} \right] \widehat S_{2,b}(x)^{-1},
\end{align}
where \rev{\(\widehat\Theta_{2,b,x}(\omega)\)} is defined when \(\widehat S_{2,b}(x)\) is invertible. For \(S\in\cS(T_x\cX)\), \(b^{-2}\widehat\Theta_{2,b,x}(m(x))(0,0_x,S)\) estimates the contraction of \(-\nabla_{\cX}^2s_x(x)\) with \(S\). Whenever \(\widehat A_{h,x}(\omega)\) is invertible, define
\begin{equation+}\label{eq:main-empirical-bias-displacement}
\widehat B_{h,b,x}(\omega) \coloneqq \frac{h^2}{2b^2} \widehat A_{h,x}(\omega)^{-1} \widehat\Theta_{2,b,x}(\omega) \left(0,0_x,\widehat M_{2,h}(x)\right)\in T_\omega\cY.
\end{equation+}
Write \(\widehat B_{h,b}(x)\coloneqq\widehat B_{h,b,x}(\widehat m_h(x))\) for the estimated leading smoothing displacement. The RRBC estimator is
\begin{equation}
\widehat m_{\mathrm{rbc}}(x) \coloneqq \Exp_{\widehat m_h(x)}^{\cY} \left[-\widehat B_{h,b}(x)\right].
\end{equation}
Geodesic reversal gives \(\Log_{m_h(x)}^{\cY}m(x)=-\Pi_{m(x)\to m_h(x)}^{\cY}B_h(x)\). This explains the negative sign in the exponential-map correction\rev{; see \Cref{fig:intrinsic-bias-correction}}.

\subsection{Leave-one-out covariance estimator}\label{subsec:linear-representation-covariance}

For \(\star\in\{\us,\rbc\}\), consider the scaled tangent-space error \(\sqrt{nh^p}\Log_{m(x)}^{\cY}\widehat m_\star(x)\). The corresponding linear representation involves the mean-zero oracle score \(\varphi_{\star,n}\), defined in \eqref{eq:supp_generic_oracle_score} of the Supplementary Material. Define the finite-bandwidth oracle covariance of the leading stochastic term by
\begin{equation+}\label{eq:rbc-asymptotic-covariance}
\Omega_{\star,n}(x) \coloneqq h^{-p}A(x)^{-1}\E[\varphi_{\star,n}((X,Y);x)^{\otimes2}]\{A(x)^{-1}\}^{*},
\end{equation+}
where \(^{*}\) denotes the adjoint.

A standard approach is to construct a sandwich covariance estimator using fitted residuals from the local polynomial regressions. Using unadjusted in-sample residuals, however, can lead to covariance estimates that understate variability in finite samples \citep{MacKinnonWhite1985,LongErvin2000}. \rev{In ordinary least squares regression, HC3 is a heteroskedasticity-consistent covariance estimator that uses squared leave-one-out (LOO) prediction residuals to adjust for leverage \citep{MacKinnon2013}. In robust bias-corrected local polynomial inference, variance estimation must also account for the variability of the bias estimate and its covariance with the original estimator \citep{Calonicoetal2018}. We construct a LOO covariance estimator directly from intrinsic displacements between the full-sample estimator and its LOO versions.} Specifically, for \(\star\in\{\us,\rbc\}\), let \(\widehat m_\star^{(-i)}(x)\) be the exact LOO \rev{estimator} computed without observation \(i\). Define
\begin{equation}\label{eq:main-loo-covariance}
\widehat\Omega_\star(x) \coloneqq nh^p\sum_{i=1}^n (\Log_{\widehat m_\star(x)}^{\cY} \widehat m_\star^{(-i)}(x))^{\otimes2}.
\end{equation}

In practice, we use the Riemannian Newton method to compute the LOO fits. Write \(\widehat Q_{h,x}^{(-i)}\) for the LOO objective, \(\widehat A_{h,x}^{(-i)}\) for its Riemannian Hessian, and \(\widehat B_{h,b,x}^{(-i)}\) for the LOO version of \eqref{eq:main-empirical-bias-displacement}. When \(\widehat A_{h,x}^{(-i)}(\widehat m_h(x))\) is invertible, the first Riemannian Newton iterate from \(\widehat m_h(x)\) is
\begin{equation+}\label{eq:main-loo-newton-update}
\widetilde m_{h,1}^{(-i)}(x)
\coloneqq
\Exp_{\widehat m_h(x)}^{\cY}\!\left[
-\{\widehat A_{h,x}^{(-i)}(\widehat m_h(x))\}^{-1}
\grad_{\widehat m_h(x)}^{\cY}\widehat Q_{h,x}^{(-i)}(\cdot)
\right].
\end{equation+}
We iterate using the stopping rule in \Cref{subsec:simulation-implementation}, reevaluating the gradient and Hessian at each current iterate, and denote the resulting fit by \(\widetilde m_h^{(-i)}(x)\). For US, set \(\widetilde m_{\us}^{(-i)}(x)\coloneqq\widetilde m_h^{(-i)}(x)\). For RRBC, recompute the LOO local quadratic fit and \(\widehat A_{h,x}^{(-i)}\) at the response center \(\widetilde m_h^{(-i)}(x)\), and set
\begin{equation+}\label{eq:main-loo-rbc-update}
\widetilde m_{\rbc}^{(-i)}(x)
\coloneqq
\Exp_{\widetilde m_h^{(-i)}(x)}^{\cY}\!\left[
-\widehat B_{h,b,x}^{(-i)}(\widetilde m_h^{(-i)}(x))
\right].
\end{equation+}
We use \(\widetilde m_\star^{(-i)}(x)\) in place of \(\widehat m_\star^{(-i)}(x)\) when evaluating \eqrefs{eq:main-loo-covariance,eq:feasible-multiplier-supremum-statistic}. The LOO displacements are computed once and reused across multiplier draws. For an approximation result, see \Cref{lem:supp-loo-newton-approximation} of the Supplementary Material.

\subsection{\texorpdfstring{\rev{Pointwise confidence sets and simultaneous confidence tubes}}{Pointwise confidence sets and simultaneous confidence tubes}}\label{subsec:uniform-intrinsic-confidence-tubes}

For \(\star\in\{\us,\rbc\}\) and a nonnegative critical value \(c\), define the intrinsic Wald set on \(\cY\) by
\[
\widehat{\cC}_n^\star(x;c) \coloneqq \Exp_{\widehat m_\star(x)}^{\cY} \left[\left\{ v\in T_{\widehat m_\star(x)}\cY: nh^p \left\langle v,\widehat\Omega_\star(x)^{-1}v \right\rangle_{\widehat m_\star(x)} \le c \right\}\right].
\]
For pointwise inference at a fixed \(x\) and \(\alpha\in(0,1)\), define
\[
\widehat{\cC}_{n,1-\alpha}^\star(x) \coloneqq \widehat{\cC}_n^\star \bigl(x;\chi^2_{q,1-\alpha}\bigr),
\]
where \(\chi^2_{q,1-\alpha}\) is the \(1-\alpha\) quantile of the chi-square distribution with \(q\) degrees of freedom.

For simultaneous inference over a nonempty deterministic evaluation set \(\cG_n\subseteq\cX_0\), we calibrate a common critical value using a Gaussian multiplier bootstrap. Here \(\cX_0\) is the compact predictor region specified in \Cref{ass:uniform-inference-conditions}. Let \(\xi_1,\ldots,\xi_n \overset{\mathrm{i.i.d.}}{\sim} N(0,1)\), independently of the data. Define
\begin{equation}\label{eq:feasible-multiplier-supremum-statistic}
\widehat T_{n,\star}^*(\cG_n) \coloneqq \sup_{x\in\cG_n} \left\| \widehat\Omega_\star(x)^{-1/2} \left(\sqrt{nh^p} \sum_{i=1}^n \xi_i\Log_{\widehat m_\star(x)}^{\cY}\widehat m_\star^{(-i)}(x)\right) \right\|_{\widehat m_\star(x)}^2.
\end{equation}
Let \(B\ge1\) denote the number of Gaussian multiplier draws. Generate \(B\) conditionally independent realizations of \eqref{eq:feasible-multiplier-supremum-statistic}, with order statistics \(\widehat T_{n,\star,(1)}^*(\cG_n)\le\cdots\le \widehat T_{n,\star,(B)}^*(\cG_n)\), and set \(\widehat T_{n,\star,(B+1)}^*(\cG_n)\coloneqq+\infty\), with \(\widehat{\cC}_n^\star(x;\infty)\coloneqq\cY\). With the critical value \(\widehat c_{1-\alpha,\star}^{*,B}(\cG_n)\coloneqq\widehat T_{n,\star,(\lceil(B+1)(1-\alpha)\rceil)}^*(\cG_n)\), we define the simultaneous confidence tube as the family of confidence sets
\[
\left\{\widehat{\cC}_n^\star\left(x;\widehat c_{1-\alpha,\star}^{*,B}(\cG_n)\right):x\in\cG_n\right\}.
\]
In each draw, the same multipliers are used across \(\cG_n\) to retain dependence across predictor locations.

\section{Asymptotic Theory}\label{sec:theory}
\subsection{Pointwise conditions and finite-bandwidth covariance}\label{subsec:pointwise-assumptions}

For the pointwise results, fix \(x\in\cX\). Each result uses either the regime with compactly supported responses \textup{(C)} or the Hadamard regime \textup{(H)} throughout. In regime \textup{(H)}, we use the following envelope functions for the logarithm and its derivatives with respect to the response center. For \(a\in\cY\) and \(R>0\), define
\[
E(a,R;y)\coloneqq\sup_{\omega\in\overline B_{\cY}(a,R)}\|\cH_\omega(y)\|_{\mathrm{op}},\qquad
J(a,R;y)\coloneqq\sup_{\substack{\omega\in\overline B_{\cY}(a,R)\\u\in T_\omega\cY,\ \|u\|_\omega\le1}}\|\nabla_u^{\cY}\cH_\omega(y)\|_{\mathrm{op}}.
\]
Set
\[
\begin{aligned}
G(a,R;y)&\coloneqq1+d_{\cY}(a,y)+E(a,R;y),\\
\cE(a,R;y)&\coloneqq G(a,R;y)+J(a,R;y),\\
\cD(a,R;y)&\coloneqq1+d_{\cY}(a,y)^2+E(a,R;y)+J(a,R;y).
\end{aligned}
\]
For response centers in the fixed neighborhoods used below, \(G\) and \(\cE\) control the logarithm and its derivatives with respect to the response center through orders one and two, respectively; \(\cD\) also bounds the squared loss.

\begin{assumption}
\label{ass:local_support_log_geometry}
There exists an open normal neighborhood \(\cV_x\) of \(x\). One of the following two response regimes holds.

\noindent\textup{(C)}
\(m(x)\) is unique, \(A(x)\) is positive definite, and there exists a compact set \(\cK_{\cY,x}\subset\cY\) such that \(\mathbb P(Y\in\cK_{\cY,x}\mid X=z)=1\) for every \(z\in\cV_x\) and \(\cK_{\cY,x}\cap\Cut_{\cY}(m(x))=\varnothing\).

\noindent\textup{(H)}
\((\cY,g_{\cY})\) is a Hadamard manifold.
There exists \(R_x>0\) such that \(\sup_{z\in\cV_x}\E[G(m(x),R_x;Y)^2\mid X=z]<\infty\).
\end{assumption}

\begin{assumption}
\label{ass:densities}
On \(\cV_x\), \(X\) has a Riemannian-volume density \(p_X\) with \(0<c_{X,x}\le p_X(z)\le C_{X,x}<\infty\) for \(\vol_{\cX}\)-almost every \(z\in\cV_x\).
The conditional distributions admit Riemannian-volume densities \(p_{Y\mid X}(y\mid z)\) satisfying the condition below for the response regime selected in \Cref{ass:local_support_log_geometry}.

\noindent\textup{(C)}
There exists an envelope \(e_x:\cY\to[0,\infty)\) with
\(\int_{\cK_{\cY,x}}e_x(y)\,\vol_{\cY}(dy)<\infty\)
such that, for \(\vol_{\cY}\)-almost every
\(y\in\cK_{\cY,x}\), the map
\(z\mapsto p_{Y\mid X}(y\mid z)\) is \(C^2\) on
\(\cV_x\) and
\[
\sup_{z\in\cV_x}\left\{
p_{Y\mid X}(y\mid z)
+\|\grad_z^{\cX}p_{Y\mid X}(y\mid\cdot)\|_z
+\|\Hess_z^{\cX}p_{Y\mid X}(y\mid\cdot)\|_{\mathrm{op}}
\right\}\le e_x(y).
\]

\noindent\textup{(H)}
For \(\vol_{\cY}\)-almost every \(y\in\cY\), the map \(z\mapsto p_{Y\mid X}(y\mid z)\) is \(C^2\) on \(\cV_x\). There exist envelopes \(e_{0,x},e_{12,x}:\cY\to[0,\infty)\) such that, almost everywhere,
\[
\begin{gathered}
\sup_{z\in\cV_x}p_{Y\mid X}(y\mid z)\le e_{0,x}(y),\quad \sup_{z\in\cV_x}\left\{\|\grad_z^{\cX}p_{Y\mid X}(y\mid\cdot)\|_z+\|\Hess_z^{\cX}p_{Y\mid X}(y\mid\cdot)\|_{\mathrm{op}}\right\}\le e_{12,x}(y),\\
\int_{\cY}\{\cD(m(x),R_x;y)e_{0,x}(y) +G(m(x),R_x;y)e_{12,x}(y)\}\,\vol_{\cY}(dy)<\infty.
\end{gathered}
\]
\end{assumption}

\begin{assumption}
\label{ass:kernels}
The kernel \(K_1:[0,\infty)\to[0,\infty)\) is bounded, has compact support, and satisfies \( \int_0^\infty r^{p-1}K_1(r)\,dr>0. \) For RRBC, \(K_2\) satisfies the same conditions.
\end{assumption}

\begin{assumption}
\label{ass:bandwidth}
The common bandwidth conditions are \(h\to0\) and \(nh^p\to\infty\). In addition, the US regime requires \(\sqrt{nh^p}\,h^2\to0\), whereas the RRBC regime requires \(b\asymp h\) and \(\sqrt{nh^p}\,h^2=O(1)\).
\end{assumption}

\Cref{ass:local_support_log_geometry,ass:densities,ass:kernels} under \textup{(C)} or \textup{(H)}, together with \(h\to0\) and \(nh^p\to\infty\), form the common pointwise conditions. Adding the US or RRBC restrictions in \Cref{ass:bandwidth} gives the corresponding pointwise regime. Under \textup{(H)}, \(m(z)\) exists uniquely for every \(z\in\cV_x\) \citep[Proposition~4.3]{Sturm2003}. \Cref{ass:local_support_log_geometry} ensures that the logarithm map and parallel transport are well defined and smooth on the local domains used in the analysis. Together, \Cref{ass:local_support_log_geometry,ass:densities} provide the regularity and integrability needed to differentiate the conditional score with respect to the predictor.

\begin{remark}[Sufficient conditions for regime \textup{(C)}]
\label{rem:scope-geometric-assumptions}
For a sectional-curvature upper bound \(\Delta\ge0\) on \(\cY\), regime \textup{(C)} is ensured if there exist \(c_x\in\cY\) and \(\rho_x>0\) such that \(\operatorname{supp}(\mathbb P_{Y\mid X=z})\subset B_{\cY}(c_x,\rho_x)\) for every \(z\in\cV_x\) and \(\rho_x<\tfrac12\min(\inj(\cY),\pi/\sqrt{\Delta})\), where \(\inj(\cY)\coloneqq\inf_{\omega\in\cY}\inj_{\cY}(\omega)\) and \(\pi/\sqrt{\Delta}=+\infty\) when \(\Delta=0\). Indeed, \citet[Theorem~2.1]{Afsari2011} gives a unique conditional Fr\'echet mean \(m(z)\in B_{\cY}(c_x,\rho_x)\) for every \(z\in\cV_x\). Completeness makes \(\cK_{\cY,x}=\overline B_{\cY}(c_x,\rho_x)\) compact, while \(d_{\cY}(m(x),y)\le2\rho_x<\inj(\cY)\) for every \(y\in\cK_{\cY,x}\); hence \(\cK_{\cY,x}\cap\Cut_{\cY}(m(x))=\varnothing\). \citet[Section~2.2.3, equation~(2.19)]{Afsari2011} gives \(A(x)\succ0\). Thus the response conditions in regime \textup{(C)} hold, while density regularity remains separate.
\end{remark}

Under the common pointwise conditions, with \(b\asymp h\) for RRBC, the eigenvalues of \(\Omega_{\star,n}(x)\) are bounded above and away from zero for all sufficiently large \(n\); see \Cref{lem:supp_oracle_residualized_score_covariance_consistency}.

\subsection{Local linear expansion}\label{subsec:local-linear-expansion}

\begin{theorem}
\label{thm:local_linear_expansion_bias}
Under the common pointwise conditions:
\begin{enumerate}
\item[\textup{(i)}] \textit{Stochastic deviation.}
\[
\Omega_{\mathrm{us},n}(x)^{-1/2}\sqrt{nh^p} \left\{ \Log_{m(x)}^{\cY}\widehat m_h(x)-B_h(x) \right\} \dto N(0,I_{T_{m(x)}\cY}) \quad\text{in }T_{m(x)}\cY.
\]

\item[\textup{(ii)}] \textit{Smoothing displacement.} With \(\beta_h(x)\) defined in \eqref{eq:explicit-bias-expansion},
\[
B_h(x) = h^2\beta_h(x)+o(h^2), \qquad \beta_h(x)=O(1).
\]
\end{enumerate}
Hence, \(d_{\cY}(\widehat m_h(x),m(x)) =O_p((nh^p)^{-1/2}+h^2)\).
\end{theorem}

\subsection{Pointwise inference}\label{subsec:theory-confidence-sets}

\begin{corollary}
\label{cor:undersmoothing-limit}
Under the pointwise US regime,
\[
\Omega_{\mathrm{us},n}(x)^{-1/2}\sqrt{nh^p} \Log_{m(x)}^{\cY}\widehat m_{\mathrm{us}}(x) \dto N(0,I_{T_{m(x)}\cY}) \quad\text{in }T_{m(x)}\cY.
\]
\end{corollary}

\begin{theorem}
\label{thm:rbc-limit}
Under the pointwise RRBC regime,
\[
\Omega_{\mathrm{rbc},n}(x)^{-1/2}\sqrt{nh^p} \Log_{m(x)}^{\cY}\widehat m_{\mathrm{rbc}}(x) \dto N(0,I_{T_{m(x)}\cY}) \quad\text{in }T_{m(x)}\cY.
\]
\end{theorem}

\begin{theorem}
\label{thm:pointwise-confidence-validity}
Let \(\star\in\{\mathrm{us},\mathrm{rbc}\}\). Under the corresponding pointwise regime,
\[
nh^p \left\langle \Log_{\widehat m_\star(x)}^{\cY}m(x), \widehat\Omega_\star(x)^{-1} \Log_{\widehat m_\star(x)}^{\cY}m(x) \right\rangle_{\widehat m_\star(x)} \dto \chi_q^2.
\]
Consequently,
\[
\sup_{\alpha\in(0,1)} \left| \mathbb P \bigl( m(x)\in \widehat{\cC}_{n,1-\alpha}^{\star}(x) \bigr) -(1-\alpha) \right| \to0.
\]
\end{theorem}

\subsection{Simultaneous inference}\label{subsec:uniform-inference-theory}

For simultaneous inference, we impose uniform versions of the pointwise regularity conditions, together with additional smoothness and modified bandwidth conditions.

\begin{assumption}
\label{ass:uniform-inference-conditions}
Let \(\cX_0\subset\cX\) be compact. There exists a relatively compact open neighborhood \(\cO_{\cX}\supset\cX_0\) on which \(X\) has a Riemannian-volume density satisfying \(0<c_{X,0}\le p_X(z)\le C_{X,0}<\infty\) for \(\vol_{\cX}\)-almost every \(z\in\cO_{\cX}\). The kernels satisfy \Cref{ass:kernels} and have bounded variation on \([0,\infty)\).

For some \(\gamma\in(0,1]\), the Hessians of \(p_{Y\mid X}\) with respect to the predictor are locally \(\gamma\)-H\"older in operator norm after parallel transport of both arguments. Fix either response regime \textup{(C)} or \textup{(H)} throughout \(\cX_0\). The response geometry, density, and H\"older envelope conditions are as follows.

\noindent\textup{(C)}
\Cref{ass:local_support_log_geometry,ass:densities} hold at every \(x\in\cX_0\). For each \(x\in\cX_0\), there is a predictor neighborhood of \(x\) contained in \(\cV_x\cap\cO_{\cX}\), chosen independently of \(y\), on which the \(\gamma\)-H\"older constants of the transported predictor Hessians are bounded, for \(\vol_{\cY}\)-almost every \(y\in\cK_{\cY,x}\), by a \(\vol_{\cY}\)-integrable envelope on \(\cK_{\cY,x}\).

\noindent\textup{(H)}
The response manifold \((\cY,g_{\cY})\) is Hadamard. There exist \(o\in\cY\), \(r_0>1\), envelopes \(e_0,e_{12}:\cY\to[0,\infty)\), and \(R>R_m\), where
\[
R_m^2\coloneqq\sup_{z\in\cO_{\cX}} \E[d_{\cY}(o,Y)^2\mid X=z]<\infty, \qquad \sup_{z\in\cO_{\cX}}\E[\cE(o,R;Y)^{2r_0}\mid X=z]<\infty.
\]
\Cref{ass:densities}~\textup{(H)} holds on \(\cO_{\cX}\) with common envelopes \(e_0,e_{12}\) and \((m(x),R_x)\) replaced by \((o,R)\). There exists an envelope \(e_\gamma:\cY\to[0,\infty)\) such that each point of \(\cO_{\cX}\) has a neighborhood, chosen independently of \(y\), on which the \(\gamma\)-H\"older constants of the transported predictor Hessians are bounded by \(e_\gamma(y)\) for \(\vol_{\cY}\)-almost every \(y\in\cY\). This envelope satisfies
\[
\int_{\cY}G(o,R;y)e_\gamma(y)\,\vol_{\cY}(dy)<\infty.
\]

Finally, assume the bandwidth conditions \(h\to0\), \(nh^p\to\infty\), and \(b\asymp h\) for RRBC, together with the following rate conditions:
\[
\begin{cases}
\dfrac{(\log n)^5}{nh^p}\to0,&\textup{(C)},\\
\dfrac{n^{1/r_0}(\log n)^3}{nh^p}\to0,&\textup{(H)},
\end{cases}
\quad\text{and}\quad
\begin{cases}
\sqrt{nh^p\log n}\,h^2\to0,&\mathrm{US},\\
\sqrt{nh^p\log n}\,h^{2+\gamma}\to0,&\mathrm{RRBC}.
\end{cases}
\]
\end{assumption}

The bounded variation condition on the kernels in \Cref{ass:uniform-inference-conditions} is used to establish uniform bounds for the empirical and multiplier processes. The H\"older condition controls the remainder in the uniform expansion of the smoothing displacement. We call the US and RRBC conditions in \Cref{ass:uniform-inference-conditions} the uniform US and RRBC regimes, respectively. The bandwidth choice \(h\asymp n^{-1/(p+4)}\), with \(b\asymp h\), satisfies the uniform RRBC rate conditions under \textup{(C)}, and under \textup{(H)} if and only if \(r_0>(p+4)/4\).

Define the supremum of the pointwise Wald statistics evaluated at the true conditional Fr\'echet means:
\[
T_{n,\star}(\cG_n) \coloneqq \sup_{x\in\cG_n} nh^p \left\| \widehat\Omega_\star(x)^{-1/2} \Log_{\widehat m_\star(x)}^{\cY}m(x) \right\|_{\widehat m_\star(x)}^2.
\]

\begin{theorem}
\label{thm:uniform-multiplier-validity}
Let \(\star\in\{\mathrm{us},\mathrm{rbc}\}\). Suppose the corresponding uniform regime holds. Then
\[
\sup_{t\in\R} \left| \mathbb P(T_{n,\star}(\cG_n)\le t) - \mathbb P^* (\widehat T_{n,\star}^*(\cG_n)\le t) \right| \pto0,
\]
where \(\mathbb P^*\) denotes probability conditional on the data. If \(B=B_n\to\infty\), then
\[
\sup_{\alpha\in(0,1)} \left| \mathbb P \left( m(x)\in \widehat{\cC}_n^\star \left(x;\widehat c_{1-\alpha,\star}^{*,B_n}(\cG_n)\right) \text{ for every }x\in\cG_n \right) -(1-\alpha) \right| \to0.
\]
\end{theorem}

\begin{remark}
\label{rem:finite-grid-calibration}
To avoid optimization over the continuum \(\cX_0\) in each multiplier draw, we can calibrate on a finite grid.
Under the conditions of \Cref{thm:uniform-multiplier-validity}, suppose \(\cG_n\) is finite and set \(\rho_n\coloneqq\sup_{x\in\cX_0}\min_{z\in\cG_n}d_{\cX}(x,z)\). If \(\rho_n=o\left(h/(\log n)^2\right)\) and \(B_n\to\infty\), the same coverage conclusion holds over \(\cX_0\) using the grid-based critical value \(\widehat c_{1-\alpha,\star}^{*,B_n}(\cG_n)\); see \Cref{cor:supp-grid-calibrated-continuum-tube} in the Supplementary Material. Only the supremum used to obtain the critical value is evaluated on the grid.
\end{remark}

We also consider inference for smooth transformations \rev{and comparisons of conditional Fr\'echet means}. \rev{Such inference} can remain valid even when the full tangent-space covariance is singular, provided that the covariance of the transformed estimates or their comparisons is nondegenerate. The required conditions and inference results are given in \Cref{thm:smooth-functional-inference,thm:uniform-smooth-functional-transfer} and \Cref{cor:supp-projected-rbc-uniform-inference}. The response-side extension in \Cref{subsec:supp-smooth-pushforward-laws} of the Supplementary Material covers the fiber-supported directional model used in \Cref{sec:ocean-currents}.

\section{Simulation Studies}\label{sec:simulation}

\rev{In the simulation studies}, we compare five procedures using the same \rev{simulated datasets and} multiplier draws: \rev{uncorrected local linear Wald inference} (\textsc{Naive}), \rev{bias-corrected Wald inference with the \textsc{Naive} covariance} (\textsc{BC + Naive Cov}), \rev{local linear inference with undersmoothing} (\textsc{US}), \textsc{RRBC} with LOO covariance (\textsc{RRBC LOO}), and \textsc{RRBC} with residual-based covariance (\textsc{RRBC Residual}). The \textsc{Naive} and \textsc{US} procedures use LOO covariance. The residual-based construction is given in \Cref{subsec:supp-residual-covariance} of the Supplementary Material. We use \(b=1.5h\) in all numerical analyses except the bandwidth-ratio sensitivity analysis. Sensitivity analyses are reported in \Cref{subsec:supp-simulation-sensitivity} of the Supplementary Material.

\subsection{Methods and implementation}\label{subsec:simulation-implementation}

The main simulations use the uniform kernel \(K_1(r)=K_2(r)=\tfrac12\mathds{1}(0\le r\le1)\). For each dataset and evaluation point, we select an initial bandwidth \(h_0\) using a regularized plug-in approximation to the asymptotic MSE (AMSE) of the uncorrected local linear estimator under squared geodesic distance. We then apply the full-sample support and directional effective sample size (ESS) rule in \Cref{subsec:supp-simulation-bandwidths}, increasing \(h\) from \(h_0\) with \(b/h\) fixed. For each LOO local linear fit, we allow at most ten Newton updates, stopping earlier when the score norm is at most \(10^{-12}\).

\textsc{Naive}, \textsc{BC + Naive Cov}, \rev{\textsc{RRBC LOO}, and \textsc{RRBC Residual}} use the same selected local linear bandwidth \(h\). The \textsc{BC + Naive Cov} procedure combines the \textsc{RRBC} center with the \textsc{Naive} covariance and observation-level multiplier contributions, transported to the tangent space at that center. \rev{\textsc{US} instead uses \(h_{\mathrm{us}}=h\,n^{-1/\{(p+3)(p+4)\}}\), where \(p\) is the predictor dimension.}

We conduct 1,000 Monte Carlo replications for each of the six designs at each sample size \(n\in\{250,500,1000,2000\}\). In these simulations, both pointwise and simultaneous 95\% critical values are computed using 1,999 multiplier draws. The pointwise calibration therefore uses a multiplier critical value rather than the chi-square critical value in \Cref{subsec:uniform-intrinsic-confidence-tubes}. For simultaneous inference, we take the maximum of the pointwise Wald statistics over the entire evaluation grid. We report empirical coverage on these finite grids.

\subsection{Sphere-valued responses}\label{subsec:sim-sphere-smooth-short}

Let \(X\sim\operatorname{Unif}([0,1])\), with responses on the round \(\mathbb S^2\). Define the curve
\[
m_0(x)=\bigl(\sin\theta(x)\cos\phi(x), \sin\theta(x)\sin\phi(x),\cos\theta(x)\bigr)^\top,
\]
where
\[
\theta(x)=\frac\pi2+\frac\pi8\sin(2\pi x),\qquad \phi(x)=-\frac\pi8+\frac{9\pi}{16}(x-1/2) +\frac\pi{16}\sin(2\pi x).
\]
We consider two tangent-noise designs. In the uniform-disk design, \(\varepsilon\mid X=x\sim\operatorname{Unif}\{u\in\R^2:\|u\|\le0.2\}\). In the beta-disk design,
\[
\varepsilon\mid X=x\mathrel{\overset d=} 3R_{\pi x/2}\Lambda(x)\sqrt{U_{\mathrm{rad}}} (\cos\vartheta,\sin\vartheta)^\top,
\]
where \(U_{\mathrm{rad}}\sim\operatorname{Beta}(1,5)\), \(\vartheta\sim\operatorname{Unif}(0,2\pi)\), and \(X\) are independent. \rev{The matrix} \(R_{\pi x/2}\) denotes counterclockwise planar rotation through angle \(\pi x/2\). The scale matrix is
\[
\Lambda(x)=\diag(s_1(x),s_2(x)),\quad s_1(x)=\frac\pi{24}\{1+\frac14\sin(2\pi x)\},\quad s_2(x)=\frac\pi{48}\{1+\frac14\cos(2\pi x)\}.
\]
\rev{Using the frame \(E(\cdot)\) defined in \Cref{sec:supp-sphere-simulation}, we generate} \(Y=\Exp_{m_0(X)}^{\mathbb S^2}\left[E(m_0(X))\varepsilon\right]\). Both designs have the unique conditional Fr\'echet mean \(m_0(x)\).

\begin{figure}[!tbp]
 \centering
 \includegraphics[width=\SphereDGPWidth]{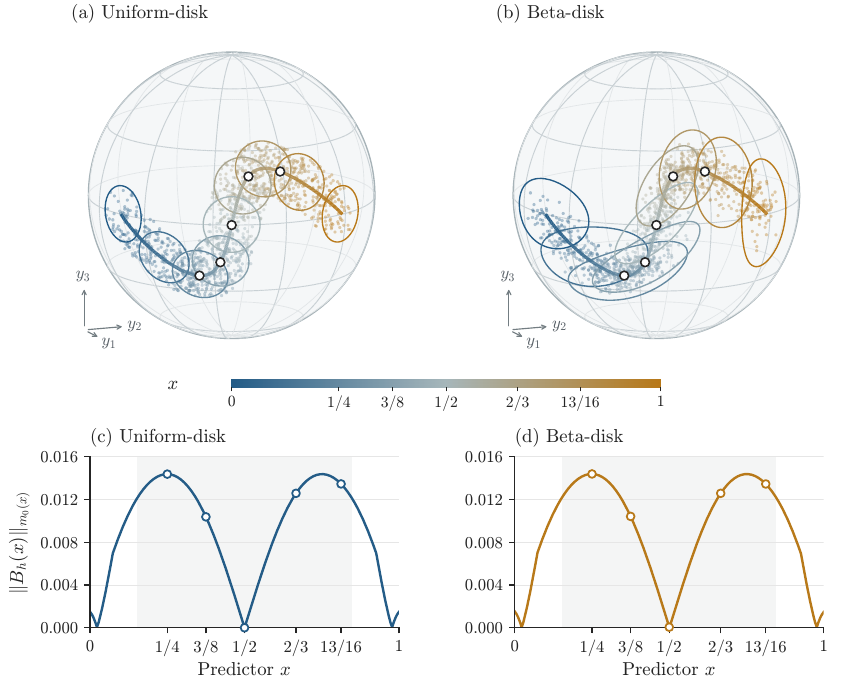}
 \caption{\rev{Panels (a) and (b) show the sphere designs, and panels (c) and (d) show the magnitude of the population smoothing displacement.} \rev{Points illustrate observations. Open circles mark evaluation points, and surrounding curves delimit conditional supports.} Displacements are computed using the uniform kernel with \(h=(1-2^{-1/2})1000^{-1/5}\). Shading indicates the inference-grid range.}
 \label{fig:sim-sphere-dgp-bias}
\end{figure}

Simultaneous coverage is evaluated over \(\mathcal G_{67}=\{j/96:j=15,\ldots,81\}\). Pointwise coverage is displayed at \(\{1/4,3/8,1/2,2/3,13/16\}\). At the reference bandwidth used in \Cref{fig:sim-sphere-dgp-bias}, the population smoothing displacement is close to zero at \(x=1/2\) and \rev{larger at the other four evaluation points}.

\begin{figure}[!tbp]
 \centering
 \includegraphics[width=\textwidth]{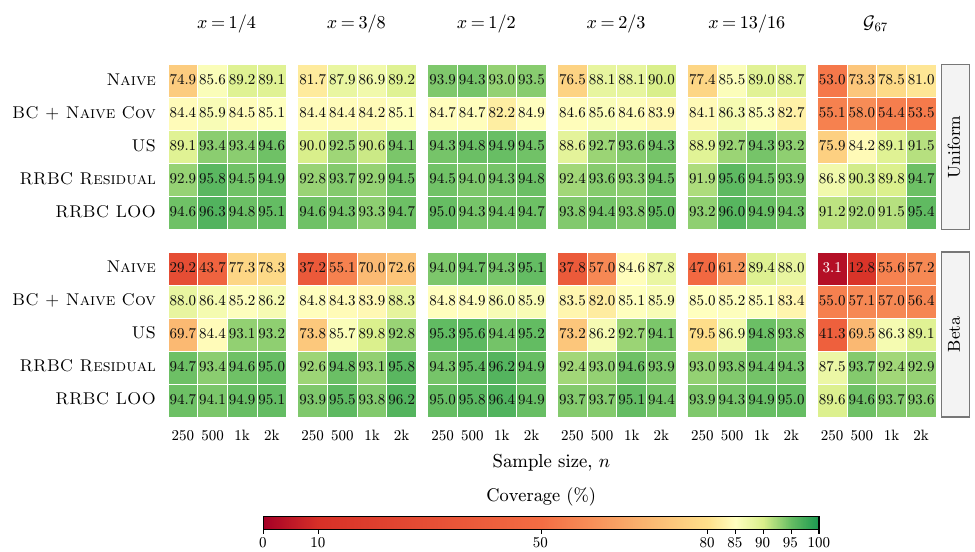}
\caption{Wald coverage (\%) for the sphere designs. \rev{The nominal coverage level is 95\%. Pointwise coverage is reported at the five evaluation points, and simultaneous coverage is reported over $\mathcal G_{67}$.}}
 \label{fig:sim-sphere-results}
\end{figure}

\rev{\Cref{fig:sim-sphere-results} shows that} \textsc{Naive} coverage varies markedly with the evaluation point. For the beta-disk design at \(n=250\), it is 94.0\% at \(x=1/2\) but only 29.2--47.0\% at the other four \rev{evaluation} points. \textsc{RRBC LOO} attains 93.7--95.0\% at all five points. This pattern is consistent with smoothing-induced miscentering as a source of undercoverage. Across both sphere designs and all sample sizes, \textsc{RRBC LOO} simultaneous coverage ranges from 89.6\% to 95.4\%.

\subsection{SPD-to-SPD regression}\label{subsec:sim-spd2spd}

Both predictor and response lie in the space \(\Sym_{++}^2\) of \(2\times2\) symmetric positive-definite matrices with the affine-invariant Riemannian metric (AIRM). Let \(E_1,E_2,E_3\) be the basis in \Cref{sec:supp-spd-target}. Draw \(a\sim N(0,\sigma_x^2I_3)\), with \(\sigma_x=1/4\), and set \(X=\exp(\sum_{j=1}^3a_jE_j)\). For \(x\in\Sym_{++}^2\), define
\[
m_0(x)=\exp\left(\frac12\log x+\frac14(\log x)^2\right).
\]
Here \(\exp\) and \(\log\) denote the matrix exponential and logarithm.

Given \(X\), generate \(Y=m_0(X)^{1/2}\exp(S)m_0(X)^{1/2}\), where \(S=\sum_{j=1}^3s_jE_j\) and \rev{\(s=(s_1,s_2,s_3)^\top\in\R^3\)} is independent of \(X\). With \(\sigma_\varepsilon=1/8\) and \(D=\diag(\sqrt{9/5},\sqrt{3/5},\sqrt{3/5})\), we consider four noise distributions:
\begin{itemize}
\item Isotropic uniform (I-U): \(s\sim\operatorname{Unif}\{u\in\R^3:\|u\|\le\sqrt5\,\sigma_\varepsilon\}\).
\item Isotropic Gaussian (I-G): \(s\sim N(0,\sigma_\varepsilon^2I_3)\).
\item Anisotropic uniform (A-U): \(s\sim\operatorname{Unif}\{u\in\R^3:\|D^{-1}u\|\le\sqrt5\,\sigma_\varepsilon\}\).
\item Anisotropic Gaussian (A-G): \(s\sim N(0,\sigma_\varepsilon^2D^2)\).
\end{itemize}
Within each isotropic or anisotropic pair, the uniform and Gaussian distributions of \(s\) have the same covariance. All four designs have the unique conditional Fr\'echet mean \(m_0(x)\).

The full predictor grid is
\[
\cG_{125}=\left\{\exp\left(\sum_{j=1}^3a_jE_j\right):a\in\{-1/8,-1/16,0,1/16,1/8\}^3\right\}.
\]
Pointwise coverage is shown at the seven \rev{evaluation points} with coordinates \(a=0,\pm e_j/8\), \(j=1,2,3\), where \(e_j\) are the standard basis vectors of \(\R^3\). Simultaneous coverage is evaluated over \(\cG_{125}\).

\begin{figure}[!tp]
 \centering
 \includegraphics[width=\textwidth]{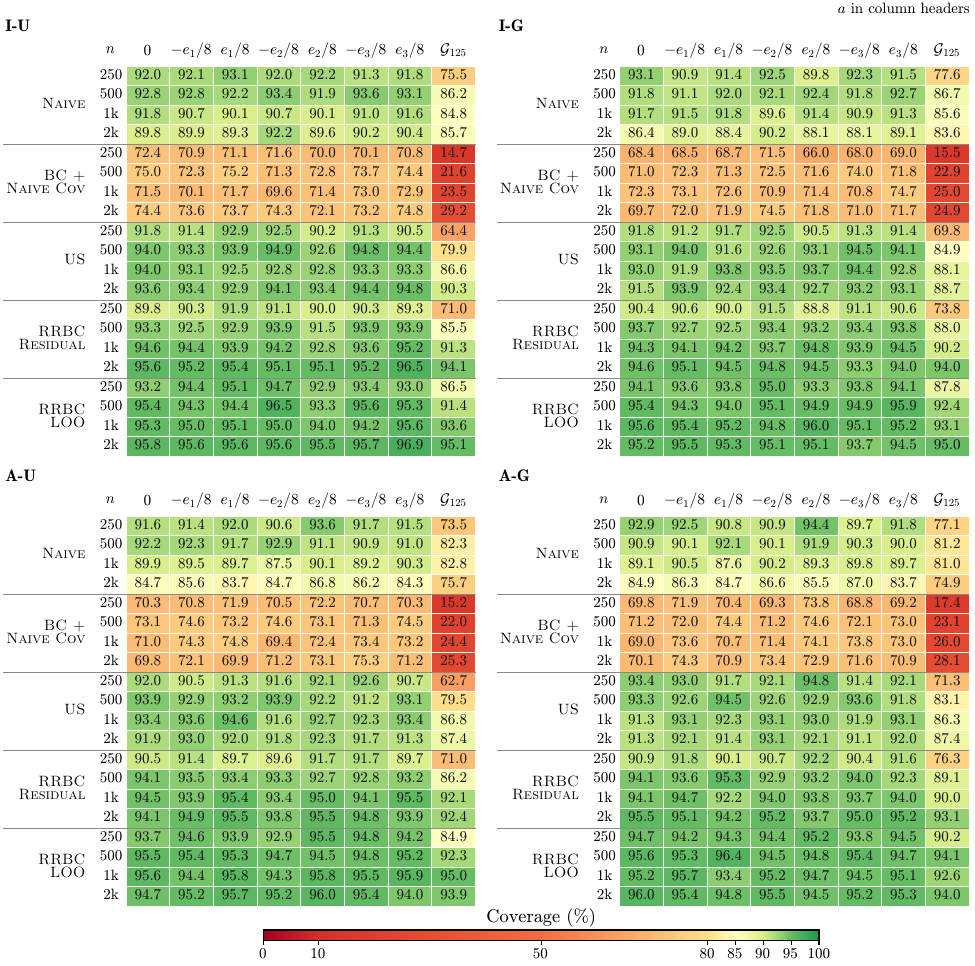}
\caption{Wald coverage (\%) for the SPD designs. \rev{The nominal coverage level is 95\%. Pointwise coverage is reported at the seven evaluation points, and simultaneous coverage is reported over $\mathcal G_{125}$.}}
 \label{fig:sim-spd2spd-results}
\end{figure}

\rev{\Cref{fig:sim-spd2spd-results} shows that at} \(n=250\), \textsc{BC + Naive Cov} has simultaneous coverage of 14.7--17.4\%, compared with 84.9--90.2\% for \textsc{RRBC LOO}. This comparison shows that correcting the center without accounting for the variability of the bias-corrected estimator can lead to severe undercoverage.

In the SPD designs, \rev{we also examine RRBC with an unresidualized covariance estimator that uses response logarithms without subtracting fitted tangent means (\Cref{subsubsec:supp-covariance-comparison} of the Supplementary Material).} At \(n=250\), RRBC Wald sets based on this covariance estimator attain higher empirical simultaneous coverage: 94.6--97.8\%, compared with 71.0--76.3\% for \rev{\textsc{RRBC Residual}} and 84.9--90.2\% for \rev{\textsc{RRBC LOO}}. This higher coverage \rev{is accompanied by} larger medians of the grid-averaged intrinsic volumes. At \(n=2000\), \rev{\textsc{RRBC LOO}} gives coverage closer to 95\% (93.9--95.1\%), while RRBC Wald inference based on the unresidualized covariance estimator remains conservative \rev{at 97.9--99.0\%; see \Cref{fig:supp-covariance-comparison}}.

\subsection{Undersmoothing, intrinsic volume, and kernel sensitivity}
\label{subsec:sim-us-volume}
\label{subsec:sim-kernel-sensitivity}

\rev{\Cref{fig:sim-sphere-results,fig:sim-spd2spd-results} show that} \textsc{US} simultaneous Wald coverage improves with \(n\) in all six designs, but \textsc{RRBC LOO} is closer to 95\% in all 24 design--sample-size settings. At \(n=2000\), simultaneous coverage is 87.4--91.5\% for \textsc{US} and 93.6--95.4\% for \textsc{RRBC LOO}.

Intrinsic volume refers to Riemannian area on \(\mathbb S^2\) and AIRM volume on \(\Sym_{++}^2\). For \rev{the Wald sets forming each simultaneous confidence tube}, we report the median across replications of the intrinsic volume averaged over the full grid. \rev{\Cref{fig:supp-procedure-sizes} compares the median intrinsic volumes across procedures.} The \textsc{RRBC LOO}/\textsc{US} ratios of these medians range from 1.06 to 1.17 for spherical area and from 1.38 to 1.62 for AIRM volume. At \(n=2000\), the \(q\)th roots of these ratios are 1.030--1.034 for the sphere (\(q=2\)) and 1.114--1.117 for SPD (\(q=3\))\rev{; see \Cref{tab:rrbc-us-volume-root-ratios}}.

For the Epanechnikov kernel, \(K_1(r)=K_2(r)=\tfrac34(1-r^2)\mathds{1}(0\le r\le1)\), we repeat the six designs with \(b=1.5h\) and the same observations and Gaussian multipliers. Bandwidths are reselected for each kernel. \rev{\Cref{fig:supp-epan-sphere-wald,fig:supp-epan-spd-wald} report Wald coverage with the Epanechnikov kernel.} For \textsc{RRBC LOO}, simultaneous Wald coverage ranges from 82.5\% to 94.8\% and is lower than with the uniform kernel in 23 of the 24 settings. \rev{The corresponding} medians of the grid-averaged intrinsic volumes \rev{are also smaller with the Epanechnikov kernel in all 24 settings; see \Cref{tab:epan-uniform-size-ratios}.} Further coverage comparisons are reported in \Cref{subsubsec:supp-epan-kernel}.

\section{Real-data Analyses}\label{sec:empirical-analyses}

Both analyses use RRBC with LOO covariance and the uniform kernel of \Cref{subsec:simulation-implementation}. For each LOO local linear fit, we allow at most ten Newton updates, stopping earlier when the score norm is at most \(10^{-12}\). Initial bandwidths are selected by regularized AMSE plug-in rules for the uncorrected local linear fit, using squared AIRM distance in the cardiovascular analysis and squared angular distance for the estimated heading in the ocean analysis. Support and ESS adjustments are detailed in \Cref{sec:supp-empirical-details} of the Supplementary Material.

\subsection{Lag-1 cardiovascular low-frequency covariance}\label{sec:cardiovascular}

\rev{We use resting electrocardiogram and continuous blood pressure recordings from the publicly available Autonomic Aging database \citep{SchumannandBar2021}, available through PhysioNet \citep{Pollardetal2026}. The eligible sample contains 840 subjects. We study age-stratum differences and variation with body mass index (BMI) in the joint low-frequency (LF) variability of systolic blood pressure (SBP) and R--R (RR) intervals, which measure the time between successive R peaks in the electrocardiogram.}

For each subject, we construct a \(2\times2\) SPD low-frequency covariance matrix from paired SBP measurements and subsequent RR intervals, after log-transforming both series and multiplying them by 100:
\[
\widehat\Sigma_{i,\mathrm{LF}}^{(1)} = 2\int_{0.04}^{0.15} \operatorname{Re}(\widehat F_i(f))\,df \in\Sym_{++}^{2},
\]
where \(\widehat F_i\) is a tapered and smoothed spectral-density estimate. Here lag 1 refers to pairing each SBP measurement with the subsequent RR interval. The \(0.04\)--\(0.15\) Hz range is the standard low-frequency band for short-term heart-rate variability \citep{TaskForce1996} and contains the \(0.07\)--\(0.14\) Hz band used for spectral assessment of SBP--RR baroreflex gain \citep{Robbeetal1987}. Subjects are the sampling units, and the SPD responses retain both variances and their correlation. See \Cref{sec:supp-cardiovascular} of the Supplementary Material for construction and quality control.

Let \(G_i\in\{1,2,3\}\) index the age strata \(18\text{--}29\), \(30\text{--}59\), and \(60\text{--}92\), respectively. For \(a=1,2,3\), we target
\[
m_a(x) \coloneqq \operatorname*{argmin}_{P\in\Sym_{++}^{2}} \E\left[ d_{\mathrm{AI}}^2 \bigl(\widehat\Sigma_{i,\mathrm{LF}}^{(1)},P\bigr) \mid \mathrm{BMI}_i=x,\ G_i=a \right].
\]
We estimate \(m_a\) separately by RRBC using \(n_1=524\), \(n_2=254\), and \(n_3=62\) subjects.

We use the 29 grid points \(x_j=19+0.5(j-1)\), \(j=1,\ldots,29\), and define
\[
S(x)\coloneqq \bigl( d_{\mathrm{AI}}^2(m_1(x),m_2(x)), d_{\mathrm{AI}}^2(m_1(x),m_3(x)), d_{\mathrm{AI}}^2(m_2(x),m_3(x)) \bigr)^{\top}.
\]
We test joint within-stratum constancy, \(H_0^{\mathrm{(i)}}:m_a(x_1)=\cdots=m_a(x_{29})\) for \(a=1,2,3\); \mbox{equality} across strata, \(H_0^{\mathrm{(ii)}}:m_1(x_j)=m_2(x_j)=m_3(x_j)\) for \(j=1,\ldots,29\); and constancy of pairwise separation across BMI, \(H_0^{\mathrm{(iii)}}:S(x_1)=\cdots=S(x_{29})\). All three tests are based on maxima of Wald statistics and use \(19{,}999\) subject-level multiplier draws and the corresponding add-one \(p\)-values \(p_1,p_2,p_3\). Since \(H_0^{\mathrm{(ii)}}\subset H_0^{\mathrm{(iii)}}\), we apply Holm's method \citep{Holm1979} to \(p_1,p_2,\max(p_2,p_3)\). The modified third \(p\)-value accommodates possible degeneracy of the separation-based Wald statistic under equality across strata. Joint calibration is described in \Cref{subsec:supp-age-contrasts}.

\begin{figure}[t]
    \centering
    \includegraphics[width=0.96\textwidth]{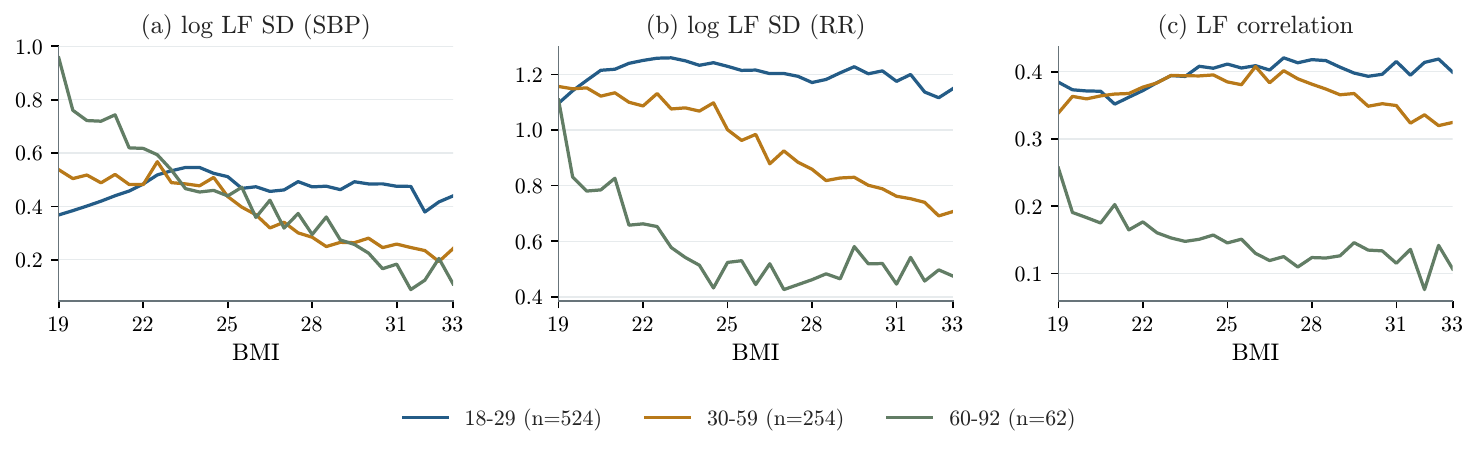}
\caption{Age-stratum RRBC estimates of log LF standard deviations of \(100\log\mathrm{SBP}\) and subsequent \(100\log\mathrm{RR}\), and LF correlation.}
    \label{fig:cardiovascular-spd}
\end{figure}

\rev{\Cref{fig:cardiovascular-spd} displays the estimated log LF standard deviations and LF correlations across BMI and age strata.} At BMI 26, the middle and oldest strata have similar estimated LF SBP standard deviations (1.45 and 1.43), whereas their estimated LF RR standard deviations are 2.68 and 1.56 and their LF correlations are 0.41 and 0.13. These standard deviations are reported on the \(100\log\mathrm{SBP}\) and \(100\log\mathrm{RR}\) scales. At this BMI, the estimated LF RR standard deviations and LF SBP--RR correlations differ between the two strata, despite similar estimated LF SBP standard deviations.

\rev{\Cref{tab:cardiovascular-intrinsic-tests} shows that} the tests reject equality across strata, but do not reject within-stratum constancy or constancy of pairwise AIRM separation across BMI.

\begin{table}[t]
    \centering
\caption{Tests based on maxima of Wald statistics for the cardiovascular covariance matrices.}
    \label{tab:cardiovascular-intrinsic-tests}

\begingroup
\footnotesize
\setlength{\tabcolsep}{2.5pt}
\begin{tabularx}{\textwidth}{@{}>{\raggedright\arraybackslash}X>{\raggedright\arraybackslash}Xrr@{}}
\toprule
Scientific question & Intrinsic contrast & Max Wald & Holm-adjusted $p$ \\
\midrule
Within-age BMI variation & Log maps between adjacent BMI-specific means & 17.184 & 0.09140 \\
Equality across age strata & Log maps between adjacent age-stratum means & 64.218 & 0.00015 \\
BMI variation in pairwise age separation & Adjacent differences in squared AIRM distances & 5.959 & 0.83645 \\
\bottomrule
\end{tabularx}
\endgroup

\end{table}

\subsection{Tropical Pacific surface-current headings}\label{sec:ocean-currents}

\rev{We compare tropical Pacific surface-current headings between El Ni\~no and La Ni\~na phases and assess north--south mirror symmetry within each phase.} We analyze quality-controlled six-hourly interpolated trajectories from NOAA's Global Drifter Program \citep{LumpkinandCenturioni2019}, using observations from 1990--2024 and retaining at most one drogue-attached observation per deployment near day seven. Using NOAA's ERSSTv6 Relative Oceanic Ni\~no Index to identify El Ni\~no--Southern Oscillation (ENSO) phases, we retain 1,107 El Ni\~no and 1,879 La Ni\~na deployments \citep{NOAACPC2026}. Details of data acquisition, sample selection, and episode definitions are given in \Cref{subsec:supp-ocean-acquisition} of the Supplementary Material.

We take the predictor to be the spherical position \(X=P\in\mathbb S^2\). The normalized horizontal velocity satisfies \(U\in T_P\mathbb S^2\), and the response is \(Y=(P,U)\in\mathrm{UT}\mathbb S^2\), \rev{where \(\mathrm{UT}\mathbb S^2\) denotes the unit tangent bundle of \(\mathbb S^2\), consisting of pairs \((p,u)\) with \(p\in\mathbb S^2\) and \(u\in T_p\mathbb S^2\) a unit vector.} Thus \(Y\mid X=x\) is supported on the unit-tangent fiber over \(x\). The response-side extension for this fiber-supported model is given in \Cref{subsec:supp-smooth-pushforward-laws} of the Supplementary Material.

We equip \(\mathrm{UT}\mathbb S^2\) with the standard Sasaki metric \citep{Sasaki1962,KlingenbergandSasaki1975}, which gives equal geometric weight to horizontal displacement and parallel-transported directional change. Explicit geometric formulas are given in \Cref{subsec:supp-ocean-sasaki} of the Supplementary Material.

For each ENSO phase \(g\), let \(m_{S,g}(x)\) denote the conditional Fr\'echet mean under the Sasaki metric. For a nonpolar point \(p\in\mathbb S^2\), let \(e_E(p)\) and \(e_N(p)\) denote the eastward and northward unit tangent vectors, and define
\[
H(p,u) = \bigl( \langle u,e_E(p)\rangle, \langle u,e_N(p)\rangle \bigr) \in\mathbb S^1.
\]
We target the phase-specific heading \(\theta_g(x)\coloneqq H(m_{S,g}(x))\). Thus estimation is performed on the full position--direction response in \(\mathrm{UT}\mathbb S^2\), while inference concerns the heading \(\theta_g(x)\).

\rev{At each grid point, we obtain a preliminary local Fr\'echet estimate under the Sasaki metric separately for each phase. We use the differential of \(H\) at that estimate to form the local linear AMSE plug-in criterion for initial bandwidth selection, holding the estimate fixed as the candidate bandwidth varies.}

The grid contains 17 longitudes from \(170^\circ\mathrm E\) to \(110^\circ\mathrm W\) at five-degree intervals and the five latitudes \(-5^\circ,-2^\circ,0^\circ,2^\circ,5^\circ\), giving 85 locations per phase. We model deployments as independent sampling units and use \(B=19{,}999\) deployment-level multiplier draws, sharing each deployment's multiplier across locations and drawing independently for the two phase samples. Confidence arcs are simultaneous over all 85 locations separately within each phase.

We test equality of the phase-specific heading fields and north--south mirror symmetry within each phase. Reflection preserves the eastward component and reverses the northward component. We calibrate one spatial maximum per hypothesis and apply Holm correction to the three multiplier \(p\)-values. The contrast families, including comparisons of equatorial headings with their reflections, are defined in \Cref{subsec:supp-ocean-heading-inference} of the Supplementary Material.

\begin{figure}[!htbp]
    \centering
    \includegraphics[width=0.98\textwidth]{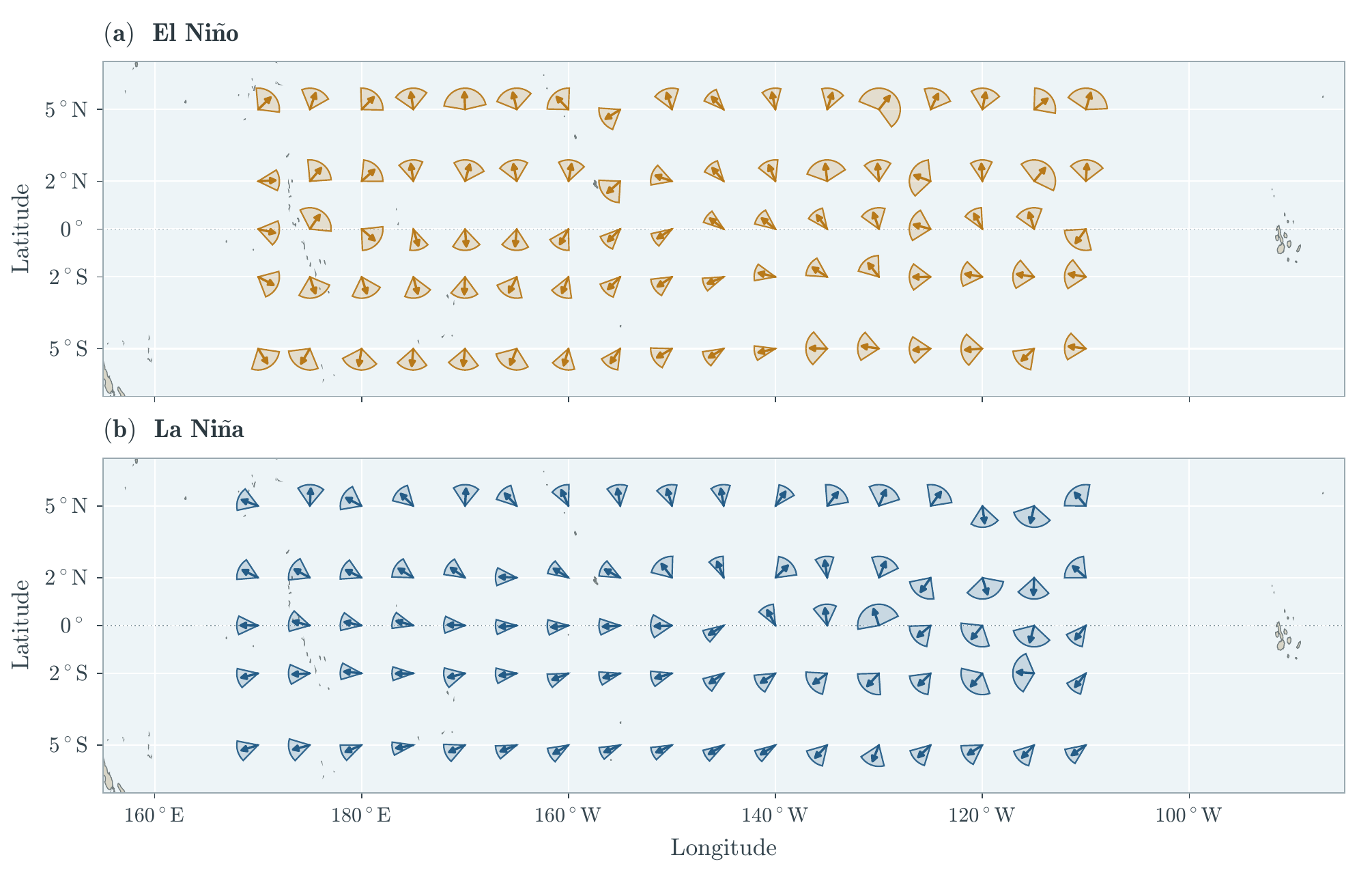}
\caption{Full-sample tropical Pacific headings at $5^\circ$S, $2^\circ$S, the equator, $2^\circ$N, and $5^\circ$N. Arrows show RRBC estimates. Sectors are 95\% simultaneous confidence arcs over all 85 locations within each phase. Latitude spacing is enlarged and arrows have fixed display length. \rev{Coastlines are from Natural Earth.}}
    \label{fig:ocean-heading-uniform}
\end{figure}

All three null hypotheses are rejected (Holm-adjusted \(p=0.00015\) for each), indicating phase-associated differences in the heading fields and departures from north--south mirror symmetry in both phases. \rev{\Cref{fig:ocean-heading-uniform} displays the estimated headings and simultaneous confidence arcs.} At \(170^\circ\mathrm E\) on the equator, the estimated phase-specific headings differ by \(167.9^\circ\). This location also has the largest pointwise Wald statistic for the phase comparison.

\section{Discussion}\label{sec:discussion}

The leading smoothing bias need not be determined by the conditional Fr\'echet mean function alone. For \rev{local linear regression with} Euclidean responses, the conditional score at a fixed response center is a difference of means, so the leading bias depends on second derivatives of the regression function. On a curved manifold, the expected logarithm at that center need not equal the logarithm of the conditional Fr\'echet mean. The bias expansion involves second derivatives of the conditional score with respect to the predictor and the inverse Hessian \(A(x)^{-1}\).

In the simulations, procedures sharing the same RRBC center yield different empirical coverage rates and intrinsic volumes. The LOO construction accounts for the effects of deleting each observation on both the original fit and \rev{the estimated leading smoothing displacement}. The first-order theory does not determine which covariance estimator gives more accurate finite-sample coverage.

\rev{First-order validity of the confidence sets} at bandwidths of the MSE-optimal order does not establish optimality for coverage error. The numerical assessment of data-dependent bandwidth rules motivates further theory on joint bandwidth selection for the local linear and local quadratic fits. Higher-order coverage error expansions would help determine how these bandwidths and the covariance construction should be chosen together for pointwise and simultaneous inference.

\section*{Acknowledgements}
Chang Jun Im was supported by the National Research Foundation of Korea
grant funded by the Korea government (MSIT) (No.~RS-2025-00515381).

\putbib[ref]
\end{bibunit}

\clearpage
\begin{bibunit}[apalike]

% Distinguish duplicate citation and bibliography anchors across bibunits.
\makeatletter
\def\@extra@b@citeb{@supp}
\gdef\@extra@binfo{@supp}
\makeatother

\switchtosupplementnumbering

\begin{center}
  \begin{minipage}{0.94\textwidth}
    \centering
    {\Large\bfseries Supplementary Material for\par
    \vspace{2mm}
    \PaperTitle\par}
  \end{minipage}

  \vspace{0.6cm}
  {\normalsize \PaperAuthorsSuppDisplay\par}
  \vspace{0.2cm}
  {\itshape\small \PaperAffiliationsSupp\par}
\end{center}

\vspace{0.35cm}
\phantomsection
\addcontentsline{toc}{part}{Appendices}
\begin{arxivsupplementformat}
% !TeX root = arxiv_audited.tex

\section{Notation and conventions}
\label{sec:supp-notation}

\subsubsection*{Basic notation and empirical processes.}
{We use the abbreviations \(m_x\coloneqq m(x)\), \(m_{h,x}\coloneqq m_h(x)\), and \(\widehat m_{h,x}\coloneqq\widehat m_h(x)\) for the conditional Fr\'echet mean, the localized Fr\'echet mean, and the local linear Fr\'echet estimator, respectively. We also write \(A_x\coloneqq A(x)\).}

Let \(W\coloneqq(X,Y)\) and \(W_i\coloneqq(X_i,Y_i)\). For an integrable function \(f\), define \(Pf\coloneqq\E[f(W)]\), \(P_nf\coloneqq n^{-1}\sum_{i=1}^n f(W_i)\), and the empirical process \(\mathbb G_nf\coloneqq\sqrt n(P_n-P)f\). For the Gaussian multipliers \(\xi_1,\ldots,\xi_n\) introduced in the Main text, define the multiplier empirical process by
\[
\mathbb G_n^\ast f\coloneqq n^{-1/2}\sum_{i=1}^n\xi_i\{f(W_i)-P_nf\}.
\]
Write \(\E^\ast\) for expectation conditional on the data. The conditional stochastic orders \(O_p^\ast\) and \(o_p^\ast\) are understood in probability with respect to the data.

For \(u,v\ge0\), \(u\lesssim v\) means \(u\le Cv\) for a deterministic \(0<C<\infty\), independent of \(n\) and the bandwidths. Constants may vary between occurrences and depend on fixed model and domain quantities, including the evaluation point in pointwise bounds. In uniform bounds, the same constant applies over the indicated indices, subject to locally stated dependencies.

{For a probability measure \(Q\), a measurable function \(f\) taking values in a finite-dimensional real inner-product space \(E\), and \(1\le r<\infty\), define \(\|f\|_{Q,r}\coloneqq\bigl(\int\|f(w)\|_E^r\,Q(dw)\bigr)^{1/r}\). For \mbox{\(E=\R\)}, this reduces to \((Q|f|^r)^{1/r}\). For real-valued functions \(f,g\), write} \(\|f\|_\infty\coloneqq\sup_s|f(s)|\) and \(e_Q(f,g)\coloneqq\|f-g\|_{Q,2}\). Let \(\cN(\cF,d,\varepsilon)\) and \(N_{[]}(\varepsilon,\cF,L_2(Q))\) denote covering and bracketing numbers, respectively, with pointwise brackets of measurable representatives \citep{VanDerVaartandWellner2023}.

For subsets \(A,B\) of a common manifold, \(A\Subset B\) means that \(\overline A\) is compact and contained in \(B\).

For a ball \(B\subset T_x\cX\), a finite-dimensional Hilbert space \(E\), and \(0<\gamma\le1\), the norm \(\|f\|_{C^{2,\gamma}(B;E)}\) is the sum of the supremum norms of \(f,Df,D^2f\) and the H\"older seminorm
\[
[D^2f]_{\gamma;B}\coloneqq\sup_{u,v\in B,\,u\ne v}\frac{\|D^2f(u)-D^2f(v)\|_{\mathrm{op}}}{\|u-v\|_x^\gamma}.
\]
{The \(C^2\) norm omits this H\"older seminorm.}
Derivatives are taken on the fixed vector space \(T_x\cX\), using \(\|\cdot\|_E\) for values and the induced multilinear operator norms for derivatives. {For ordinary derivatives, parentheses indicate evaluation at a point and square brackets application to directions.}

\subsubsection*{Rates and consequences of the bandwidth conditions.}
We use \(\nu=\mathrm{pt}\) for pointwise bounds at a fixed predictor location and \(\nu=\mathrm{unif}\) for bounds uniform over the compact predictor region \(\cX_0\). For a bandwidth \(a\in(0,1)\), \(\delta_{n,\nu}(a)\) is the stochastic scale for local kernel averages, while \(r_{n,\nu}\) also includes {the smoothing bias term} \(h^2\):
\[
\delta_{n,\nu}(a)\coloneqq\begin{cases}(na^p)^{-1/2},&\nu=\mathrm{pt},\\ \sqrt{\frac{\log(1/a)}{na^p}},&\nu=\mathrm{unif},\end{cases}\quad r_{n,\nu}\coloneqq\delta_{n,\nu}(h)+h^2.
\]
The logarithmic factor in the uniform rate accounts for the supremum over predictor locations. Under the common bandwidth conditions in \Cref{ass:bandwidth}, \(r_{n,\mathrm{pt}}\to0\). For RRBC, \(b\asymp h\), so \(\delta_{n,\nu}(b)\asymp\delta_{n,\nu}(h)\).

For uniform inference, let \(t_n=1\) in {regime} \textup{(C)} and \(t_n=n^{1/(2r_0)}\) in the Hadamard regime \textup{(H)}, where \(r_0>1\) is the moment exponent in \Cref{ass:uniform-inference-conditions}~\textup{(H)}. Under \textup{(H)}, \(t_n\) is the truncation level for the {envelope functions for the response factors}; no truncation is needed under \textup{(C)}. We use the following rate to bound uniform empirical product terms:
\[
c_n(a)\coloneqq\begin{cases}\displaystyle\sqrt{\frac{\log n}{na^p}},&\textup{(C)},\\[1mm] \displaystyle\sqrt{\frac{\log n}{na^p}}+\frac{t_n^2\log n}{na^p},&\textup{(H)}.\end{cases}
\]
The additional term under \textup{(H)} accounts for the truncated response envelopes.

For \(a=h\) and, in RRBC, also for \(a=b\asymp h\), the uniform bandwidth conditions in \Cref{ass:uniform-inference-conditions} imply
\begin{equation}\label{eq:supp-uniform-bandwidth-rate}
\log(1/a)\asymp\log n,\quad \begin{cases}(\log n)^5/(na^p)\to0,&\textup{(C)},\\ t_n^2(\log n)^3/(na^p)\to0,&\textup{(H)}.\end{cases}
\end{equation}
Consequently, the rates used in uniform covariance and multiplier approximations satisfy
\begin{equation}\label{eq:supp-uniform-rate-consequences}
\{r_{n,\mathrm{unif}}+c_n(a)\}(\log n)^2\to0.
\end{equation}
The pointwise and uniform bounds used for quadratic remainders are
\begin{equation}\label{eq:supp-rbc-rate-consequences}
\begin{aligned}
\sqrt{nh^p}\,\delta_{n,\mathrm{pt}}(h)^2&\le\sqrt{nh^p}\,r_{n,\mathrm{pt}}\delta_{n,\mathrm{pt}}(h)\to0,\\
\sqrt{nh^p\log(1/h)}\,\delta_{n,\mathrm{unif}}(h)^2&\le\sqrt{nh^p\log(1/h)}\,r_{n,\mathrm{unif}}\delta_{n,\mathrm{unif}}(h)\to0.
\end{aligned}
\end{equation}

\subsubsection*{Local polynomial notation and weights for bias correction.}
To use common notation for the local linear and local quadratic fits, write \(U_x(z)\coloneqq\Log_x^{\cX}z\in T_x\cX\) for \(z\) in the injectivity region of \(x\). For \(j\in\{1,2\}\) and \(a>0\), let \(K_{j,a}\) denote the kernel in \eqref{eq:main-predictor-kernel} with \((K_1,h)\) replaced by \((K_j,a)\). The corresponding population design operator is
\begin{equation}\label{eq:supp-local-polynomial-design-definition}
S_{j,a}(x)\coloneqq\E\left[K_{j,a}(X,x)r_{j,x}(U_x(X)/a)^{\otimes2}\right].
\end{equation}

Choose \(R_j<\infty\) with \(K_j(r)=0\) for \(r>R_j\), \(j=1,2\), and set \(R_K\coloneqq\max(R_1,R_2)\). The local polynomial design vectors are then uniformly bounded on the kernel supports:
\begin{equation}\label{eq:supp-standing-design-vector-bound}
\sup_{j\in\{1,2\}}\sup_{x\in\cX}\sup_{\|t\|_x\le R_K}\|r_{j,x}(t)\|<\infty.
\end{equation}

For the population counterpart of the local quadratic bias correction, define the bias weight
\begin{equation}\label{eq:supp-population-bias-weight}
L_{\bias,h,b}(z;x)\coloneqq\frac{h^2}{2b^2}K_{2,b}(z,x)\left\langle S_{2,b}(x)^{-1}r_{2,x}(U_x(z)/b),(0,0,M_{2,h}(x))\right\rangle_{\cR_{2,x}}.
\end{equation}

\subsubsection*{Measurability conventions.}
The following conventions fix measurable versions of the quantities used in uniform inference. All spaces carry their Borel sigma-fields. We use the regular conditional distribution \(z\mapsto\mathbb P_{Y\mid X=z}\) from \Cref{subsec:problem-setting} and the jointly Borel density versions in \Cref{ass:densities}. All kernels and envelope functions in \Cref{ass:densities,ass:kernels,ass:uniform-inference-conditions} are measurable, and the deterministic evaluation sets \(\cG_n\) are Borel.

To define the random quantities also outside the domains where their geometric operations are defined, extend logarithms, Hessians, and ordinary inverses by zero; covariance inverses and inverse square roots by the identity; and manifold-valued quantities by a fixed base point. For transport from \(T_{\omega_0}\cY\) to \(T_{\omega_1}\cY\) outside these domains, use \(E_{\omega_1}E_{\omega_0}^{-1}\), where \(E_\omega:\R^q\to T_\omega\cY\) are fixed measurable orthonormal frames. For empirical Fr\'echet minimizers, use the jointly Borel selector in \Cref{lem:supp_measurable_empirical_local_frechet_minimizer}.

The resulting Wald and multiplier maps are jointly Borel in their arguments. Their suprema over deterministic Borel evaluation sets are upper semianalytic and hence universally measurable \citep[Corollary~7.42.1 and Propositions~7.47 and~7.48]{BertsekasShreve1978}. All probabilities are evaluated on the corresponding completed probability spaces, so we use ordinary \(O_p\) and \(o_p\) for the model-specific suprema in \Cref{sec:supp_consistency_localization,sec:supp_score_expansions_covariances,sec:supp_uniform_inference}.

\section{Geometric localization and consistency}
\label{sec:supp_consistency_localization}

{We first establish existence and consistency of the global minimizers, then show that they lie in response neighborhoods where the score expansions in \Cref{sec:supp_score_expansions_covariances} are valid.}

Fix an evaluation point \(x\), and set \(\mathcal I_{\mathrm{pt}}\coloneqq\{x\}\) and \(\mathcal I_{\mathrm{unif}}\coloneqq\cX_0\). Uniform geometric and population results use \Cref{ass:uniform-inference-conditions} without its bandwidth conditions.

Since \(\cX\) and \(\cY\) are complete, Hopf--Rinow implies that they are proper metric spaces, that is, every closed and bounded subset is compact \citep[Theorem~5.7.1]{Petersen2016}.

{We use the following response-moment bounds for minimizer consistency and for the score and Hessian expansions in \Cref{sec:supp_score_expansions_covariances}.} {Under \textup{(H)}, set} \(\cK_{\cY,x}=\cY\) {and, for pointwise bounds, define}
\[
\cE_R(y)\coloneqq\cE(m_x,R_x;y),\qquad G_R(y)\coloneqq G(m_x,R_x;y),\qquad J_R(y)\coloneqq J(m_x,R_x;y).
\]
Under \textup{(C)}, take \(\cE_R=G_R=1\) and \(J_R=0\).
By \Cref{ass:local_support_log_geometry,ass:densities},
\begin{equation}\label{eq:supp-pointwise-split-moments}
 \sup_{z\in\cV_x}\E[G_R(Y)^2+J_R(Y)\mid X=z]<\infty.
\end{equation}
{Since \(\cE_R=G_R+J_R\), \eqref{eq:supp-pointwise-split-moments} implies that \(\cE_R(Y)\) has locally uniformly bounded conditional first moments. The second-moment bound for \(G_R\) controls score and Hessian fluctuations at a fixed response center. Changes in the response center and quadratic score remainders require only the first-moment bound for \(\cE_R\).}

{For uniform bounds under \textup{(H)}, use the quantities in} \Cref{ass:uniform-inference-conditions} and set
\[
 \bar\rho_{\cY}\coloneqq(R-R_m)/3,\qquad \cE_R(y)\coloneqq\cE(o,R;y),\qquad G_R(y)\coloneqq G(o,R;y),\qquad J_R(y)\coloneqq J(o,R;y).
\]
Under \textup{(C)}, take \(\cE_R=G_R=1\), \(J_R=0\), and any \(r_0>1\).
Envelopes for the quantities bounded in
\Cref{lem:supp_fixed_local_log_domain}~\textup{(ii)} and for their
pairwise products are bounded by \(C\cE_R\) and
\(C\cE_R^2\), respectively, with
\[
 \int_{\cO_{\cX}\times\cY}\cE_R(y)^{2r_0}\,\mathbb P(dz,dy)<\infty.
\]

Under \textup{(H)}, the Hadamard variance inequality
\citep[Proposition~4.4]{Sturm2003} gives
\[
 d_{\cY}(o,m(z))\le R_m,\qquad
 \overline B_{\cY}(m(z),2\bar\rho_{\cY})
 \subset\overline B_{\cY}(o,R),
 \qquad z\in\cO_{\cX}.
\]
The triangle inequality and inclusion of the center balls give
\[
 G(m_x,2\bar\rho_{\cY};y)\lesssim G_R(y),\qquad
 \cE(m_x,2\bar\rho_{\cY};y)\lesssim\cE_R(y),\qquad
 \cD(m_x,2\bar\rho_{\cY};y)\lesssim\cD(o,R;y),
\]
uniformly over \(x\in\cX_0\) and \(y\in\cY\).
Together with \(2r_0>2\) and the moment and density conditions in
\Cref{ass:uniform-inference-conditions}, these bounds verify
\Cref{ass:local_support_log_geometry,ass:densities} uniformly over
\(x\in\cX_0\) on the local predictor neighborhoods, with
\(R_x=2\bar\rho_{\cY}\) and a common bound on the envelope moments.

{The set}
\[
 {\cD_{\log}^{\cY}\coloneqq
 \{(\omega,y)\in\cY\times\cY:y\notin\Cut_{\cY}(\omega)\}}
\]
{is open}, and \((\omega,y)\mapsto\Log_\omega^{\cY}y\) is {well defined} and smooth \citep[Theorems~10.33 and~10.34]{Lee2018}.

\subsection{Geometric results and consistency of minimizers}

\begin{lemma}
\label{lem:supp_fixed_local_log_domain}
Suppose \Cref{ass:local_support_log_geometry} holds.
There exist \(\rho>0\) and a relatively compact strongly geodesically
convex normal neighborhood \(\cW_{m_x}\) of \(m_x\) such that:
\begin{enumerate}[(i)]
\item
\[
 \overline B_{\cY}(m_x,2\rho)\subset\cW_{m_x},
 \qquad
 \overline{\cW_{m_x}}\times
 \bigl(\cK_{\cY,x}\cup\overline{\cW_{m_x}}\bigr)
 \subset\cD_{\log}^{\cY}.
\]
Under \textup{(H)}, one may also require
\(\overline{\cW_{m_x}}\subset B_{\cY}(m_x,R_x)\).

\item
Fix smooth center coordinates and a smooth frame near \(\overline{\cW_{m_x}}\). Uniformly for \(\omega\in\overline{\cW_{m_x}}\) and \(y\in\cK_{\cY,x}\), the center derivatives of \(d_{\cY}(y,\omega)^2/2\) of orders one through three and of \(\Log_\omega^{\cY}y\) of orders zero through two are bounded by \(C\cE_R(y)\). {For pointwise bounds, the loss derivatives of orders one and two and the logarithm derivatives of orders zero and one are bounded by} \(CG_R(y)\).

\item
In these coordinates and frames, the maps
\[
 (\omega_0,\omega_1)\mapsto\Log_{\omega_0}^{\cY}\omega_1,
 \qquad
 (\omega_0,\omega_1,v)\mapsto
 \Pi_{\omega_0\to\omega_1}^{\cY}v
\]
have uniformly bounded derivatives in \((\omega_0,\omega_1)\)
of total order at most two for
\(\omega_0,\omega_1\in\overline{\cW_{m_x}}\) and
\(\|v\|_{\omega_0}\le1\).
\end{enumerate}
\end{lemma}
\begin{proof}
Under \textup{(C)}, openness of \(\cD_{\log}^{\cY}\) and compactness
of \(\cK_{\cY,x}\) give a neighborhood \(O_x\) of \(m_x\) with
\(\overline{O_x}\times\cK_{\cY,x}\subset\cD_{\log}^{\cY}\).
Choose \(\cW_{m_x}\) relatively compact and strongly geodesically
convex, with its closure contained in both \(O_x\) and a slightly larger
strongly geodesically convex normal neighborhood. Then
\[
\overline{\cW_{m_x}}\times
\bigl(\cK_{\cY,x}\cup\overline{\cW_{m_x}}\bigr)
\subset\cD_{\log}^{\cY}.
\]
Smoothness on these compact sets gives the asserted bounds with
\(\cE_R\equiv1\).

Under \textup{(H)}, the logarithm is globally smooth. Choose
\(\overline{\cW_{m_x}}\subset B_{\cY}(m_x,R_x)\).
The identities
\[
{\grad_\omega^{\cY}\{d_{\cY}(y,\cdot)^2/2\}}
=-\Log_\omega^{\cY}y,\qquad
\nabla_v^{\cY}\Log_\omega^{\cY}y
=-\cH_\omega(y)v,\qquad v\in T_\omega\cY,
\]
and the bounds defining \(E\) and \(J\) give the required
covariant derivative bounds. Smooth coordinate and frame coefficients
on the compact center region transfer these bounds to the stated
representations.

In either regime, shrink \(\rho\) so that
\(\overline B_{\cY}(m_x,2\rho)\subset\cW_{m_x}\).
The logarithm and parallel transport between centers are smooth on a
neighborhood of the compact domains in part~(iii). Compactness of the
unit disk bundle gives the uniform bounds in part~(iii).
\end{proof}

{The next two lemmas give coercivity and argmin consistency for the resulting signed objectives.}

\begin{lemma}
\label{lem:supp_signed_measure_frechet_coercivity}
Let \((\cY,d_{\cY})\) be proper, fix \(o\in\cY\), and let
\((\nu_t:t\in\cI)\) be finite signed Borel measures with \(\nu_t(\cY)=1\). Set
\begin{equation}\label{eq:m}
 \cM(\nu_t;o)\coloneqq \int_{\cY}\{1+d_{\cY}(o,y)^2\}\,|\nu_t|(dy).
\end{equation}
{For \(\cM(\nu_t;o)<\infty\), the function
\(\cQ_t(\omega)\coloneqq \tfrac12\int_{\cY}d_{\cY}(y,\omega)^2\,\nu_t(dy)\)
is finite and continuous, with compact sublevel sets, and}
\begin{equation}\label{eq:supp-signed-coercivity}
 \cQ_t(\omega)\ge\tfrac12d_{\cY}(o,\omega)^2-\tfrac12\cM(\nu_t;o)\{1+d_{\cY}(o,\omega)\}.
\end{equation}
Its global argmin is nonempty and compact. If
\(\overline{\cM}\coloneqq \sup_{t\in\cI}\cM(\nu_t;o)<\infty\), then, for every compact \(\cC\subset\cY\),
\begin{equation}\label{eq:supp-signed-compact-containment}
 \bigcup_{t\in\cI}\argmin_{\omega\in\cY}\cQ_t(\omega)\subset\overline B_{\cY}(o,2\overline{\cM}),\qquad
 \sup_{t\in\cI,\,\omega\in\cC}|\cQ_t(\omega)|\le\{1+\sup_{\omega\in\cC}d_{\cY}(o,\omega)^2\}\overline{\cM}.
\end{equation}
For a single \(t\), the same containment holds with radius \(2\cM(\nu_t;o)\).
\end{lemma}
\begin{proof}
Fix \(t\in\cI\) with \(\cM(\nu_t;o)<\infty\). By \eqref{eq:m},
\(1=|\nu_t(\cY)|\le|\nu_t|(\cY)\le\cM(\nu_t;o)\).
Applying \(2a\le1+a^2\) with
\(a=d_{\cY}(o,y)\) and integrating against \(|\nu_t|\) gives
\(2\int_{\cY}d_{\cY}(o,y)\,|\nu_t|(dy)\le\cM(\nu_t;o)\).
\begin{equation}\label{eq:supp-signed-absolute-bound}
\begin{aligned}
 |\cQ_t(\omega)|
 &\le\tfrac12\int_{\cY}d_{\cY}(y,\omega)^2\,|\nu_t|(dy)
 \le\int_{\cY}\{d_{\cY}(o,y)^2+d_{\cY}(o,\omega)^2\}\,|\nu_t|(dy)\\
 &\le\{1+d_{\cY}(o,\omega)^2\}\cM(\nu_t;o)<\infty.
\end{aligned}
\end{equation}
The triangle inequality gives
\begin{equation}\label{eq:supp-signed-distance-comparison}
 |d_{\cY}(y,\omega)^2-d_{\cY}(o,\omega)^2|
 \le d_{\cY}(o,y)\{2d_{\cY}(o,\omega)+d_{\cY}(o,y)\}.
\end{equation}
Since \(\nu_t(\cY)=1\), \eqref{eq:supp-signed-distance-comparison} gives
\[
\begin{aligned}
 \cQ_t(\omega)-\tfrac12d_{\cY}(o,\omega)^2
 &=\tfrac12\int_{\cY}\{d_{\cY}(y,\omega)^2-d_{\cY}(o,\omega)^2\}\,\nu_t(dy)\\
 &\ge-\tfrac12\int_{\cY}d_{\cY}(o,y)
 \{2d_{\cY}(o,\omega)+d_{\cY}(o,y)\}\,|\nu_t|(dy)\\
 &\ge-\tfrac12\cM(\nu_t;o)\{1+d_{\cY}(o,\omega)\},
\end{aligned}
\]
which is \eqref{eq:supp-signed-coercivity}.

On each bounded set of centers,
\[
 \tfrac12d_{\cY}(y,\omega)^2
 \le C\{1+d_{\cY}(o,y)^2\}.
\]
The envelope is integrable with respect to \(|\nu_t|\) by
\eqref{eq:m}, so dominated convergence makes \(\cQ_t\) continuous.
The bound \eqref{eq:supp-signed-coercivity} makes every sublevel set
bounded, hence compact by properness.
Weierstrass' theorem gives a nonempty compact global argmin.

For \(\omega\in\argmin_{\eta\in\cY}\cQ_t(\eta)\), minimality gives
\(\cQ_t(\omega)\le\cQ_t(o)\le\tfrac12\cM(\nu_t;o)\).
Together with \eqref{eq:supp-signed-coercivity}, this yields
\[
 d_{\cY}(o,\omega)^2-\cM(\nu_t;o)d_{\cY}(o,\omega)-2\cM(\nu_t;o)\le0.
\]
Since \(\cM(\nu_t;o)\ge1\), this implies \(d_{\cY}(o,\omega)\le2\cM(\nu_t;o)\). Taking the union over \(t\) {and the suprema in} \eqref{eq:supp-signed-absolute-bound} proves \eqref{eq:supp-signed-compact-containment}.
\end{proof}

\begin{lemma}
\label{lem:supp_coercive_compact_argmin_consistency}
Let \((\cY,d_{\cY})\) be proper.
{Let \(\cI=\cI_n\) be a possibly \(n\)-dependent index set.}
For \(t\in\cI\), let \(Q(\cdot;t)\)
have a unique minimizer \(m_t\), and suppose \(\widehat Q_n(\cdot;t)\)
has a nonempty argmin. Assume that the random suprema below are measurable.
Suppose that the following three conditions hold.

\textup{(i)} \emph{Uniform approximation.} For every compact \(\cC\subset\cY\),
\[
 \Delta_n(\cC)\coloneqq \sup_{t\in\cI,\,\omega\in\cC}
 |\widehat Q_n(\omega;t)-Q(\omega;t)|=o_p(1).
\]

\textup{(ii)} \emph{Uniform separation.} {For every compact \(\cC\subset\cY\) and \(\varepsilon>0\), there exists \(\gamma(\cC,\varepsilon)>0\), independent of \(n\), such that}
\[
 d_{\cY}(\omega,m_t)\ge\varepsilon
 \quad\Longrightarrow\quad
 Q(\omega;t)-Q(m_t;t)\ge\gamma(\cC,\varepsilon)
\]
{for all \(n\), \(t\in\cI_n\), and \(\omega\in\cC\).}

\textup{(iii)} \emph{Coercivity.} On measurable events \(\cA_n\) with
\(\mathbb P(\cA_n)\to1\), for some \(o\in\cY\) and
nonnegative \(C_n,B_n=O_p(1)\),
\[
 \widehat Q_n(\omega;t)\ge\tfrac12d_{\cY}(o,\omega)^2
       -C_n\{1+d_{\cY}(o,\omega)\},\qquad
 \widehat Q_n(m_t;t)\le B_n
\]
for every \(t,\omega\). Then
\[
 {D_n\coloneqq} \sup_{t\in\cI}\sup_{\omega\in\argmin\widehat Q_n(\cdot;t)}
 d_{\cY}(\omega,m_t)=o_p(1).
\]
{The deterministic analogue holds as \(n\to\infty\) or \(h\downarrow0\), provided that \textup{(i)} holds with \(o(1)\) and \textup{(iii)} holds eventually with bounded coefficients.}
\end{lemma}
\begin{proof}
{Fix \(\varepsilon>0\). On \(\cA_n\), condition \textup{(iii)} gives \(\widehat Q_n(m_t;t)\le B_n\), and minimality gives \(\widehat Q_n(\omega;t)\le\widehat Q_n(m_t;t)\le B_n\) for every minimizer \(\omega\) of \(\widehat Q_n(\cdot;t)\). For \(r=d_{\cY}(o,m_t)\) and for \(r=d_{\cY}(o,\omega)\), the coercive lower bound gives \(\tfrac12r^2\le C_nr+C_n+B_n\le\tfrac14r^2+C_n^2+C_n+B_n\). Hence, on \(\cA_n\), simultaneously over \(t\in\cI_n\) and all such minimizers \(\omega\),}
\begin{equation+}\label{eq:supp-compact-argmin-comparison}
{\max\{d_{\cY}(o,m_t)^2,d_{\cY}(o,\omega)^2\}
 \le4(C_n^2+C_n+B_n).}
\end{equation+}

{For \(R>0\), let \(\cC_R\coloneqq\overline B_{\cY}(o,R)\), which is compact by properness. On \(\cA_n\cap\{4(C_n^2+C_n+B_n)\le R^2\}\), \eqref{eq:supp-compact-argmin-comparison} places \(m_t\) and every minimizer \(\omega\) in \(\cC_R\), simultaneously over \(t\in\cI_n\). In the following decomposition, minimality makes the middle term nonpositive, while each of the other two terms is at most \(\Delta_n(\cC_R)\):}
\begingroup
\ifdefined\JASAAnonymityMode\footnotesize\fi
\begin{equation+}\label{eq:supp-compact-argmin-gap}
{\begin{aligned}
 Q(\omega;t)-Q(m_t;t)
 =\{Q(\omega;t)-\widehat Q_n(\omega;t)\}
 +\{\widehat Q_n(\omega;t)-\widehat Q_n(m_t;t)\}
 +\{\widehat Q_n(m_t;t)-Q(m_t;t)\}\le2\Delta_n(\cC_R).
\end{aligned}}
\end{equation+}
\endgroup

{Condition \textup{(ii)} requires an objective gap of at least \(\gamma(\cC_R,\varepsilon)\) at any point of \(\cC_R\) whose distance from \(m_t\) is at least \(\varepsilon\). If a minimizer on \(\cA_n\cap\{4(C_n^2+C_n+B_n)\le R^2\}\) had this property, condition \textup{(ii)} and \eqref{eq:supp-compact-argmin-gap} would give}
\[
{\gamma(\cC_R,\varepsilon)
 \le Q(\omega;t)-Q(m_t;t)
 \le2\Delta_n(\cC_R).}
\]
{On this event,
\(2\Delta_n(\cC_R)<\gamma(\cC_R,\varepsilon)\)
implies \(D_n\le\varepsilon\). Therefore,}
\[
{\begin{aligned}
 \mathbb P(D_n>\varepsilon)
 \le\mathbb P(\cA_n^c)
 +\mathbb P\bigl(4(C_n^2+C_n+B_n)>R^2\bigr)+\mathbb P\bigl(2\Delta_n(\cC_R)\ge\gamma(\cC_R,\varepsilon)\bigr).
\end{aligned}}
\]
{Since \(C_n,B_n=O_p(1)\), for any \(\eta>0\) we can choose a fixed \(R\) large enough that \(\limsup_{n\to\infty}\mathbb P\bigl(4(C_n^2+C_n+B_n)>R^2\bigr)\le\eta\). For this fixed \(R\), the first probability tends to zero because \(\mathbb P(\cA_n)\to1\), and the last tends to zero by condition \textup{(i)}, since \(\gamma(\cC_R,\varepsilon)>0\) is independent of \(n\). Hence \(\limsup_{n\to\infty}\mathbb P(D_n>\varepsilon)\le\eta\). Since \(\eta>0\) is arbitrary, this proves \(D_n=o_p(1)\).}

{For deterministic criteria, bounded coefficients and eventual validity of condition \textup{(iii)} allow a fixed \(R\) such that \eqref{eq:supp-compact-argmin-comparison} places \(m_t\) and all minimizers in \(\cC_R\) for all sufficiently large \(n\). Condition \textup{(i)} then gives \(2\Delta_n(\cC_R)<\gamma(\cC_R,\varepsilon)\) eventually, so \eqref{eq:supp-compact-argmin-gap} and condition \textup{(ii)} yield \(D_n\le\varepsilon\) eventually. Since \(\varepsilon>0\) is arbitrary, \(D_n\to0\). The same argument applies to deterministic approximations indexed by \(h\downarrow0\).}
\end{proof}

\begin{lemma}
\label{lem:supp_measurable_empirical_local_frechet_minimizer}
Let \(\cI\in\{\{x\},\cX_0\}\), fix \(\omega_0\in\cY\), and set
\(\cZ_n\coloneqq (\cX\times\cY)^n\). Define
\(\cD_{n,h}(\cI)\coloneqq \{(\mathbf z,x)\in\cZ_n\times\cI:
\widehat S_{1,h}(x;\mathbf z)\text{ is nonsingular}\}\).
Extend \(\widehat Q_{h,x}\) from \Cref{subsec:target-estimator} by
\begin{equation}\label{eq:supp-measurable-criterion-extension}
 \widehat Q_{h,x}^{\mathrm{ext}}(\omega;\mathbf z)\coloneqq
 \begin{cases}
 \widehat Q_{h,x}(\omega;\mathbf z),&(\mathbf z,x)\in\cD_{n,h}(\cI),\\
 \tfrac12d_{\cY}(\omega,\omega_0)^2,&(\mathbf z,x)\notin\cD_{n,h}(\cI).
 \end{cases}
\end{equation}
There is a jointly Borel map \(\widehat m_h:\cZ_n\times\cI\to\cY\)
such that, for every \((\mathbf z,x)\),
\[
 \widehat m_h(\mathbf z,x)\in
 \argmin_{\omega\in\cY}\widehat Q_{h,x}^{\mathrm{ext}}(\omega;\mathbf z),
 \qquad \widehat m_h(\mathbf z,x)=\omega_0
 \quad\text{off }\cD_{n,h}(\cI).
\]
\end{lemma}
\begin{proof}
{Put \(S\coloneqq\cZ_n\times\cI\), and write \(s\coloneqq(\mathbf z,x)\), where \(\mathbf z=((x_i,y_i))_{i=1}^n\). In local orthonormal frames, the empirical design operator has Borel measurable matrix entries. Hence \(\cD_{n,h}(\cI)\) is Borel, and continuity of matrix inversion implies that the local linear weights are Borel measurable on this set. It follows from \eqref{eq:supp-measurable-criterion-extension} that \(\widehat Q_{h,x}^{\mathrm{ext}}\) is real-valued and jointly Borel in \((s,\omega)\) and continuous in \(\omega\) for each fixed \(s\).}

{For \(s\in\cD_{n,h}(\cI)\), constant reproduction gives}
\[
{\nu_s\coloneqq\frac1n\sum_{i=1}^n
 \widehat L_{1,h,i}(x;\mathbf z)\delta_{y_i},\qquad
 \nu_s(\cY)=1.}
\]
{This finite signed measure has finite support and finite second moment. On this domain,}
\[
{\widehat Q_{h,x}^{\mathrm{ext}}(\omega;\mathbf z)
 =\frac12\int_{\cY}d_{\cY}(y,\omega)^2\,\nu_s(dy),}
\]
{so \Cref{lem:supp_signed_measure_frechet_coercivity} shows that its argmin is nonempty and compact. For \(s\notin\cD_{n,h}(\cI)\), the extended objective has the unique minimizer \(\omega_0\). Thus the minimum is attained for every \(s\in S\).}

{Both \(S\) and \(\cY\) are complete separable metric spaces, since \(\cI\) is a singleton or a compact subset of \(\cX\), and \(\cY\) is \(\sigma\)-compact by properness. Apply \citet[Corollary~1]{BrownandPurves1973} with parameter space \(S\), action space \(\cY\), full domain \(S\times\cY\), and objective \((s,\omega)\mapsto\widehat Q_{h,x}^{\mathrm{ext}}(\omega;\mathbf z)\). Each feasible section is \(\cY\), hence \(\sigma\)-compact, and the objective is continuous on that section. Since the minimum is attained at every parameter, the corollary yields a Borel map \(\widehat m_h:S\to\cY\) selecting an exact minimizer everywhere. Off \(\cD_{n,h}(\cI)\), this selector necessarily equals \(\omega_0\), because the minimizer there is unique.}
\end{proof}

{Under \Cref{ass:local_support_log_geometry,ass:densities}, the squared distance \(d_{\cY}(m_x,Y)^2\) is uniformly integrable over the conditional laws \(\mathbb P_{Y\mid X=z}\), \(z\in\cV_x\). More precisely,}
\begin{equation}\label{eq:supp-pointwise-response-tail}
 \lim_{T\to\infty}\sup_{z\in\cV_x}
 \E[(1+d_{\cY}(m_x,Y)^2)\mathds{1}\{d_{\cY}(m_x,Y)>T\}\mid X=z]=0.
\end{equation}
{Under \textup{(H)}, \eqref{eq:supp-pointwise-response-tail} follows from dominated convergence with the integrable envelope \((1+d_{\cY}(m_x,y)^2)e_{0,x}(y)\). Under \textup{(C)}, all conditional response distributions are supported on the fixed compact set \(\cK_{\cY,x}\). Thus, \eqref{eq:supp-pointwise-response-tail} verifies the conditional uniform-integrability hypothesis of \Cref{lem:supp-pointwise-local-truncation} for \(G(y)=1+d_{\cY}(m_x,y)^2\).}

\begin{lemma}
\label{lem:supp_conditional_moment_regularity}
{Suppose \Cref{ass:local_support_log_geometry,ass:densities} hold at \(x\). Fix an open set \(\cV\Subset\cV_x\). Use the center coordinates of \Cref{lem:supp_fixed_local_log_domain}, and let \(\Theta\) be the coordinate image of a compact subset of \(\cW_{m_x}\). Write \(\omega=\omega(\theta)\), and let \(E_\omega:\R^q\to T_\omega\cY\) be a smooth orthonormal response frame on this center neighborhood. Define}
\[
{\begin{aligned}
 T(\theta,y)&\coloneqq E_\omega^\ast\Log_\omega^{\cY}y,
 \qquad H(\theta,y)\coloneqq E_\omega^\ast\cH_\omega(y)E_\omega,\\
 M_T(z,\theta)&\coloneqq\E[T(\theta,Y)\mid X=z],
 \qquad M_H(z,\theta)\coloneqq\E[H(\theta,Y)\mid X=z].
\end{aligned}}
\]
{All predictor derivatives below are taken with \(\omega(\theta)\) and \(E_\omega\) held fixed.}

\noindent{\textup{(i) Continuity.} For every compact set \(\cC\subset\cY\), \(F\) is jointly continuous on \(\cC\times\overline{\cV}\). The map \((z,\theta)\mapsto\E[T(\theta,Y)^{\otimes2}\mid X=z]\) is jointly continuous on \(\overline{\cV}\times\Theta\).}

\noindent{\textup{(ii) Differentiation under the integral.} The map \(M_T\) is jointly \(C^2\) in the center and predictor variables. Here \(\nabla_{\cX}^0f=f\), \(\nabla_{\cX}^1f=df\), and \(\nabla_{\cX}^2f\) is the second covariant derivative on \(\cX\). For nonnegative integers \(a,b\) with \(a+b\le2\), its derivatives satisfy}
\[
{D_\theta^a\nabla_{\cX}^bM_T(z,\theta)
 =\int_{\cK_{\cY,x}}D_\theta^aT(\theta,y)\,
 \nabla_{\cX}^bp_{Y\mid X}(y\mid z)\,\vol_{\cY}(dy).}
\]
{For the conditional Hessian matrix \(M_H(z,\theta)=E_\omega^\ast\Hess_\omega^{\cY}F(\cdot;z)E_\omega\), the predictor derivatives through order two exist and satisfy, for \(b=0,1,2\),}
\[
{\nabla_{\cX}^bM_H(z,\theta)
 =\int_{\cK_{\cY,x}}H(\theta,y)\,
 \nabla_{\cX}^bp_{Y\mid X}(y\mid z)\,\vol_{\cY}(dy).}
\]
{All these derivatives are jointly continuous on \(\overline{\cV}\times\Theta\).}

\noindent{\textup{(iii) Uniform predictor regularity.} Consider a fixed finite family of the preceding local constructions satisfying the common density and \(\gamma\)-H\"older envelope conditions of \Cref{ass:uniform-inference-conditions}. Under \textup{(H)}, take the response-center regions inside the common ball \(B_{\cY}(o,R)\). In predictor normal coordinates, the maps}
\[
{u\longmapsto M_T\bigl(\Exp_z^{\cX}[u],\theta\bigr)}
\]
{have uniformly bounded \(C^{2,\gamma}\) norms, and the corresponding maps for \(M_H\) have uniformly bounded \(C^2\) norms, on each fixed compact predictor disk bundle whose exponential image lies in the corresponding predictor neighborhood. The bounds are uniform over \(\theta\in\Theta\), the predictor base points \(z\), and the finite family.}
\end{lemma}
\begin{proof}
{For differentiation, take the bounds on slightly larger compact parameter sets contained in the given coordinate domains, with \(\overline{\cV}\times\Theta\) in their interiors, and then restrict to \(\overline{\cV}\times\Theta\). For the uniform bounds under \textup{(H)}, keep the enlarged center sets inside the same ball \(B_{\cY}(o,R)\).}

{For \textup{(i)}, under \textup{(C)}, the functions \(d_{\cY}(y,\omega)^2/2\) and \(T(\theta,y)^{\otimes2}\) are continuous and bounded on the respective compact center sets times \(\cK_{\cY,x}\). Their products with \(p_{Y\mid X}(y\mid z)\) are therefore dominated by \(Ce_x(y)\), which is integrable with respect to \(\vol_{\cY}\). Under \textup{(H)}, the triangle inequality and orthonormality give}
\[
{\sup_{\omega\in\cC}\frac12d_{\cY}(y,\omega)^2
 +\sup_{\theta\in\Theta}\|T(\theta,y)^{\otimes2}\|_{\mathrm{op}}
 \le C\{1+d_{\cY}(m_x,y)^2\}.}
\]
{The right-hand side multiplied by \(e_{0,x}(y)\) is integrable by \Cref{ass:densities}. For \(\vol_{\cY}\)-almost every \(y\), the integrands are jointly continuous in their parameters. Dominated convergence proves both assertions in \textup{(i)}.}

{For \textup{(ii)}, under \textup{(H)}, \Cref{lem:supp_fixed_local_log_domain} gives}
\[
{\begin{aligned}
 \sup_{\theta\in\Theta}\bigl\{\|T(\theta,y)\|
 +\|D_\theta T(\theta,y)\|+\|H(\theta,y)\|_{\mathrm{op}}\bigr\}
 &\le CG(m_x,R_x;y),\\
 \sup_{\theta\in\Theta}\|D_\theta^2T(\theta,y)\|
 &\le C\cE(m_x,R_x;y).
\end{aligned}}
\]
{Because \(a+b\le2\), the only required product involving the second bound is \(D_\theta^2T(\theta,y)p_{Y\mid X}(y\mid z)\). Since \(G(m_x,R_x;y)+\cE(m_x,R_x;y)\le C\cD(m_x,R_x;y)\), all integrands needed for the stated derivatives are dominated by the integrable function}
\[
{C\bigl\{\cD(m_x,R_x;y)e_{0,x}(y)
 +G(m_x,R_x;y)e_{12,x}(y)\bigr\}.}
\]
{Under \textup{(C)}, the geometric factors are bounded and the corresponding products are dominated by \(Ce_x(y)\).}

{Ordinary second predictor derivatives of the density equal its covariant second derivatives plus smooth connection coefficients times its first derivatives, so the same density envelopes apply in local coordinates. On the enlarged compact parameter sets, the mean value theorem and the derivative bounds also dominate the relevant difference quotients. Successive differentiation under the integral gives all partial derivatives of \(M_T\) of total order at most two, and dominated convergence makes them jointly continuous. Thus \(M_T\) is jointly \(C^2\). The same argument gives the predictor derivatives of \(M_H\) through order two and their joint continuity. The predictor connection coefficients are bounded on the compact sets and independent of \(y\); incorporating them yields the stated covariant formulas.}

{The first two center derivatives of \(d_{\cY}(y,\omega)^2/2\) are bounded by \(CG(m_x,R_x;y)\) under \textup{(H)}, and by a constant under \textup{(C)}, by \Cref{lem:supp_fixed_local_log_domain}. Differentiating its conditional integral and including the response connection terms therefore gives}
\[
{\Hess_\omega^{\cY}F(\cdot;z)
 =\int_{\cK_{\cY,x}}\cH_\omega(y)
 p_{Y\mid X}(y\mid z)\,\vol_{\cY}(dy).}
\]
{Its matrix representation in \(E_\omega\) is the stated identity for \(M_H\).}

{For \textup{(iii)}, under \textup{(H)}, the common bound \(\|T(\theta,y)\|+\|H(\theta,y)\|_{\mathrm{op}}\le CG(o,R;y)\) holds on all chosen center regions. The common integrability condition \(\int_{\cY}G(o,R;y)\{e_0(y)+e_{12}(y)\}\,\vol_{\cY}(dy)<\infty\) and the formulas in \textup{(ii)} give uniform bounds for both conditional moments and their predictor derivatives through order two.}

{Holding \(\theta\) fixed, integrate the density H\"older bound in \Cref{ass:uniform-inference-conditions}. For sufficiently close \(\zeta,\zeta'\) in each predictor neighborhood, this gives}
\[
{\begin{aligned}
 \left\|\nabla_{\cX}^2M_T(\zeta,\theta)
 -(\Pi_{\zeta\to\zeta'}^{\cX})^\ast
   \nabla_{\cX}^2M_T(\zeta',\theta)\right\|_{\mathrm{op}}\le C d_{\cX}(\zeta,\zeta')^\gamma
 \int_{\cY}G(o,R;y)e_\gamma(y)\,\vol_{\cY}(dy).
\end{aligned}}
\]
{Here the pullback by \(\Pi_{\zeta\to\zeta'}^{\cX}\) acts on both predictor arguments. The integral is finite by \Cref{ass:uniform-inference-conditions}. Under \textup{(C)}, bounded geometric factors and the integrable density and H\"older envelopes give the same bounds. Taking maxima over the fixed finite family makes the constants uniform.}

{For each fixed predictor base point \(z\), the second-order covariant chain rule expresses the second derivatives of \(u\mapsto M_T(\Exp_z^{\cX}[u],\theta)\) and \(u\mapsto M_H(\Exp_z^{\cX}[u],\theta)\) in terms of their first and second predictor derivatives. The Hessian term has two factors of \(d\Exp_z^{\cX}\), and the first-derivative term has the covariant derivative of \(d\Exp_z^{\cX}\). On a fixed compact predictor disk bundle, these smooth coefficients are uniformly bounded and locally Lipschitz, uniformly over the predictor base points. The first predictor derivatives are locally Lipschitz by the second-derivative bounds. Expressing the transported second derivatives in smooth predictor frames adds only locally Lipschitz coefficient terms. Since \(0<\gamma\le1\), the covariant H\"older bound for \(M_T\) therefore yields a uniform local \(\gamma\)-H\"older bound for the second derivatives of its exponential pullbacks. Compactness gives a finite cover on which these bounds are uniform; for pairs outside these local comparison neighborhoods, uniform boundedness controls the H\"older quotient. This proves the asserted \(C^{2,\gamma}\) bounds for the pullbacks of \(M_T\). The bounds through order two give the asserted \(C^2\) bounds for the pullbacks of \(M_H\).}
\end{proof}
Under \textup{(H)}, dominated differentiation and Hessian comparison
\citep[Theorem~11.7]{Lee2018} give,
on the fixed center region,
\begin{equation}\label{eq:supp-hadamard-hessian-lower-bound}
 \Hess_\omega^{\cY}F(\cdot;z)=\E[\cH_\omega(Y)\mid X=z]\succeq I.
\end{equation}
In particular, \(A_x\succeq I\).

\subsection{Conditional Fr\'echet means and local neighborhoods}
\label{subsec:supp_uniform_version}

Under \Cref{ass:uniform-inference-conditions} without its bandwidth conditions, let \(\cO_{\cX}\) be an open
neighborhood of \(\cX_0\) on which the common density versions are defined.
Choose \(\rho_X>0\) so that
\begin{equation}\label{eq:supp_predictor_buffer_containment}
 \overline{\cX}_0^{\,\rho_X}
 \coloneqq\{z:d_{\cX}(z,\cX_0)\le\rho_X\}
 \subset\cO_{\cX},\qquad
 c_{X,0}\le p_X(z)\le C_{X,0}\quad\vol_{\cX}\text{-a.e.}
\end{equation}
Compactness of \(\cX_0\) and openness of \(\cO_{\cX}\) permit this choice.
Shrink \(\rho_X\) so that \(\rho_X<\iota_X/2\), where
\(\iota_X\coloneqq \inf_{x\in\cX_0}\inj_{\cX}(x)>0\).
Write \(\cX_0^{\rho_X}\coloneqq \{z:d_{\cX}(z,\cX_0)<\rho_X\}\).

All sufficiently small bandwidths \(a\in\{h,b\}\) satisfy
\[
    x\in\cX_0,\qquad d_{\cX}(z,x)\le R_Ka
    \quad\Longrightarrow\quad
    z\in\overline{\cX}_0^{\,\rho_X}\subset\cO_{\cX},
\]
and \(z\) lies in the injectivity region of \(x\), so
\(\Log_x^{\cX}z\) is defined on all local kernel supports.

\begin{proposition}
\label{prop:supp_smoothness_conditional_frechet_mean_map}
Fix \(x_0\in\cX\), and suppose \Cref{ass:local_support_log_geometry,ass:densities} hold at \(x_0\). Then there exist an open neighborhood
\(\cN_{x_0}\subset\cV_{x_0}\) of \(x_0\) and a relatively compact open
neighborhood \(\cO_{m_{x_0}}\subset\cY\) of \(m_{x_0}\) such that
\[
    \overline{\cO_{m_{x_0}}}\times\cK_{\cY,x_0}
    \subset\cD_{\log}^{\cY},
\]
and, for every \(z\in\cN_{x_0}\), the function \(F(\cdot;z)\) has a unique global
minimizer \(m_z\in\cO_{m_{x_0}}\). {Moreover, \(m\in C^2(\cN_{x_0},\cO_{m_{x_0}})\).}
\end{proposition}
\begin{proof}
By \Cref{lem:supp_fixed_local_log_domain}, choose normal coordinates
centered at \(x_0\) and \(m_{x_0}\), with the response coordinate
neighborhood having compact closure in the fixed logarithm domain.
Let \(\widetilde\Phi(v,u)\) denote the coordinate representation of
\({\grad_\omega^{\cY}F(\cdot;z)}
=-\E[\Log_\omega^{\cY}Y\mid X=z]\), where \(u=0\) and \(v=0\)
correspond to \(x_0\) and \(m_{x_0}\), respectively.
By \Cref{lem:supp_conditional_moment_regularity}~\textup{(ii)},
\(\widetilde\Phi\) is \(C^2\).

Since \(m_{x_0}\) is a critical point, \(\widetilde\Phi(0,0)=0\).
{In \(D_v\widetilde\Phi\), \(u\) is held fixed.} The Christoffel correction vanishes there, so \({(D_v\widetilde\Phi)(0,0)}\)
is the coordinate representation of the positive definite Hessian \(A_{x_0}\)
\citep[Proposition~2.2.6]{Petersen2016}.
Thus \({(D_v\widetilde\Phi)(0,0)}\) is invertible.
The implicit function theorem
\citep[cf.][Theorem~C.40]{Lee2013} therefore gives neighborhoods
\(\cN\subset\cV_{x_0}\), \(\cO\ni m_{x_0}\), and a \(C^2\) map
\(\mu:\cN\to\cO\) such that \(\mu(z)\) is the unique critical point
of \(F(\cdot;z)\) in \(\cO\).
We may choose \(\cO\) relatively compact with
\(\overline{\cO}\times\cK_{\cY,x_0}\subset\cD_{\log}^{\cY}\).

Let \(M(z)\coloneqq\argmin_\omega F(\omega;z)\).
Apply \Cref{lem:supp_signed_measure_frechet_coercivity}
to the conditional distribution \(\mathbb P_{Y\mid X=z}\).
The local conditional moment bound implies that, for \(z\) near \(x_0\),
the nonempty compact sets \(M(z)\) lie in a common compact set
\(\cC\subset\cY\) containing \(m_{x_0}\).
By \Cref{lem:supp_conditional_moment_regularity}~\textup{(i)},
\(\Delta(z)\coloneqq
\sup_{\omega\in\cC}|F(\omega;z)-F(\omega;x_0)|\to0\)
as \(z\to x_0\).

For every \(\omega\in M(z)\),
\(F(\omega;z)\le F(m_{x_0};z)\), and hence
\begin{equation}\label{eq:supp-conditional-argmin-gap}
F(\omega;x_0)-F(m_{x_0};x_0)
\le
|F(\omega;x_0)-F(\omega;z)|+
|F(m_{x_0};z)-F(m_{x_0};x_0)|\le 2\Delta(z).
\end{equation}
Since \(m_{x_0}\) is the unique minimizer of \(F(\cdot;x_0)\),
continuity and compactness give
\[
\eta_{\cO}
\coloneqq
\inf_{\omega\in\cC\setminus\cO}
\{F(\omega;x_0)-F(m_{x_0};x_0)\}
>0.
\]
Shrink \(\cN\), if necessary, so that
\(2\Delta(z)<\eta_{\cO}\) for \(z\in\cN\).
Equation \eqref{eq:supp-conditional-argmin-gap} gives \(M(z)\subset\cO\).

Every global minimizer in the open set \(\cO\) is a critical point,
while \(\mu(z)\) is the unique critical point there.
Hence \(M(z)=\{\mu(z)\}\).
Taking \(\cN_{x_0}=\cN\) and
\(\cO_{m_{x_0}}=\cO\) proves the result.
\end{proof}

{The next lemma chooses finitely many predictor and response neighborhoods on which the preceding local bounds hold uniformly.}

\begin{lemma}
\label{lem:supp_uniform_finite_shrunken_cover}
{Assume \Cref{ass:uniform-inference-conditions} without its bandwidth conditions. Use its H\"older exponent \(\gamma\), and fix \(\rho_X\) as in \eqref{eq:supp_predictor_buffer_containment}. There exist \(x_1,\ldots,x_L\in\cX_0\), a finite open cover \((\cU_\ell)_{\ell\le L}\) of \(\cX_0\), predictor neighborhoods \(\cN_\ell,\cV_\ell\subset\cX\), response neighborhoods \(\cW_\ell^{\cY}\subset\cY\), and constants \(\rho_Y,c,\delta_L>0\) with the following {\normalcolor four} properties. Set \(\cX_\ell\coloneqq\overline{\cU_\ell}^{\,\cX}\cap\cX_0\) and \(\cK_\ell\coloneqq\cK_{\cY,x_\ell}\). All these choices, including the frames below, are fixed independently of \(n\) and \(h\).}

\noindent{\textup{(i) Predictor neighborhoods and conditional regularity.} The sets \(\cN_\ell\) are relatively compact and strongly geodesically convex, and}
\begin{equation}\label{eq:supp-matched-predictor-domains}
{\begin{aligned}
 \cX_\ell&\subset\cN_\ell\Subset\cV_\ell\Subset\cX_0^{\rho_X},\qquad
 \mathbb P(Y\in\cK_\ell\mid X=z)=1,\quad z\in\cV_\ell.
\end{aligned}}
\end{equation}
{On a neighborhood of \(\overline{\cV_\ell}^{\,\cX}\), the map \(m\) is \(C^2\), and the fixed conditional density version satisfies the density and H\"older envelope conditions of \Cref{ass:uniform-inference-conditions}. The predictor neighborhoods satisfy}
\begin{equation}\label{eq:supp-matched-predictor-radius}
{\min_{\ell\le L}d_{\cX}(\cX_\ell,\cV_\ell^c)>0,\qquad
 \Exp_x^{\cX}[u]\in\cN_\ell\quad
 (x\in\cX_\ell,\ \|u\|_x\le\delta_L).}
\end{equation}

\noindent{\textup{(ii) Response neighborhoods and Hessian bounds.} The sets \(\cW_\ell^{\cY}\) are relatively compact and strongly geodesically convex. For \(x\in\cX_\ell\) and \(\omega\in\overline{\cW_\ell^{\cY}}\),}
\begin{equation}\label{eq:supp-matched-response-domains}
{\begin{aligned}
 m(\overline{\cV_\ell}^{\,\cX})&\Subset\cW_\ell^{\cY},\qquad
 \overline{\cW_\ell^{\cY}}\times
 (\cK_\ell\cup\overline{\cW_\ell^{\cY}})\subset\cD_{\log}^{\cY},\\
 \overline B_{\cY}(m_x,2\rho_Y)&\subset\cW_\ell^{\cY},\qquad
 \Hess_\omega^{\cY}F(\cdot;x)\succeq2cI_{T_\omega\cY}.
\end{aligned}}
\end{equation}

\noindent{\textup{(iii) Response frames.} For each \(\ell\), choose a smooth orthonormal response frame \(E_\ell(\omega):\R^q\to T_\omega\cY\) on a neighborhood of \(\overline{\cW_\ell^{\cY}}\). The maps \(E_{z,\ell}\coloneqq E_\ell(m_z)\) form a \(C^2\) family of isometries on a neighborhood of \(\overline{\cV_\ell}^{\,\cX}\), with \(E_{z,\ell}^\ast E_{z,\ell}=I_q\).}

\noindent{\textup{(iv) Uniform geometric bounds.} The derivative bounds in \Cref{lem:supp_fixed_local_log_domain}~\textup{(ii) {\normalcolor and} (iii)} hold uniformly on these neighborhoods in fixed response coordinates and the chosen frames. The quantities in \Cref{lem:supp_fixed_local_log_domain}~\textup{(ii)} are bounded by \(C\cE_R(y)\), while the squared-distance loss \(d_{\cY}(y,\omega)^2/2\) and pairwise products of those quantities are bounded by \(C\cE_R(y)^2\), with a common constant over the finite family.}
\end{lemma}
\begin{proof}
{Fix \(x_0\in\cX_0\). By \Cref{prop:supp_smoothness_conditional_frechet_mean_map} and \Cref{lem:supp_conditional_moment_regularity}~\textup{(ii)}, shrink the support and density neighborhood \(\cV_{x_0}\Subset\cX_0^{\rho_X}\) so that \(m\) is \(C^2\), the map \((\omega,z)\mapsto\Hess_\omega^{\cY}F(\cdot;z)\) is continuous in smooth response frames, and the density and H\"older envelope conditions hold on a neighborhood of \(\overline{\cV_{x_0}}^{\,\cX}\). Under \textup{(C)}, \(A_{x_0}\succ0\), compactness of \(\cK_{\cY,x_0}\), and openness of \(\cD_{\log}^{\cY}\) permit a relatively compact strongly geodesically convex response neighborhood \(\cW_{x_0}^{\cY}\), with closure in a normal chart, and a further shrinking of \(\cV_{x_0}\) such that}
\begin{equation}\label{eq:supp-matched-local-construction}
{m(\overline{\cV_{x_0}}^{\,\cX})\Subset\cW_{x_0}^{\cY},\quad \overline{\cW_{x_0}^{\cY}}\times (\cK_{\cY,x_0}\cup\overline{\cW_{x_0}^{\cY}})\subset\cD_{\log}^{\cY},\quad \inf_{\substack{z\in\overline{\cV_{x_0}}^{\,\cX}\\ \omega\in\overline{\cW_{x_0}^{\cY}}}} \lambda_{\min}(\Hess_\omega^{\cY}F(\cdot;z))>0.}
\end{equation}
{Under \textup{(H)}, take \(\cW_{x_0}^{\cY}\coloneqq B_{\cY}(o,(R_m+R)/2)\), whose closure lies in \(B_{\cY}(o,R)\). The bound \(d_{\cY}(o,m_z)\le R_m\) and \eqref{eq:supp-hadamard-hessian-lower-bound} give \eqref{eq:supp-matched-local-construction}. Choose a relatively compact strongly geodesically convex predictor neighborhood \(\cN_{x_0}\) and an open set \(\cU_{x_0}\) satisfying}
\[
{x_0\in\cU_{x_0}\Subset\cN_{x_0}\Subset\cV_{x_0}\Subset\cX_0^{\rho_X}.}
\]

{Compactness of \(\cX_0\) gives a finite subcover \((\cU_\ell)_{\ell\le L}\), with the associated neighborhoods indexed by \(\ell\). This gives \eqref{eq:supp-matched-predictor-domains} and the asserted regularity near \(\overline{\cV_\ell}^{\,\cX}\). Compact containment makes the distances to the neighborhood boundaries positive and permits the common choices}
\begin{gather}
{0<\delta_L<\min\left\{\iota_X,
 \min_{\ell\le L}d_{\cX}(\cX_\ell,\cN_\ell^c)\right\},\qquad
0<2\rho_Y<\min_{\ell\le L}
 d_{\cY}(m(\cX_\ell),(\cW_\ell^{\cY})^c),}\notag\\
{2c\coloneqq\min_{\ell\le L}
 \inf_{\substack{z\in\overline{\cV_\ell}^{\,\cX}\\
                  \omega\in\overline{\cW_\ell^{\cY}}}}
 \lambda_{\min}(\Hess_\omega^{\cY}F(\cdot;z))>0.}
\label{eq:supp-matched-common-constants}
\end{gather}
{For \(x\in\cX_\ell\) and \(\|u\|_x\le\delta_L\),}
\[
{d_{\cX}(x,\Exp_x^{\cX}[u])=\|u\|_x\le\delta_L
 <d_{\cX}(\cX_\ell,\cN_\ell^c).}
\]
{Thus \eqrefs{eq:supp-matched-local-construction,eq:supp-matched-common-constants} give \eqrefs{eq:supp-matched-predictor-radius,eq:supp-matched-response-domains}, proving \textup{(i) {\normalcolor and} (ii)}.}

{For \textup{(iii)}, apply Gram--Schmidt orthonormalization to the coordinate frame on a response chart containing \(\overline{\cW_\ell^{\cY}}\), using a global normal chart under \textup{(H)}. This gives a smooth orthonormal frame \(E_\ell(\omega):\R^q\to T_\omega\cY\) on a neighborhood of \(\overline{\cW_\ell^{\cY}}\). By \Cref{prop:supp_smoothness_conditional_frechet_mean_map}, \(m\) is \(C^2\) near \(\overline{\cV_\ell}^{\,\cX}\). Since \(m(\overline{\cV_\ell}^{\,\cX})\Subset\cW_\ell^{\cY}\), the composition \(E_{z,\ell}=E_\ell(m_z)\) is defined and \(C^2\) on an open neighborhood of \(\overline{\cV_\ell}^{\,\cX}\). Orthonormality gives \(E_{z,\ell}^\ast E_{z,\ell}=I_q\).}

{Finally, apply \Cref{lem:supp_fixed_local_log_domain}~\textup{(ii) {\normalcolor and} (iii)} on the finite family of response neighborhoods, with the fixed charts and frames just chosen. Under \textup{(C)}, the response supports \(\cK_\ell\) and center closures are compact. Under \textup{(H)}, the inclusions \(\overline{\cW_\ell^{\cY}}\subset B_{\cY}(o,R)\) give the common envelope \(C\cE_R(y)\). The logarithm and parallel transport between centers are smooth near the compact sets of center pairs and unit vectors allowed by \eqref{eq:supp-matched-response-domains}, giving the uniform derivative bounds in \textup{(iv)}. The triangle inequality bounds \(d_{\cY}(y,\omega)^2/2\) by \(C\cE_R(y)^2\), and the pairwise product bounds follow from the individual \(C\cE_R(y)\) bounds. Taking maxima over the finite family gives a common constant. Every neighborhood, frame, and constant was chosen from the fixed model and domains, independently of \(n\) and \(h\).}
\end{proof}

For uniform arguments, use the predictor neighborhoods, response neighborhoods, and corresponding frames constructed in
\Cref{lem:supp_uniform_finite_shrunken_cover}.

For pointwise arguments and \((j,a)\in\{(1,h),(2,b)\}\),
\begin{equation}\label{eq:supp-pointwise-response-support}
 \mathbb P\left(\bigcap_{i=1}^n
 \bigl(\{K_{j,a}(X_i,x)=0\}\cup\{Y_i\in\cK_{\cY,x}\}\bigr)\right)=1.
\end{equation}
For small \(a\), nonzero kernels imply \(X_i\in\cV_x\).
The fixed version of the conditional distribution in \Cref{ass:local_support_log_geometry}
gives \eqref{eq:supp-pointwise-response-support}.

Define the uniform response support event
\begin{equation}
\label{eq:supp_uniform_response_support_event}
    \cE_{n,\cX_0}^{\mathrm{supp}}
    \coloneqq
    \bigcap_{i=1}^n\bigcap_{\ell=1}^L
    \left(\{X_i\notin\cV_\ell\}\cup\{Y_i\in\cK_\ell\}\right).
\end{equation}
By \Cref{lem:supp_uniform_finite_shrunken_cover}~\textup{(i)},
\(\mathbb P(X_i\in\cV_\ell,Y_i\notin\cK_\ell)
=\E[\mathds{1}\{X_i\in\cV_\ell\}
\mathbb P(Y_i\notin\cK_\ell\mid X_i)]=0\)
for every \(i\) and \(\ell\). Hence
\(\mathbb P(\cE_{n,\cX_0}^{\mathrm{supp}})=1\) for every \(n\).

\begin{proposition}
\label{prop:supp_derived_uniform_identification_hessian_nondegeneracy}
Under \Cref{ass:uniform-inference-conditions} without its bandwidth conditions, for every compact
\(\cC\subset\cY\) and every \(\varepsilon>0\),
\begin{equation}\label{eq:supp-uniform-identification-gap}
    \inf_{\substack{x\in\cX_0,\ \omega\in\cC\\
      d_{\cY}(\omega,m_x)\ge\varepsilon}}
    \{F(\omega;x)-F(m_x;x)\}
    >
    0 .
\end{equation}
\end{proposition}

\begin{proof}
By \Cref{lem:supp_conditional_moment_regularity}~\textup{(i)} and
continuity of \(m\), \(F(\omega;x)-F(m_x;x)\) is continuous on
\(\cX_0\times\cC\) and positive wherever
\(d_{\cY}(\omega,m_x)\ge\varepsilon\). Compactness gives the assertion,
with the infimum over the empty set taken to be positive infinity.
\end{proof}

\subsection{Approximation of the objective functions and consistency of the centers}

For {a Borel set} \(A{\subseteq}\cY\), define {the signed response measure and the empirical objective with population weights by}
\[
 \nu_{h,x}(A)\coloneqq\E[L_{1,h}(X;x)\mathds{1}\{Y\in A\}],\qquad
 Q_{n,h,x}^{\circ}(\omega)\coloneqq
 {\tfrac12P_n[L_{1,h}(X;x)d_{\cY}(Y,\omega)^2]}.
\]
In the following proofs, take \(o=m_x\) pointwise. Uniformly, use any
fixed \(o\in\cY\) under \textup{(C)} and the reference point in
\Cref{ass:uniform-inference-conditions} under \textup{(H)}.

\begin{lemma}
\label{lem:supp_uniform_compact_criterion_approximation}
For \(\nu=\mathrm{pt}\), suppose
\Cref{ass:local_support_log_geometry,ass:densities,ass:kernels} hold.
For \(\nu=\mathrm{unif}\), impose
\Cref{ass:uniform-inference-conditions} with its bandwidth conditions
omitted. Let \(h\to0\). For every compact \(\cC\subset\cY\):

\noindent\textup{(i) Population approximation.}
\[
 \sup_{x\in\mathcal I_\nu}\sup_{\omega\in\cC}
 |Q_{h,x}(\omega)-F(\omega;x)|\to0.
\]

\noindent\textup{(ii) Empirical approximation.}
Additionally assume \(nh^p\to\infty\) pointwise, or
\eqref{eq:supp-uniform-bandwidth-rate} with \(a=h\) uniformly.
For \(\widehat Q_{h,x}^{\mathrm{ext}}\) from
\Cref{lem:supp_measurable_empirical_local_frechet_minimizer},
\[
 \sup_{x\in\mathcal I_\nu}\sup_{\omega\in\cC}
 |\widehat Q_{h,x}^{\mathrm{ext}}(\omega)-F(\omega;x)|=o_p(1).
\]
\end{lemma}
\begin{proof}
Exact constant reproduction and
\Cref{prop:supp_selector_weight_design_moment_master}~\textup{(ii)} give
\begin{equation}\label{eq:supp_population_criterion_approximation}
\begin{aligned}
 \sup_{\omega\in\cC}|Q_{h,x}(\omega)-F(\omega;x)|
 &\le\E[|L_{1,h}(X;x)|]
 \sup_{\substack{\omega\in\cC\\d_{\cX}(z,x)\le R_1h}}
 |F(\omega;z)-F(\omega;x)|\to0.
\end{aligned}
\end{equation}
Here \Cref{lem:supp_conditional_moment_regularity}~\textup{(i)} supplies
uniform continuity on compact sets of predictors and response centers.
Under the uniform conditions, apply \eqref{eq:supp_population_criterion_approximation}
for \(x\in\cX_\ell\). For sufficiently small \(h\), the kernel supports
lie in \(\cV_\ell\). The uniform weight bound and finite cover make the
convergence uniform over \(\cX_0\).

For \textup{(ii)}, work on the design and response support events, where \(\widehat Q_{h,x}^{\mathrm{ext}}=\widehat Q_{h,x}\). {We use the decomposition}
\begin{equation}\label{eq:supp-criterion-three-term-decomposition}
{\begin{aligned}
 &\sup_{x\in\mathcal I_\nu}\sup_{\omega\in\cC}
 |\widehat Q_{h,x}^{\mathrm{ext}}(\omega)-F(\omega;x)|\\
 &\quad\leq\sup_{\substack{x\in\mathcal I_\nu\\\omega\in\cC}}|\widehat Q_{h,x}(\omega)-Q_{n,h,x}^{\circ}(\omega)|+\sup_{\substack{x\in\mathcal I_\nu\\\omega\in\cC}}|Q_{n,h,x}^{\circ}(\omega)-Q_{h,x}(\omega)|
 +\sup_{\substack{x\in\mathcal I_\nu\\\omega\in\cC}}|Q_{h,x}(\omega)-F(\omega;x)|.
\end{aligned}}
\end{equation}
{The last term tends to zero by part~\textup{(i)}. To control the first term,} the loss envelope and \Cref{lem:supp_localized_support_count} give
\[
\begin{gathered}
 \sup_{\omega\in\cC}d_{\cY}(y,\omega)^2
 \le C_{\cC}\{1+d_{\cY}(o,y)^2\},\\
 \sup_{x\in\mathcal I_\nu}h^{-p}P_n\left[
 \mathds{1}\{d_{\cX}(X,x)\le R_1h\}\{1+d_{\cY}(o,Y)^2\}\right]=O_p(1).
\end{gathered}
\]
{The weight-replacement bound in \eqref{eq:supp-local-linear-weight-replacement-envelope} and the empirical design bounds} {in \Cref{lem:supp_empirical_design_concentration}} {imply}
\begin{equation}\label{eq:supp-criterion-feasible-weight-error}
\begin{aligned}
 &\sup_{x\in\mathcal I_\nu}\sup_{\omega\in\cC}
 |\widehat Q_{h,x}(\omega)-Q_{n,h,x}^{\circ}(\omega)|\\
 &\quad\le C_{\cC}
 \sup_{x\in\mathcal I_\nu}\|\widehat S_{1,h}(x)^{-1}-S_{1,h}(x)^{-1}\|_{\mathrm{op}}
 \sup_{x\in\mathcal I_\nu}h^{-p}P_n\left[
 \mathds{1}\{d_{\cX}(X,x)\le R_1h\}\{1+d_{\cY}(o,Y)^2\}\right]\\
 &\quad=O_p(\delta_{n,\nu}(h)).
\end{aligned}
\end{equation}

Pointwise, \eqref{eq:supp-pointwise-response-tail} and
\Cref{lem:supp-pointwise-local-truncation}, with
\(G(y)=1+d_{\cY}(m_x,y)^2\), give
\(Q_{n,h,x}^{\circ}(\omega)-Q_{h,x}(\omega)=o_p(1)\) for fixed
\(\omega\). Moreover,
\[
 \tfrac12|d_{\cY}(y,\omega)^2-d_{\cY}(y,\omega')^2|
 \le d_{\cY}(\omega,\omega')\{d_{\cY}(m_x,y)+C_{\cC}\},
 \qquad \omega,\omega'\in\cC.
\]
{\(P_n[|L_{1,h}|\{d_{\cY}(m_x,Y)+C_{\cC}\}]=O_p(1)\) and \(P[|L_{1,h}|\{d_{\cY}(m_x,Y)+C_{\cC}\}]=O(1)\).} A finite \(\delta\)-net {argument on} \(\cC\), first letting \(n\to\infty\) and then \(\delta\downarrow0\), gives \(\sup_{\omega\in\cC}|Q_{n,h,x}^{\circ}(\omega)-Q_{h,x}(\omega)|=o_p(1)\).

Uniformly, the conditional support property in
\eqref{eq:supp-matched-predictor-domains} gives, for \(x\in\cX_\ell\) and small \(h\),
\begin{equation}\label{eq:supp-criterion-carrier-restriction}
 \tfrac12L_{1,h}(X;x)d_{\cY}(Y,\omega)^2\mathds{1}\{Y\in\cK_\ell\}
 =\tfrac12L_{1,h}(X;x)d_{\cY}(Y,\omega)^2\quad\text{a.s.}
\end{equation}
For \(\omega,\omega'\in\cC\),
\begin{equation}\label{eq:supp-criterion-loss-envelope}
\begin{aligned}
 \tfrac12d_{\cY}(y,\omega)^2\mathds{1}\{y\in\cK_\ell\}
 &\le C_{\cC}\cE_R(y)^2,\\
 \tfrac12|d_{\cY}(y,\omega)^2-d_{\cY}(y,\omega')^2|\mathds{1}\{y\in\cK_\ell\}
 &\le C_{\cC}\cE_R(y)d_{\cY}(\omega,\omega').
\end{aligned}
\end{equation}
{By} \eqref{eq:supp-population-design-bounds}{, the coordinates of \(S_{1,h}(x)^{-1}e_{1,x}\) are uniformly bounded.} {Cover \(\cC\) by finitely many response coordinate charts.} By \eqref{eq:supp-criterion-loss-envelope}, \Cref{lem:supp_local_polynomial_kernel_factors_are_localized} applies to \((z,y)\mapsto\tfrac12L_{1,h}(z;x)d_{\cY}(y,\omega)^2\mathds{1}\{y\in\cK_\ell\}\) with {\((a,M_a,H,m)=(h,h^{-p},1,0)\)} under \textup{(C)}, or {\((a,M_a,H,m)=(h,h^{-p},\cE_R,2)\)} under \textup{(H)}. Apply \Cref{lem:supp_localized_bracketing_maximal_inequality} in the former case and \Cref{lem:supp_localized_finite_moment_maximal_inequality} in the latter. Using \eqref{eq:supp-criterion-carrier-restriction} and taking the maximum over the finitely many predictor neighborhoods and response charts gives
\begin{equation}\label{eq:supp-criterion-oracle-empirical-error}
 \sup_{x\in\mathcal I_\nu}\sup_{\omega\in\cC}
 |Q_{n,h,x}^{\circ}(\omega)-Q_{h,x}(\omega)|
 =\begin{cases}
 o_p(1),&\nu=\mathrm{pt},\\
 O_p(\delta_{n,\mathrm{unif}}(h)),&\nu=\mathrm{unif},\ \textup{(C)},\\
 O_p(c_n(h)),&\nu=\mathrm{unif},\ \textup{(H)}.
 \end{cases}
\end{equation}

Substituting \eqrefs{eq:supp_population_criterion_approximation,eq:supp-criterion-feasible-weight-error,eq:supp-criterion-oracle-empirical-error} into \eqref{eq:supp-criterion-three-term-decomposition} proves \textup{(ii)}, since the design and support events have probability tending to one.
\end{proof}

\begin{proposition}
\label{prop:supp_population_smoothed_target_consistency}
\label{prop:supp_derived_uniqueness_population_local_linear_target}
For \(\nu=\mathrm{pt}\), suppose
\Cref{ass:local_support_log_geometry,ass:densities,ass:kernels} hold.
For \(\nu=\mathrm{unif}\), impose
\Cref{ass:uniform-inference-conditions} without its bandwidth conditions.
For all sufficiently small \(h>0\), \(Q_{h,x}(\cdot)\) has a unique global
minimizer \(m_{h,x}\) for every \(x\in\mathcal I_\nu\), and
\[
 \sup_{x\in\mathcal I_\nu}d_{\cY}(m_{h,x},m_x)\to0
 \qquad(h\downarrow0).
\]
For each such \(h\), the map \(x\mapsto m_{h,x}\) is Borel.
\end{proposition}
\begin{proof}
For small \(h\), \Cref{lem:supp_kernel_change_population_design}
gives nonsingular population designs.
Exact reproduction and
\Cref{prop:supp_selector_weight_design_moment_master}~\textup{(ii)} give
\[
 \nu_{h,x}(\cY)=1,\qquad
 \sup_{x\in\mathcal I_\nu}\cM(\nu_{h,x};o)
 \le\sup_{x\in\mathcal I_\nu}
 \E[|L_{1,h}(X;x)|\{1+d_{\cY}(o,Y)^2\}]=O(1).
\]
Use the local conditional second moment of the distance pointwise.
Uniformly, use the compact union \(\bigcup_{\ell=1}^L\cK_\ell\) containing the conditional response supports under \textup{(C)} and the conditional
\(2r_0\)-moment bound under \textup{(H)}. {By \Cref{lem:supp_signed_measure_frechet_coercivity}, \(Q_{h,x}(\cdot)\) has a nonempty compact argmin for all sufficiently small \(h\). The conditional Fr\'echet means \(m_x\) are bounded over \(\mathcal I_\nu\), trivially in the pointwise case and by {\Cref{lem:supp_uniform_finite_shrunken_cover}~\textup{(ii)}} in the uniform case. Together with the preceding moment bound, \Cref{eq:supp-signed-coercivity,eq:supp-signed-absolute-bound} give a coercive lower bound and a uniform upper bound at \(m_x\). Thus condition \textup{(iii)} of \Cref{lem:supp_coercive_compact_argmin_consistency} holds with bounded deterministic coefficients. For every compact set \(\cC\subset\cY\), \Cref{lem:supp_uniform_compact_criterion_approximation}~\textup{(i)} gives}
\[
{{\sup_{x\in\mathcal I_\nu,\,\omega\in\cC}}
 |Q_{h,x}(\omega)-F(\omega;x)|\to0
 \qquad(h\downarrow0).}
\]
{This verifies condition \textup{(i)} of \Cref{lem:supp_coercive_compact_argmin_consistency} in its deterministic form. Condition \textup{(ii)} follows pointwise from continuity of \(F(\cdot;x)\) and uniqueness of its minimizer \(m_x\), and uniformly from \Cref{eq:supp-uniform-identification-gap} in \Cref{prop:supp_derived_uniform_identification_hessian_nondegeneracy}. Applying the deterministic version of \Cref{lem:supp_coercive_compact_argmin_consistency} as \(h\downarrow0\) gives}
\begin{equation}\label{eq:supp_uniform_population_argmin_localization}
 \sup_{x\in\mathcal I_\nu}\sup_{\omega\in\argmin_{\omega\in\cY}Q_{h,x}(\omega)}
 d_{\cY}(\omega,m_x)\to0.
\end{equation}

Pointwise choose a relatively compact strongly geodesically convex
neighborhood \(\cW_x\) within the fixed logarithm region, on which
\(\Hess_\omega^{\cY}F(\cdot;x)\) has a positive lower bound.
Uniformly use \(\cW_\ell^{\cY}\) from
\Cref{lem:supp_uniform_finite_shrunken_cover}~\textup{(ii)}.
Set \(A_{h,x}(\omega)\coloneqq \Hess_\omega^{\cY}Q_{h,x}(\cdot)\). In fixed response
frames, conditioning and constant reproduction give
\begin{equation}\label{eq:supp-population-hessian-local-approximation}
\begin{aligned}
 A_{h,x}(\omega)-\Hess_\omega^{\cY}F(\cdot;x)
 &=\E[L_{1,h}(X;x)
 \{\Hess_\omega^{\cY}F(\cdot;X)-\Hess_\omega^{\cY}F(\cdot;x)\}],\\
 \|A_{h,x}(\omega)-\Hess_\omega^{\cY}F(\cdot;x)\|_{\mathrm{op}}
 &\le\E[|L_{1,h}(X;x)|]
 \sup_{d_{\cX}(z,x)\le R_1h}
 \|\Hess_\omega^{\cY}F(\cdot;z)-\Hess_\omega^{\cY}F(\cdot;x)\|_{\mathrm{op}}.
\end{aligned}
\end{equation}
By {\Cref{lem:supp_conditional_moment_regularity}~\textup{(ii)}} and the
bound on the absolute weights, the last quantity in
\eqref{eq:supp-population-hessian-local-approximation} tends to zero uniformly over
\(\omega\in\overline{\cW_x}\) pointwise, and over all corresponding
\((\ell,x,\omega)\) uniformly. Hence the positive lower bound on
\(\cW_x\) persists for small \(h\). Uniformly,
\eqrefs{eq:supp-matched-response-domains,eq:supp-population-hessian-local-approximation}
give, for some \(h_0>0\),
\begin{equation}\label{eq:supp_population_hessian_uniform_lower_bound}
 \inf_{\substack{x\in\cX_\ell, \enspace\omega\in\overline{\cW_\ell^{\cY}}}}
 \lambda_{\min}(A_{h,x}(\omega))\ge c,
 \qquad \ell=1,\ldots,L,\quad 0<h\le h_0.
\end{equation}
For sufficiently small \(h\),
\eqref{eq:supp_uniform_population_argmin_localization} places every global
minimizer in \(\cW_x\) pointwise and in \(\cW_\ell^{\cY}\), uniformly over
\(\ell\le L\) and \(x\in\cX_\ell\).
The Hessian lower bounds above imply strict geodesic convexity on these
neighborhoods, so \(\argmin_{\omega\in\cY}Q_{h,x}(\omega)\) is a singleton.

The map \((x,\omega)\mapsto Q_{h,x}(\omega)\) is jointly Borel and
continuous in \(\omega\), and
\Cref{lem:supp_signed_measure_frechet_coercivity} makes its sublevel sets compact.
The measurable selection argument in the proof of
\Cref{lem:supp_measurable_empirical_local_frechet_minimizer}, with parameter
\(x\), makes its unique minimizer \(x\mapsto m_{h,x}\) Borel.
\end{proof}

\begin{proposition}
\label{prop:supp_pointwise_empirical_smoothed_target_consistency}
\label{prop:supp_uniform_crude_localization_local_frechet_centers}
For \(\nu=\mathrm{pt}\), impose
\Cref{ass:local_support_log_geometry,ass:densities,ass:kernels},
\(h\to0\), and \(nh^p\to\infty\).
For \(\nu=\mathrm{unif}\), impose
\Cref{ass:uniform-inference-conditions} without the bandwidth conditions
used to control smoothing bias, and retain \(\log(1/h)\asymp\log n\).
Then
\[
 \sup_{x\in\mathcal I_\nu}d_{\cY}(\widehat m_{h,x},m_x)=o_p(1),\qquad
 \sup_{x\in\mathcal I_\nu}d_{\cY}(\widehat m_{h,x},m_{h,x})=o_p(1).
\]
\end{proposition}
\begin{proof}
On the design and response support events, put
\(\widehat\nu_{n,h,x}\coloneqq n^{-1}\sum_i\widehat L_{1,h,i}(x)\delta_{Y_i}\).
\Cref{prop:supp_selector_weight_design_moment_master,lem:supp_localized_support_count,lem:supp_empirical_design_concentration}
give
\[
 \widehat\nu_{n,h,x}(\cY)=1,\qquad
 \sup_{x\in\mathcal I_\nu}\cM(\widehat\nu_{n,h,x};o)
 \le\sup_{x\in\mathcal I_\nu}
 P_n[|\widehat L_{1,h}|\{1+d_{\cY}(o,Y)^2\}]=O_p(1).
\]
Here nonzero weights restrict observations to the compact union
\(\bigcup_{\ell=1}^L\cK_\ell\) containing the conditional response supports under uniform \textup{(C)}. Under uniform
\textup{(H)}, use \(1+d_{\cY}(o,Y)^2\le\cE_R(Y)^2\).
Equations \eqrefs{eq:supp-signed-coercivity,eq:supp-signed-compact-containment}
verify \Cref{lem:supp_coercive_compact_argmin_consistency}~\textup{(iii)}
with
\[
 C_n\coloneqq \tfrac12\sup_{x\in\mathcal I_\nu}\cM(\widehat\nu_{n,h,x};o)=O_p(1),\quad
 B_n\coloneqq \{1+\sup_{x\in\mathcal I_\nu}d_{\cY}(o,m_x)^2\}
 \sup_{x\in\mathcal I_\nu}\cM(\widehat\nu_{n,h,x};o)=O_p(1).
\]

{Here the conditional Fr\'echet means \(m_x\) are bounded over \(\mathcal I_\nu\), trivially in the pointwise case and by {\Cref{lem:supp_uniform_finite_shrunken_cover}~\textup{(ii)}} in the uniform case. This justifies the bound \(B_n=O_p(1)\). For every compact set \(\cC\subset\cY\), \Cref{lem:supp_uniform_compact_criterion_approximation}~\textup{(ii)} gives}
\[
{
 {\sup_{x\in\mathcal I_\nu,\,\omega\in\cC}}
 \left|\widehat Q_{h,x}^{\mathrm{ext}}(\omega)-F(\omega;x)\right|
 =o_p(1).}
\]
{This verifies condition \textup{(i)} of \Cref{lem:supp_coercive_compact_argmin_consistency}. The separation condition in \Cref{lem:supp_coercive_compact_argmin_consistency}~\textup{(ii)} follows pointwise from continuity and uniqueness, and uniformly from \Cref{prop:supp_derived_uniform_identification_hessian_nondegeneracy}. \Cref{lem:supp_coercive_compact_argmin_consistency} therefore gives the first conclusion. The second follows from \Cref{prop:supp_population_smoothed_target_consistency} and the triangle inequality.}
\end{proof}

Pointwise, choose \(\rho\) and \(\cW_{m_x}\) from
\Cref{lem:supp_fixed_local_log_domain}~\textup{(i)}.
Uniformly, use the finite cover in
\Cref{lem:supp_uniform_finite_shrunken_cover}~\textup{(i) and (ii)}
and \(\rho_Y\) in \eqref{eq:supp-matched-response-domains}.
Under the hypotheses of
\Cref{prop:supp_pointwise_empirical_smoothed_target_consistency},
combine its empirical consistency with
\Cref{prop:supp_population_smoothed_target_consistency} to obtain,
with probability tending to one,
\begin{equation}\label{eq:supp-localized-center-containment}
 \{m_x,m_{h,x},\wh m_{h,x}\}\subset
 \begin{cases}
  \overline B_{\cY}(m_x,\rho)\subset\cW_{m_x},&\nu=\mathrm{pt},\\
  \overline B_{\cY}(m_x,\rho_Y)\subset\cW_\ell^{\cY},
   &\nu=\mathrm{unif},\ x\in\cX_\ell,\ \ell=1,\ldots,L.
 \end{cases}
\end{equation}
The uniform inclusion holds simultaneously for all \(x\) and \(\ell\).

By \Cref{lem:supp_empirical_design_concentration}, the containment in \eqref{eq:supp-localized-center-containment} holds jointly with nonsingularity of \(\widehat S_{1,h}(x)\) on \(\mathcal I_\nu\) and, uniformly, \(\inf_{x\in\cX_0}\lambda_{\min}(\widehat S_{1,h}(x))\ge c_{S,1}\). For RRBC with \(b\asymp h\), the same high-probability assertion includes nonsingularity of \(\widehat S_{2,b}(x)\) on \(\mathcal I_\nu\) and, in the uniform regime, {the} spectral lower bound \(c_{S,2}\).

For sufficiently small bandwidths, \eqrefs{eq:supp-pointwise-response-support,eq:supp_uniform_response_support_event} and \eqref{eq:supp-matched-predictor-radius} place responses with nonzero local linear weights in \(\cK_{\cY,x}\) pointwise and in \(\cK_\ell\) for \(x\in\cX_\ell\) uniformly, almost surely. For RRBC, the same holds for local quadratic weights. Strong geodesic convexity keeps the minimizing geodesics joining the centers in \eqref{eq:supp-localized-center-containment} inside these neighborhoods. The bounds for response logarithms and maps between centers hold pointwise by \Cref{lem:supp_fixed_local_log_domain}~\textup{(ii) and (iii)} and uniformly by {\Cref{lem:supp_uniform_finite_shrunken_cover}~\textup{(iv)}}.

\subsection{\texorpdfstring{{Regularity of the framed logarithm and Hessian}}{Regularity of the framed logarithm and Hessian}}
{The following Hessian and logarithm regularity bounds are used in the population expansions in}
\Cref{sec:supp_score_expansions_covariances}
{and the comparison of grid and continuum suprema in}
\Cref{cor:supp-grid-calibrated-continuum-tube}{.}

\begin{lemma}
\label{lem:supp_population_center_hessian_patchwise_moduli}
Assume \Cref{ass:uniform-inference-conditions} with its bandwidth conditions omitted, and use the cover and corresponding frames from \Cref{lem:supp_uniform_finite_shrunken_cover}. {Write} \(\overline A_\ell(x)\coloneqq E_{x,\ell}^\ast A_xE_{x,\ell}\). {For} \(x,x'\in\cX_\ell\),
\begin{equation}\label{eq:supp-population-hessian-modulus}
 \|\overline A_\ell(x)-\overline A_\ell(x')\|_{\mathrm{op}}
 +\|\overline A_\ell(x)^{-1}-\overline A_\ell(x')^{-1}\|_{\mathrm{op}}
 \lesssim d_{\cX}(x,x').
\end{equation}
\end{lemma}
\begin{proof}
Differentiate
\(\overline A_\ell(x)=E_{x,\ell}^\ast
\E[\cH_{m_x}(Y)\mid X=x]E_{x,\ell}\), varying the center, frame,
and conditioning predictor together. Under \textup{(H)},
\[
 \|(d\overline A_\ell)_x[v]\|_{\mathrm{op}}
 \lesssim\|v\|_x\int_{\cK_\ell}
 \{\cE_Re_0+G_Re_{12}\}\,\vol_{\cY}(dy)\lesssim\|v\|_x,\qquad x\in\cN_\ell.
\]
Center and frame derivatives give the first term in the integrand, and the density derivative gives the second. Under \textup{(C)}, bounded geometric derivatives and the integrable density envelope associated with each predictor neighborhood give the same bound. The minimizing predictor geodesic stays in \(\cN_\ell\) by \eqref{eq:supp-matched-predictor-domains}, so integration gives the first term of \eqref{eq:supp-population-hessian-modulus}. The lower bound in \eqref{eq:supp-matched-response-domains} and \Cref{lem:supp_covariance_transfer_spectral_stability} give the {second term of \eqref{eq:supp-population-hessian-modulus}}.
\end{proof}

{For \(s\in T_{m_x}\cY\) near zero, define the transported logarithm on the response neighborhoods from {\Cref{lem:supp_uniform_finite_shrunken_cover}~\textup{(ii)}} by}
\begin{equation}
 \mathcal L_x(s,y)\coloneqq \Pi_{\Exp_{m_x}^{\cY}\left[s\right]\to m_x}^{\cY}
 \Log_{\Exp_{m_x}^{\cY}\left[s\right]}^{\cY}y.
\end{equation}

\begin{lemma}
\label{lem:supp_uniform_density_pullback_regularity}
Assume \Cref{ass:uniform-inference-conditions} with its bandwidth conditions omitted. Use the cover, frames \(E_{x,\ell}\), and radius \(\delta_L\) from \Cref{lem:supp_uniform_finite_shrunken_cover}. For \(y\in\cK_\ell\), {the value and first derivative of \(s\mapsto\mathcal L_x(s,y)\) at \(s=0_{m_x}\), expressed in the frame \(E_{x,\ell}\), are}
\[
 L_{\ell,x}^{(0)}(y)\coloneqq E_{x,\ell}^\ast\Log_{m_x}^{\cY}y,\qquad
 L_{\ell,x}^{(1)}(y)\coloneqq -E_{x,\ell}^\ast\cH_{m_x}(y)E_{x,\ell}.
\]
{Equivalently, \(L_{\ell,x}^{(1)}(y)=E_{x,\ell}^\ast\bigl(\left.D_s\mathcal L_x(s,y)\right|_{s=0_{m_x}}\bigr)E_{x,\ell}\), with \(x,y\) held fixed under differentiation in \(s\).} In this lemma and its proof, use the Euclidean norm on \(\mathbb R^q\) and the Frobenius norm on \(\mathbb R^{q\times q}\), suppressing the corresponding norm subscripts. Then
\begin{equation}\label{eq:supp-transported-log-frame-bounds}
 \|L_{\ell,x}^{(r)}(y)\|\lesssim\cE_R(y),\qquad
 \|L_{\ell,x}^{(r)}(y)-L_{\ell,x'}^{(r)}(y)\|
 \lesssim d_{\cX}(x,x')\cE_R(y),
\end{equation}
uniformly over \(\ell\le L\), \(r=0,1\), \(x,x'\in\cX_\ell\), and
\(y\in\cK_\ell\). Moreover,
\begin{equation}\label{eq:supp-transported-log-predictor-regularity}
{\begin{aligned}
 \max_{\ell\le L}\sup_{x\in\cX_\ell}
 \left\|u\mapsto\E[L_{\ell,x}^{(0)}(Y)\mid X=\Exp_x^{\cX}\left[u\right]]
 \right\|_{C^{2,\gamma}(B_{T_x\cX}(0,\delta_L);\mathbb R^q)}&\lesssim1,\\
 \max_{\ell\le L}\sup_{x\in\cX_\ell}
 \left\|u\mapsto\E[L_{\ell,x}^{(1)}(Y)\mid X=\Exp_x^{\cX}\left[u\right]]
 \right\|_{C^2(B_{T_x\cX}(0,\delta_L);\mathbb R^{q\times q})}&\lesssim1.
\end{aligned}}
\end{equation}
\end{lemma}
\begin{proof}
{The domain inclusion in} \eqref{eq:supp-matched-response-domains}, {the bounds in} \Cref{lem:supp_fixed_local_log_domain}, and orthonormality give the first bound in \eqref{eq:supp-transported-log-frame-bounds}. The chain rule, differentiating both \(m_z\) and \(E_{z,\ell}\), gives
\begin{equation}\label{eq:supp-transported-log-basepoint-derivative}
 \|(dL_{\ell,\cdot}^{(r)}(y))_z[v]\|
 \lesssim\|v\|_z\cE_R(y),\qquad
 z\in\cN_\ell,\quad y\in\cK_\ell,\quad r=0,1.
\end{equation}
Integrating \eqref{eq:supp-transported-log-basepoint-derivative} along
the minimizing geodesic in \(\cN_\ell\) proves the second bound in
\eqref{eq:supp-transported-log-frame-bounds}.

{For \eqref{eq:supp-transported-log-predictor-regularity}, fix \(\ell,x\) and hold \(m_x,E_{x,\ell}\) fixed when differentiating the conditioning predictor. Apply \Cref{lem:supp_conditional_moment_regularity}~\textup{(ii) and (iii)} to the conditional moments of \(\Log_{m_x}^{\cY}Y\) for \(r=0\) and \(\cH_{m_x}(Y)\) for \(r=1\), obtaining the \(C^{2,\gamma}\) and \(C^2\) bounds, respectively. Equation \eqref{eq:supp-matched-predictor-radius} keeps the exponential images in \(\cN_\ell\). The fixed frame isometries preserve these bounds, which are uniform over the finite cover.}
\end{proof}

\section{Local expansions and linear representations}
\label{sec:supp_score_expansions_covariances}

{We derive linear representations for the local linear and RRBC estimators in terms of the mean-zero oracle score in \eqref{eq:supp_generic_oracle_score}. The population displacements are retained until the final corollary, where the US and RRBC bandwidth restrictions make them negligible.}

Unless otherwise stated, suprema over \(x\) are over \(\mathcal I_\nu\).
For \(\nu=\mathrm{unif}\), impose the geometric, density, kernel,
and moment conditions of \Cref{ass:uniform-inference-conditions},
together with \eqref{eq:supp-uniform-bandwidth-rate}
for \(a=h\) and, for RRBC, \(a=b\asymp h\).
Bias restrictions are used only when removing the population displacement.
The common conditions below refer
to the common pointwise conditions in \Cref{subsec:pointwise-assumptions}
or the preceding uniform conditions.
For RRBC, retain \(b\asymp h\) throughout this section.

\subsection{Regression identities, design, and weights}

In expectations and sums weighted by the kernel, set the integrand or summand to
zero before evaluating logarithms if \(K_{j,a}(X,x)=0\), or if
\(Y\notin\cK_{\cY,x}\) pointwise or \(Y\notin\cK_\ell\) for
\(x\in\cX_\ell\) uniformly. The response centers lie in
\(\overline{\cW_{m_x}}\) pointwise and
\(\overline{\cW_\ell^{\cY}}\) uniformly. See
\Cref{lem:supp_fixed_local_log_domain} and
\eqrefs{eq:supp-matched-response-domains,eq:supp-pointwise-response-support,eq:supp_uniform_response_support_event}.

For US, omit all terms involving
\((j,a)=(2,b)\). For \((j,a)=(1,h)\) and, {for RRBC,} \((j,a)=(2,b)\), define
\begin{equation}\label{eq:supp-regression-coefficients}
 C_{j,a,x}(\omega)\coloneqq{\E}[K_{j,a}(X,x)\Log_\omega^{\cY}Y\otimes r_{j,x}(U_x(X)/a)],\qquad \Theta_{j,a,x}(\omega)=C_{j,a,x}(\omega)S_{j,a}^{-1}(x).
\end{equation}
Both are linear maps from \(\cR_{j,x}\) to \(T_\omega\cY\). {At a fixed \(\omega\), define the population counterparts of \(\widehat A_{h,x}(\omega)\) and \(\widehat B_{h,b,x}(\omega)\) by}
\begin{equation+}\label{eq:main-population-response-quantities}
{\begin{aligned}
 A_{h,x}(\omega)&\coloneqq\Hess_\omega^{\cY}Q_{h,x}(\cdot)
 =\E[L_{1,h}(X;x)\cH_\omega(Y)],\\
 B_{h,b,x}(\omega)&=\frac{h^2}{2b^2}A_{h,x}(\omega)^{-1}
 \Theta_{2,b,x}(\omega)(0,0,M_{2,h}(x)).
\end{aligned}}
\end{equation+}
The empirical bias weight is
\begin{equation}\label{eq:supp-empirical-bias-weight}
 \widehat L_{\bias,h,b,i}(x)
 \coloneqq \frac{h^2}{2b^2}K_{2,b}(X_i,x)
 \langle\widehat S_{2,b}^{-1}r_{2,x}(U_x(X_i)/b),
                      (0,0,\widehat M_{2,h})\rangle.
\end{equation}
Define the population regression residual by
\begin{equation}\label{eq:supp-regression-residual}
 \varepsilon_{j,a,x}(\omega;W)
 \coloneqq \Log_\omega^{\cY}Y
       -\Theta_{j,a,x}(\omega)r_{j,x}(U_x(X)/a).
\end{equation}
Equations \eqrefs{eq:supp-local-polynomial-design-definition,eq:supp-regression-coefficients,eq:supp-regression-residual}
give, for every admissible center,
\begin{equation}\label{eq:supp-residual-normal-equations}
 {P[K_{j,a}(X,x)\varepsilon_{j,a,x}(\omega;W)\otimes r_{j,x}(U_x(X)/a)]}=C_{j,a,x}(\omega)-\Theta_{j,a,x}(\omega)S_{j,a}(x)=0.
\end{equation}

The population score equation at \(m_{h,x}\) gives
\[
 {P[L_{1,h}\Log_{m_{h,x}}^{\cY}Y]}
 =\Theta_{1,h,x}(m_{h,x})e_{1,x}=0.
\]
Define the oracle score by
\begin{equation}\label{eq:supp_generic_oracle_score}
 \varphi_{\star,n}(W;x)
 \coloneqq h^p\{L_{1,h}(X;x)\varepsilon_{1,h,x}(m_x;W)
 -\mathds{1}\{\star=\rbc\}L_{\bias,h,b}(X;x)
                 \varepsilon_{2,b,x}(m_x;W)\}.
\end{equation}
By \eqref{eq:supp-residual-normal-equations},
\({P\varphi_{\star,n}(\cdot;x)}=0\).

{Let \(\lambda_x\) denote the Lebesgue measure induced by \(g_{\cX,x}\), and \(J_x\) the normal-coordinate volume Jacobian. On a sufficiently small normal-coordinate ball whose image lies in \(\cV_x\), define}
\[
{\pi_{X,x}(u)\coloneqq p_X(\Exp_x^{\cX}u)J_x(u).}
\]
{The local predictor-density bounds and smooth positivity of \(J_x\) give}
\begin{equation}\label{eq:supp-design-chart-density-bounds}
{0<c\le\pi_{X,x}(u)\le C
\qquad\lambda_x\text{-a.e. near }0_x.}
\end{equation}
{Under the uniform predictor-density and neighborhood conditions in}
\Cref{ass:uniform-inference-conditions},
{the constants and coordinate radius are uniform over \(\cX_0\).}

\begin{lemma}
\label{lem:supp_kernel_change_population_design}
Fix \(j\in\{1,2\}\). Under the marginal density condition in
\Cref{ass:densities} and \Cref{ass:kernels}, for sufficiently small \(a\)
and bounded measurable \(g\) on the kernel support,
\begin{equation}\label{eq:pscnew_design_change_of_variables_new}
 \E[K_{j,a}(X,x)g(U_x(X)/a)]
 =\int K_j(\|t\|_x)g(t)\pi_{X,x}(at)\,d\lambda_x(t).
\end{equation}
Moreover, for some \(c_{S,j},C>0\),
\begin{equation}\label{eq:supp-population-design-bounds}
 2c_{S,j}I\preceq S_{j,a}(x)\preceq CI.
\end{equation}
Under the uniform
predictor density and neighborhood conditions of
\Cref{ass:uniform-inference-conditions}, the constants and bandwidth
threshold are uniform over \(\cX_0\).
\end{lemma}
\begin{proof}
For sufficiently small \(a\), the kernel support lies in the normal
neighborhood. Substitute \(u=at\) in the normal coordinate integral
to obtain \eqref{eq:pscnew_design_change_of_variables_new}.

{To bound \(S_{j,a}(x)\), define} \(G_{j,x}\coloneqq \int K_j(\|t\|_x)r_{j,x}(t)^{\otimes2}\,d\lambda_x(t)\). For \(0\ne v\in\cR_{j,x}\), the polynomial \(t\mapsto\langle v,r_{j,x}(t)\rangle\) is nonzero, so its zero set has \(\lambda_x\)-measure zero. By \Cref{ass:kernels},
\[
 \langle G_{j,x}v,v\rangle
 =\int K_j(\|t\|_x)\langle v,r_{j,x}(t)\rangle^2\,d\lambda_x(t)>0.
\]
Applying \eqref{eq:pscnew_design_change_of_variables_new} to
\(g(t)=\langle v,r_{j,x}(t)\rangle^2\) and using
\eqref{eq:supp-design-chart-density-bounds} yield
\[
 cG_{j,x}\preceq S_{j,a}(x)\preceq CG_{j,x}.
\]
Since \(G_{j,x}\succ0\), \eqref{eq:supp-population-design-bounds}
follows with \(2c_{S,j}=c\lambda_{\min}(G_{j,x})>0\),
after enlarging \(C\).

In orthonormal frames, the matrix of \(G_{j,x}\) is independent of \(x\).
The uniform density and predictor neighborhood conditions make the preceding comparison uniform.
\end{proof}

Under \Cref{ass:local_support_log_geometry,ass:densities,ass:kernels},
\eqrefs{eq:supp-standing-design-vector-bound,eq:supp-regression-coefficients,eq:supp-population-design-bounds}
give, for \(j\in\{1,2\}\) and sufficiently small \(a\), uniformly over
\(\omega\in\overline{\cW_{m_x}}\),
\begin{equation}\label{eq:supp-population-regression-coefficient-bounds}
\begin{aligned}
 \|C_{j,a,x}(\omega)\|_{\mathrm{op}}
  +\|\Theta_{j,a,x}(\omega)\|_{\mathrm{op}}
 &\le (1+\|S_{j,a}(x)^{-1}\|_{\mathrm{op}})
 \E\left[|K_{j,a}(X,x)|d_{\cY}(\omega,Y)
             \|r_{j,x}(U_x(X)/a)\|\right]\\
 &\lesssim a^{-p}\E\left[
 (1+d_{\cY}(m_x,Y))\mathds{1}\{d_{\cX}(X,x)\le R_ja\}\right]
 \lesssim1.
\end{aligned}
\end{equation}
Under \Cref{ass:uniform-inference-conditions} without its bandwidth conditions,
the bound and bandwidth threshold are uniform over \(\ell\),
\(x\in\cX_\ell\), and \(\omega\in\overline{\cW_\ell^{\cY}}\).

\begin{lemma}
\label{lem:supp_localized_support_count}
Fix \(R<\infty\) and let \(a\to0\). Under the pointwise predictor density
condition when \(\nu=\mathrm{pt}\), and under
the uniform predictor density, compactness, and neighborhood conditions of
\Cref{ass:uniform-inference-conditions} and
\(\log(1/a)/(na^p)\to0\) when \(\nu=\mathrm{unif}\),
\[
    \sup_{z\in\mathcal I_\nu}
    P_n\mathds{1}\{d_{\cX}(X,z)\le Ra\}
    =
    O_p(a^p).
\]
For \(a=h\), or \(a=b\) with \(b\asymp h\), the same bound holds when
each indicator is multiplied by \(\cE_R(Y_i)\) or \(G_R(Y_i)^2\)
under the common pointwise conditions. Uniformly it holds with
\(\cE_R(Y_i)^k\), \(k\in\{1,2\}\), under
\Cref{ass:uniform-inference-conditions}.
\end{lemma}

\begin{proof}
{For all sufficiently small \(a\), the local density and Jacobian bounds in normal coordinates, followed by \(u=at\), give, uniformly over \(z\in\mathcal I_\nu\),}
\[
 \mathbb P(d_{\cX}(X,z)\le Ra)
 =\int_{\|u\|_z\le Ra}
\pi_{X,z}(u)\,d\lambda_z(u)
 \lesssim \int_{\|u\|_z\le Ra}\!d\lambda_z(u)
 =a^p\int_{\|t\|_z\le R}d\lambda_z(t)
 \lesssim a^p.
\]

Pointwise, Markov's inequality and the preceding population bound
give the unweighted assertion. For the weighted assertions, take
\(a=h\) or \(a=b\asymp h\). For sufficiently small \(a\), the
predictor ball lies in \(\cV_x\).
Since \(\cE_R=G_R+J_R\) and \(G_R\le1+G_R^2\),
\eqref{eq:supp-pointwise-split-moments} and conditioning give
\[
 {\E\left[P_n\!\left[
 \mathds{1}\{d_{\cX}(X,x)\le Ra\}
 \{\cE_R(Y)+G_R(Y)^2\}
 \right]\right]}\lesssim \mathbb P(d_{\cX}(X,x)\le Ra)\lesssim a^p.
\]
Markov's inequality and nonnegativity give both weighted pointwise bounds.

{For \(\nu=\mathrm{unif}\), consider the class}
\[
    \left\{
        \mathds{1}\{d_{\cX}(\cdot,z)\le Ra\}:
        z\in\cX_0
    \right\}.
\]
{The indicator kernel is of bounded variation. Applying}
\Cref{lem:supp_local_polynomial_kernel_factors_are_localized}
{with \((M_a,H,m)=(1,1,0)\) on each fixed predictor
neighborhood, and taking the finite union, verifies}
\eqref{eq:supp-bounded-class-bounds}{.}
\Cref{lem:supp_localized_bracketing_maximal_inequality} {therefore gives}
\[
    \sup_{z\in\cX_0}
    \left|
        (P_n-P)\mathds{1}\{d_{\cX}(X,z)\le Ra\}
    \right|
    =O_p\left(
        \sqrt{a^p\log(1/a)/n}
        +\log(1/a)/n
    \right)=o_p(a^p).
\]
Adding the population bound gives the uniform bound for the local observation counts.
Uniformly, \(\cE_R\ge1\), so it suffices to treat \(k=2\).
Under uniform \textup{(C)},
\(\cE_R=1\) and the result for the local observation counts applies. {Under uniform \textup{(H)}, apply}
\eqref{eq:supp-localized-finite-moment-rate}
{with \((M_a,H,m)=(1,\cE_R,2)\) to the restrictions of the class}
\[
{\left\{(z,y)\mapsto\cE_R(y)^2
 \mathds{1}\{d_{\cX}(z,x)\le Ra\}:x\in\cX_0\right\}}
\]
{to the fixed predictor neighborhoods. The indicator kernel is of
bounded variation, and the response factor is independent of the center.
Taking the maximum over the fixed finite cover gives}
\[
{\sup_{x\in\cX_0}\left|(P_n-P)\left[
 \cE_R^2\mathds{1}\{d_{\cX}(X,x)\le Ra\}\right]\right|
 =O_p(a^pc_n(a))=o_p(a^p).}
\]
{The conditional second-moment and small-ball bounds give}
\[
{\sup_{x\in\cX_0}P\left[
 \cE_R^2\mathds{1}\{d_{\cX}(X,x)\le Ra\}\right]\lesssim a^p.}
\]
{Adding this population bound proves the weighted assertion.
The same argument applies to \(b\asymp h\).}
\end{proof}
\begin{lemma}
\label{lem:supp_empirical_design_concentration}
Under the predictor density and kernel conditions of
\Cref{lem:supp_kernel_change_population_design}, fix
\(j\in\{1,2\}\), \(\nu\in\{\mathrm{pt},\mathrm{unif}\}\), and
\(a\to0\) with \(\delta_{n,\nu}(a)\to0\).
For \(\nu=\mathrm{unif}\), use the uniform predictor conditions of that
lemma and the bounded variation condition in
\Cref{ass:uniform-inference-conditions}. Then
\begin{equation}
\label{eq:supp_empirical_design_rates}
{\sup_{z\in\mathcal I_\nu}
\left\{
\|\widehat S_{j,a}(z)-S_{j,a}(z)\|_{\mathrm{op}}
+\|\widehat S_{j,a}(z)^{-1}-S_{j,a}(z)^{-1}\|_{\mathrm{op}}
\right\}}
=O_p(\delta_{n,\nu}(a)).
\end{equation}
In the uniform regime, with \(c_{S,j}>0\) from
\Cref{lem:supp_kernel_change_population_design},
\(\mathbb P(\inf_{x\in\cX_0}\lambda_{\min}(\widehat S_{j,a}(x))
\ge c_{S,j})\to1\).
{Consequently, for \((j,a)=(1,h)\), the local linear weights satisfy}
\begin{equation}
\label{eq:supp_raw_weight_L1_bounds}
{
    \sup_{z\in\mathcal I_\nu}P_n[|\wh L_{1,h}-L_{1,h}|]
    =O_p(\delta_{n,\nu}(h)),\qquad
    \sup_{z\in\mathcal I_\nu}P_n[|\wh L_{1,h}|]
    =O_p(1).
}
\end{equation}
\end{lemma}

\begin{proof}
{Choose a finite cover of \(\mathcal I_\nu\) by predictor neighborhoods with smooth orthonormal frames.}
On each neighborhood \(V\), let \(r^V_{j,z,k}(t)\) denote the
\(k\)th coordinate of \(r_{j,z}(t)\). For
\(w\in\cX\), set
\begin{gather}
f^V_{z,kk'}(w)
\coloneqq K_{j,a}(w,z)
r^V_{j,z,k}(U_z(w)/a)r^V_{j,z,k'}(U_z(w)/a),\\
[\widehat S_{j,a}(z)-S_{j,a}(z)]^V_{kk'}
=(P_n-P)f^V_{z,kk'},\qquad
|f^V_{z,kk'}(w)|
\lesssim a^{-p}\mathds{1}\{d_{\cX}(w,z)\le R_ja\}.
\end{gather}
At a fixed \(z\), independence and the marginal density bound give
\[
\begin{aligned}
 {\E}\bigl[|(P_n-P)f^V_{z,kk'}|^2\bigr]
 &=n^{-1}\operatorname{Var}(f^V_{z,kk'}(X))
 \le n^{-1}{P[(f^V_{z,kk'})^2]}\\
 &\lesssim n^{-1}a^{-2p}
       \mathbb P(d_{\cX}(X,z)\le R_ja)
 \lesssim (na^p)^{-1}.
\end{aligned}
\]
Chebyshev's inequality therefore gives
\((P_n-P)f^V_{z,kk'}=O_p(\delta_{n,\mathrm{pt}}(a))\).

{On each fixed predictor neighborhood, the coordinate products are bounded polynomial factors.}
\Cref{lem:supp_local_polynomial_kernel_factors_are_localized}
{with \((M_a,H,m)=(a^{-p},1,0)\) verifies}
\eqref{eq:supp-bounded-class-bounds} {for each coordinate class.
Applying} \Cref{lem:supp_localized_bracketing_maximal_inequality} {gives}
\[
 \sup_{z\in\cX_0\cap V}|(P_n-P)f^V_{z,kk'}|
 =O_p\!\left(a^{-p}\sqrt{a^p\log(1/a)/n}\right)
 =O_p(\delta_{n,\mathrm{unif}}(a)).
\]
Finitely many neighborhoods with fixed frames and finitely many coordinates, together with norm
equivalence, give
\[
\begin{aligned}
 \sup_{z\in\mathcal I_\nu}
 \|\widehat S_{j,a}(z)-S_{j,a}(z)\|_{\mathrm{op}}
 \lesssim \max_V\max_{k,k'}\sup_{z\in\mathcal I_\nu\cap V}
       |(P_n-P)f^V_{z,kk'}|
 =O_p(\delta_{n,\nu}(a))
 =o_p(1).
\end{aligned}
\]

By \eqref{eq:supp-population-design-bounds},
\(S_{j,a}(z)\succeq2c_{S,j}I\).
The preceding concentration bound and
\Cref{lem:supp_covariance_transfer_spectral_stability}
give \eqref{eq:supp_empirical_design_rates} and
\(\mathbb P(\inf_{z\in\mathcal I_\nu}
\lambda_{\min}(\widehat S_{j,a}(z))\ge c_{S,j})\to1\).

{For \((j,a)=(1,h)\), the weight definitions in}
\eqref{eq:main-population-local-linear-weight}{, the kernel bound, and}
\eqref{eq:supp-standing-design-vector-bound} {give}
\begin{equation}
\label{eq:supp-local-linear-weight-replacement-envelope}
{
h^p|\wh L_{1,h,i}(z)-L_{1,h}(X_i;z)|
\lesssim \|\wh S_{1,h}(z)^{-1}-S_{1,h}(z)^{-1}\|_{\mathrm{op}}
\mathds{1}\{d_{\cX}(X_i,z)\le R_1h\}
}
\end{equation}
{and}
\[
{
h^p|\wh L_{1,h,i}(z)|
\lesssim \|\wh S_{1,h}(z)^{-1}\|_{\mathrm{op}}
\mathds{1}\{d_{\cX}(X_i,z)\le R_1h\}.
}
\]
{By} \eqref{eq:supp-population-design-bounds} {and}
\eqref{eq:supp_empirical_design_rates}{,}
{\(\sup_{z\in\mathcal I_\nu}\|\wh S_{1,h}(z)^{-1}\|_{\mathrm{op}}=O_p(1)\).
Averaging the two envelopes and applying the unweighted local observation
count bound in} \Cref{lem:supp_localized_support_count}
{and the inverse rates in} \eqref{eq:supp_empirical_design_rates}
{proves} \eqref{eq:supp_raw_weight_L1_bounds}{.}
\end{proof}

\begin{proposition}
\label{prop:supp_selector_weight_design_moment_master}
Suppose the marginal predictor density condition in \Cref{ass:densities} and
\Cref{ass:kernels} hold. Fix \(j\in\{1,2\}\) and a sufficiently
small bandwidth \(a\), and let \(v(x),\wh v(x)\in\cR_{j,x}\), with
\(\wh v(x)\) possibly random.

\noindent\textup{(i) Exact reproduction.}
The following identities hold whenever the corresponding design
operator is nonsingular:
\begin{align}
{P\left[K_{j,a}(X,x)
\left\langle S_{j,a}(x)^{-1}r_{j,x}(U_x(X)/a),v(x)\right\rangle_{\cR_{j,x}}
r_{j,x}(U_x(X)/a)\right]}&=v(x),\label{eq:supp-population-polynomial-reproduction}\\
{P_n\left[K_{j,a}(X,x)
\left\langle\widehat S_{j,a}(x)^{-1}r_{j,x}(U_x(X)/a),\widehat v(x)\right\rangle_{\cR_{j,x}}
r_{j,x}(U_x(X)/a)\right]}
&=\widehat v(x).\label{eq:supp-empirical-polynomial-reproduction}
\end{align}

\noindent\textup{(ii) Localized bounds.}
\begin{gather}\label{eq:supp-local-polynomial-weight-envelopes}
\begin{aligned}
&\left|K_{j,a}(z,x)
\left\langle S_{j,a}(x)^{-1}r_{j,x}(U_x(z)/a),v(x)\right\rangle_{\cR_{j,x}}\right|
\lesssim a^{-p}\|v(x)\|_{\cR_{j,x}}
\mathds{1}\{d_{\cX}(z,x)\le R_ja\},\\
&\left|K_{j,a}(X_i,x)
\left\langle\widehat S_{j,a}(x)^{-1}r_{j,x}(U_x(X_i)/a),\widehat v(x)\right\rangle_{\cR_{j,x}}\right|\lesssim a^{-p}\|\widehat S_{j,a}(x)^{-1}\|_{\mathrm{op}}
\|\widehat v(x)\|_{\cR_{j,x}}
\mathds{1}\{d_{\cX}(X_i,x)\le R_ja\}.
\end{aligned}
\end{gather}
Consequently, for \(k\in\{0,1,2\}\),
\begin{equation}\label{eq:supp-local-polynomial-weight-moments}
\E\left[\left|K_{j,a}(X,x)
\left\langle S_{j,a}(x)^{-1}r_{j,x}(U_x(X)/a),v(x)\right\rangle_{\cR_{j,x}}\right|
\|U_x(X)\|_x^k\right]\lesssim a^k\|v(x)\|_{\cR_{j,x}}.
\end{equation}
Under the uniform design conditions, these bounds hold uniformly over \(\cX_0\).
\end{proposition}

\begin{proof}
Self-adjointness of \(S_{j,a}(x)^{-1}\) gives, for \(t\in T_x\cX\),
\[
 \langle S_{j,a}(x)^{-1}r_{j,x}(t),v(x)\rangle r_{j,x}(t)
 =\langle r_{j,x}(t),S_{j,a}(x)^{-1}v(x)\rangle r_{j,x}(t)
 =r_{j,x}(t)^{\otimes2}S_{j,a}(x)^{-1}v(x).
\]
By \eqref{eq:supp-local-polynomial-design-definition}, the population expression equals
\[
{P\left[K_{j,a}(X,x)\{r_{j,x}(U_x(X)/a)\otimes
r_{j,x}(U_x(X)/a)\}\right]}S_{j,a}(x)^{-1}v(x)
=S_{j,a}(x)S_{j,a}(x)^{-1}v(x)=v(x).
\]
Replacing \(\E\) by \(P_n\) in \eqref{eq:supp-local-polynomial-design-definition}
gives \eqref{eq:supp-empirical-polynomial-reproduction}, since
\(\widehat S_{j,a}(x)\widehat S_{j,a}(x)^{-1}\widehat v(x)=\widehat v(x)\).

For \(\|t\|_x\le R_j\), Cauchy--Schwarz and
\eqrefs{eq:supp-standing-design-vector-bound,eq:supp-population-design-bounds} give
\[
 \bigl|\langle S_{j,a}(x)^{-1}r_{j,x}(t),v(x)\rangle\bigr|
 \le\|S_{j,a}(x)^{-1}\|_{\mathrm{op}}
       \|r_{j,x}(t)\|\|v(x)\|
 \lesssim \|v(x)\|.
\]
The same argument gives
\[
 \bigl|\langle\widehat S_{j,a}(x)^{-1}r_{j,x}(t),\widehat v(x)\rangle\bigr|
 \lesssim \|\widehat S_{j,a}(x)^{-1}\|_{\mathrm{op}}\|\widehat v(x)\|.
\]
Together with
\(\lvert K_{j,a}(z,x)\rvert
\lesssim a^{-p}\mathds{1}\{d_{\cX}(z,x)\le R_ja\}\) from
\Cref{ass:kernels}, these bounds prove
\eqref{eq:supp-local-polynomial-weight-envelopes}.

For \(k\in\{0,1,2\}\),
\eqref{eq:supp-local-polynomial-weight-envelopes} and the marginal density bound yield
\[
\begin{aligned}
 \E\!\left[\left|K_{j,a}(X,x)
 \left\langle S_{j,a}(x)^{-1}r_{j,x}(U_x(X)/a),v(x)\right\rangle\right|
 \|U_x(X)\|_x^k\right]
 &\lesssim a^{-p}\|v(x)\|
 \E\!\left[\mathds{1}\{d_{\cX}(X,x)\le R_ja\}\|U_x(X)\|_x^k\right]\\
 &\lesssim a^{-p}\|v(x)\|(R_ja)^k
 \mathbb P(d_{\cX}(X,x)\le R_ja)\\
 &\lesssim a^k\|v(x)\|.
\end{aligned}
\]
Under the uniform design conditions, the bounds on the inverses and on the probabilities of small balls,
including their bandwidth thresholds, hold uniformly over \(\cX_0\).
\end{proof}

\begin{lemma}
\label{lem:supp-pointwise-local-truncation}
Fix \(x\), \(0<R_\ast<\infty\), and \(h\to0\) with \(nh^p\to\infty\).
Assume \(\mathbb P(d_{\cX}(X,x)\le R_\ast h)\lesssim h^p\) and, for some
neighborhood \(\cV_x\) of \(x\) and measurable \(G\ge0\),
\(\sup_{z\in\cV_x}\E[G(Y)\mathds{1}\{G(Y)>T\}\mid X=z]\to0\)
as \(T\to\infty\). For any deterministic measurable \(f_h\) valued in a
fixed finite-dimensional Hilbert space with
\(\|f_h(X,Y)\|\lesssim \mathds{1}\{d_{\cX}(X,x)\le R_\ast h\}G(Y)\),
\begin{equation}
 h^{-p}(P_n-P)f_h=o_p(1),\qquad
 (nh^p)^{-1}\max_i\mathds{1}\{d_{\cX}(X_i,x)\le R_\ast h\}G(Y_i)=o_p(1).
\end{equation}
\end{lemma}
\begin{proof}
For small \(h\), the support lies in \(\cV_x\). For fixed \(T>0\),
independence and the envelope bound give
\[
\begin{aligned}
 {\E\left[\left\|h^{-p}(P_n-P)
 [f_h\mathds{1}\{G(Y)\le T\}]\right\|^2\right]}
 &\lesssim T^2(nh^{2p})^{-1}
 \mathbb P(d_{\cX}(X,x)\le R_\ast h)
 \lesssim T^2(nh^p)^{-1},\\
 {\E\left[\left\|h^{-p}(P_n-P)
 [f_h\mathds{1}\{G(Y)>T\}]\right\|\right]}
 &\lesssim h^{-p}{\E}\left[
 \mathds{1}\{d_{\cX}(X,x)\le R_\ast h\}G(Y)\mathds{1}\{G(Y)>T\}\right]\\
 &\lesssim \sup_{z\in\cV_x}
 {\E}[G(Y)\mathds{1}\{G(Y)>T\}\mid X=z].
\end{aligned}
\]
Chebyshev's and Markov's inequalities, first as \(n\to\infty\) and
then as \(T\to\infty\), prove the first assertion. For \(\varepsilon>0\),
the union bound gives
\[
\begin{aligned}
 \mathbb P\!\left(\max_i
 \mathds{1}\{d_{\cX}(X_i,x)\le R_\ast h\}G(Y_i)>\varepsilon nh^p\right)
 &\le n\mathbb P\left(
 d_{\cX}(X,x)\le R_\ast h,\ G(Y)>\varepsilon nh^p\right)\\
 &\lesssim \varepsilon^{-1}\sup_{z\in\cV_x}
 {\E}[G(Y)\mathds{1}\{G(Y)>\varepsilon nh^p\}\mid X=z]\to 0.
\end{aligned}
\]
\end{proof}

\begin{lemma}
\label{lem:supp_local_linear_second_moment}
Let \(h\to0\).

\noindent\textup{(i) Population bound.}
Under the conditions of \Cref{lem:supp_kernel_change_population_design}
with \(j=1\) and \(a=h\), \(\|M_{2,h}(x)\|_{\mathrm{op}}\lesssim 1\) for
sufficiently small \(h\), uniformly over \(\cX_0\) under the uniform
predictor conditions of that lemma.

\noindent\textup{(ii) Empirical concentration.}
Under the conditions of \Cref{lem:supp_empirical_design_concentration}
with \(j=1\), \(a=h\), and \(\nu\in\{\mathrm{pt},\mathrm{unif}\}\),
\begin{equation}
    \sup_{z\in\mathcal I_\nu}
    \|\wh M_{2,h}(z)-M_{2,h}(z)\|_{\mathrm{op}}
    =O_p(\delta_{n,\nu}(h)).
\end{equation}
\end{lemma}

\begin{proof}
For \textup{(i)}, \eqref{eq:supp-local-polynomial-weight-moments} with
\((j,a,k,v)=(1,h,2,e_{1,x})\) gives
\[
 \|M_{2,h}(x)\|_{\mathrm{op}}
 \le h^{-2}{\E}\left[|L_{1,h}(X;x)|\|U_x(X)\|_x^2\right]
 \lesssim 1.
\]

For \textup{(ii)}, at fixed \(x\), independence, finite-dimensional norm equivalence,
\eqref{eq:supp-local-polynomial-weight-envelopes}, and the marginal
density bound give
\[
\begin{aligned}
 {\E\left[\left\|(P_n-P)\left[
 L_{1,h}(X;x)(U_x(X)/h)^{\otimes2}\right]\right\|_{\mathrm{op}}^2\right]}
 &\lesssim n^{-1}{\E}\left[
 |L_{1,h}(X;x)|^2\|U_x(X)/h\|_x^4\right]\\
 &\lesssim (nh^{2p})^{-1}
 \mathbb P(d_{\cX}(X,x)\le R_1h)
 \lesssim (nh^p)^{-1}.
\end{aligned}
\]
Chebyshev's inequality gives \(O_p(\delta_{n,\mathrm{pt}}(h))\).

For the uniform bound, the coordinates of
\(S_{1,h}(x)^{-1}e_{1,x}\) in the fixed orthonormal frames
are deterministic and uniformly bounded by
\Cref{lem:supp_kernel_change_population_design},
so they factor out of \(P_n-P\).
{For the remaining kernel-weighted polynomial coordinate classes,}
\Cref{lem:supp_local_polynomial_kernel_factors_are_localized}
{with \((a,M_a,H,m)=(h,h^{-p},1,0)\) verifies}
\eqref{eq:supp-bounded-class-bounds}{.
Applying} \Cref{lem:supp_localized_bracketing_maximal_inequality}
{coordinatewise and using finite-dimensional norm equivalence gives}
\[
 \sup_{x\in\cX_0}\left\|(P_n-P)\left[
 L_{1,h}(X;x)(U_x(X)/h)^{\otimes2}\right]\right\|_{\mathrm{op}}
 =O_p\left(h^{-p}\sqrt{h^p\log(1/h)/n}\right)
 =O_p(\delta_{n,\mathrm{unif}}(h)).
\]
Finally, the definitions {of \(\widehat M_{2,h}\) in \eqref{eq:main-empirical-local-linear-second-moment} and \(M_{2,h}\) in \eqref{eq:main-local-linear-second-moment}}, the bound
\(\|U_x(X_i)/h\|_x\le R_1\) on the kernel support, and
\Cref{lem:supp_empirical_design_concentration} yield
\[
\begin{aligned}
 &\sup_{x\in\mathcal I_\nu}
 \|\wh M_{2,h}(x)-M_{2,h}(x)\|_{\mathrm{op}}\\
 &\quad\le R_1^2
 \sup_{x\in\mathcal I_\nu}{P_n[|\wh L_{1,h}-L_{1,h}|]}
 +\sup_{x\in\mathcal I_\nu}
 \left\|(P_n-P)\left[
 L_{1,h}(X;x)(U_x(X)/h)^{\otimes2}\right]\right\|_{\mathrm{op}}=O_p(\delta_{n,\nu}(h)).
\end{aligned}
\]
\end{proof}

{For the population increment bounds below, assume}
\Cref{ass:uniform-inference-conditions}
{without its bandwidth conditions.
Fix \(\ell\le L\) in} \Cref{lem:supp_uniform_finite_shrunken_cover}
{and compare tensors in a fixed smooth orthonormal predictor frame on a neighborhood
of \(\cX_\ell\).}

{For sufficiently small \(h>0\) and \(x,x'\in\cX_\ell\) with
\(d_{\cX}(x,x')\le h\), there is a nonnegative measurable \(w_1\) such that}
\begin{equation}
\label{eq:supp_population_weight_basepoint_modulus}
{h^p|L_{1,h}(z;x)-L_{1,h}(z;x')|\le w_1(z),}
\end{equation}
{and}
\begin{equation}
\label{eq:supp_population_second_moment_basepoint_modulus}
{\|M_{2,h}(x)-M_{2,h}(x')\|_{\mathrm{op}}
\lesssim h^{-1}d_{\cX}(x,x').}
\end{equation}
{For RRBC, let \(h,b>0\) be sufficiently small with
\(0<\underline\tau\le h/b\le\overline\tau<\infty\);
for \(x,x'\in\cX_\ell\) with \(d_{\cX}(x,x')\le\min\{h,b\}\),
there is a nonnegative measurable \(w_b\) such that}
\begin{equation}
\label{eq:supp_population_bias_weight_basepoint_modulus}
{h^p|L_{\bias,h,b}(z;x)-L_{\bias,h,b}(z;x')|\le w_b(z).}
\end{equation}
{Choose the envelopes to vanish outside the unions of their corresponding
kernel supports at \(x,x'\), which lie in \(\cO_{\cX}\) for sufficiently small
bandwidths. Each envelope \(w\), on its stated distance range, satisfies}
\[
{\E[w(X)\cE_R(Y)]+\E[w(X)^2\cE_R(Y)^2]
\lesssim h^{p-1}d_{\cX}(x,x').}
\]
{Since \(\cE_R\ge1\), this also bounds the unweighted first and second
moments. All constants are uniform over \(\ell\le L\).}

{For \((j,a)=(1,h)\) and, for RRBC, \((2,b)\), with
\(d_{\cX}(x,x')\le a\), apply the first-moment oscillation bound in}
\Cref{lem:supp_local_polynomial_kernel_factors_are_localized}
{to the entries of
\(K_{j,a}(z,x)r_{j,x}(U_x(z)/a)^{\otimes2}\), with
\((M_a,H,m,\delta)=(a^{-p},1,0,a^{-1}d_{\cX}(x,x'))\).
The lower bound in} \Cref{lem:supp_kernel_change_population_design}
{and the inverse perturbation bound in}
\Cref{lem:supp_covariance_transfer_spectral_stability} {give}
\begin{equation}
\label{eq:supp-population-design-inverse-modulus}
{\|S_{j,a}(x)-S_{j,a}(x')\|_{\mathrm{op}}
+\|S_{j,a}(x)^{-1}-S_{j,a}(x')^{-1}\|_{\mathrm{op}}
\lesssim a^{-1}d_{\cX}(x,x').}
\end{equation}
{The weight formula in} \eqref{eq:main-population-local-linear-weight}{,}
\Cref{prop:supp_selector_weight_design_moment_master}~\textup{(ii)}{,}
\eqref{eq:supp-population-design-inverse-modulus}{, and the kernel oscillation
bounds give} \eqref{eq:supp_population_weight_basepoint_modulus}
{and the unweighted envelope moments for \(w_1\).
Apply the first-moment oscillation bound in}
\Cref{lem:supp_local_polynomial_kernel_factors_are_localized}
{to the entries of \(z\mapsto L_{1,h}(z;x)(U_x(z)/h)^{\otimes2}\),
with \((a,M_a,H,m,\delta)=(h,h^{-p},1,0,h^{-1}d_{\cX}(x,x'))\).}
Equation \eqref{eq:main-local-linear-second-moment}
{and Jensen's inequality give}
\eqref{eq:supp_population_second_moment_basepoint_modulus}{.
The bias-weight formula in} \eqref{eq:supp-population-bias-weight}{,}
\eqrefs{eq:supp_population_second_moment_basepoint_modulus,eq:supp-population-design-inverse-modulus}{,}
\Cref{lem:supp_local_linear_second_moment}~\textup{(i)}{, and the kernel oscillation
bounds, with \(h\asymp b\), give}
\eqref{eq:supp_population_bias_weight_basepoint_modulus}
{and the unweighted envelope moments for \(w_b\).
Conditioning on \(X\) and using the uniform conditional second-moment bound
for \(\cE_R(Y)\) gives the displayed weighted moment bound.}

{For \(u\in T_x\cX\) near zero, write the conditional score in predictor normal coordinates as}
\[
    \psi_x(u)\coloneqq  s_x(\Exp_x^{\cX}\left[u\right])
    =
    -\int_{\cK_{\cY,x}}
 \Log_{m_x}^{\cY}y\,
 p_{Y\mid X}(y\mid \Exp_x^{\cX}\left[u\right])
    \,\vol_{\cY}(dy),
\]
whenever \(\Exp_x^{\cX}\left[u\right]\) belongs to the current predictor neighborhood.

\begin{lemma}
\label{lem:supp_lq_bias_weight_moment_bounds}
Let \(h,b\to0\) with \(b\asymp h\).
Suppose \Cref{ass:local_support_log_geometry,ass:densities,ass:kernels}
hold. For \(\nu=\mathrm{unif}\), impose
\Cref{ass:uniform-inference-conditions} without its bandwidth conditions.
Then
\begin{align}
{\sup_{x\in\mathcal I_\nu}
 \left\|\E[L_{\bias,h,b}\Log_{m_x}Y]-h^2A_x\beta_h(x)\right\|}
 &{=\begin{cases}
 o(h^2),&\nu=\mathrm{pt},\\
 O(h^2b^\gamma),&\nu=\mathrm{unif},
 \end{cases}}\label{eq:supp-bias-weight-population-expansion}\\
{\sup_{x\in\mathcal I_\nu}
 \left\|\E[L_{\bias,h,b}\cH_{m_x}(Y)]\right\|_{\mathrm{op}}}
 &{=O(h^2).}\label{eq:supp-bias-weight-population-hessian}
\end{align}
\end{lemma}

\begin{proof}
{With \((j,a)=(2,b)\) and \(v(x)=\frac{h^2}{2b^2}(0,0,M_{2,h}(x))\), \eqrefs{eq:supp-population-bias-weight,eq:main-local-linear-second-moment,eq:supp-population-polynomial-reproduction} give}
\begin{equation}\label{eq:supp-bias-weight-reproduction}
{\E[L_{\bias,h,b}\,(1,U_x(X))]=0,\quad \E[L_{\bias,h,b}U_x(X)^{\otimes2}]=h^2M_{2,h}(x)=\E[L_{1,h}U_x(X)^{\otimes2}].}
\end{equation}
{Equation \eqref{eq:supp-local-polynomial-weight-moments} with \(k=2\) and \Cref{lem:supp_local_linear_second_moment}~\textup{(i)} give}
\[
{\sup_{x\in\mathcal I_\nu}\E[|L_{\bias,h,b}|\|U_x(X)\|_x^2]\lesssim h^2\sup_{x\in\mathcal I_\nu}\|M_{2,h}(x)\|_{\mathrm{op}}=O(h^2).}
\]

{With \(m_x\) and its response frame fixed under predictor differentiation, set}
\[
{M_{H,x}(u)\coloneqq\E[\cH_{m_x}(Y)\mid X=\Exp_x^{\cX}[u]].}
\]
{\Cref{lem:supp_conditional_moment_regularity}~\textup{(ii)} and \eqref{eq:supp-transported-log-predictor-regularity} give \(C^2\) regularity pointwise and common \(C^{2,\gamma}\) and \(C^2\) bounds for \(\psi_x\) and \(M_{H,x}\), respectively, in the uniform regime. Since \(D^2\psi_x(0)=\nabla_{\cX}^2s_x(x)\), \eqrefs{eq:supp-bias-weight-reproduction,eq:explicit-bias-expansion} yield}
\[
{-\tfrac12\E[L_{\bias,h,b}D^2\psi_x(0)[U_x(X),U_x(X)]]=-\tfrac12\E[L_{1,h}D^2\psi_x(0)[U_x(X),U_x(X)]]=h^2A_x\beta_h(x).}
\]
{Condition on \(X\). The moment identities and the second-order integral Taylor remainder then give}
\[
{\begin{aligned}
\sup_{x\in\mathcal I_\nu}\left\|\E[L_{\bias,h,b}\Log_{m_x}^{\cY}Y]-h^2A_x\beta_h(x)\right\|&\lesssim h^2\sup_{x\in\mathcal I_\nu}\sup_{\|u\|_x\le R_2b}\|D^2\psi_x(u)-D^2\psi_x(0)\|_{\mathrm{op}}\\
&=\begin{cases}o(h^2),&\nu=\mathrm{pt},\\O(h^2b^\gamma),&\nu=\mathrm{unif},\end{cases}
\end{aligned}}
\]
{proving \eqref{eq:supp-bias-weight-population-expansion}. For the Hessian, Taylor's formula and \eqref{eq:supp-bias-weight-reproduction} give}
\[
{\begin{aligned}
\left\|\E[L_{\bias,h,b}\cH_{m_x}(Y)]\right\|_{\mathrm{op}}&=\left\|\E[L_{\bias,h,b}\{M_{H,x}(U_x(X))-M_{H,x}(0)-DM_{H,x}(0)[U_x(X)]\}]\right\|_{\mathrm{op}}\\
&\lesssim\E[|L_{\bias,h,b}|\|U_x(X)\|_x^2]=O(h^2).
\end{aligned}}
\]
{This proves \eqref{eq:supp-bias-weight-population-hessian}.}
\end{proof}

\subsection{\texorpdfstring{{Population score expansion and smoothing displacement}}{Population score expansion and smoothing displacement}}

{To expand the smoothing displacement \(B_h(x)\), we write the population score equation in the fixed tangent space \(T_{m_x}\cY\).} For sufficiently small \(s\in T_{m_x}\cY\), define
\[
 \widetilde\Phi_{h,x}(s)
 \coloneqq\grad_s\{Q_{h,x}(\Exp_{m_x}^{\cY}\left[s\right])\}.
\]
Here \(\grad_s\) is the gradient on the fixed linear space
\(T_{m_x}\cY\) with its inner product at \(m_x\). {We hold \(x\) fixed under differentiation in \(s\), and write \((D_s^j\widetilde\Phi_{h,x})(s_0)\) for the \(j\)th derivative evaluated at \(s=s_0\).}
The chain rule gives
\[
 \widetilde\Phi_{h,x}(s)
 =(d\Exp_{m_x}^{\cY})_s^\ast
   \left.{\grad_\omega^{\cY}Q_{h,x}(\cdot)}
   \right|_{\omega=\Exp_{m_x}^{\cY}\left[s\right]},
\]
where the adjoint uses the Riemannian inner products at
\(m_x\) and \(\Exp_{m_x}^{\cY}\left[s\right]\).

At the origin, \((d\Exp_{m_x}^{\cY})_0=I\) and normal coordinates give
\[
 \widetilde\Phi_{h,x}(0)
 =\E[L_{1,h}(X;x)\psi_x(U_x(X))],\qquad
 {\left.D_s\widetilde\Phi_{h,x}(s)\right|_{s=0_{m_x}}}=A_{h,x}(m_x).
\]
The derivative identity uses the vanishing response Christoffel symbols
at the origin.

By \Cref{prop:supp_population_smoothed_target_consistency,lem:supp_fixed_local_log_domain},
\(B_h(x)=\Log_{m_x}^{\cY}m_{h,x}\) and the segment \(tB_h(x)\),
\(0\le t\le1\), lie in the smooth pullback domain for small \(h\).
This holds uniformly under the uniform conditions by
\eqref{eq:supp_uniform_population_argmin_localization} and
{\Cref{lem:supp_uniform_finite_shrunken_cover}~\textup{(ii)}}.
The first-order condition at \(m_{h,x}\) and Taylor's formula
in integral form give
\begin{equation}
\label{eq:supp_smoothing_first_order_linearization}
 0=\widetilde\Phi_{h,x}(B_h(x))
 =\widetilde\Phi_{h,x}(0)+\bar A_{h,x}B_h(x),
\end{equation}
where
\begin{equation}
\label{eq:supp_mean_value_hessian_smoothing_path}
 \bar A_{h,x}
 \coloneqq
 \int_0^1{\left.D_s\widetilde\Phi_{h,x}(s)\right|_{s=tB_h(x)}}\,dt
 \in\mathcal L(T_{m_x}\cY,T_{m_x}\cY).
\end{equation}

\begin{lemma}
\label{lem:supp_conditional_score_smoothness}
\noindent\textup{(i) Conditional score.} Under \Cref{ass:local_support_log_geometry,ass:densities},
\(\psi_x\) is \(C^2\) near zero and \(\psi_x(0)=0\). {In normal coordinates, \(D^2\psi_x(0)=\nabla_{\cX}^2s_x(x)\).}

Under \Cref{ass:uniform-inference-conditions} without its bandwidth
conditions,
\begin{equation}\label{eq:supp-uniform-conditional-score-regularity}
 \sup_{x\in\cX_0}
 \|\psi_x\|_{C^{2,\gamma}(B_{T_x\cX}(0,\delta_L);T_{m_x}\cY)}\lesssim 1.
\end{equation}

\noindent\textup{(ii) Hessian comparisons.} Suppose
\Cref{ass:local_support_log_geometry,ass:densities,ass:kernels}
hold, and let \(h\to0\).
For all sufficiently small \(h\),
\[
    \|A_{h,x}(m_x)-A_x\|_{\mathrm{op}}
    =
    O(h^2),
    \qquad
    \|\bar A_{h,x} -A_{h,x}(m_x)\|_{\mathrm{op}}
    \lesssim \|B_h(x)\|_{m_x}.
\]
Under \Cref{ass:uniform-inference-conditions} without its bandwidth
conditions, both bounds hold uniformly over \(\cX_0\).
\end{lemma}

\begin{proof}
\noindent\textup{(i).} By
\Cref{lem:supp_conditional_moment_regularity}~\textup{(ii)} at the fixed
center \(m_x\), \(\psi_x\) is \(C^2\) near zero.

The first-order condition gives \(\psi_x(0)=s_x(x)=0\). Since
\((d\Exp_x^{\cX})_0=\operatorname{Id}_{T_x\cX}\) and the codomain
\(T_{m_x}\cY\) carries the trivial connection, the chain rule gives
\(D\psi_x(0)=(ds_x)_x\). In \(x\)-centered normal coordinates,
the Christoffel symbols of \(\cX\) vanish at the origin, so the ordinary
second derivative there equals the second covariant derivative. Hence
\(D^2\psi_x(0)=\nabla_{\cX}^2s_x(x)\).

Taylor's formula with integral remainder gives
\[
 \psi_x(u)={D\psi_x(0)[u]}+\tfrac12{D^2\psi_x(0)[u,u]}+R_{\psi,x}(u),
\]
where
\[
 R_{\psi,x}(u)\coloneqq
 \int_0^1 (1-t)
 {\{D^2\psi_x(tu)-D^2\psi_x(0)\}[u,u]}\,dt.
\]
Continuity of \(D^2\psi_x\) gives
\[
\begin{aligned}
\|R_{\psi,x}(u)\|_{m_x}&\le\|u\|_x^2\int_0^1(1-t)\|D^2\psi_x(tu)-D^2\psi_x(0)\|_{\mathrm{op}}\,dt\\
&\le\tfrac12\|u\|_x^2\sup_{\|v\|_x\le\|u\|_x}\|D^2\psi_x(v)-D^2\psi_x(0)\|_{\mathrm{op}}=o(\|u\|_x^2).
\end{aligned}
\]
Under \Cref{ass:uniform-inference-conditions}, the identity in these frames is
\[
 \E[L_{\ell,x}^{(0)}(Y)\mid X=\Exp_x^{\cX}\left[u\right]]
 =-E_{x,\ell}^\ast\psi_x(u).
\]
Since \(E_{x,\ell}\) is fixed in \(u\) and isometric,
\eqref{eq:supp-transported-log-predictor-regularity} with \(r=0\)
gives \eqref{eq:supp-uniform-conditional-score-regularity}.
The H\"older bound and the same integral yield
\[
 \|R_{\psi,x}(u)\|_{m_x}
 \lesssim \|u\|_x^{2+\gamma}\int_0^1(1-t)t^\gamma\,dt\lesssim \|u\|_x^{2+\gamma},
 \qquad x\in\cX_0,\quad\|u\|_x\le\delta_L.
\]

\noindent\textup{(ii).} Set
\({M_{H,x}(u)}\coloneqq
\E[\cH_{m_x}(Y)\mid X=\Exp_x^{\cX}\left[u\right]]\).
{By
\Cref{lem:supp_conditional_moment_regularity}~\textup{(ii)}, this map is
\(C^2\) near zero in a fixed response frame, with
\(M_{H,x}(0)=A_x\).}
Under the uniform conditions,
\[
 \E[L_{\ell,x}^{(1)}(Y)\mid X=\Exp_x^{\cX}\left[u\right]]
 =-E_{x,\ell}^\ast {M_{H,x}(u)}E_{x,\ell},
\]
so \eqref{eq:supp-transported-log-predictor-regularity} with \(r=1\)
gives common \(C^2\) bounds.
Taylor's theorem gives
\[
\|{M_{H,x}(u)}-{M_{H,x}(0)}-D{M_{H,x}(0)}[u]\|_{\mathrm{op}}
\lesssim \|u\|_x^2.
\]
Condition on \(X\). Constant and linear reproduction in
\eqref{eq:supp-population-polynomial-reproduction} and the bound in
\eqref{eq:supp-local-polynomial-weight-moments} with \(k=2\) give
\[
\begin{aligned}
 A_{h,x}(m_x)-A_x
 &=\E\Bigl[L_{1,h}\bigl\{
       {M_{H,x}(U_x(X))}-{M_{H,x}(0)}
       -D{M_{H,x}(0)}[U_x(X)]\bigr\}\Bigr],\\
 \|A_{h,x}(m_x)-A_x\|_{\mathrm{op}}
 &\lesssim \E[|L_{1,h}(X;x)|\|U_x(X)\|_x^2]=O(h^2).
\end{aligned}
\]

For small \(h\), the kernel support lies in the predictor neighborhood.
The conditional envelope bounds and
\eqref{eq:supp-local-polynomial-weight-moments} with \(k=0\) give
\[
 \E[|L_{1,h}(X;x)|\cE_R(Y)]
 =\E\bigl[|L_{1,h}(X;x)|\E[\cE_R(Y)\mid X]\bigr]\lesssim \E[|L_{1,h}(X;x)|]\lesssim 1.
\]
By \Cref{lem:supp_fixed_local_log_domain}~\textup{(ii)} and smoothness
of the exponential map, the third derivatives in \(s\) of the squared distance loss expressed in normal coordinates have norms \(\lesssim\cE_R(y)\) on
the fixed local domain. Differentiation under the integral and the
fundamental theorem of calculus therefore give
\[
\begin{aligned}
 \|{(D_s^2\widetilde\Phi_{h,x})(s)}\|_{\mathrm{op}}
 &\lesssim \E[|L_{1,h}(X;x)|\cE_R(Y)]\lesssim 1,\\
 \|{(D_s\widetilde\Phi_{h,x})(s)-(D_s\widetilde\Phi_{h,x})(0)}\|_{\mathrm{op}}
 &\le\int_0^1
       \|{(D_s^2\widetilde\Phi_{h,x})(ts)[s,\cdot]}\|_{\mathrm{op}}\,dt
 \lesssim \|s\|_{m_x}.
\end{aligned}
\]
Using \eqref{eq:supp_mean_value_hessian_smoothing_path} and
\({\left.D_s\widetilde\Phi_{h,x}(s)\right|_{s=0_{m_x}}}=A_{h,x}(m_x)\), we obtain
\[
 \|\bar A_{h,x}-A_{h,x}(m_x)\|_{\mathrm{op}}
 \le\int_0^1
       \|{(D_s\widetilde\Phi_{h,x})(tB_h(x))}
          -{(D_s\widetilde\Phi_{h,x})(0)}\|_{\mathrm{op}}\,dt
 \lesssim \|B_h(x)\|_{m_x}.
\]

Under the uniform conditions,
{\Cref{lem:supp_uniform_finite_shrunken_cover}~\textup{(iv)}} and the
uniform envelope and weight bounds make these estimates uniform.
\end{proof}

\begin{proposition}
\label{prop:supp_smoothing_bias_expansion}
Let \(h\to0\). For \(\nu=\mathrm{pt}\), assume
\Cref{ass:local_support_log_geometry,ass:densities,ass:kernels}.
For \(\nu=\mathrm{unif}\), assume \Cref{ass:uniform-inference-conditions}
without its bandwidth conditions. Then
\begin{equation}\label{eq:supp-smoothing-bias-expansion}
\begin{aligned}
 \sup_{x\in\mathcal I_\nu}
 \{\|\beta_h(x)\|_{m_x}+h^{-2}d_{\cY}(m_{h,x},m_x)\}&=O(1),\\
 \sup_{x\in\mathcal I_\nu}
 \|B_h(x)-h^2\beta_h(x)\|_{m_x}
 &=\begin{cases}
 o(h^2),&\nu=\mathrm{pt},\\
 O(h^{2+\gamma}),&\nu=\mathrm{unif}.
 \end{cases}
\end{aligned}
\end{equation}
\end{proposition}

\begin{proof}
With \(R_{\psi,x}\) as defined in the proof of
\Cref{lem:supp_conditional_score_smoothness}, set
\(R_{\psi,h}(x)\coloneqq
\E[L_{1,h}(X;x)R_{\psi,x}(U_x(X))]\).
Conditioning on \(X\), applying
\Cref{lem:supp_conditional_score_smoothness}~\textup{(i)} and
\Cref{prop:supp_selector_weight_design_moment_master}~\textup{(i)}
with \((j,a,v)=(1,h,e_{1,x})\), and using
\eqref{eq:explicit-bias-expansion} give
\begin{equation}
\label{eq:supp_population_score_bias_expansion}
\begin{aligned}
\widetilde\Phi_{h,x}(0)
&=\E[L_{1,h}(X;x)\psi_x(U_x(X))]\\
&=\tfrac12\E[L_{1,h}(X;x)\nabla_{\cX}^2s_x(x)(U_x(X),U_x(X))]
  +R_{\psi,h}(x)\\
&=-h^2A_x\beta_h(x)+R_{\psi,h}(x).
\end{aligned}
\end{equation}
\Cref{lem:supp_local_linear_second_moment}~\textup{(i)},
\Cref{lem:supp_conditional_score_smoothness}~\textup{(i)}, and
invertibility of \(A_x\) give
\[
 \sup_{x\in\mathcal I_\nu}\|\beta_h(x)\|_{m_x}=O(1),
\]
with the uniform inverse bound supplied by
\Cref{lem:supp_uniform_finite_shrunken_cover}~\textup{(ii)}.
Kernel localization and
\Cref{prop:supp_selector_weight_design_moment_master}~\textup{(ii)},
with \(k=2\), give
\[
 \sup_{x\in\mathcal I_\nu}\|R_{\psi,h}(x)\|_{m_x}
 \lesssim h^2\sup_{x\in\mathcal I_\nu}
 \sup_{0<\|u\|_x\le R_1h}
 \frac{\|R_{\psi,x}(u)\|_{m_x}}{\|u\|_x^2}=\begin{cases}
 o(h^2),&\nu=\mathrm{pt},\\
 O(h^{2+\gamma}),&\nu=\mathrm{unif}.
 \end{cases}\]
Thus \eqref{eq:supp_population_score_bias_expansion} gives
\(\sup_{x\in\mathcal I_\nu}\|\widetilde\Phi_{h,x}(0)\|_{m_x}=O(h^2)\).

\Cref{prop:supp_population_smoothed_target_consistency} gives
\(\sup_{x\in\mathcal I_\nu}\|B_h(x)\|_{m_x}=o(1)\).
By \Cref{lem:supp_conditional_score_smoothness}~\textup{(ii)},
\[
 \sup_{x\in\mathcal I_\nu}\|\bar A_{h,x}-A_x\|_{\mathrm{op}}
 \lesssim \sup_{x\in\mathcal I_\nu}\|B_h(x)\|_{m_x}+h^2=o(1).
\]
Use positive definiteness of \(A_x\) and, in the uniform case,
the spectral lower bound in
\Cref{lem:supp_uniform_finite_shrunken_cover}~\textup{(ii)}.
Together with the preceding comparison,
\Cref{lem:supp_covariance_transfer_spectral_stability} gives
\(\sup_{x\in\mathcal I_\nu}\|\bar A_{h,x}^{-1}\|_{\mathrm{op}}=O(1)\).
Hence \eqref{eq:supp_smoothing_first_order_linearization} and
\(\sup_{x\in\mathcal I_\nu}\|\widetilde\Phi_{h,x}(0)\|_{m_x}=O(h^2)\) yield
\[
 \sup_{x\in\mathcal I_\nu}\|B_h(x)\|_{m_x}
 =O(1)O(h^2)=O(h^2).
\]
By \eqrefs{eq:supp_smoothing_first_order_linearization,eq:supp_population_score_bias_expansion},
\Cref{lem:supp_conditional_score_smoothness}~\textup{(ii)}, and the preceding bounds,
\[
\begin{aligned}
 \sup_{x\in\mathcal I_\nu}\|B_h(x)-h^2\beta_h(x)\|_{m_x}
 &=\sup_{x\in\mathcal I_\nu}\Bigl\|
 \bar A_{h,x}^{-1}\bigl[h^2\{A_x-\bar A_{h,x}\}\beta_h(x)
 -R_{\psi,h}(x)\bigr]\Bigr\|_{m_x}\\
 &\lesssim \sup_{x\in\mathcal I_\nu}\|R_{\psi,h}(x)\|_{m_x}
 +h^2\sup_{x\in\mathcal I_\nu}\|\bar A_{h,x}-A_x\|_{\mathrm{op}}\\
 &\lesssim \sup_{x\in\mathcal I_\nu}\|R_{\psi,h}(x)\|_{m_x}
 +h^2\left\{\sup_{x\in\mathcal I_\nu}\|B_h(x)\|_{m_x}+h^2\right\}\\
 &\lesssim \sup_{x\in\mathcal I_\nu}\|R_{\psi,h}(x)\|_{m_x}+h^4=\begin{cases}o(h^2),&\nu=\mathrm{pt},\\
O(h^{2+\gamma}),&\nu=\mathrm{unif}
 \end{cases}.
\end{aligned}
\]
Finally, \(d_{\cY}(m_{h,x},m_x)=\|B_h(x)\|_{m_x}\) on the fixed
logarithm region gives the bounds in \eqref{eq:supp-smoothing-bias-expansion}.
\end{proof}

\subsection{Local linear expansion}
\label{sec:supp_local_linear_oracle_limit}

Use \(\widehat A_{h,x}(\omega)\) from \Cref{subsec:rrbc}
and its population counterpart in
\eqref{eq:main-population-response-quantities}.
For comparison on \(T_{m_x}\cY\), define
\[
 A_{h,x}^\parallel(\omega)
 \coloneqq\Pi_{\omega\to m_x}A_{h,x}(\omega)\Pi_{m_x\to\omega},\qquad
 \widehat A_{h,x}^\parallel(\omega)
 \coloneqq\Pi_{\omega\to m_x}\widehat A_{h,x}(\omega)\Pi_{m_x\to\omega}.
\]

\begin{lemma}
\label{lem:supp_feasible_ll_hessian_replacement}
Under the common conditions for \(\nu\in\{\mathrm{pt},\mathrm{unif}\}\),
\begin{align}
{\sup_{x\in\mathcal I_\nu}\|A_{h,x}^\parallel(m_{h,x})-A_x\|_{\mathrm{op}}}
 &{=O(h^2),}\label{eq:supp-population-hessian-comparison}\\
{\sup_{x\in\mathcal I_\nu}\|\widehat A_{h,x}^\parallel(m_{h,x})-A_{h,x}^\parallel(m_{h,x})\|_{\mathrm{op}}}
 &{=O_p(\delta_{n,\nu}(h)).}\label{eq:supp-stochastic-hessian-comparison}
\end{align}
Moreover, for some \(c>0\),
\(\mathbb P(\inf_{x\in\mathcal I_\nu}
\lambda_{\min}(\widehat A_{h,x}^\parallel(m_{h,x}))\ge c)\to1\).
\end{lemma}
\begin{proof}
On the fixed response regions, use
\Cref{lem:supp_fixed_local_log_domain}~\textup{(ii) and (iii)} pointwise
and {\Cref{lem:supp_uniform_finite_shrunken_cover}~\textup{(iv)}} uniformly
to obtain
\begin{equation}\label{eq:supp-transported-hessian-lipschitz}
 \|\Pi_{\omega\to m_x}\cH_\omega(y)\Pi_{m_x\to\omega}
  -\Pi_{\omega'\to m_x}\cH_{\omega'}(y)\Pi_{m_x\to\omega'}\|_{\mathrm{op}}
 \lesssim d_{\cY}(\omega,\omega')\cE_R(y).
\end{equation}
The conditional envelope bounds and
\eqref{eq:supp-local-polynomial-weight-moments} with \(k=0\) give
\(\sup_x\E[|L_{1,h}|\cE_R]\lesssim 1\).
Thus \eqref{eq:supp-transported-hessian-lipschitz},
\Cref{lem:supp_conditional_score_smoothness}~\textup{(ii)}, and
\Cref{prop:supp_smoothing_bias_expansion} yield
\[
\begin{aligned}
 \sup_x\|A_{h,x}^\parallel(m_{h,x})-A_x\|_{\mathrm{op}}
 &\lesssim \sup_xd_{\cY}(m_{h,x},m_x)
       \sup_x\E[|L_{1,h}|\cE_R]
       +\sup_x\|A_{h,x}(m_x)-A_x\|_{\mathrm{op}}\\
 &=O(h^2)O(1)+O(h^2)=O(h^2).
\end{aligned}
\]

At a deterministic center \(\omega_x\) in the fixed response region,
\[
\begin{aligned}
 \widehat A_{h,x}(\omega_x)-A_{h,x}(\omega_x)
 &=P_n[(\widehat L_{1,h}-L_{1,h})\cH_{\omega_x}(Y)]
   +(P_n-P)[L_{1,h}\cH_{\omega_x}(Y)].
\end{aligned}
\]
{\crefplainname{Equation} \eqref{eq:supp-local-linear-weight-replacement-envelope}}
and {\Cref{lem:supp_localized_support_count,lem:supp_empirical_design_concentration}} give
\[
\begin{aligned}
 \sup_x\|P_n[(\widehat L_{1,h}-L_{1,h})\cH_{\omega_x}(Y)]\|_{\mathrm{op}}
 &\lesssim \sup_x\|\widehat S_{1,h}(x)^{-1}-S_{1,h}(x)^{-1}\|_{\mathrm{op}}
       \sup_x h^{-p}P_n[\mathds{1}\{d_{\cX}(X,x)\le R_1h\}G_R(Y)]\\
 &=O_p(\delta_{n,\nu}(h))O_p(1)=O_p(\delta_{n,\nu}(h)).
\end{aligned}
\]
Pointwise, independence, finite-dimensional norm equivalence,
\eqref{eq:supp-local-polynomial-weight-envelopes}, and
\eqref{eq:supp-pointwise-split-moments} yield
\[
\begin{aligned}
 {\E\left[\left\|(P_n-P)[L_{1,h}\cH_{\omega_x}(Y)]\right\|_{\mathrm{op}}^2\right]}
 &\lesssim n^{-1}{\E}[|L_{1,h}|^2\|\cH_{\omega_x}(Y)\|_{\mathrm{op}}^2]\\
 &\lesssim (nh^{2p})^{-1}{\E}[
       \mathds{1}\{d_{\cX}(X,x)\le R_1h\}G_R(Y)^2]\\
 &\lesssim (nh^p)^{-1}.
\end{aligned}
\]
Chebyshev's inequality gives \(O_p(\delta_{n,\mathrm{pt}}(h))\).
{Uniformly, the localized response centers, bounded design coefficients,
and framed Hessian envelopes verify the conditions for the function classes.} Apply
\Cref{lem:supp_local_polynomial_kernel_factors_are_localized,lem:supp_localized_finite_moment_maximal_inequality}
with {\((a,M_a,H,m)=(h,h^{-p},\cE_R,1)\)} to obtain
\(O_p(\delta_{n,\mathrm{unif}}(h))\) for each framed coordinate.
{Conjugation by parallel transport} preserves the operator norm, so taking
\(\omega_x=m_{h,x}\) proves \eqref{eq:supp-stochastic-hessian-comparison}.

By \Cref{ass:local_support_log_geometry},
\eqref{eq:supp-hadamard-hessian-lower-bound}, and
\Cref{lem:supp_uniform_finite_shrunken_cover}~\textup{(ii)},
choose \(c>0\) with \(A_x\succeq2cI\) uniformly over \(\mathcal I_\nu\).

Equations \eqrefs{eq:supp-population-hessian-comparison,eq:supp-stochastic-hessian-comparison} give
\[
\begin{aligned}
 \sup_x\|\widehat A_{h,x}^\parallel(m_{h,x})-A_x\|_{\mathrm{op}}
 &\le\sup_x\|\widehat A_{h,x}^\parallel(m_{h,x})-A_{h,x}^\parallel(m_{h,x})\|_{\mathrm{op}}+\sup_x\|A_{h,x}^\parallel(m_{h,x})-A_x\|_{\mathrm{op}}\\
 &=O_p(\delta_{n,\nu}(h))+O(h^2)=O_p(r_{n,\nu})=o_p(1).
\end{aligned}
\]
Weyl's inequality gives the stated probability limit and the inverse bound \(c^{-1}\).
\end{proof}

For either procedure, define the oracle process and its empirically
centered Gaussian multiplier process by
\begin{equation}\label{eq:supp-generic-oracle-linear-process}
 \mathfrak Z_{n,\star}(x)\coloneqq h^{-p/2}A_x^{-1}\mathbb G_n\varphi_{\star,n}(\cdot;x),
 \qquad
 \mathfrak Z_{n,\star}^{\ast o}(x)
 \coloneqq h^{-p/2}A_x^{-1}\mathbb G_n^\ast\varphi_{\star,n}(\cdot;x).
\end{equation}
Since \(P\varphi_{\star,n}=0\), the first process equals
\((nh^p)^{-1/2}\sum_iA_x^{-1}\varphi_{\star,n}(W_i;x)\).

\begin{lemma}
\label{lem:supp_fixed_center_oracle_score_replacement}
Under the common conditions, for \((j,a,L)=(1,h,L_{1,h})\) and,
{for RRBC,} \((j,a,L)=(2,b,L_{\bias,h,b})\),
\begin{equation}\label{eq:supp-oracle-score-recentering-rate}
 \sup_{x\in\mathcal I_\nu}\|P_n[L\{\Pi_{m_{h,x}\to m_x}
       \varepsilon_{j,a,x}(m_{h,x})-\varepsilon_{j,a,x}(m_x)\}]\|
 =O_p(h^2\delta_{n,\nu}(h)).
\end{equation}
\end{lemma}
\begin{proof}
Put \(\gamma_x(t)\coloneqq \Exp_{m_x}\left[tB_h(x)\right]\), \(0\le t\le1\).
By \Cref{prop:supp_smoothing_bias_expansion},
\(\sup_x\|B_h(x)\|_{m_x}=O(h^2)\).
{For \(z\) on the kernel support and \(y\) in the response domain, expand the residual using}
\eqrefs{eq:supp-regression-coefficients,eq:supp-regression-residual}{. Since \(S_{j,a}(x)\) does not depend on \(\omega\), we apply \(\nabla_v^{\cY}\Log_\omega^{\cY}y=-\cH_\omega(y)v\), \(v\in T_\omega\cY\), along \(\gamma_x\), holding \(y\) fixed, to obtain}
\[
\begin{aligned}
 &\left\|\Pi_{m_{h,x}\to m_x}\varepsilon_{j,a,x}(m_{h,x};(z,y))
       -\varepsilon_{j,a,x}(m_x;(z,y))\right\|_{m_x}\\
 &=\left\|\Pi_{m_{h,x}\to m_x}\Log_{m_{h,x}}y-\Log_{m_x}y-\{\Pi_{m_{h,x}\to m_x}\Theta_{j,a,x}(m_{h,x})-\Theta_{j,a,x}(m_x)\}r_{j,x}(U_x(z)/a)\right\|_{m_x}\\
 &=\left\|-\int_0^1\Pi_{\gamma_x(t)\to m_x}\cH_{\gamma_x(t)}(y)
             \Pi_{m_x\to\gamma_x(t)}B_h(x)\,dt-\{\Pi_{m_{h,x}\to m_x}\Theta_{j,a,x}(m_{h,x})-\Theta_{j,a,x}(m_x)\}r_{j,x}(U_x(z)/a)\right\|_{m_x}\\
 &\le\|B_h(x)\|_{m_x}\int_0^1\|\cH_{\gamma_x(t)}(y)\|_{\mathrm{op}}\,dt\\
 &\qquad+\|B_h(x)\|_{m_x}\|S_{j,a}(x)^{-1}\|_{\mathrm{op}}
       \|r_{j,x}(U_x(z)/a)\|\E\Bigl[|K_{j,a}(X,x)|\|r_{j,x}(U_x(X)/a)\|
       \int_0^1\|\cH_{\gamma_x(t)}(Y)\|_{\mathrm{op}}\,dt\Bigr]\\
 &\lesssim\|B_h(x)\|_{m_x}\left\{G_R(y)+
       \E\Bigl[|K_{j,a}(X,x)|G_R(Y)\|r_{j,x}(U_x(X)/a)\|\Bigr]\right\}\\
 &\lesssim\|B_h(x)\|_{m_x}\{G_R(y)+1\}
 \lesssim h^2G_R(y).
\end{aligned}
\]
{The Hessian envelopes follow from}
\Cref{lem:supp_fixed_local_log_domain}{. The remaining bounds use the population design bounds in} \eqref{eq:supp-population-design-bounds}{,}
\eqref{eq:supp-standing-design-vector-bound}{, the conditional first-moment bound for \(G_R\), and \(G_R\ge1\). The same estimates also give \(\sup_x\|\Pi_{m_{h,x}\to m_x}\Theta_{j,a,x}(m_{h,x})-\Theta_{j,a,x}(m_x)\|_{\mathrm{op}}\lesssim h^2\).}

Contracting \eqref{eq:supp-residual-normal-equations} with
\(S_{1,h}^{-1}e_{1,x}\) for \(j=1\), or
\(\frac{h^2}{2b^2}S_{2,b}^{-1}(0,0,M_{2,h})\) for \(j=2\), gives
\(P[L\varepsilon_{j,a,x}(\omega)]=0\). Hence
\[
 P[L\{\Pi_{m_{h,x}\to m_x}\varepsilon_{j,a,x}(m_{h,x})
                         -\varepsilon_{j,a,x}(m_x)\}]
=\Pi_{m_{h,x}\to m_x}P[L\varepsilon_{j,a,x}(m_{h,x})]
             -P[L\varepsilon_{j,a,x}(m_x)]=0.
\]
Pointwise, independence,
\eqrefs{eq:supp-pointwise-split-moments,eq:supp-local-polynomial-weight-envelopes},
and \Cref{lem:supp_local_linear_second_moment}~\textup{(i)} yield
\[
\begin{aligned}
 &{\E\left[\left\|P_n[L\{\Pi_{m_{h,x}\to m_x}\varepsilon_{j,a,x}(m_{h,x})  -\varepsilon_{j,a,x}(m_x)\}]\right\|_{m_x}^2\right]}\\
 &={\E\left[\left\|(P_n-P)[L\{\Pi_{m_{h,x}\to m_x}\varepsilon_{j,a,x}(m_{h,x})
                             -\varepsilon_{j,a,x}(m_x)\}]\right\|_{m_x}^2\right]}\\
 &\lesssim n^{-1}\E\left[|L|^2
       \|\Pi_{m_{h,x}\to m_x}\varepsilon_{j,a,x}(m_{h,x})
                              -\varepsilon_{j,a,x}(m_x)\|_{m_x}^2\right]\\
 &\lesssim h^4(na^{2p})^{-1}\E[
       \mathds{1}\{d_{\cX}(X,x)\le R_ja\}G_R(Y)^2]
 \lesssim h^4(na^p)^{-1}.
\end{aligned}
\]
Chebyshev's inequality and \(a\asymp h\) prove the pointwise case.

{For the uniform bound, the \(O(h^2)\) bounds on \(B_h(x)\) and
\(\Pi_{m_{h,x}\to m_x}\Theta_{j,a,x}(m_{h,x})-\Theta_{j,a,x}(m_x)\) give uniformly bounded normalized coefficients. The uniform
geometric bounds in}
\Cref{lem:supp_uniform_finite_shrunken_cover}~\textup{(iv)}
{provide a common \(C\cE_R\) envelope and Lipschitz bound
for the response factors in fixed frames. Apply}
\Cref{lem:supp_local_polynomial_kernel_factors_are_localized,lem:supp_localized_finite_moment_maximal_inequality}
{coordinatewise to the two \(h^{-2}\)-normalized terms below,
with \((M_a,H,m)=(a^{-p},\cE_R,1)\).
The mean-zero identity and Fubini's theorem then give}
\[
{\begin{aligned}
 &\sup_{x\in\cX_0}\left\|P_n\Bigl[L\bigl(\Pi_{m_{h,x}\to m_x}\varepsilon_{j,a,x}(m_{h,x};W)-\varepsilon_{j,a,x}(m_x;W)\bigr)\Bigr]\right\|_{m_x}\\
 &\le h^2\sup_{\substack{x\in\cX_0\\0\le t\le1}}\left\|(P_n-P)\Bigl[L\Pi_{\gamma_x(t)\to m_x}\cH_{\gamma_x(t)}(Y)\Pi_{m_x\to\gamma_x(t)}\bigl(h^{-2}B_h(x)\bigr)\Bigr]\right\|_{m_x}\\
 &\quad+h^2\sup_{x\in\cX_0}\left\|(P_n-P)\Bigl[Lh^{-2}\bigl(\Pi_{m_{h,x}\to m_x}\Theta_{j,a,x}(m_{h,x})-\Theta_{j,a,x}(m_x)\bigr)r_{j,x}(U_x(X)/a)\Bigr]\right\|_{m_x}\\
 &=O_p\!\left(h^2\sqrt{\frac{\log n}{na^p}}\right)=O_p\!\left(h^2\delta_{n,\mathrm{unif}}(h)\right).
\end{aligned}}
\]
{Here \(a\asymp h\), and}
\eqref{eq:supp-uniform-bandwidth-rate}
{gives \(\log n\asymp\log(1/h)\).}
\end{proof}

\begin{lemma}
\label{lem:supp_local_empirical_frechet_linearization_mh}
Under the common conditions,
\(\sup_{x\in\mathcal I_\nu} d_{\cY}(\widehat m_{h,x},m_{h,x})=O_p(\delta_{n,\nu}(h))\).
\begin{equation}\label{eq:supp-native-local-linear-representation}
 \sup_{x\in\mathcal I_\nu}\|\Log_{m_{h,x}}\widehat m_{h,x}-A_{h,x}(m_{h,x})^{-1}
             P_n[L_{1,h}\varepsilon_{1,h,x}(m_{h,x})]\|
 =O_p(\delta_{n,\nu}(h)^2).
\end{equation}
\end{lemma}
\begin{proof}
Work on the localization and design events, whose probabilities tend to one.
Apply \eqref{eq:supp-exact-regression-coefficient-identity} to
\(\widehat\Theta_{1,h,x}(m_{h,x})-\Theta_{1,h,x}(m_{h,x})\),
then apply both sides to \(e_{1,x}\). Since
\(\Theta_{1,h,x}(m_{h,x})e_{1,x}=0\), this gives
\begin{equation}\label{eq:supp-score-design-exact-replacement}
\begin{aligned}
 &P_n[\widehat L_{1,h}\Log_{m_{h,x}}Y]\\
 &\quad =P_n[L_{1,h}\varepsilon_{1,h,x}(m_{h,x})]+(P_n-P)[K_{1,h}(X,x)\varepsilon_{1,h,x}(m_{h,x})\otimes r_{1,x}(U_x(X)/h)]
       (\widehat S_{1,h}^{-1}-S_{1,h}^{-1})e_{1,x}.
\end{aligned}
\end{equation}
{\crefplainname{Equation}} \eqref{eq:supp-residual-normal-equations}
{gives centering, and the design and regression coefficient bounds in}
\eqrefs{eq:supp-population-design-bounds,eq:supp-population-regression-coefficient-bounds}
{bound the coefficients. Pointwise, use the second-moment bound in}
\eqref{eq:supp-pointwise-split-moments}{. Uniformly, expanding the residual
gives a finite sum of kernel-weighted predictor polynomials with response factors
\(1\) or framed logarithm coordinates. The localized centers, bounded coefficients,
and response envelopes verify the conditions of}
\Cref{lem:supp_local_polynomial_kernel_factors_are_localized,lem:supp_localized_finite_moment_maximal_inequality}
{with \((a,M_a,H,m)=(h,h^{-p},\cE_R,1)\). Thus}
\begin{equation}
 \sup_{x\in\mathcal I_\nu}\|P_n[L_{1,h}\varepsilon_{1,h,x}(m_{h,x})]\|
 \lesssim \sup_{x\in\mathcal I_\nu}
 \|(P_n-P)[K_{1,h}(X,x)\varepsilon_{1,h,x}(m_{h,x})
                     \otimes r_{1,x}(U_x(X)/h)]\|=O_p(\delta_{n,\nu}(h)).
\end{equation}
{For RRBC, the same argument gives}
\begin{equation}\label{eq:supp-centered-pilot-score-design-rate}
 \sup_{x\in\mathcal I_\nu}\|(P_n-P)[K_{2,b}(X,x)\varepsilon_{2,b,x}(m_{h,x})
                     \otimes r_{2,x}(U_x(X)/b)]\|
 =O_p(\delta_{n,\nu}(h)).
\end{equation}
Thus {\eqref{eq:supp-score-design-exact-replacement} and \Cref{lem:supp_empirical_design_concentration}} yield
\[
\begin{aligned}
 &\sup_{x\in\mathcal I_\nu}\|P_n[\widehat L_{1,h}\Log_{m_{h,x}}Y]
             -P_n[L_{1,h}\varepsilon_{1,h,x}(m_{h,x})]\|\\
 &\quad\le
 \sup_{x\in\mathcal I_\nu}\|(P_n-P)[K_{1,h}(X,x)\varepsilon_{1,h,x}(m_{h,x})
             \otimes r_{1,x}(U_x(X)/h)]\|
 \sup_{x\in\mathcal I_\nu}\|\widehat S_{1,h}^{-1}-S_{1,h}^{-1}\|_{\mathrm{op}}\\
 &\quad=O_p(\delta_{n,\nu}(h))O_p(\delta_{n,\nu}(h))=O_p(\delta_{n,\nu}(h)^2).
\end{aligned}
\]

On the localized domains in \eqref{eq:supp-localized-center-containment},
the first-order condition at \(\widehat m_{h,x}\),
the bound on the second derivative of \(\omega\mapsto\Log_\omega Y\) in
\Cref{lem:supp_fixed_local_log_domain}, and the bound on the weighted first moment
in \Cref{lem:supp_localized_support_count} give
\[
 \sup_{x\in\mathcal I_\nu}\|\widehat A_{h,x}(m_{h,x})\Log_{m_{h,x}}\widehat m_{h,x}
              -P_n[\widehat L_{1,h}\Log_{m_{h,x}}Y]\|
 =O_p\!\left(\left\{\sup_{x\in\mathcal I_\nu}d_{\cY}(\widehat m_{h,x},m_{h,x})\right\}^{2}\right).
\]
{The centered score bound,}
{\eqref{eq:supp-score-design-exact-replacement}, and \Cref{lem:supp_empirical_design_concentration}}
give \(\sup_{x\in\mathcal I_\nu}\|P_n[\widehat L_{1,h}\Log_{m_{h,x}}Y]\|=O_p(\delta_{n,\nu}(h))\).
The quadratic Taylor bound,
\Cref{lem:supp_feasible_ll_hessian_replacement}, and consistency from
\Cref{prop:supp_pointwise_empirical_smoothed_target_consistency}
then imply that, for every
\(\eta>0\), there is \(M>0\) such that, with
probability at least \(1-\eta\) for all sufficiently large \(n\),
\(M\sup_{x\in\mathcal I_\nu}d_{\cY}(\widehat m_{h,x},m_{h,x})\le1/2\) and
\[
 \sup_{x\in\mathcal I_\nu}d_{\cY}(\widehat m_{h,x},m_{h,x})
 \le M\delta_{n,\nu}(h)
       +M\left\{\sup_{x\in\mathcal I_\nu}d_{\cY}(\widehat m_{h,x},m_{h,x})\right\}^2\le M\delta_{n,\nu}(h)
       +\tfrac12\sup_{x\in\mathcal I_\nu}d_{\cY}(\widehat m_{h,x},m_{h,x}).
\]
Hence \(\sup_{x\in\mathcal I_\nu}d_{\cY}(\widehat m_{h,x},m_{h,x})\le2M\delta_{n,\nu}(h)\)
on this event, proving the distance rate.
Thus \eqref{eq:supp-stochastic-hessian-comparison}, bounded population inverses,
and the isometry of parallel transport yield
\[
\begin{aligned}
&\sup_{x\in\mathcal I_\nu}\|\Log_{m_{h,x}}\widehat m_{h,x}-A_{h,x}(m_{h,x})^{-1}P_n[L_{1,h}\varepsilon_{1,h,x}(m_{h,x})]\|\\
&\quad\lesssim\sup_{x\in\mathcal I_\nu}\|\widehat A_{h,x}(m_{h,x})-A_{h,x}(m_{h,x})\|_{\mathrm{op}}\sup_{x\in\mathcal I_\nu}d_{\cY}(\widehat m_{h,x},m_{h,x})\\
&\qquad+\sup_{x\in\mathcal I_\nu}\|\widehat A_{h,x}(m_{h,x})\Log_{m_{h,x}}\widehat m_{h,x}-P_n[\widehat L_{1,h}\Log_{m_{h,x}}Y]\|\\
&\qquad+\sup_{x\in\mathcal I_\nu}\|P_n[\widehat L_{1,h}\Log_{m_{h,x}}Y]-P_n[L_{1,h}\varepsilon_{1,h,x}(m_{h,x})]\|\\
&\quad=O_p(\delta_{n,\nu}(h)^2).
\end{aligned}
\]
\end{proof}

The bounds on derivatives with respect to the center in
\Cref{lem:supp_fixed_local_log_domain}~\textup{(ii) and (iii)} also give
\eqref{eq:supp-transported-hessian-lipschitz}, with \(m_x\) replaced by \(m_{h,x}\).
The center rate in \Cref{lem:supp_local_empirical_frechet_linearization_mh}, \Cref{lem:supp_localized_support_count},
\eqref{eq:supp-stochastic-hessian-comparison}, and the isometry of parallel transport
give the first bound below. With the operators transported to \(T_{m_x}\cY\), the same argument,
\eqref{eq:supp-population-hessian-comparison}, and
\Cref{lem:supp_covariance_transfer_spectral_stability} give the second:
\begin{equation}\label{eq:supp-fitted-center-hessian-rates}
\begin{aligned}
 \sup_{x\in\mathcal I_\nu}\|\Pi_{\widehat m_{h,x}\to m_{h,x}}\widehat A_{h,x}(\widehat m_{h,x})
             \Pi_{m_{h,x}\to\widehat m_{h,x}}-A_{h,x}(m_{h,x})\|_{\mathrm{op}}
 &=O_p(\delta_{n,\nu}(h)),\\
 \sup_{x\in\mathcal I_\nu}\{\|\widehat A_{h,x}^\parallel(\widehat m_{h,x})-A_x\|_{\mathrm{op}}
       +\|(\widehat A_{h,x}^\parallel(\widehat m_{h,x}))^{-1}-A_x^{-1}\|_{\mathrm{op}}\}
 &=O_p(r_{n,\nu}).
\end{aligned}
\end{equation}

\subsection{\texorpdfstring{{Bias correction and linear representations with population displacements}}{Bias correction and linear representations with population displacements}}
\label{sec:supp_rrbc_bias_matching}
\label{sec:supp_intrinsic_oracle_representations}

{Write} {\(\widehat m_{\star,x}\coloneqq\widehat m_\star(x)\)} {for} \(\star\in\{\us,\rbc\}\). {Define the population displacements by}
\begin{equation}\label{eq:main-population-centers}
{B_{\us,n}(x)=B_h(x),\quad B_{\rbc,n}(x)=\Log_{m(x)}^{\cY}
 \Exp_{m_h(x)}^{\cY}\left[-B_{h,b,x}(m_h(x))\right].}
\end{equation}

\begin{lemma}
\label{lem:supp_rbc_bias_matching_linearization_mean}
{Under the common conditions, let} \(\iota(M)\coloneqq(0,0,M)\) {for \(M\in\cS(T_x\cX)\)}. Pointwise, take \(\omega\in\overline{\cW_{m_x}}\) from \Cref{lem:supp_fixed_local_log_domain}. Uniformly, use the response domains in \eqref{eq:supp-matched-response-domains}. Then
\begin{equation}\label{eq:supp-quadratic-block-bound}
 \sup_{x,\omega}\|\Theta_{2,b,x}(\omega)\circ\iota\|_{\mathrm{op}}=O(b^2),
 \qquad
 \sup_x\|{\widehat\Theta_{2,b,x}(\widehat m_{h,x})}\circ\iota\|_{\mathrm{op}}=O_p(r_{n,\nu}).
\end{equation}
Moreover, \(\sup_x\|B_{h,b,x}(m_{h,x})\|_{m_{h,x}}=O(h^2)\), and
{\(B_{\rbc,n}(x)\) satisfies}
\begin{equation}\label{eq:supp-exact-bias-comparison}
 B_{\rbc,n}(x)=B_h(x)-A_x^{-1}{\E[L_{\bias,h,b}\Log_{m_x}Y]}+O(h^4),
\end{equation}
with the remainder uniform in \(x\).
Consequently, \(\|B_{\rbc,n}(x)\|_{m_x}=o(h^2)\) pointwise and
\(\sup_x\|B_{\rbc,n}(x)\|_{m_x}=O(h^{2+\gamma})\) uniformly.
\end{lemma}
\begin{proof}
By \Cref{lem:supp_conditional_moment_regularity}~\textup{(ii) and (iii)},
the conditional means of the logarithm map have bounded second predictor derivatives
on the stated regions. Exact reproduction in
\eqref{eq:supp-population-polynomial-reproduction} cancels the constant
and linear Taylor terms in \(\Theta_{2,b,x}(\omega)\circ\iota\).
The integral Taylor remainder and
\eqref{eq:supp-local-polynomial-weight-moments} with \(k=2\)
therefore give
\begin{equation}\label{eq:supp-population-quadratic-block-rate}
\begin{aligned}
 \sup_{x,\omega}\|\Theta_{2,b,x}(\omega)\circ\iota\|_{\mathrm{op}}
 &\lesssim b^2
 \sup_{\substack{x,\omega\\\|u\|_x\le R_2b}}
 \left\|D_u^2{\E}[\Log_\omega Y\mid X=\Exp_x^{\cX}\left[u\right]]\right\|_{\mathrm{op}}
 =O(b^2).
\end{aligned}
\end{equation}
By \eqref{eq:main-population-response-quantities},
\Cref{lem:supp_local_linear_second_moment}~\textup{(i)}, and
\Cref{lem:supp_feasible_ll_hessian_replacement},
\[
 \sup_x\|B_{h,b,x}(m_{h,x})\|_{m_{h,x}}\lesssim\frac{h^2}{2b^2}
 \sup_x\|A_{h,x}(m_{h,x})^{-1}\|_{\mathrm{op}}
 \sup_x\|\Theta_{2,b,x}(m_{h,x})\circ\iota\|_{\mathrm{op}}
 \sup_x\|M_{2,h}(x)\|_{\mathrm{op}}=O(h^2).
\]

For the empirical bound, work on the localization and design events,
whose probabilities tend to one.
Since \(\widehat S_{2,b}(x)\) {does not depend on \(\omega\)},
\eqref{eq:supp-exact-regression-coefficient-identity} gives
\begin{equation}\label{eq:supp-quadratic-block-decomposition}
\begin{aligned}
&\bigl\{\Pi_{\widehat m_{h,x}\to m_{h,x}}{\widehat\Theta_{2,b,x}(\widehat m_{h,x})}
          -\Theta_{2,b,x}(m_{h,x})\bigr\}\circ\iota\\
&\quad=\Biggl\{
 (P_n-P)\!\left[
 K_{2,b}(X,x)\varepsilon_{2,b,x}(m_{h,x})
 \otimes r_{2,x}(U_x(X)/b)\right]\\
 &\quad\quad+
 P_n\Bigl[
 K_{2,b}(X,x)
 \bigl\{\Pi_{\widehat m_{h,x}\to m_{h,x}}\Log_{\widehat m_{h,x}}Y
                  -\Log_{m_{h,x}}Y\bigr\}\otimes r_{2,x}(U_x(X)/b)\Bigr]
 \Biggr\}\widehat S_{2,b}^{-1}\circ\iota .
\end{aligned}
\end{equation}
Hence
\begin{equation}\label{eq:supp-empirical-quadratic-block-rate}
\begin{aligned}
&\sup_x\|{\widehat\Theta_{2,b,x}(\widehat m_{h,x})}\circ\iota\|_{\mathrm{op}}\\
&\quad\le\sup_x\|\Theta_{2,b,x}(m_{h,x})\circ\iota\|_{\mathrm{op}}+\sup_x\Bigl\|\bigl\{\Pi_{\widehat m_{h,x}\to m_{h,x}}{\widehat\Theta_{2,b,x}(\widehat m_{h,x})}-\Theta_{2,b,x}(m_{h,x})\bigr\}\circ\iota\Bigr\|_{\mathrm{op}}\\
&\quad\lesssim b^2+\Biggl\{\sup_x\Bigl\|(P_n-P)\!\left[K_{2,b}(X,x)\varepsilon_{2,b,x}(m_{h,x})\otimes r_{2,x}(U_x(X)/b)\right]\Bigr\|_{\mathrm{op}}\\
&\qquad+\sup_xd_{\cY}(\widehat m_{h,x},m_{h,x})\sup_xP_n\!\left[|K_{2,b}(X,x)|G_R(Y)\|r_{2,x}(U_x(X)/b)\|\right]\Biggr\}\sup_x\|\widehat S_{2,b}(x)^{-1}\circ\iota\|_{\mathrm{op}}\\
&\quad=O(b^2)+\bigl\{O_p(\delta_{n,\nu}(h))+O_p(\delta_{n,\nu}(h))O_p(1)\bigr\}O_p(1)\\
&\quad=O_p(b^2+\delta_{n,\nu}(h))=O_p(r_{n,\nu}).
\end{aligned}
\end{equation}
The first inequality in \eqref{eq:supp-empirical-quadratic-block-rate}
uses the isometry of parallel transport and the triangle inequality.
The second inequality in \eqref{eq:supp-empirical-quadratic-block-rate}
follows from \eqrefs{eq:supp-population-quadratic-block-rate,eq:supp-quadratic-block-decomposition}
and the first center derivative bound in
\Cref{lem:supp_fixed_local_log_domain} on the stated response domains.
The rate bounds follow from
{\eqrefs{eq:supp-population-design-bounds,eq:supp-centered-pilot-score-design-rate}}
and {\Cref{lem:supp_localized_support_count,lem:supp_empirical_design_concentration,lem:supp_local_empirical_frechet_linearization_mh}},
using \(\|r_{2,x}(U_x(X)/b)\|\lesssim1\) on
\(\{K_{2,b}(X,x)\ne0\}\) and \(b\asymp h\).

For the population displacement, {\Cref{lem:supp_lq_bias_weight_moment_bounds}}
{gives} \(\sup_x\|{\E[L_{\bias,h,b}\Log_{m_x}Y]}\|_{m_x}=O(h^2)\) and
\(\sup_x\|{\E[L_{\bias,h,b}\cH_{m_x}(Y)]}\|_{\mathrm{op}}=O(h^2)\).
The conditional first moment of \(\cE_R\) and
\eqref{eq:supp-local-polynomial-weight-moments} give
\(\sup_x{\E[|L_{\bias,h,b}|\cE_R]}=O(1)\).
Taylor's theorem in the response center and
\Cref{lem:supp_fixed_local_log_domain,prop:supp_smoothing_bias_expansion} yield
\[
\begin{aligned}
 &\sup_x\bigl\|
 \Pi_{m_{h,x}\to m_x}{\E[L_{\bias,h,b}\Log_{m_{h,x}}Y]}
 -{\E[L_{\bias,h,b}\Log_{m_x}Y]}\bigr\|_{m_x}\\
 &\quad\lesssim
 \sup_x\|{\E[L_{\bias,h,b}\cH_{m_x}(Y)]}\|_{\mathrm{op}}
 \sup_x\|B_h(x)\|_{m_x}
 +\sup_x{\E[|L_{\bias,h,b}|\cE_R]}
       \left\{\sup_x\|B_h(x)\|_{m_x}\right\}^2\\
 &\quad=O(h^2)O(h^2)+O(1)O(h^4)=O(h^4).
\end{aligned}
\]
Thus \eqref{eq:supp-population-hessian-comparison}, \Cref{lem:supp_covariance_transfer_spectral_stability}, and {the definition of \(B_{h,b,x}\)} {in \eqref{eq:main-population-response-quantities}} give, uniformly in \(x\),
\begin{equation}\label{eq:supp-population-correction-transport}
{\Pi_{m_{h,x}\to m_x}B_{h,b,x}(m_{h,x})=A_x^{-1}\E[L_{\bias,h,b}\Log_{m_x}Y]+O(h^4).}
\end{equation}
Apply \Cref{lem:supp_local_logarithmic_additivity} with
\((a,z,v)=(m_x,m_{h,x},-B_{h,b,x}(m_{h,x}))\) to obtain
\begin{equation}\label{eq:supp-population-correction-additivity}
 B_{\rbc,n}(x)=B_h(x)-\Pi_{m_{h,x}\to m_x}B_{h,b,x}(m_{h,x})+O(h^4),
\end{equation}
with a uniform remainder because
\(\|B_h(x)\|_{m_x}+\|B_{h,b,x}(m_{h,x})\|_{m_{h,x}}=O(h^2)\).
{Substituting \eqref{eq:supp-population-correction-transport} into \eqref{eq:supp-population-correction-additivity} proves \eqref{eq:supp-exact-bias-comparison}.}
{Finally, \Cref{lem:supp_lq_bias_weight_moment_bounds,prop:supp_smoothing_bias_expansion}, together with \(\sup_x\|A_x^{-1}\|_{\mathrm{op}}<\infty\), give}
\[
\begin{aligned}
 \sup_x\|B_{\rbc,n}(x)\|_{m_x}
 &\lesssim\sup_x\|B_h(x)-h^2\beta_h(x)\|_{m_x}
 +\sup_x\|{\E[L_{\bias,h,b}\Log_{m_x}Y]}-h^2A_x\beta_h(x)\|_{m_x}+h^4\\
 &=\begin{cases}
 o(h^2),&\nu=\mathrm{pt},\\
 O(h^{2+\gamma}),&\nu=\mathrm{unif}
 \end{cases}.
\end{aligned}
\]
\end{proof}

{The next lemma isolates the stochastic term contributed by estimating the leading smoothing displacement.}

\begin{lemma}
\label{lem:supp_feasible_to_oracle_rbc_replacement}
Under the common conditions,
\begin{equation}\label{eq:supp-correction-increment}
 \sup_x\Bigl\|\Pi_{\widehat m_{h,x}\to m_{h,x}}\widehat B_{h,b}(x)
                    -B_{h,b,x}(m_{h,x})-A_{h,x}(m_{h,x})^{-1}
                     P_n[L_{\bias,h,b}\varepsilon_{2,b,x}(m_{h,x})]\Bigr\|
 =O_p(r_{n,\nu}\delta_{n,\nu}(h)).
\end{equation}
In particular,
\begin{equation}\label{eq:supp_unscaled_feasible_bias_vector_size}
 \sup_x\|\widehat B_{h,b}(x)\|=O_p(r_{n,\nu}).
\end{equation}
\end{lemma}
\begin{proof}
{We first isolate the leading stochastic term at {fixed \(m_{h,x}\)}.} By \Cref{lem:supp_empirical_design_concentration,lem:supp_local_linear_second_moment,lem:supp_local_empirical_frechet_linearization_mh}, \(\sup_xd_{\cY}(\widehat m_{h,x},m_{h,x})=O_p(\delta_{n,\nu}(h))\), \(\sup_x\{\|\widehat M_{2,h}-M_{2,h}\| +\|\widehat S_{2,b}^{-1}-S_{2,b}^{-1}\|\}=O_p(\delta_{n,\nu}(h))\), and {\(\sup_{x\in\mathcal I_\nu}\|S_{2,b}(x)^{-1}\|_{\mathrm{op}}=O(1)\) and \(\sup_{x\in\mathcal I_\nu}\|\widehat S_{2,b}(x)^{-1}\|_{\mathrm{op}}=O_p(1)\)}. Also \(\sup_x\|A_{h,x}(m_{h,x})^{-1}\|_{\mathrm{op}}=O(1)\). {\crefplainname{Equations} \eqrefs{eq:supp-residual-normal-equations,eq:supp-centered-pilot-score-design-rate,eq:supp-exact-regression-coefficient-identity,eq:supp-quadratic-block-bound} give}
\begin{align}
&\sup_x\Bigl\|
 P_n[\widehat L_{\bias,h,b}\Log_{m_{h,x}}Y]
 -P[L_{\bias,h,b}\Log_{m_{h,x}}Y]
 -P_n[L_{\bias,h,b}\varepsilon_{2,b,x}(m_{h,x})]
 \Bigr\|\\
&\le\frac{h^2}{2b^2}
 \sup_x\Bigl\|
 P_n[K_{2,b}(X,x)\varepsilon_{2,b,x}(m_{h,x})
       \otimes r_{2,x}(U_x(X)/b)]
 \Bigr\|\!
 \sup_x\Bigl\|
 \widehat S_{2,b}^{-1}(0,0,\widehat M_{2,h})
 -S_{2,b}^{-1}(0,0,M_{2,h})
 \Bigr\|\\
&\quad+\frac{h^2}{2b^2}
 \sup_x\|\Theta_{2,b,x}(m_{h,x})\circ\iota\|_{\mathrm{op}}
 \sup_x\|\widehat M_{2,h}-M_{2,h}\|\\
&=O_p\!\left(\delta_{n,\nu}(h)^2+b^2\delta_{n,\nu}(h)\right)
 =O_p(r_{n,\nu}\delta_{n,\nu}(h)).
\end{align}

{To control the effect of {replacing \(m_{h,x}\) by \(\widehat m_{h,x}\)}, repeat the fixed-center Hessian calculation in the proof of \Cref{lem:supp_feasible_ll_hessian_replacement}, with \(L_{1,h}\) replaced by \(L_{\bias,h,b}\) and \(a=b\).} Equations \eqrefs{eq:supp-population-bias-weight,eq:supp-population-design-bounds} and \Cref{lem:supp_local_linear_second_moment}~\textup{(i)}, with \(b\asymp h\), bound the weight coefficients over \(\mathcal I_\nu\). Pointwise, the calculation uses the \(G_R^2\) bound in \eqref{eq:supp-pointwise-split-moments}. {The predictor-polynomial bounds and the framed Hessian bounds in} \Cref{lem:supp_uniform_density_pullback_regularity} {verify the factor bounds of} \Cref{lem:supp_local_polynomial_kernel_factors_are_localized}. Apply \Cref{lem:supp_localized_finite_moment_maximal_inequality} with {\((a,M_a,H,m)=(b,b^{-p},\cE_R,1)\)}. Since \(\delta_{n,\nu}(b)\asymp\delta_{n,\nu}(h)\), this yields
\[
 \sup_x\|(P_n-P)[L_{\bias,h,b}\cH_{m_x}(Y)]\|_{\mathrm{op}}
 =O_p(\delta_{n,\nu}(h)).
\]
The design and second moment rates and
\Cref{lem:supp_localized_support_count} give
\[
 \sup_xP_n[|\widehat L_{\bias,h,b}-L_{\bias,h,b}|G_R]
 =O_p(\delta_{n,\nu}(h)),\qquad
 \sup_xP_n[|\widehat L_{\bias,h,b}|\cE_R]=O_p(1).
\]
Thus \eqrefs{eq:supp-bias-weight-population-hessian,eq:supp-transported-hessian-lipschitz},
the isometry of parallel transport, and \(d_{\cY}(m_{h,x},m_x)=O(h^2)\) yield
\begin{equation}\label{eq:supp-bias-weight-empirical-hessian-rate}
\begin{aligned}
 \sup_x\|P_n[\widehat L_{\bias,h,b}\cH_{m_{h,x}}(Y)]\|_{\mathrm{op}}
 =O(h^2)+O_p(\delta_{n,\nu}(h))=O_p(r_{n,\nu}).
\end{aligned}
\end{equation}
Equation \eqref{eq:supp-bias-weight-empirical-hessian-rate},
the center rate in
\Cref{lem:supp_local_empirical_frechet_linearization_mh},
and Taylor's theorem with
\Cref{lem:supp_fixed_local_log_domain}~\textup{(ii) and (iii)}
give, uniformly in \(x\in\mathcal I_\nu\),
\[
\begin{aligned}
&P_n\!\left[\widehat L_{\bias,h,b}
 \left\{\Pi_{\widehat m_{h,x}\to m_{h,x}}\Log_{\widehat m_{h,x}}Y
                   -\Log_{m_{h,x}}Y\right\}\right]\\*
&\quad=
 -P_n[\widehat L_{\bias,h,b}\cH_{m_{h,x}}(Y)]
  \Log_{m_{h,x}}\widehat m_{h,x}
 +O_p(\delta_{n,\nu}(h)^2)
  =O_p(r_{n,\nu}\delta_{n,\nu}(h)).
\end{aligned}
\]

{For the inverse Hessian \(\widehat A_{h,x}(\widehat m_{h,x})^{-1}\), the} first bound in \eqref{eq:supp-fitted-center-hessian-rates} and \eqref{eq:supp-inverse-difference-identity} give
\[
 \sup_{x\in\mathcal I_\nu}\Bigl\|
 \Pi_{\widehat m_{h,x}\to m_{h,x}}
 \widehat A_{h,x}(\widehat m_{h,x})^{-1}
 \Pi_{m_{h,x}\to\widehat m_{h,x}}-A_{h,x}(m_{h,x})^{-1}\Bigr\|_{\mathrm{op}}
 =O_p(\delta_{n,\nu}(h)).
\]
The bound at the fitted center \eqref{eq:supp-quadratic-block-bound} and
\(\sup_x\|\widehat M_{2,h}\|=O_p(1)\) give
\[
 \sup_x\bigl\|P_n[\widehat L_{\bias,h,b}\Log_{\widehat m_{h,x}}Y]\bigr\|
 \lesssim\frac{h^2}{2b^2}
 \sup_x\|{\widehat\Theta_{2,b,x}(\widehat m_{h,x})}\circ\iota\|_{\mathrm{op}}
 \sup_x\|\widehat M_{2,h}\|
 =O_p(r_{n,\nu}).
\]

The {definitions of \(\widehat B_{h,b,x}\)} {in \eqref{eq:main-empirical-bias-displacement}} {and \(B_{h,b,x}\)} {in \eqref{eq:main-population-response-quantities}}, the preceding bounds, and the isometry of parallel transport yield
\[
\begin{aligned}
&\sup_x\Bigl\|
 \Pi_{\widehat m_{h,x}\to m_{h,x}}\widehat B_{h,b}(x)
 -B_{h,b,x}(m_{h,x})
 -A_{h,x}(m_{h,x})^{-1}
 P_n[L_{\bias,h,b}\varepsilon_{2,b,x}(m_{h,x})]
 \Bigr\|\\*
&\lesssim \sup_x\Bigl\|
 P_n[\widehat L_{\bias,h,b}\Log_{m_{h,x}}Y]
 -P[L_{\bias,h,b}\Log_{m_{h,x}}Y]
 -P_n[L_{\bias,h,b}\varepsilon_{2,b,x}(m_{h,x})]
 \Bigr\|\\
&\quad+\sup_x\Bigl\|
 P_n\!\left[\widehat L_{\bias,h,b}
 \left\{\Pi_{\widehat m_{h,x}\to m_{h,x}}\Log_{\widehat m_{h,x}}Y
                    -\Log_{m_{h,x}}Y\right\}\right]
 \Bigr\|\\
&\quad+
 \sup_x\Bigl\|
 \Pi_{\widehat m_{h,x}\to m_{h,x}}\widehat A_{h,x}(\widehat m_{h,x})^{-1}
 \Pi_{m_{h,x}\to\widehat m_{h,x}}
 -A_{h,x}(m_{h,x})^{-1}
 \Bigr\|_{\mathrm{op}}
 \sup_x\Bigl\|
 P_n[\widehat L_{\bias,h,b}\Log_{\widehat m_{h,x}}Y]
 \Bigr\|\\
&=O_p(r_{n,\nu}\delta_{n,\nu}(h))
 +O_p(r_{n,\nu}\delta_{n,\nu}(h))
 +O_p(\delta_{n,\nu}(h))O_p(r_{n,\nu})=O_p(r_{n,\nu}\delta_{n,\nu}(h)).
\end{aligned}
\]

{\(A_{h,x}(m_{h,x})^{-1}P_n[L_{\bias,h,b}\varepsilon_{2,b,x}(m_{h,x})]=O_p(\delta_{n,\nu}(h))\)} by {\eqrefs{eq:supp-residual-normal-equations,eq:supp-centered-pilot-score-design-rate}}. Since \(r_{n,\nu}\to0\) {and} {\(B_{h,b,x}(m_{h,x})=O(h^2)\)},
\begin{align}
&\sup_x\|\Pi_{\widehat m_{h,x}\to m_{h,x}}\widehat B_{h,b}(x)
                   -B_{h,b,x}(m_{h,x})\|
 =O_p(\delta_{n,\nu}(h)),\\
&\sup_x\|\widehat B_{h,b}(x)\|
 \le\sup_x\|\Pi_{\widehat m_{h,x}\to m_{h,x}}\widehat B_{h,b}(x)
                   -B_{h,b,x}(m_{h,x})\|
     +\sup_x\|B_{h,b,x}(m_{h,x})\|\\
&\qquad=O_p(\delta_{n,\nu}(h))+O(h^2)=O_p(r_{n,\nu}).
\end{align}
\end{proof}

\begin{proposition}
\label{prop:supp_intrinsic_oracle_representations}
Under the common conditions, with \(B_{\star,n}\) as in
\eqref{eq:main-population-centers}, both procedures satisfy
\begin{equation}\label{eq:supp-exact-centered-alr}
 \sup_{x\in\mathcal I_\nu}
 \|\sqrt{nh^p}\{\Log_{m_x}\widehat m_{\star,x}-B_{\star,n}(x)\}
                      -\mathfrak Z_{n,\star}(x)\|
 =O_p(\sqrt{nh^p}\,r_{n,\nu}\delta_{n,\nu}(h)).
\end{equation}
\end{proposition}
\begin{proof}
For US, apply \Cref{lem:supp_local_logarithmic_additivity} to
\(m_x,m_{h,x}\), and \(\Log_{m_{h,x}}\widehat m_{h,x}\).
\[
 \sup_{x\in\mathcal I_\nu}
 \bigl\|\Log_{m_x}\widehat m_{h,x}-B_h(x)
 -\Pi_{m_{h,x}\to m_x}\Log_{m_{h,x}}\widehat m_{h,x}\bigr\|_{m_x}
 =O_p(h^2\delta_{n,\nu}(h)+\delta_{n,\nu}(h)^2).
\]
By {\Cref{lem:supp_local_empirical_frechet_linearization_mh}},
\[
 \Log_{m_x}\widehat m_{\us,x}-B_{\us,n}(x)=\Pi_{m_{h,x}\to m_x}A_{h,x}(m_{h,x})^{-1}
 P_n[L_{1,h}\varepsilon_{1,h,x}(m_{h,x})]
 +O_p(r_{n,\nu}\delta_{n,\nu}(h)).
\]

For RRBC, put \(v\coloneqq \Log_{m_{h,x}}\widehat m_{h,x}\) {and} \(w\coloneqq \Pi_{\widehat m_{h,x}\to m_{h,x}}\widehat B_{h,b}(x)-B_{h,b,x}(m_{h,x})\). On the common normal neighborhoods,
\[
 \Log_{m_x}\widehat m_{\rbc,x}
 =\Log_{m_x}\Exp_{\Exp_{m_{h,x}}\left[v\right]}
 \left[-\Pi_{m_{h,x}\to\Exp_{m_{h,x}}\left[v\right]}
              \{B_{h,b,x}(m_{h,x})+w\}\right].
\]
As a function of \((v,w)\), the right-hand side is smooth and equals
\(B_{\rbc,n}(x)\) at \((0,0)\).
Since \(d_{\cY}(m_{h,x},m_x)+\|B_{h,b,x}(m_{h,x})\|=O(h^2)\), its derivatives
at \((0,0)\) with respect to \(v\) and \(w\) are
\(\Pi_{m_{h,x}\to m_x}+O(h^2)\) and
\(-\Pi_{m_{h,x}\to m_x}+O(h^2)\), respectively,
and its second derivatives are uniformly bounded.
Taylor's formula gives
\[
 \Log_{m_x}\widehat m_{\rbc,x}-B_{\rbc,n}(x)
 =\Pi_{m_{h,x}\to m_x}(v-w)
       +O_p(h^2\delta_{n,\nu}(h)+\delta_{n,\nu}(h)^2).
\]
By {\Cref{lem:supp_local_empirical_frechet_linearization_mh} and \eqref{eq:supp-correction-increment}},
\[
\begin{aligned}
 &\Log_{m_x}\widehat m_{\rbc,x}-B_{\rbc,n}(x)\\
 &\quad=\Pi_{m_{h,x}\to m_x}A_{h,x}(m_{h,x})^{-1}
 P_n[L_{1,h}\varepsilon_{1,h,x}(m_{h,x})
       -L_{\bias,h,b}\varepsilon_{2,b,x}(m_{h,x})]+O_p(r_{n,\nu}\delta_{n,\nu}(h)).
\end{aligned}
\]
{\crefplainname{Equation} \eqref{eq:supp-population-hessian-comparison} and \Cref{lem:supp_fixed_center_oracle_score_replacement}} then give, for both procedures {and uniformly in \(x\in\mathcal I_\nu\)},
\begin{equation}\label{eq:supp-unscaled-centered-score-alr}
\begin{aligned}
 \Log_{m_x}\widehat m_{\star,x}-B_{\star,n}(x)
 &=h^{-p}A_x^{-1}P_n\varphi_{\star,n}(\cdot;x)
       +O_p(r_{n,\nu}\delta_{n,\nu}(h)).
\end{aligned}
\end{equation}
Multiply \eqref{eq:supp-unscaled-centered-score-alr} by \(\sqrt{nh^p}\) and use
\eqref{eq:supp-generic-oracle-linear-process}.
\end{proof}

\begin{corollary}
\label{cor:supp-bias-free-score-representations}
Under the assumptions of \Cref{prop:supp_intrinsic_oracle_representations},
{for \(\nu=\mathrm{pt}\),} if \(\sqrt{nh^p}\|B_{\star,n}(x)\|\to0\), then
\begin{equation}\label{eq:supp_pointwise_intrinsic_oracle_representation}
 \sqrt{nh^p}\Log_{m_x}\widehat m_{\star,x}
 =\mathfrak Z_{n,\star}(x)+o_p(1).
\end{equation}
{For \(\nu=\mathrm{unif}\), if} \(\sqrt{nh^p\log(1/h)}\sup_x\|B_{\star,n}(x)\|\to0\), then
\begin{equation}\label{eq:supp_uniform_intrinsic_oracle_representation}
 \sqrt{\log(1/h)}\sup_x
 \|\sqrt{nh^p}\Log_{m_x}\widehat m_{\star,x}-\mathfrak Z_{n,\star}(x)\|
 =o_p(1).
\end{equation}
For pointwise or uniform inference, the respective condition on \(B_{\star,n}\) stated in this corollary may replace the
bandwidth bias restrictions. All other assumptions remain unchanged.
\end{corollary}
\begin{proof}
By {\Cref{prop:supp_intrinsic_oracle_representations} and \eqref{eq:supp-rbc-rate-consequences}},
\[
\begin{aligned}
 \bigl\|\sqrt{nh^p}
 \{\Log_{m_x}\widehat m_{\star,x}-B_{\star,n}(x)\}
 -\mathfrak Z_{n,\star}(x)\bigr\|
 &=O_p(r_{n,\mathrm{pt}})=o_p(1),\\
 \sqrt{\log(1/h)}\sup_{x\in\cX_0}
 \bigl\|\sqrt{nh^p}
 \{\Log_{m_x}\widehat m_{\star,x}-B_{\star,n}(x)\}
 -\mathfrak Z_{n,\star}(x)\bigr\|
 &=O_p\!\left(r_{n,\mathrm{unif}}\log(1/h)\right)=o_p(1).
\end{aligned}
\]
The assumed bias conditions and the triangle inequality give
\eqrefs{eq:supp_pointwise_intrinsic_oracle_representation,eq:supp_uniform_intrinsic_oracle_representation}.
The US and RRBC regimes in the main text satisfy these bias conditions by
{\eqref{eq:main-population-centers}},
{\Cref{prop:supp_smoothing_bias_expansion,lem:supp_rbc_bias_matching_linearization_mean}}, and their
bandwidth restrictions.
\end{proof}

\section{Oracle Gaussian approximation and validity of inference}
\label{sec:supp_uniform_inference}

{{We establish pointwise Gaussian limits for the standardized oracle process from} \Cref{sec:supp_score_expansions_covariances} {and Gaussian approximations for suprema of its norm.} The inference results then reduce feasible calibration to the covariance and multiplier approximations verified in \Cref{sec:supp-covariance-constructions}.}

\subsection{Oracle covariance and pointwise approximation}

Under the common conditions of \Cref{sec:supp_score_expansions_covariances},
\eqrefs{eq:supp-residual-normal-equations,eq:supp_generic_oracle_score} give
\(P\varphi_{\star,n}(\cdot;x)=0\).
Fix \(R_\ast\ge R_1\); for RRBC, choose it so that
\(R_2b\le R_\ast h\) for all sufficiently large \(n\), using \(b\asymp h\).
Equations
\eqrefs{eq:supp-standing-design-vector-bound,eq:supp-regression-residual,eq:supp-population-regression-coefficient-bounds,eq:supp-local-polynomial-weight-envelopes}
and
\Cref{lem:supp_local_linear_second_moment}~\textup{(i)} give
\begin{equation}\label{eq:supp_centered_oracle_score_envelope}
\begin{aligned}
 \|\varphi_{\star,n}((z,y);x)\|
 &\le h^p|L_{1,h}(z;x)|
       \|\varepsilon_{1,h,x}(m_x;(z,y))\|+\mathds{1}\{\star=\rbc\}
       h^p|L_{\bias,h,b}(z;x)|
       \|\varepsilon_{2,b,x}(m_x;(z,y))\|\\
 &\lesssim (1+d_{\cY}(m_x,y))
       \mathds{1}\{d_{\cX}(z,x)\le R_\ast h\}.
\end{aligned}
\end{equation}
Under the uniform conditions, for all sufficiently large \(n\),
\eqref{eq:supp_centered_oracle_score_envelope} holds for every
\(x\in\cX_0\) with the same constant.
Uniformly, conditional moments and support probabilities give
\begin{equation+}\label{eq:supp_centered_oracle_score_moments}
 \sup_{x\in\cX_0}{\E[\|\varphi_{\star,n}(W;x)\|^k]}
 \lesssim h^p,\qquad 2\le k\le 2r_0.
\end{equation+}

Independence and \(P\varphi_{\star,n}=0\) give
\(\operatorname{Cov}(\mathfrak Z_{n,\star}(x))=\Omega_{\star,n}(x)\) defined in
\eqref{eq:rbc-asymptotic-covariance}.

Pointwise, take \(L=1\), \(\cX_1=\{x\}\), and an orthonormal {frame} \(E_{x,1}:\R^q\to T_{m_x}\cY\). Uniformly, use the frames from {\Cref{lem:supp_uniform_finite_shrunken_cover}~\textup{(iii)}}. For \(x\in\cX_\ell\) and \(x'\in\cX_{\ell'}\), define
\begin{equation}\label{eq:supp-oracle-cross-covariance}
{\zeta_{\star,n,\ell}(W;x)\coloneqq h^{-p/2}E_{x,\ell}^\ast A_x^{-1}\varphi_{\star,n}(W;x),\quad C^{\ell,\ell'}_{\star,n}(x,x')\coloneqq \E[\zeta_{\star,n,\ell}(W;x)\zeta_{\star,n,\ell'}(W;x')^\top].}
\end{equation}
Thus \(C^{\ell,\ell}_{\star,n}(x,x)=E_{x,\ell}^\ast\Omega_{\star,n}(x)E_{x,\ell}\).

\begin{lemma}
\label{lem:supp_oracle_residualized_score_covariance_consistency}
Under the common pointwise conditions, with \(b\asymp h\) for RRBC,
\(cI\preceq\Omega_{\star,n}(x)\preceq CI\) for all sufficiently large
\(n\), and
\begin{equation}\label{eq:supp-finite-bandwidth-standardized-clt}
 \Omega_{\star,n}(x)^{-1/2}\mathfrak Z_{n,\star}(x)
 \dto N(0,I_{T_{m_x}\cY}).
\end{equation}
The covariance bounds hold uniformly over \(\cX_0\) under
\Cref{ass:uniform-inference-conditions}.
For subsequent inference on smooth transformations, let \(L:T_{m_x}\cY\to\mathbb R^d\) be a fixed
linear map with \(\liminf_n\lambda_{\min}(L\Omega_{\star,n}(x)L^\ast)>0\).
The projected limit \eqref{eq:supp-projected-standardized-clt}
also holds when \eqref{eq:supp-pointwise-response-tail}
replaces {the existence and regularity requirements for the
conditional response density in}
\Cref{ass:densities}{.
All other assumptions, including those on \(p_X\), are retained.}
Nondegeneracy is required only for
\(L\Omega_{\star,n}(x)L^\ast\).
\end{lemma}
\begin{proof}
By \eqrefs{eq:supp-design-chart-density-bounds,eq:supp_centered_oracle_score_envelope}
and \Cref{ass:local_support_log_geometry},
\begin{align}
 h^{-p}{\E[\|\varphi_{\star,n}\|^2]}
 &\lesssim h^{-p}\E\!\left[
 \mathds{1}\{d_{\cX}(X,x)\le R_\ast h\}
 \{1+d_{\cY}(m_x,Y)^2\}\right]
  \lesssim h^{-p}\mathbb P(d_{\cX}(X,x)\le R_\ast h)\lesssim1,\notag\\
  {\|\Omega_{\star,n}(x)\|_{\mathrm{op}}}
  &{\le\|A_x^{-1}\|_{\mathrm{op}}^2h^{-p}\E[\|\varphi_{\star,n}\|^2]
  \lesssim1.}\label{eq:supp-oracle-covariance-upper-bound}
\end{align}

The conditional density and smooth logarithm chart exclude concentration
of \(\Log_{m_x}Y\mid X=x\) on any proper affine hyperplane, so
\(\operatorname{Cov}(\Log_{m_x}Y\mid X=x)\succ0\).
With \(m_x\) fixed, {\Cref{lem:supp_conditional_moment_regularity}~\textup{(i) and (ii)}}
{make} \(z\mapsto\operatorname{Cov}(\Log_{m_x}Y\mid X=z)\) continuous.
Thus this covariance is bounded below by \(cI\) on the kernel supports
for small \(h\). Conditioning \eqref{eq:supp_generic_oracle_score} gives
\[
 \operatorname{Cov}(\varphi_{\star,n}(W;x)\mid X=z)
 =h^{2p}\{L_{1,h}(z;x)-\mathds{1}\{\star=\rbc\}L_{\bias,h,b}(z;x)\}^2
 \operatorname{Cov}(\Log_{m_x}Y\mid X=z).
\]
Reproduction in
\eqrefs{eq:supp-population-polynomial-reproduction,eq:supp-bias-weight-reproduction},
Cauchy--Schwarz, and the support bound yield
\[
\begin{aligned}
 1
 =\left(\E[L_{1,h}-\mathds{1}\{\star=\rbc\}L_{\bias,h,b}]\right)^2
 &\le\mathbb P(d_{\cX}(X,x)\le R_\ast h)
 \E[\{L_{1,h}-\mathds{1}\{\star=\rbc\}L_{\bias,h,b}\}^2]\\
 &\lesssim h^p\E[\{L_{1,h}-\mathds{1}\{\star=\rbc\}L_{\bias,h,b}\}^2].
\end{aligned}
\]
Since \(P\varphi_{\star,n}=0\), total covariance and boundedness of
\(A_x\) give
\begin{equation}\label{eq:supp-oracle-covariance-lower-bound}
{\begin{aligned}
 \Omega_{\star,n}(x)
 &\succeq h^{-p}A_x^{-1}
 \E[\operatorname{Cov}(\varphi_{\star,n}\mid X)]\{A_x^{-1}\}^\ast\\
 &\succeq ch^p
 \E[\{L_{1,h}-\mathds{1}\{\star=\rbc\}L_{\bias,h,b}\}^2]
 A_x^{-1}\{A_x^{-1}\}^\ast\succeq cI.
\end{aligned}}
\end{equation}
Joint continuity of the conditional moments in these frames, compactness of
\(\cX_0\), and the uniform support and moment bounds make both spectral
bounds uniform under \Cref{ass:uniform-inference-conditions}.

For every \(\varepsilon>0\),
\eqrefs{eq:supp-pointwise-response-tail,eq:supp_centered_oracle_score_envelope}
and the support bound give, for sufficiently large \(n\),
\[
\begin{aligned}
 &h^{-p}\E\!\left[\|\varphi_{\star,n}\|^2
 \mathds{1}\{\|\varphi_{\star,n}\|>\varepsilon\sqrt{nh^p}\}\right]\\
 &\quad\lesssim\sup_{z\in\cV_x}\E\!\left[
 \{1+d_{\cY}(m_x,Y)^2\}
 \mathds{1}\{d_{\cY}(m_x,Y)>c\varepsilon\sqrt{nh^p}\}\mid X=z
 \right]\to 0.
\end{aligned}
\]
Here \(nh^p\to\infty\). The standardized oracle sum has covariance
\(I\), and \(\Omega_{\star,n}(x)^{-1/2}A_x^{-1}\) is bounded.
Lindeberg--Feller and Cram\'er--Wold therefore give
\eqref{eq:supp-finite-bandwidth-standardized-clt}.

The upper bound and Lindeberg argument use only the retained support
and moment conditions and \eqref{eq:supp-pointwise-response-tail}.
For fixed \(L\) as stated, the projected covariance inverse square root
is bounded, so the same argument gives
\begin{equation}\label{eq:supp-projected-standardized-clt}
 \{L\Omega_{\star,n}(x)L^\ast\}^{-1/2}L\mathfrak Z_{n,\star}(x)
 \dto N(0,I_d),
\end{equation}
without requiring nondegeneracy of the full tangent-space covariance.
\end{proof}

\subsection{Uniform oracle Gaussian approximation}

{For uniform Gaussian approximation, we express the norm of the standardized oracle process as a supremum over scalar projections. Use} the finite cover and the \(C^2\) frame isometries \(E_{x,\ell}:\R^q\to T_{m_x}\cY\) from \Cref{lem:supp_uniform_finite_shrunken_cover}. {The score functions below are set to zero on} \(N_\ell\coloneqq\cV_\ell\times(\cY\setminus\cK_\ell)\). By \Cref{lem:supp_uniform_finite_shrunken_cover}~\textup{(i)}, \(\mathbb P(N_\ell)=0\). {By} \eqrefs{eq:supp-oracle-covariance-upper-bound,eq:supp-oracle-covariance-lower-bound}{, \(\Omega_{\star,n}\) is uniformly positive definite for all sufficiently large \(n\).} For \(\star\in\{\us,\rbc\}\), put
\begin{equation}\label{eq:supp-oracle-score-class-definition}
\begin{aligned}
    f_{\star,n,\ell,x,u}(w)
    &\coloneqq
    \begin{cases}
    h^{-p/2}
    \left\langle
 E_{x,\ell}u,
 \Omega_{\star,n}(x)^{-1/2}A_x^{-1}
 \varphi_{\star,n}(w;x)
    \right\rangle_{m_x},
    & w\notin N_\ell,\\
    0,
    & w\in N_\ell,
    \end{cases}\\
    \cF_{\star,n}
    &\coloneqq \bigcup_{\ell=1}^L
    \left\{
 f_{\star,n,\ell,x,u}:
 x\in\cX_{\ell},\
 u\in\mathbb S^{q-1}
    \right\}.
\end{aligned}
\end{equation}
{Write
\(\left.\cF_{\star,n}\right|_{\cG_n}\) for the subclass obtained by
restricting \(x\) to \(\cG_n\).}
On the event of probability one \(\cE_{n,\cX_0}^{\mathrm{supp}}\),
centering and orthonormality give
\[
 \sup_{f\in\cF_{\star,n}}\mathbb G_n f
 =\sup_{x\in\cX_0}\|\Omega_{\star,n}(x)^{-1/2}\mathfrak Z_{n,\star}(x)\|_{m_x}.
\]

{Under}
\Cref{ass:uniform-inference-conditions}
{without its bandwidth conditions, let \(h\to0\), with \(b\asymp h\) for RRBC. Fix \(\ell\). Square brackets denote coordinates in the fixed smooth orthonormal predictor and response frames and the induced orthonormal design bases. Put}
\[
{[\Log_\omega y]_\alpha
 \coloneqq\langle E_\ell(\omega)e_\alpha,\Log_\omega y\rangle_\omega,
 \qquad\chi_\ell(z,y)\coloneqq\mathds{1}\{z\in\cV_\ell\}\mathds{1}\{y\in\cK_\ell\}.}
\]
{We embed \(f_{\star,n,\ell,x,u}\), \(\star\in\{\us,\rbc\}\), in the family \(\{f_{h,\vartheta}:\vartheta\in\Theta\}\), where}
\[
{f_{h,\vartheta}(z,y)\coloneqq
 h^{-p/2}\chi_\ell(z,y)\sum_{j=1}^J
 K_j\!\left(\frac{d_{\cX}(z,x)}{ht_j}\right)
 Q_{j,h}(z,y;\vartheta),}
\]
\[
{Q_{j,h}(z,y;\vartheta)\coloneqq
 \sum_{|\gamma|\le4}
 \left[\frac{U_x(z)}{ht_j}\right]^\gamma
 \left\{c_{j,\gamma,0}
 +\sum_{\alpha=1}^q c_{j,\gamma,\alpha}[\Log_\omega y]_\alpha\right\}.}
\]
{Here \(J=1\) for US and \(J=2\) for RRBC, with \(t_1=1\). For RRBC, let \(t_2\in[\underline t,\overline t]\), with fixed \(0<\underline t<1<\overline t<\infty\) and \(b/h\in[\underline t,\overline t]\) eventually. The multi-index \(\gamma\) belongs to \(\mathbb N_0^p\). The compact parameter set \(\Theta\) consists of \(x\in\cX_\ell\), the stated ratios, \(\omega\in\overline{\cW_\ell^{\cY}}\) in fixed coordinates, and independently varying coefficients \(c_{j,\gamma,\alpha}\in[-C,C]\), for a fixed \(C\) chosen below. Each summand is set to zero outside its prescribed response domain and kernel support before evaluating its factors, as in}
\eqref{eq:supp-kernel-factor-class}{.}

{After factoring out its kernel, each local polynomial weight has degree at most \(j\), while its residual is the logarithm vector \(\Log_\omega y\) minus another polynomial of degree at most \(j\). Their product therefore has predictor degree at most \(2j\le4\), while remaining affine in \(\Log_\omega y\).}

{Put}
\[
{\beta_\alpha\coloneqq\left\langle E_{x,\ell}u,
 \Omega_{\star,n}(x)^{-1/2}A_x^{-1}E_{x,\ell}e_\alpha\right\rangle_{m_x}.}
\]
{In the next display only, write \(r_1=r_{1,x}(U_x(z)/h)\) and \(r_2=r_{2,x}(U_x(z)/b)\). The weight definitions in \eqrefs{eq:supp-population-bias-weight,eq:main-population-local-linear-weight} and the residual definition in \eqref{eq:supp-regression-residual} give, for all sufficiently large \(n\),}
\begin{equation}\label{eq:supp-oracle-score-fixed-frame-expansion}
{\begin{aligned}
 &f_{\star,n,\ell,x,u}(z,y)\\
 &=h^{-p/2}\chi_\ell(z,y)\Bigg[
 K_1\!\left(\frac{d_{\cX}(z,x)}h\right)
 \sum_{\alpha,k}\beta_\alpha[S_{1,h}(x)^{-1}e_{1,x}]_k[r_1]_k\Bigl\{[\Log_{m_x}y]_\alpha
 -\sum_s[\Theta_{1,h,x}(m_x)]_{\alpha s}[r_1]_s\Bigr\}\\
 &\quad-\frac{\mathds{1}\{\star=\rbc\}}2
 \left(\frac hb\right)^{p+2}K_2\!\left(\frac{d_{\cX}(z,x)}b\right)\\
 &\qquad\times\sum_{\alpha,k}\beta_\alpha[S_{2,b}(x)^{-1}(0,0,M_{2,h}(x))]_k[r_2]_k\Bigl\{[\Log_{m_x}y]_\alpha
 -\sum_s[\Theta_{2,b,x}(m_x)]_{\alpha s}[r_2]_s\Bigr\}\Bigg].
\end{aligned}}
\end{equation}
{Here \(1\le\alpha\le q\), and \(k,s\) range over the coordinates of \(\cR_{j,x}\) in terms containing \(r_j\). For US, omit the entire \(K_2\) contribution. By}
\eqrefs{eq:supp-population-design-bounds,eq:supp-population-regression-coefficient-bounds},
\Cref{lem:supp_local_linear_second_moment}~\textup{(i)},
\Cref{lem:supp_uniform_finite_shrunken_cover}~\textup{(ii)}, {and}
\eqrefs{eq:supp-oracle-covariance-upper-bound,eq:supp-oracle-covariance-lower-bound},
{the coefficient factors in}
\eqref{eq:supp-oracle-score-fixed-frame-expansion}
{satisfy, uniformly for all sufficiently large \(n\),}
\[
{\begin{aligned}
 |\beta_\alpha|
 &\le\|\Omega_{\star,n}(x)^{-1/2}\|_{\mathrm{op}}
       \|A_x^{-1}\|_{\mathrm{op}}\lesssim1,\\
 |[S_{1,h}(x)^{-1}e_{1,x}]_k|
 &\le\|S_{1,h}(x)^{-1}\|_{\mathrm{op}}\lesssim1,\\
 |[S_{2,b}(x)^{-1}(0,0,M_{2,h}(x))]_k|
 &\le\|S_{2,b}(x)^{-1}\|_{\mathrm{op}}
       \|M_{2,h}(x)\|_{\mathrm{HS}}\le\sqrt p\,\|S_{2,b}(x)^{-1}\|_{\mathrm{op}}
       \|M_{2,h}(x)\|_{\mathrm{op}}\lesssim1,\\
 |[\Theta_{j,a,x}(m_x)]_{\alpha s}|
 &\le\|\Theta_{j,a,x}(m_x)\|_{\mathrm{op}}\lesssim1,
 \qquad (h/b)^{p+2}\lesssim1.
\end{aligned}}
\]
{These bounds and the fixed design bases yield a common \(C\), independent of \(n,x,u\), for which the choices \(\omega=m_x\), \(t_2=b/h\) for RRBC, and the coefficients in}
\eqref{eq:supp-oracle-score-fixed-frame-expansion}
{give}
\begin{equation}\label{eq:supp-oracle-score-kernel-class-selection}
{\left\{f_{\star,n,\ell,x,u}:x\in\cX_\ell,\ u\in\mathbb S^{q-1}\right\}
 \subseteq\left\{f_{h,\vartheta}:\vartheta\in\Theta\right\}.}
\end{equation}

{We verify the envelope and increment conditions in}
\eqrefs{eq:supp-kernel-factor-bounds,eq:supp-kernel-factor-increment-bound}
{for this larger family. Choose \(C_0>1+\overline t\max_{j\le J}R_j\) and take \(n\) large enough that \(C_0h<\delta_L\). On the enlarged predictor balls, the normalized logarithm coordinates are bounded, with derivatives with respect to the predictor location \(x\) of order \(O(h^{-1})\) and derivatives with respect to the bandwidth ratios of order \(O(1)\). By}
\Cref{lem:supp_uniform_finite_shrunken_cover}~\textup{(iv)},
{the coordinates \([\Log_\omega y]_\alpha\) and their first derivatives with respect to the response center \(\omega\) have envelope \(C\cE_R\). Thus, for every \(1\le j\le J\) and \(\vartheta\in\Theta\), on \(\mathcal D_{j,h}(\vartheta)\),}
\[
 {|Q_{j,h}(z,y;\vartheta)|\lesssim\cE_R(y).}
\]
{For \(\vartheta,\vartheta'\in\Theta\) with \(d_h(\vartheta,\vartheta')\le1\), on \(\mathcal D_{j,h}(\vartheta)\cap\mathcal D_{j,h}(\vartheta')\),}
\[
 {|Q_{j,h}(z,y;\vartheta)-Q_{j,h}(z,y;\vartheta')|
 \lesssim\cE_R(y)d_h(\vartheta,\vartheta').}
\]
{Consequently, this family has the form in}
\eqref{eq:supp-kernel-factor-class}
{with \(\chi_j=\chi_\ell\), \((x_{j1},K_{j1},t_{j1})=(x,K_j,t_j)\), and}
\[
{(a,M_a,H,m,s_\nu)=(h,h^{-p/2},\cE_R,1,1).}
\]
{The uniform moment condition gives}
\eqref{eq:supp-response-envelope-moment}{. The oracle scores are indexed through a deterministic Borel map \((x,u)\mapsto\vartheta\in\Theta\), since their parameter coordinates are obtained from Borel integrals by operations continuous on the stated domains. Taking the finite union over \(\ell\) and restricting \(x\) to \(\cG_n\) preserve these bounds, omitting empty frame domains.}

\begin{remark}
\label{rem:supp-unstandardized-oracle-score-class}
Let \(\mathcal G_{\star,n}\) be defined by omitting
\(\Omega_{\star,n}(x)^{-1/2}\) in
\eqref{eq:supp-oracle-score-class-definition}.
{Under the corresponding uniform regime,
\(\mathcal G_{\star,n}\) is a jointly Borel parametrized class of the form}
\eqref{eq:supp-kernel-factor-class}
{with \((a,M_a,H,m)=(h,h^{-p/2},\cE_R,1)\).
It has envelope
\(Ch^{-p/2}\cE_R(y)\mathds{1}\{z\in\cO_{\cX}\}\) and satisfies
\(Pg=0\) and \(\sup_{g\in\mathcal G_{\star,n}}Pg^2\lesssim1\).
The class conditions follow from}
\eqrefs{eq:supp-oracle-score-fixed-frame-expansion,eq:supp-oracle-score-kernel-class-selection}
{with \(\beta_\alpha=\langle E_{x,\ell}u,A_x^{-1}E_{x,\ell}e_\alpha\rangle_{m_x}\).}
The envelope, zero means, and second-moment bound follow from
\eqref{eq:supp_centered_oracle_score_envelope}, \eqref{eq:supp-residual-normal-equations} and
\eqref{eq:supp-oracle-covariance-upper-bound}, respectively.
{
They require no positive lower bound on
\(\lambda_{\min}(\Omega_{\star,n}(x))\).}
Moreover,
\[
 \sup_x\|\mathfrak Z_{n,\star}(x)\|
 =O_p(\sqrt{\log(1/h)}),\qquad
 \sup_x\|\mathfrak Z_{n,\star}^{\ast o}(x)\|
 =O_p^\ast(\sqrt{\log(1/h)}).
\]
The maximal bounds follow from
\Cref{lem:supp_localized_finite_moment_maximal_inequality}
and the representation of the norm as a supremum over unit directions.
\end{remark}

For \(\star\in\{\us,\rbc\}\), use the oracle multiplier process in
\eqref{eq:supp-generic-oracle-linear-process}.
Let \(\{G_{\star,n}f:f\in\cF_{\star,n}\}\), with
\(\cF_{\star,n}\) as in \eqref{eq:supp-oracle-score-class-definition},
be a separable centered Gaussian process with
\(\E[G_{\star,n}f\,G_{\star,n}g]=P(fg)\).
{Define}
\[
 S_{n,\star}^{o}(\cG_n)\coloneqq \sup_{f\in\left.\cF_{\star,n}\right|_{\cG_n}}\mathbb G_n f,\quad
 S_{n,\star}^{G}(\cG_n)\coloneqq \sup_{f\in\left.\cF_{\star,n}\right|_{\cG_n}}G_{\star,n}f,\quad
 S_{n,\star}^{\ast o}(\cG_n)\coloneqq \sup_{f\in\left.\cF_{\star,n}\right|_{\cG_n}}\mathbb G_n^\ast f.
\]
Omit \((\cG_n)\) when \(\cG_n=\cX_0\).

\begin{theorem}
\label{thm:supp_oracle_cck_gaussian_approximation_uniform_wald}
Fix \(\star\in\{\us,\rbc\}\) and suppose the corresponding uniform regime
holds.
{Then}
\begin{equation}
\sup_{t\in\R}\left|\mathbb P(S_{n,\star}^{o}(\cG_n)\le t)-\mathbb P(S_{n,\star}^{G}(\cG_n)\le t)\right|\to0,\quad
\sup_{t\in\R}\left|\mathbb P^\ast(S_{n,\star}^{\ast o}(\cG_n)\le t)-\mathbb P(S_{n,\star}^{G}(\cG_n)\le t)\right|\pto0.
\end{equation}
Moreover,
\begin{equation}
\label{eq:supp-oracle-supremum-size}
S_{n,\star}^{o}(\cG_n)
=O_p(\sqrt{\log(1/h)}),\quad
S_{n,\star}^{G}(\cG_n)=O_p(\sqrt{\log(1/h)}),\quad
S_{n,\star}^{\ast o}(\cG_n)
=O_p^\ast (\sqrt{\log(1/h)}).
\end{equation}
{The distribution function approximations also hold after squaring.
The constants in the bounds do not depend on \(\cG_n\). In particular,}
\begin{equation}
\label{eq:supp_oracle_gaussian_anticoncentration}
\sup_{t\in\R}\mathbb P(|S_{n,\star}^{G}(\cG_n)-t|\le\varepsilon)
\lesssim \varepsilon\sqrt{\log(1/h)},\qquad \varepsilon>0.
\end{equation}
\end{theorem}
\begin{proof}
{By}
\eqrefs{eq:supp-oracle-covariance-upper-bound,eq:supp-oracle-covariance-lower-bound}
{and}
\eqref{eq:supp_centered_oracle_score_envelope}{, the score class has envelope}
\[
{C_Bh^{-p/2}\cE_R(y)\mathds{1}\{z\in\cO_{\cX}\},}
\]
{for a fixed \(C_B>0\). For \(f=f_{\star,n,\ell,x,u}\),}
\eqrefs{eq:supp-residual-normal-equations,eq:rbc-asymptotic-covariance}
{give}
\[
{Pf=0,\qquad
Pf^2=u^\top E_{x,\ell}^\ast\Omega_{\star,n}(x)^{-1/2}
\Omega_{\star,n}(x)\Omega_{\star,n}(x)^{-1/2}E_{x,\ell}u=1.}
\]
The nonempty class
\(\left.\cF_{\star,n}\right|_{\cG_n}\) is symmetric by
\eqref{eq:supp-oracle-score-class-definition}.
{The class inclusion in}
\eqref{eq:supp-oracle-score-kernel-class-selection}
{and the bandwidth conditions in}
\eqref{eq:supp-uniform-bandwidth-rate}
{therefore verify the hypotheses of}
\Cref{prop:cck-multiplier-bounded-envelope}
{for this class with
\((a,M_a,H,m)=(h,h^{-p/2},\cE_R,1)\).}
Choose \(G_{\star,n}\) from the application of the proposition to the entire class and use its restrictions.
With \(\log n\asymp\log(1/h)\), {this gives the distributional approximations, including after squaring, and the order and anti-concentration bounds for \(S_{n,\star}^G(\cG_n)\).
The bounds on the empirical and multiplier suprema follow from these distributional approximations and the bound on \(S_{n,\star}^G(\cG_n)\), conditionally in probability for the multiplier supremum.}
\end{proof}

\subsection{Inference from the linear representation}
\label{sec:supp-modular-inference}

Fix \(\star\in\{\us,\rbc\}\). {Let \(\widehat V_\star(x)\) denote a generic covariance estimator on \(T_{\widehat m_\star(x)}\cY\), assumed positive semidefinite and measurable with respect to the data. Its parallel transport to \(T_{m_x}\cY\) is}
\[
\widehat V_\star^{\parallel}\coloneqq\Pi_{\widehat m_\star\to m_x}\widehat V_\star\Pi_{m_x\to\widehat m_\star}.
\]

\begin{proposition}
\label{prop:supp-generic-pointwise-wald}
Suppose \eqref{eq:supp_pointwise_intrinsic_oracle_representation} holds,
\(cI\preceq\Omega_{\star,n}(x)\preceq CI\), and the oracle CLT
\eqref{eq:supp-finite-bandwidth-standardized-clt} holds. If
\begin{equation}\label{eq:supp-generic-pointwise-covariance}
 \|\widehat V_\star^{\parallel}(x)-\Omega_{\star,n}(x)\|_{\mathrm{op}}
 =o_p(1),
\end{equation}
then \(\widehat V_\star(x)\succ0\) with probability tending to one and
\[
 nh^p\left\langle\Log_{\widehat m_\star(x)}m_x,
 \widehat V_\star(x)^{-1}\Log_{\widehat m_\star(x)}m_x\right\rangle
 \dto\chi_q^2.
\]
\end{proposition}
\begin{proof}
\Cref{lem:supp_covariance_transfer_spectral_stability} gives
\(\widehat V_\star\succ0\) with probability tending to one and
\(\|\Omega_{\star,n}^{1/2}(\widehat V_\star^{\parallel})^{-1}
\Omega_{\star,n}^{1/2}-I\|_{\mathrm{op}}=o_p(1)\).
Since \(\Pi_{\widehat m_\star\to m_x}\Log_{\widehat m_\star}m_x
=-\Log_{m_x}\widehat m_\star\),
\eqrefs{eq:supp_pointwise_intrinsic_oracle_representation,eq:supp-finite-bandwidth-standardized-clt}
and Slutsky's theorem give the claim.
\end{proof}

{For uniform inference, let \(\cG_n\) be a nonempty deterministic
standard Borel space, with deterministic target \(m_x\in\cY\) and a centered
oracle field \(\mathfrak Z_{n,\star}(x)\in T_{m_x}\cY\) with covariance
\(\Omega_{\star,n}(x)\). Assume joint Borel measurability of the target,
estimators, and fields. For the multiplier assertion, also let
\(\mathfrak Z_{n,\star}^{\ast o}\) and \(\widetilde{\mathfrak Z}_{n,\star}^{\ast}\)
be fields in \(T_{m_x}\cY\). Set}
\[
{\begin{aligned}
S_n^o&\coloneqq\sup_{x\in\cG_n}\|\Omega_{\star,n}^{-1/2}\mathfrak Z_{n,\star}\|,\quad
\widehat S_n\coloneqq\sup_{x\in\cG_n}\|\{\widehat V_\star^\parallel\}^{-1/2}\sqrt{nh^p}\Log_{m_x}\widehat m_\star\|,\\
S_n^{\ast o}&\coloneqq\sup_{x\in\cG_n}\|\Omega_{\star,n}^{-1/2}\mathfrak Z_{n,\star}^{\ast o}\|,\quad
\widehat S_n^\ast\coloneqq\sup_{x\in\cG_n}\|\{\widehat V_\star^\parallel\}^{-1/2}\widetilde{\mathfrak Z}_{n,\star}^\ast\|.
\end{aligned}}
\]

\begin{theorem}
\label{thm:supp-modular-inference-transfer}
{Under the preceding setup, suppose \(cI\preceq\Omega_{\star,n}(x)\preceq CI\)
uniformly over \(x\in\cG_n\) for all sufficiently large \(n\), and}
\[
{\sup_{x\in\cG_n}\|\sqrt{nh^p}\Log_{m_x}\widehat m_\star(x)
                  -\mathfrak Z_{n,\star}(x)\|
=o_p(\{\log(1/h)\}^{-1/2}).}
\]
{Assume}
\begin{equation}\label{eq:supp-generic-uniform-covariance}
{\sup_{x\in\cG_n}
\|\widehat V_\star^\parallel(x)-\Omega_{\star,n}(x)\|_{\mathrm{op}}
=o_p(\{\log(1/h)\}^{-1}).}
\end{equation}
{Let \(S_n^G\) be a nonnegative Gaussian supremum satisfying}
\[
{S_n^G=O_p(\sqrt{\log(1/h)}),\qquad
\sup_t\mathbb P(|S_n^G-t|\le\varepsilon)
\lesssim\varepsilon\sqrt{\log(1/h)},\qquad\varepsilon>0.}
\]
{For the multiplier assertion, additionally assume that
\(\mathfrak Z_{n,\star}^{\ast o}\) and \(\widetilde{\mathfrak Z}_{n,\star}^{\ast}\)
are conditionally centered Gaussian fields and}
\begin{equation}\label{eq:supp-generic-generator-replacement}
{\sup_{x\in\cG_n}\|
\widetilde{\mathfrak Z}_{n,\star}^\ast(x)-\mathfrak Z_{n,\star}^{\ast o}(x)\|
=o_p^\ast(\{\log(1/h)\}^{-1/2}).}
\end{equation}
{Consider the following distributional approximations:}
\begin{equation}\label{eq:supp-generic-oracle-law}
{\sup_t|\mathbb P(S_n^o\le t)-\mathbb P(S_n^G\le t)|\to0,\qquad
\sup_t|\mathbb P^\ast(S_n^{\ast o}\le t)-\mathbb P(S_n^G\le t)|\pto0.}
\end{equation}
{Then \(\inf_{x\in\cG_n}\lambda_{\min}(\widehat V_\star^\parallel(x))\ge c/2\)
with probability tending to one, and the following implications hold.}

{\textup{(i)} The unconditional approximation in}
\eqref{eq:supp-generic-oracle-law}
{implies the corresponding approximation in}
\eqref{eq:supp-generic-feasible-laws}{.}

{\textup{(ii)} Under the additional multiplier assumptions,
the conditional approximation in}
\eqref{eq:supp-generic-oracle-law}
{implies the corresponding approximation in}
\eqref{eq:supp-generic-feasible-laws}{:}
\begin{equation}\label{eq:supp-generic-feasible-laws}
{\sup_t|\mathbb P(\widehat S_n\le t)-\mathbb P(S_n^G\le t)|\to0,\qquad
\sup_t|\mathbb P^\ast(\widehat S_n^\ast\le t)-\mathbb P(S_n^G\le t)|\pto0.}
\end{equation}
{Each implication also holds after squaring the statistics.
When both approximations in}
\eqref{eq:supp-generic-feasible-laws}
{hold,}
\Cref{lem:supp_uniform_random_quantile_calibration}
{gives generalized quantile and Monte Carlo calibration.}
\end{theorem}
\begin{proof}
{The oracle distributional approximations and the bound on \(S_n^G\) give}
\(S_n^o=O_p(\sqrt{\log(1/h)})\) and
\(S_n^{\ast o}=O_p^\ast(\sqrt{\log(1/h)})\). The covariance upper bound yields
\[
 \sup_{x\in\cG_n}\|\mathfrak Z_{n,\star}\|\le\sqrt C\,S_n^o,\qquad
 \sup_{x\in\cG_n}\|\mathfrak Z_{n,\star}^{\ast o}\|\le\sqrt C\,S_n^{\ast o}.
\]
By \Cref{lem:supp_covariance_transfer_spectral_stability} and
\eqref{eq:supp-generic-uniform-covariance}, the stated eigenvalue bound
holds, and, on an event with probability tending to one,
\[
{\sup_{x\in\cG_n}\|\{\widehat V_\star^\parallel\}^{-1/2}
                      -\Omega_{\star,n}^{-1/2}\|_{\mathrm{op}}
 \lesssim\sup_{x\in\cG_n}\|\widehat V_\star^\parallel-\Omega_{\star,n}\|_{\mathrm{op}},\quad \sup_{x\in\cG_n}\|\{\widehat V_\star^\parallel\}^{-1/2}\|_{\mathrm{op}}
 \le\sqrt{2/c}.}
\]
The linear representation,
\eqrefs{eq:supp-generic-uniform-covariance,eq:supp-generic-generator-replacement},
and the preceding bounds give
\[
\begin{aligned}
 |\widehat S_n-S_n^o|
 &\lesssim\sup_{x\in\cG_n}\|\sqrt{nh^p}\Log_{m_x}\widehat m_\star
                                  -\mathfrak Z_{n,\star}\|+\sup_{x\in\cG_n}\|\widehat V_\star^\parallel-\Omega_{\star,n}\|_{\mathrm{op}}
                \sup_{x\in\cG_n}\|\mathfrak Z_{n,\star}\|
 =o_p(\{\log(1/h)\}^{-1/2}),\\
 |\widehat S_n^\ast-S_n^{\ast o}|
 &\lesssim\sup_{x\in\cG_n}\|\widetilde{\mathfrak Z}_{n,\star}^\ast-\mathfrak Z_{n,\star}^{\ast o}\|+\sup_{x\in\cG_n}\|\widehat V_\star^\parallel-\Omega_{\star,n}\|_{\mathrm{op}}
                \sup_{x\in\cG_n}\|\mathfrak Z_{n,\star}^{\ast o}\|=o_p^\ast(\{\log(1/h)\}^{-1/2}).
\end{aligned}
\]

Choose deterministic \(\varepsilon_n=o(\{\log(1/h)\}^{-1/2})\) such that
\[
 \mathbb P(|\widehat S_n-S_n^o|>\varepsilon_n)\to0,\qquad
 \mathbb P^\ast(|\widehat S_n^\ast-S_n^{\ast o}|>\varepsilon_n)\pto0.
\]
The anti-concentration bound gives
\[
 \sup_t\mathbb P(t-\varepsilon_n<S_n^G\le t+\varepsilon_n)
 \lesssim\varepsilon_n\sqrt{\log(1/h)}\to0.
\]
Equation \eqref{eq:supp-cdf-perturbation-bound}, its conditional version,
and \eqref{eq:supp-generic-oracle-law} prove
\eqref{eq:supp-generic-feasible-laws}. {Nonnegativity gives the same distributional approximations for the squared statistics. Anti-concentration implies that \(S_n^G\) and \((S_n^G)^2\) have continuous distribution functions.}
\Cref{lem:supp_uniform_random_quantile_calibration} gives the generalized
quantile and rank conclusions before and after squaring.
\end{proof}

For US and RRBC under their uniform regimes and the conditions on the population displacements in \Cref{cor:supp-bias-free-score-representations},
\Cref{lem:supp_oracle_residualized_score_covariance_consistency},
\eqref{eq:supp_uniform_intrinsic_oracle_representation}, and
\Cref{thm:supp_oracle_cck_gaussian_approximation_uniform_wald}
{verify the spectral bounds, linear representation, oracle distributional approximations, and the order and anti-concentration bounds for \(S_n^G\), with}
\(S_n^G=S_{n,\star}^G(\cG_n)\). Thus only \eqrefs{eq:supp-generic-uniform-covariance,eq:supp-generic-generator-replacement}
remain to be verified for the feasible construction.

\section{Covariance constructions and Gaussian calibration}
\label{sec:supp-covariance-constructions}

{We verify the covariance and multiplier approximations for the residual-based construction and then transfer them to the LOO construction by comparing the observation-level contributions. }The covariance targets \(\Omega_{\star,n}\) and \(C^{\ell,\ell'}_{\star,n}\) are defined in \eqref{eq:rbc-asymptotic-covariance} and \eqref{eq:supp-oracle-cross-covariance}, respectively, {and the required approximation bounds are \eqrefs{eq:supp-generic-pointwise-covariance,eq:supp-generic-uniform-covariance,eq:supp-generic-generator-replacement}.}

For arrays of vectors \(a_i(x),b_i(x)\in\mathbb R^q\) that are measurable
with respect to the data, the Cauchy--Schwarz inequality gives
\begin{equation}\label{eq:supp-column-second-moment-transfer}
\begin{aligned}
 &\sup_{x,x'}\left\|\frac1{nh^p}\sum_i
       \{a_i(x)a_i(x')^\top-b_i(x)b_i(x')^\top\}\right\|_{\mathrm{op}}^2\\
 &\quad\le2\sup_x\frac1{nh^p}\sum_i\|a_i(x)-b_i(x)\|^2
       \sup_x\frac1{nh^p}\sum_i(\|a_i(x)\|^2+\|b_i(x)\|^2).
\end{aligned}
\end{equation}

\subsection{Residual-based covariance and Gaussian multiplier process}
\label{subsec:supp-residual-covariance}

For \((j,a)=(1,h)\) and, for RRBC with \(b\asymp h\), \((j,a)=(2,b)\), let \(\widehat\varepsilon_{j,i}(x)\) denote the empirical version of \eqref{eq:supp-regression-residual}, evaluated at \(\omega=\widehat m_{h,x}\) and \(W=W_i\). {Define the residual-based contributions, their covariance estimator, and the Gaussian multiplier process by}
\begin{equation}\label{eq:supp-auxiliary-influence}
\begin{aligned}
 \widehat\psi^{\mathrm A}_{\star,i}(x)
 &\coloneqq h^p\Pi_{\widehat m_{h,x}\to\widehat m_{\star,x}}^{\cY}
       \widehat A_{h,x}(\widehat m_{h,x})^{-1}\{\widehat L_{1,h,i}(x)\widehat\varepsilon_{1,i}(x)
 -\mathds{1}\{\star=\rbc\}\widehat L_{\bias,h,b,i}(x)
                         \widehat\varepsilon_{2,i}(x)\},\\
 \widehat\Omega^{\mathrm A}_\star(x)
 &\coloneqq(nh^p)^{-1}\sum_i(\widehat\psi^{\mathrm A}_{\star,i}(x))^{\otimes2},\\
 \widetilde{\mathfrak Z}_{n,\star}^{\ast,\mathrm A}(x)
 &\coloneqq(nh^p)^{-1/2}\sum_i\xi_i
       \Pi_{\widehat m_{\star,x}\to m_x}^{\cY}
       \widehat\psi^{\mathrm A}_{\star,i}(x).
\end{aligned}
\end{equation}
The empirical version of \eqref{eq:supp-residual-normal-equations} gives
\(\sum_i\widehat\psi^{\mathrm A}_{\star,i}(x)=0\).

Write \(\widehat\Omega_\star^{\mathrm A,\parallel}\) for {the transport of \(\widehat\Omega_\star^{\mathrm A}\)} to \(T_{m_x}\cY\). In the frames of \eqref{eq:supp-oracle-cross-covariance}, define
\begin{equation}\label{eq:supp-auxiliary-coordinate-kernel}
 \widehat\zeta^{\mathrm A}_{\star,n,i,\ell}(x)
 \coloneqq h^{-p/2}E_{x,\ell}^\ast\Pi_{\widehat m_{\star,x}\to m_x}^{\cY}
                         \widehat\psi^{\mathrm A}_{\star,i}(x),\qquad
 \widehat C^{\mathrm A,\ell,\ell'}_{\star,n}(x,x')
 \coloneqq P_n[\widehat\zeta^{\mathrm A}_{\star,n,\ell}(x)
              \widehat\zeta^{\mathrm A}_{\star,n,\ell'}(x')^\top].
\end{equation}
{Set the residual-based quantities defined in} \eqrefs{eq:supp-auxiliary-influence,eq:supp-auxiliary-coordinate-kernel} {to zero off the localization event.}

\begin{proposition}
\label{prop:supp-score-replacement}
Under the common conditions {of}
\Cref{sec:supp_score_expansions_covariances},
with \(b\asymp h\) for RRBC:

\noindent\textup{(i) Covariances.}
\begin{align}
 \sup_{x\in\mathcal I_\nu}
 \|\widehat\Omega_\star^{\mathrm A,\parallel}(x)-\Omega_{\star,n}(x)\|_{\mathrm{op}}
 &=\begin{cases}
 o_p(1),&\nu=\mathrm{pt},\\
 O_p(r_{n,\mathrm{unif}}+c_n(h)),&\nu=\mathrm{unif},
 \end{cases}
 \label{eq:supp-auxiliary-covariance-replacement}\\
 \max_{\ell,\ell'\le L}\sup_{\substack{x\in\cX_\ell\\x'\in\cX_{\ell'}}}
 \|\widehat C^{\mathrm A,\ell,\ell'}_{\star,n}(x,x')-C^{\ell,\ell'}_{\star,n}(x,x')\|_{\mathrm{op}}
 &=O_p(r_{n,\mathrm{unif}}+c_n(h)),\qquad \nu=\mathrm{unif}.
 \label{eq:supp-auxiliary-cross-covariance-replacement}
\end{align}
\noindent\textup{(ii) Uniform multipliers.}
\begin{align}
 \sup_{x\in\cX_0}
 \|\widetilde{\mathfrak Z}_{n,\star}^{\ast,\mathrm A}(x)
                         -\mathfrak Z_{n,\star}^{\ast o}(x)\|
 &=O_p^\ast(r_{n,\mathrm{unif}}\sqrt{\log(1/h)}),
 \label{eq:supp-unstandardized-feasible-multiplier-replacement}\\
 \sup_{x\in\cX_0}
 \{\|\mathfrak Z_{n,\star}^{\ast o}(x)\|
               +\|\widetilde{\mathfrak Z}_{n,\star}^{\ast,\mathrm A}(x)\|\}
 &=O_p^\ast(\sqrt{\log(1/h)}).
 \label{eq:supp-unstandardized-multiplier-size}
\end{align}
\end{proposition}

\begin{proof}
\noindent{\textit{Step 1. Replacement of the fitted contributions.}}
Work on the localization event. Use the first derivative bounds for \(\omega\mapsto\Pi_{\omega\to m_x}\Log_\omega Y\) in \Cref{lem:supp_fixed_local_log_domain}~\textup{(ii) and (iii)}, with uniform bounds from {\Cref{lem:supp_uniform_finite_shrunken_cover}~\textup{(iv)}}. Combine these with \eqref{eq:supp-exact-regression-coefficient-identity}, \Cref{lem:supp_localized_support_count}, the {score and design bounds} and center rate established in the proof of \Cref{lem:supp_local_empirical_frechet_linearization_mh}, and \Cref{prop:supp_smoothing_bias_expansion} to obtain, for \((j,a)=(1,h)\) and, for RRBC, \((2,b)\),
\begin{equation}\label{eq:supp-empirical-coefficient-transport-rate}
 \sup_x\|\Pi_{\widehat m_{h,x}\to m_x}\widehat\Theta_{j,a,x}(\widehat m_{h,x})
                    -\Theta_{j,a,x}(m_x)\|_{\mathrm{op}}
 =O_p(r_{n,\nu}).
\end{equation}

By \eqrefs{eq:supp-population-design-bounds,eq:supp-population-regression-coefficient-bounds}, \Cref{lem:supp_local_linear_second_moment}~\textup{(i)}, and the Hessian lower bounds in \Cref{ass:local_support_log_geometry} pointwise and \Cref{lem:supp_uniform_finite_shrunken_cover}~\textup{(ii)} uniformly, choose fixed compact convex coefficient sets containing {fixed-radius neighborhoods of the oracle values}. {To compare the oracle and fitted contributions through the same smooth map, let \(\eta=(s,J,(B_j,T_j)_{j=1,2},M)\) range over these sets in the fixed frames. Here \(s\) specifies the response center \({\Exp_{m_x}^{\cY}[s]}\), while \(J,B_j,T_j,M\) correspond to the {transported inverse Hessian}, inverse design operators, regression coefficient operators, and local linear second-moment operator, respectively.} {Restrict \(s\) to a small ball about zero whose exponential image lies in the prescribed response neighborhood.} For US, omit \(j=2\) and \(M\). {Use \(\iota\) from \Cref{lem:supp_rbc_bias_matching_linearization_mean}. For \(u\in\R^q\) with \(\|u\|=1\), define}
\begin{equation}\label{eq:supp-auxiliary-score-map}
\begin{aligned}
&H_{\star,\ell}((v,y);x,u,\eta)\\
&\quad\coloneqq h^{p/2}\Bigl[K_{1,h}(v,x)\langle B_1r_{1,x}(U_x(v)/h),e_{1,x}\rangle\bigl\langle E_{x,\ell}u,J\{\mathcal L_x(s,y)-T_1r_{1,x}(U_x(v)/h)\}\bigr\rangle\\
&\qquad-\mathds{1}\{\star=\rbc\}\frac{h^2}{2b^2}K_{2,b}(v,x)\langle B_2r_{2,x}(U_x(v)/b),\iota(M)\rangle\bigl\langle E_{x,\ell}u,J\{\mathcal L_x(s,y)-T_2r_{2,x}(U_x(v)/b)\}\bigr\rangle\Bigr].
\end{aligned}
\end{equation}
Summands are zero off their kernel support and corresponding response domain.

The oracle and estimated coefficient blocks, indexed by the same pairs, are
\begin{equation}\label{eq:supp-auxiliary-score-blocks}
\begin{aligned}
\eta_{\star,\ell}(x)&\coloneqq\bigl(0,A_x^{-1},(S_{j,a}^{-1},\Theta_{j,a,x}(m_x))_{(j,a)},M_{2,h}\bigr),\\
\widehat\eta_{\star,\ell}(x)&\coloneqq\Bigl(\Log_{m_x}^{\cY}\widehat m_{h,x},\Pi_{\widehat m_{\star,x}\to m_x}^{\cY}\!\!\Pi_{\widehat m_{h,x}\to\widehat m_{\star,x}}^{\cY}\!\!\widehat A_{h,x}(\widehat m_{h,x})^{-1}\Pi_{m_x\to\widehat m_{h,x}}^{\cY},\\
&\qquad\bigl(\widehat S_{j,a}^{-1},\Pi_{\widehat m_{h,x}\to m_x}^{\cY}\!\!\widehat\Theta_{j,a,x}(\widehat m_{h,x})\bigr)_{(j,a)},\widehat M_{2,h}\Bigr).
\end{aligned}
\end{equation}
Equations
\eqrefs{eq:supp_generic_oracle_score,eq:supp-auxiliary-coordinate-kernel,eq:supp-auxiliary-score-map,eq:supp-auxiliary-score-blocks}
give
\begin{equation}\label{eq:supp-auxiliary-score-map-identities}
 H_{\star,\ell}(W_i;x,u,\widehat\eta_{\star,\ell}(x))
 =u^\top\widehat\zeta^{\mathrm A}_{\star,n,i,\ell}(x),\qquad
 H_{\star,\ell}(w;x,u,\eta_{\star,\ell}(x))
 =u^\top\zeta_{\star,n,\ell}(w;x).
\end{equation}

The composition of the three parallel transports in the estimated \(J\) block is smooth
in the centers and equals \(I\) when they coincide. The rates for the center and estimated Hessian therefore give an \(O_p(r_{n,\nu})\) error in this block relative
to \(A_x^{-1}\). Thus
{\eqrefs{eq:supp-fitted-center-hessian-rates,eq:supp-empirical-coefficient-transport-rate}}
and {\Cref{lem:supp_empirical_design_concentration,lem:supp_local_linear_second_moment}} give
\begin{equation}\label{eq:supp-d3-nuisance-error}
 \max_{\ell\le L}\sup_{x\in\cX_\ell}\|\widehat\eta_{\star,\ell}(x)-\eta_{\star,\ell}(x)\|
 =O_p(r_{n,\nu}).
\end{equation}
{By} \eqref{eq:supp-d3-nuisance-error}, {the joining segments
lie in these sets with probability tending to one.
On this event, the fundamental theorem of calculus in \(\eta\),
with \(w,x,u\) fixed, gives}
\begin{equation}\label{eq:supp-d3-integral-score-difference}
\begin{aligned}
 &H_{\star,\ell}(w;x,u,\widehat\eta_{\star,\ell}(x))
       -H_{\star,\ell}(w;x,u,\eta_{\star,\ell}(x))\\
 &\quad=\int_0^1{(D_\eta H_{\star,\ell})}
       \bigl(w;x,u,(1-t)\eta_{\star,\ell}(x)
                         +t\widehat\eta_{\star,\ell}(x)\bigr)\bigl[\widehat\eta_{\star,\ell}(x)
                         -\eta_{\star,\ell}(x)\bigr]\,dt.
\end{aligned}
\end{equation}

In the following bounds, replace \(G_R\) by \(\cE_R\) in the uniform case.
{The coefficient blocks are bounded on these sets, and the design vectors
are bounded on the kernel supports by}
\eqref{eq:supp-standing-design-vector-bound}.
{Each differentiated summand contains at most one factor
\(\mathcal L_x\) or \(D_s\mathcal L_x\).
The product rule, kernel scaling, and the logarithm bounds in}
\Cref{lem:supp_fixed_local_log_domain}~\textup{(ii) and (iii)} {give}
\begin{equation}
\label{eq:supp-auxiliary-derivative-envelope}
 \bigl|{(D_\eta H_{\star,\ell})}((v,y);x,u,\eta)[d]\bigr|
 \lesssim h^{-p/2}G_R(y)
 \mathds{1}\{d_{\cX}(v,x)\le Ch\}\|d\|.
\end{equation}
Equations
\eqrefs{eq:supp-auxiliary-score-map-identities,eq:supp-d3-nuisance-error,eq:supp-d3-integral-score-difference,eq:supp-auxiliary-derivative-envelope}
and \Cref{lem:supp_localized_support_count}
{give}
\begin{equation}\label{eq:supp-auxiliary-column-second-moments}
\begin{aligned}
 &\left\{\max_{\ell\le L}\sup_{x\in\cX_\ell}P_n\|\widehat\zeta^{\mathrm A}_{\star,n,\ell}(x)
                                     -\zeta_{\star,n,\ell}(x)\|^2\right\}^{1/2}\\
 &\quad\lesssim\max_{\ell\le L}\sup_{x\in\cX_\ell}\|\widehat\eta_{\star,\ell}(x)-\eta_{\star,\ell}(x)\|
 \left\{\sup_xh^{-p}P_n[
 \mathds{1}\{d_{\cX}(X,x)\le Ch\}G_R(Y)^2]\right\}^{1/2}
 =O_p(r_{n,\nu}).
\end{aligned}
\end{equation}
Since \(A_x^{-1}\) is uniformly bounded,
\eqrefs{eq:supp_centered_oracle_score_envelope,eq:supp-auxiliary-column-second-moments} and
\Cref{lem:supp_localized_support_count} give
\[
\begin{aligned}
 &\max_{\ell\le L}\sup_{x\in\cX_\ell}P_n\left[
 \|\widehat\zeta^{\mathrm A}_{\star,n,\ell}(x)\|^2
 +\|\zeta_{\star,n,\ell}(x)\|^2\right]\\
 &\quad\lesssim
 \max_{\ell\le L}\sup_{x\in\cX_\ell}P_n\|\widehat\zeta^{\mathrm A}_{\star,n,\ell}(x)
                    -\zeta_{\star,n,\ell}(x)\|^2
 +\sup_xh^{-p}P_n[
 G_R(Y)^2\mathds{1}\{d_{\cX}(X,x)\le R_\ast h\}]=O_p(1).
\end{aligned}
\]

\noindent{\textit{Step 2. Covariance approximation.}}
Uniformly, apply
\Cref{lem:supp_local_polynomial_kernel_factors_are_localized,lem:supp_localized_finite_moment_maximal_inequality}
with {\((a,M_a,H,m)=(h,h^{-p},\cE_R,2)\)} {to products of score coordinates}. This gives
\[
 \max_{\ell,\ell'\le L}\sup_{\substack{x\in\cX_\ell\\x'\in\cX_{\ell'}}}\|(P_n-P)[\zeta_{\star,n,\ell}(x)
                  \zeta_{\star,n,\ell'}(x')^\top]\|_{\mathrm{op}}
 =O_p(c_n(h)).
\]
Apply \eqref{eq:supp-column-second-moment-transfer} with
\(a_i(x,\ell)=h^{p/2}\widehat\zeta^{\mathrm A}_{\star,n,i,\ell}(x)\) and
\(b_i(x,\ell)=h^{p/2}\zeta_{\star,n,\ell}(W_i;x)\).
Equation \eqref{eq:supp-auxiliary-column-second-moments}, the preceding
total second-moment bound, and the empirical-product bound give
\begin{equation}\label{eq:supp-feasible-cross-covariance-replacement}
\begin{aligned}
 &\max_{\ell,\ell'\le L}\sup_{\substack{x\in\cX_\ell\\x'\in\cX_{\ell'}}}\|\widehat C^{\mathrm A,\ell,\ell'}_{\star,n}(x,x')
                         -C^{\ell,\ell'}_{\star,n}(x,x')\|_{\mathrm{op}}\\
 &\le\max_{\ell,\ell'\le L}\sup_{\substack{x\in\cX_\ell\\x'\in\cX_{\ell'}}}\Bigl\|\widehat C^{\mathrm A,\ell,\ell'}_{\star,n}(x,x')
        -P_n[\zeta_{\star,n,\ell}(x)\zeta_{\star,n,\ell'}(x')^\top]
       \Bigr\|_{\mathrm{op}}+\max_{\ell,\ell'\le L}\sup_{\substack{x\in\cX_\ell\\x'\in\cX_{\ell'}}}\|(P_n-P)[\zeta_{\star,n,\ell}(x)
                                   \zeta_{\star,n,\ell'}(x')^\top]\|_{\mathrm{op}}\\
  &=O_p(r_{n,\mathrm{unif}})+O_p(c_n(h))
 =O_p(r_{n,\mathrm{unif}}+c_n(h)).
\end{aligned}
\end{equation}
Taking \(x'=x\) and \(\ell'=\ell\) proves \eqref{eq:supp-auxiliary-covariance-replacement}
uniformly.

Pointwise, \Cref{lem:supp-pointwise-local-truncation} and
\eqref{eq:supp-pointwise-response-tail} give
\[
  (P_n-P)[\zeta_{\star,n,\ell}(x)\zeta_{\star,n,\ell}(x)^\top]=o_p(1).
\]
The inequality in \eqref{eq:supp-feasible-cross-covariance-replacement} and
\eqref{eq:supp-auxiliary-column-second-moments}, with
\(r_{n,\mathrm{pt}}\to0\), prove
\eqref{eq:supp-auxiliary-covariance-replacement} pointwise.

\noindent{\textit{Step 3. Multiplier approximation.}}
{Equation} \eqref{eq:supp-auxiliary-derivative-envelope} and the logarithm bounds through order two in \Cref{lem:supp_fixed_local_log_domain}~\textup{(ii) and (iii)} control, respectively, the envelope and increments of the smooth factors of the class of derivatives \(D_\eta H[d]\), \(\|d\|\le1\). Apply \Cref{lem:supp_local_polynomial_kernel_factors_are_localized,lem:supp_localized_finite_moment_maximal_inequality} with {\((a,M_a,m)=(h,h^{-p/2},1)\)} to obtain
\begin{equation}\label{eq:supp-auxiliary-derivative-multiplier-bound}
 \max_{\ell\le L}\sup_{\substack{x\in\cX_\ell,\ \|u\|=1\\\eta,\ \|d\|\le1}}
 |\mathbb G_n^\ast {(D_\eta H_{\star,\ell})}(\cdot;x,u,\eta)[d]|
 =O_p^\ast(\sqrt{\log(1/h)}).
\end{equation}
By empirical centering, the representation of the norm as a supremum over unit directions, and
\eqrefs{eq:supp-d3-nuisance-error,eq:supp-d3-integral-score-difference,eq:supp-auxiliary-derivative-multiplier-bound},
\[
\begin{aligned}
 \sup_x\|\widetilde{\mathfrak Z}_{n,\star}^{\ast,\mathrm A}(x)
                            -\mathfrak Z_{n,\star}^{\ast o}(x)\|
 &=\max_{\ell\le L}\sup_{\substack{x\in\cX_\ell\\\|u\|=1}}
 \big|\mathbb G_n^\ast\{H_{\star,\ell}(\cdot;x,u,\widehat\eta_{\star,\ell}(x))
                    -H_{\star,\ell}(\cdot;x,u,\eta_{\star,\ell}(x))\}\big|\\
 &\le\max_{\ell\le L}\sup_{x\in\cX_\ell}\|\widehat\eta_{\star,\ell}(x)-\eta_{\star,\ell}(x)\|
 {\sup_{\substack{\ell\le L,\ \eta,\ \|d\|\le1\\x\in\cX_\ell,\ \|u\|=1}}}
 |\mathbb G_n^\ast {(D_\eta H_{\star,\ell})}(\cdot;x,u,\eta)[d]|\\
 &=O_p^\ast(r_{n,\mathrm{unif}}\sqrt{\log(1/h)}).
\end{aligned}
\]
The preceding bound,
\Cref{rem:supp-unstandardized-oracle-score-class},
and the triangle inequality give
\eqref{eq:supp-unstandardized-multiplier-size}.
\end{proof}

\subsection{Leave-one-out covariance and multiplier construction}
\label{sec:supp_loo_covariance}

Write \(P_n^{(-i)}\coloneqq P_n-n^{-1}\delta_{W_i}\). {This normalization gives the same fits as division by \(n-1\), because the common factor cancels against the inverse design operators.} Superscripts \((-i)\) replace \(P_n\) by \(P_n^{(-i)}\), holding the explicitly specified response centers fixed. Apply \Cref{lem:supp_measurable_empirical_local_frechet_minimizer} with \eqref{eq:supp-measurable-criterion-extension} to the normalized retained sample to select jointly Borel local linear minimizers.

For the covariance estimator in \eqref{eq:main-loo-covariance}, define
\begin{equation}\label{eq:supp-loo-contributions}
 d_{\star,i}(x)
 \coloneqq
 -\Log_{\widehat m_\star(x)}^{\cY}
       \widehat m_\star^{(-i)}(x)
 \in T_{\widehat m_\star(x)}\cY.
\end{equation}
{The minus sign aligns these contributions with the oracle score. It leaves the covariance and the squared multiplier statistic unchanged.}

Let \(\chi_{i,x}\coloneqq \mathds{1}\{d_{\cX}(X_i,x)\le R_\ast h\}\), {with \(R_\ast\) chosen so that all kernel supports used by the procedure
lie within distance \(R_\ast h\) of \(x\).}
All \(x\)-suprema are over \(\mathcal I_\nu\).
Pointwise take \(o=m_x\). Uniformly use the reference
point in \Cref{sec:supp_consistency_localization}.
In \textup{(H)}, use
\[
 (E_i,J_i)\coloneqq
 \begin{cases}
 (G(m_x,R_x;Y_i),J(m_x,R_x;Y_i)),&\nu=\mathrm{pt},\\
 (\cE(o,R;Y_i),J(o,R;Y_i)),&\nu=\mathrm{unif}.
 \end{cases}
\]
In \textup{(C)}, take \(E_i=J_i=1\), absorbing the derivative bounds in
{\Cref{lem:supp_fixed_local_log_domain} \crefplainname{and} \Cref{lem:supp_uniform_finite_shrunken_cover}~\textup{(iv)}}
into constants.

\begin{lemma}
\label{lem:supp_loo_localization}
Under the corresponding pointwise or uniform regime,
\begin{equation}\label{eq:supp-loo-local-moments}
 \sup_x(nh^p)^{-1}\sum_i\chi_{i,x}(E_i^2+J_i)=O_p(1).
\end{equation}
Use the fixed strongly geodesically convex response neighborhoods
\(\cW_x\) pointwise and \(\cW_\ell^{\cY}\), \(x\in\cX_\ell\), uniformly,
as in \eqref{eq:supp-population-hessian-local-approximation}.
For some deterministic \(c>0\), with probability tending to one,
the following hold simultaneously for \(x\in\mathcal I_\nu\) and
\(1\le i\le n\).
\begin{enumerate}[(i)]
\item \(\widehat S_{1,h}^{(-i)}(x)\) and, for RRBC,
\(\widehat S_{2,b}^{(-i)}(x)\) are nonsingular.
\item \(\widehat Q_{h,x}(\cdot)\) and
\(\widehat Q_{h,x}^{(-i)}(\cdot)\) have unique global minimizers in
\(\cW_x\) pointwise and in every \(\cW_\ell^{\cY}\) with
\(x\in\cX_\ell\) uniformly. For \(\omega\) in these response domains,
\begin{equation}\label{eq:supp-loo-hessian-lower-bound}
 \widehat A_{h,x}(\omega)\succeq cI,\qquad
 \widehat A_{h,x}^{(-i)}(\omega)\succeq cI.
\end{equation}
\item For RRBC, \(\widehat m_{\rbc,x}\) and
\(\widehat m_{\rbc,x}^{(-i)}\), \(1\le i\le n\), lie in a common
normal neighborhood of \(m_x\) for each \(x\).
\end{enumerate}
\end{lemma}

\begin{proof}
\noindent{\textit{Step 1. Local moments and deletion design bounds.}}
Pointwise, \eqref{eq:supp-pointwise-split-moments} gives
\[
 {\E}\!\left[(nh^p)^{-1}\sum_i\chi_{i,x}(E_i^2+J_i)\right]
 \lesssim h^{-p}\mathbb P(d_{\cX}(X,x)\le R_\ast h)\lesssim1.
\]
Markov's inequality proves \eqref{eq:supp-loo-local-moments} pointwise.
{Uniformly, \(J_i\le E_i\) and \(E_i\ge1\) imply
\(E_i^2+J_i\le2E_i^2\). The weighted local observation bound in}
\Cref{lem:supp_localized_support_count}
{therefore gives} \eqref{eq:supp-loo-local-moments}{.
Under \textup{(H)},
\(\sup_{z\in\cO_{\cX}}\E[\cE_R(Y)^{2r_0}\mid X=z]<\infty\);
under \textup{(C)}, \(E_i=J_i=1\).}
The argument in \eqref{eq:supp-localized-truncation-probability},
with \(V\) replaced by \(\cO_{\cX}\), also gives
\(\mathbb P(\exists i:X_i\in\cO_{\cX},E_i>t_n)=o(1)\)
in the uniform regime, with \(t_n\) from \Cref{sec:supp-notation}.
Since \(E_i\ge1\), \eqref{eq:supp-loo-local-moments} implies
\begin{equation}\label{eq:supp-loo-local-first-and-max}
\begin{gathered}
 \sup_x(nh^p)^{-1}\sum_i\chi_{i,x}E_i=O_p(1),\\
 \sup_{x,i}\frac{\chi_{i,x}E_i}{nh^p}
 \le(nh^p)^{-1/2}
 \left\{\sup_x(nh^p)^{-1}\sum_i\chi_{i,x}E_i^2\right\}^{1/2}
 =O_p((nh^p)^{-1/2})=o_p(1).
\end{gathered}
\end{equation}

Take \((j,a)=(1,h)\), and also \((2,b)\) for RRBC. Then
\[
 \widehat S_{j,a}^{(-i)}(x)
 =\widehat S_{j,a}(x)
   -n^{-1}K_{j,a}(X_i,x)r_{j,x}(U_x(X_i)/a)^{\otimes2}.
\]
The bounds \(K_{j,a}(X_i,x)\le Ca^{-p}\chi_{i,x}\) and
\(\|r_{j,x}(U_x(X_i)/a)\|\le C\) whenever
\(K_{j,a}(X_i,x)\ne0\), together with \(a\asymp h\), imply
\(\|\widehat S_{j,a}^{(-i)}(x)-\widehat S_{j,a}(x)\|_{\mathrm{op}}
\le C\chi_{i,x}(nh^p)^{-1}\).
By \Cref{lem:supp_empirical_design_concentration}, with probability
tending to one, \(\inf_x\lambda_{\min}(\widehat S_{j,a}(x))
\ge c_{S,j}>0\) simultaneously for the required pairs \((j,a)\).
{Since \(\chi_{i,x}\le1\) and \(nh^p\to\infty\), the preceding perturbation bound ensures that} \Cref{lem:supp_covariance_transfer_spectral_stability} {applies with \(A=\widehat S_{j,a}(x)\), \(B=\widehat S_{j,a}^{(-i)}(x)\), and \(c=c_{S,j}/2\) for all sufficiently large \(n\). This yields, on the same event and simultaneously for the required pairs \((j,a)\) and all \(x,i\),}
\begin{equation}\label{eq:supp-loo-design-path}
\begin{gathered}
 \|\widehat S_{j,a}^{(-i)}-\widehat S_{j,a}\|_{\mathrm{op}}
 +\|(\widehat S_{j,a}^{(-i)})^{-1}-\widehat S_{j,a}^{-1}\|_{\mathrm{op}}
 \le C\chi_{i,x}(nh^p)^{-1},\qquad
 \sup_{x,i}\|(\widehat S_{j,a}^{(-i)})^{-1}\|_{\mathrm{op}}
 \le2c_{S,j}^{-1}.
\end{gathered}
\end{equation}

On this design event, \eqrefs{eq:supp-loo-local-first-and-max,eq:supp-loo-design-path}
and the empirical version of \eqref{eq:supp-regression-coefficients}
give, with \(\omega\) in the response
domains of \Cref{lem:supp_loo_localization},
\begin{equation}\label{eq:supp-loo-regression-coefficient-bounds}
 \sup_{x,i,\omega}\left\{
 \|\widehat\Theta_{j,a,x}(\omega)\|_{\mathrm{op}}
 +\|\widehat\Theta_{j,a,x}^{(-i)}(\omega)\|_{\mathrm{op}}\right\}
\lesssim\sup_x(nh^p)^{-1}\sum_k\chi_{k,x}E_k=O_p(1).
\end{equation}

\noindent{\textit{Step 2. Localization of all deleted minimizers.}}
Fix a compact \(\cC\subset\cY\). The triangle inequality gives
\(\sup_{\omega\in\cC}d_{\cY}(Y,\omega)^2
\le C_{\cC}\{1+d_{\cY}(o,Y)^2\}\).
Pointwise, \eqref{eq:supp-pointwise-response-tail} and
\Cref{lem:supp-pointwise-local-truncation} imply
\(\sup_i\chi_{i,x}\{1+d_{\cY}(o,Y_i)^2\}/(nh^p)=o_p(1)\).
Uniformly under \textup{(H)}, on
\(\{E_i\le t_n\text{ for every }i\text{ with }X_i\in\cO_{\cX}\}\),
\[
 \sup_{x,i}\frac{\chi_{i,x}\{1+d_{\cY}(o,Y_i)^2\}}{nh^p}
 \lesssim\frac{t_n^2}{nh^p}\longrightarrow0
\]
by \eqref{eq:supp-uniform-bandwidth-rate}. Under \textup{(C)}, compactness
of \(\cK_{\cY,x}\) pointwise and of
\(\bigcup_{\ell=1}^L\cK_\ell\) uniformly, using
\Cref{lem:supp_uniform_finite_shrunken_cover}~\textup{(i)} in the uniform
regime, gives
\(\sup_{x,i}\chi_{i,x}\{1+d_{\cY}(o,Y_i)^2\}/(nh^p)
\le C/(nh^p)\to0\).

Apply \eqref{eq:supp-exact-regression-coefficient-identity} with
\(P=P_n\), \(Q=P_n^{(-i)}\), and response
\(d_{\cY}(Y,\omega)^2/2\), and evaluate the coefficient difference at
\(e_{1,x}\). Equation \eqref{eq:supp-loo-design-path} then gives
\[
 \sup_{x,i,\omega\in\cC}
 |\widehat Q_{h,x}^{(-i)}(\omega)-\widehat Q_{h,x}(\omega)|
 \le C_{\cC}\sup_{x,i}
 \frac{\chi_{i,x}\{1+d_{\cY}(o,Y_i)^2\}}{nh^p}
 +\frac{C_{\cC}}{nh^p}\sup_x\frac1{nh^p}\sum_k
 \chi_{k,x}\{1+d_{\cY}(o,Y_k)^2\}.
\]
Since \(\sup_x(nh^p)^{-1}\sum_k\chi_{k,x}
\{1+d_{\cY}(o,Y_k)^2\}=O_p(1)\) by
\eqref{eq:supp-loo-local-moments},
\begin{equation}\label{eq:supp-loo-criterion-path}
\begin{gathered}
 \sup_{x,i}\frac{\chi_{i,x}\{1+d_{\cY}(o,Y_i)^2\}}{nh^p}=o_p(1),\qquad
 \sup_{x,i,\omega\in\cC}|\widehat Q_{h,x}^{(-i)}(\omega)
                          -\widehat Q_{h,x}(\omega)|=o_p(1).
\end{gathered}
\end{equation}
Constant reproduction and
\eqrefs{eq:supp-loo-local-moments,eq:supp-loo-design-path} give
\begin{equation}\label{eq:supp-loo-weighted-distance-moments}
\begin{aligned}
 P_n^{(-i)}\widehat L_{1,h}^{(-i)}&=1,\\
 \sup_{x,i}P_n^{(-i)}\!\left[
 |\widehat L_{1,h}^{(-i)}|\{1+d_{\cY}(o,Y)^2\}\right]
 &\lesssim\sup_x(nh^p)^{-1}
 \sum_k\chi_{k,x}\{1+d_{\cY}(o,Y_k)^2\}=O_p(1).
\end{aligned}
\end{equation}
Replacing \((P_n^{(-i)},\widehat L_{1,h}^{(-i)},\sup_{x,i})\) by
\((P_n,\widehat L_{1,h},\sup_x)\) gives
\eqref{eq:supp-loo-weighted-distance-moments} without deletion.
{On the same design and response-support events,} \Cref{lem:supp_signed_measure_frechet_coercivity} and \eqref{eq:supp-loo-weighted-distance-moments} {show that both \(\widehat Q_{h,x}\) and \(\widehat Q_{h,x}^{(-i)}\) attain their global minima. Apply} \Cref{lem:supp_coercive_compact_argmin_consistency} {with index \((x,i)\), population criterion \(F(\cdot;x)\), and empirical criterion \(\widehat Q_{h,x}^{(-i)}\). Its condition (i) follows from} \eqref{eq:supp-loo-criterion-path} and \Cref{lem:supp_uniform_compact_criterion_approximation}~\textup{(ii)}{, while condition (ii) follows pointwise from continuity of \(F(\cdot;x)\) and uniqueness of \(m_x\), and uniformly from} \eqref{eq:supp-uniform-identification-gap}{. Condition (iii) follows from} \Cref{lem:supp_signed_measure_frechet_coercivity} and \eqref{eq:supp-loo-weighted-distance-moments}{, using boundedness of the centers \(m_x\), which is immediate pointwise and follows from} \Cref{lem:supp_uniform_finite_shrunken_cover}~\textup{(ii)} {uniformly. Hence,}
\begin{equation}\label{eq:supp-loo-global-argmin-localization}
 \sup_{x,i}\sup_{\omega\in\argmin\widehat Q_{h,x}^{(-i)}(\cdot)}
 d_{\cY}(\omega,m_x)=o_p(1).
\end{equation}
Applying the argument to \(\widehat Q_{h,x}\) also gives
\(\sup_x\sup_{\omega\in\argmin\widehat Q_{h,x}(\cdot)}
d_{\cY}(\omega,m_x)=o_p(1)\).

\noindent{\textit{Step 3. Hessian lower bounds and uniqueness.}}
In the Hessian comparisons below, take \(\omega\in\cW_x\) pointwise.
Uniformly take \(x\in\cX_\ell\), \(\omega\in\cW_\ell^{\cY}\),
for every \(\ell=1,\ldots,L\), and include the supremum over \(\ell\).
Recall
\[
{\widehat A_{h,x}(\omega)=P_n[\widehat L_{1,h}\cH_\omega(Y)],\quad
\widehat A_{h,x}^{(-i)}(\omega)=P_n^{(-i)}[\widehat L_{1,h}^{(-i)}\cH_\omega(Y)],\quad
A_{h,x}(\omega)=P[L_{1,h}\cH_\omega(Y)].}
\]
At the same {\(\omega\)},
\[
\widehat A_{h,x}^{(-i)}(\omega)-\widehat A_{h,x}(\omega)
=-n^{-1}\widehat L_{1,h,i}(x)\cH_\omega(Y_i)+P_n^{(-i)}[(\widehat L_{1,h}^{(-i)}-\widehat L_{1,h})\cH_\omega(Y)].
\]
Since \(\|\cH_\omega(Y_i)\|_{\mathrm{op}}\lesssim E_i\) when
\(\chi_{i,x}=1\),
\eqrefs{eq:supp-loo-local-first-and-max,eq:supp-loo-design-path} imply
\[
 \sup_{x,i,\omega}\|\widehat A_{h,x}^{(-i)}(\omega)
                      -\widehat A_{h,x}(\omega)\|_{\mathrm{op}}
 \lesssim\sup_{x,i}\frac{\chi_{i,x}E_i}{nh^p}
 +(nh^p)^{-1}\sup_x(nh^p)^{-1}\sum_k\chi_{k,x}E_k=o_p(1).
\]

Pointwise, for each fixed \(\omega\in\overline{\cW_x}\), the decomposition
in the proof of \Cref{lem:supp_feasible_ll_hessian_replacement} gives
\(\|\widehat A_{h,x}(\omega)-A_{h,x}(\omega)\|_{\mathrm{op}}=o_p(1)\).
On \(T_{m_x}\cY\), \eqref{eq:supp-transported-hessian-lipschitz} gives
\[
\begin{aligned}
 &\bigl\|[\widehat A_{h,x}^{\parallel}(\omega)-A_{h,x}^{\parallel}(\omega)]
 -[\widehat A_{h,x}^{\parallel}(\omega')-A_{h,x}^{\parallel}(\omega')]\bigr\|_{\mathrm{op}}\\
 &\quad\lesssim d_{\cY}(\omega,\omega')
 \{P_n[|\widehat L_{1,h}|\cE_R(Y)]+P[|L_{1,h}|\cE_R(Y)]\}=d_{\cY}(\omega,\omega')O_p(1).
\end{aligned}
\]
Here \(P_n[|\widehat L_{1,h}|\cE_R]=O_p(1)\) by \eqrefs{eq:supp-local-polynomial-weight-envelopes,eq:supp-loo-local-moments} and \Cref{lem:supp_empirical_design_concentration}. The bound \(P[|L_{1,h}|\cE_R]\lesssim1\) is established in the proof of \Cref{lem:supp_feasible_ll_hessian_replacement}. A finite \(\varepsilon\)-net {argument on} \(\overline{\cW_x}\), first letting \(n\to\infty\) and then \(\varepsilon\downarrow0\), gives \(\sup_\omega\|\widehat A_{h,x}(\omega)-A_{h,x}(\omega)\|_{\mathrm{op}}=o_p(1)\). Uniformly, the proof of \Cref{lem:supp_feasible_ll_hessian_replacement}, with \(\omega\) ranging over \(\overline{\cW_\ell^{\cY}}\) on each \(\cX_\ell\), gives \(\sup_{\ell,x,\omega}\|\widehat A_{h,x}(\omega)-A_{h,x}(\omega)\|_{\mathrm{op}} =o_p(1)\).

By
\eqrefs{eq:supp-population-hessian-local-approximation,eq:supp_population_hessian_uniform_lower_bound},
choose a deterministic \(c>0\) such that
\(A_{h,x}(\omega)\succeq3cI\) for all sufficiently small \(h\).
The decomposition
\[
 \widehat A_{h,x}^{(-i)}(\omega)
 =A_{h,x}(\omega)
 +\{\widehat A_{h,x}(\omega)-A_{h,x}(\omega)\}
 +\{\widehat A_{h,x}^{(-i)}(\omega)-\widehat A_{h,x}(\omega)\}
\]
therefore proves \eqref{eq:supp-loo-hessian-lower-bound} simultaneously
on the specified \(\omega\)-domains.
Thus \(\widehat Q_{h,x}(\cdot)\) and
\(\widehat Q_{h,x}^{(-i)}(\cdot)\) are strictly geodesically convex on
\(\cW_x\) pointwise and on \(\cW_\ell^{\cY}\) for
\(x\in\cX_\ell\) uniformly.
Equation \eqref{eq:supp-loo-global-argmin-localization} and the
\(\widehat Q_{h,x}\) argmin bound place all their global minimizers in
these neighborhoods, so each objective function has a unique global minimizer.
By \Cref{lem:supp_measurable_empirical_local_frechet_minimizer},
\(\argmin\widehat Q_{h,x}(\cdot)=\{\widehat m_{h,x}\}\) and
\(\argmin\widehat Q_{h,x}^{(-i)}(\cdot)=\{\widehat m_{h,x}^{(-i)}\}\)
simultaneously over \(x,i\), with probability tending to one.

\noindent{\textit{Step 4. Localization of the RRBC estimators.}}
{For RRBC, \textup{(iii)} follows from \(\sup_x\|\widehat B_{h,b}\|+\sup_{x,i}\|\widehat B_{h,b}^{(-i)}\|=o_p(1)\) and consistency of the local linear fits.} For the LOO fits, \eqref{eq:supp-loo-hessian-lower-bound} and \(b\asymp h\) give
\[
 \sup_{x,i}\|\widehat B_{h,b}^{(-i)}\|
 \lesssim
 \sup_{x,i}\|\widehat\Theta_{2,b,x}^{(-i)}
 (\widehat m_{h,x}^{(-i)})\circ\iota\|_{\mathrm{op}}
 \sup_{x,i}\|\widehat M_{2,h}^{(-i)}\|_{\mathrm{op}},
\]
where \(\iota\) is as in {\Cref{lem:supp_rbc_bias_matching_linearization_mean}}.
{We prove} \eqref{eq:supp-loo-deleted-pilot-bounds}.
Apply \eqref{eq:supp-exact-regression-coefficient-identity}
with \(P=P_n\), \(Q=P_n^{(-i)}\), and response
\(\Log_\omega Y\), first for fixed \(\omega\) in the stated
response domains. Then evaluate at \(\omega=\widehat m_{h,x}\).
Equations \eqrefs{eq:supp-loo-design-path,eq:supp-loo-regression-coefficient-bounds} give
\[
 \sup_{x,i}\left\|
 \{\widehat\Theta_{2,b,x}^{(-i)}(\widehat m_{h,x})
 -\widehat\Theta_{2,b,x}(\widehat m_{h,x})\}\circ\iota
 \right\|_{\mathrm{op}}
 \lesssim\sup_{x,i}\frac{\chi_{i,x}E_i}{nh^p}
 +(nh^p)^{-1}\sup_x(nh^p)^{-1}\sum_k\chi_{k,x}E_k=o_p(1).
\]
The derivative of
\(\omega\mapsto\Pi_{\omega\to\widehat m_{h,x}}\Log_\omega Y_k\)
is bounded by \(CE_k\) on the geodesic from
\(\widehat m_{h,x}\) to \(\widehat m_{h,x}^{(-i)}\), when
\(\chi_{k,x}=1\), pointwise by \Cref{lem:supp_fixed_local_log_domain}
and uniformly over \(x,i,k\) by
{\Cref{lem:supp_uniform_finite_shrunken_cover}~\textup{(iv)}}.
Hence
\[
\begin{aligned}
 \sup_{x,i}\left\|
 \left\{\Pi_{\widehat m_{h,x}^{(-i)}\to\widehat m_{h,x}}
 \widehat\Theta_{2,b,x}^{(-i)}(\widehat m_{h,x}^{(-i)})
 -\widehat\Theta_{2,b,x}^{(-i)}(\widehat m_{h,x})\right\}\circ\iota
 \right\|_{\mathrm{op}}\lesssim\sup_{x,i}d_{\cY}(\widehat m_{h,x}^{(-i)},\widehat m_{h,x})
 \sup_x(nh^p)^{-1}\sum_k\chi_{k,x}E_k=o_p(1)
\end{aligned}
\]
by \eqref{eq:supp-loo-local-first-and-max} and the consistency of
\(\widehat m_{h,x}^{(-i)}\) and \(\widehat m_{h,x}\).
Since parallel transport preserves norms, the two comparisons above and
\eqref{eq:supp-quadratic-block-bound} give the first bound below.
The second follows from \eqref{eq:supp-loo-weighted-distance-moments},
since \(\|U_x(X)/h\|_x\le R_1\) on
\(\{K_{1,h}(X,x)\ne0\}\):
\begin{equation}\label{eq:supp-loo-deleted-pilot-bounds} \sup_{x,i}\|\widehat\Theta_{2,b,x}^{(-i)}
 (\widehat m_{h,x}^{(-i)})\circ\iota\|_{\mathrm{op}}
 =o_p(1),\qquad
 \sup_{x,i}\|\widehat M_{2,h}^{(-i)}\|_{\mathrm{op}}
 \le R_1^2\sup_{x,i}P_n^{(-i)}|\widehat L_{1,h}^{(-i)}|=O_p(1).
\end{equation}

Equation \eqref{eq:supp-loo-hessian-lower-bound} gives
\(\sup_{x,i}\|\widehat A_{h,x}^{(-i)}
(\widehat m_{h,x}^{(-i)})^{-1}\|_{\mathrm{op}}\le c^{-1}\).
Using \(h^2/b^2=O(1)\), \eqref{eq:supp-loo-deleted-pilot-bounds}, and
\(\|\widehat M_{2,h}^{(-i)}\|_{\mathrm{HS}}
\le\sqrt p\|\widehat M_{2,h}^{(-i)}\|_{\mathrm{op}}\), we obtain
\begin{equation}\label{eq:supp-loo-rbc-localization}
\begin{aligned}
 \sup_{x,i}\|\widehat B_{h,b}^{(-i)}\|
 &\lesssim\frac{h^2}{2b^2}
 \sup_{x,i}\|\widehat A_{h,x}^{(-i)}
 (\widehat m_{h,x}^{(-i)})^{-1}\|_{\mathrm{op}}
 \sup_{x,i}\|\widehat\Theta_{2,b,x}^{(-i)}
 (\widehat m_{h,x}^{(-i)})\circ\iota\|_{\mathrm{op}}
 \sup_{x,i}\|\widehat M_{2,h}^{(-i)}\|_{\mathrm{op}}=o_p(1),\\
 \sup_{x,i}d_{\cY}(\widehat m_{\rbc,x}^{(-i)},m_x)
 &\le\sup_{x,i}d_{\cY}(\widehat m_{h,x}^{(-i)},m_x)
 +\sup_{x,i}\|\widehat B_{h,b}^{(-i)}\|=o_p(1).
\end{aligned}
\end{equation}
{By} \eqref{eq:supp_unscaled_feasible_bias_vector_size} and \Cref{prop:supp_pointwise_empirical_smoothed_target_consistency}{,}
\[
{\sup_x d_{\cY}(\widehat m_{\rbc,x},m_x)\le\sup_x d_{\cY}(\widehat m_{h,x},m_x)+\sup_x\|\widehat B_{h,b}(x)\|=o_p(1).}
\]
Thus, with probability tending to one, simultaneously over \(x,i\),
\(\widehat m_{h,x}\), \(\widehat m_{h,x}^{(-i)}\),
\(\widehat m_{\rbc,x}\), and \(\widehat m_{\rbc,x}^{(-i)}\)
belong to \(\overline B_{\cY}(m_x,\rho)\subset\cW_{m_x}\)
pointwise, and to
\(\overline B_{\cY}(m_x,\rho_Y)\subset\cW_\ell^{\cY}\)
for every \(\ell\) with \(x\in\cX_\ell\) uniformly, as in
\eqref{eq:supp-localized-center-containment}.
\end{proof}

Under the conditions of \Cref{lem:supp_loo_localization}, \eqrefs{eq:supp-local-polynomial-weight-envelopes,eq:supp-loo-local-moments,eq:supp-loo-design-path,eq:supp-loo-deleted-pilot-bounds} {also give}
\begin{equation}\label{eq:supp-loo-weighted-derivative-moments}
 \sup_{x,i}P_n^{(-i)}\!\left[
 (|\widehat L_{1,h}^{(-i)}|+|\widehat L_{\bias,h,b}^{(-i)}|)
 \cE_R\right]=O_p(1).
\end{equation}
Pointwise under \textup{(H)},
\(\cE_R(Y_i)=E_i+J_i\le E_i^2+J_i\). Uniformly under
\textup{(H)}, \(\cE_R(Y_i)=E_i\). Under \textup{(C)},
\(\cE_R(Y_i)=1\).
On the design events of \Cref{lem:supp_empirical_design_concentration},
\eqref{eq:supp-local-polynomial-weight-envelopes} and
\Cref{lem:supp_localized_support_count,lem:supp_local_linear_second_moment}
give
\[
 \sup_xP_n[(|\widehat L_{1,h}|+|\widehat L_{\bias,h,b}|)\cE_R]\lesssim
 \left\{1+\sup_x\|\widehat M_{2,h}(x)\|_{\mathrm{op}}\right\}
 \sup_x\frac1{nh^p}\sum_k\chi_{k,x}\cE_R(Y_k)=O_p(1).
\]
For US, omit \(|\widehat L_{\bias,h,b}^{(-i)}|\) in
\eqref{eq:supp-loo-weighted-derivative-moments}, and use
\(\sup_xP_n[|\widehat L_{1,h}|\cE_R]=O_p(1)\).

\begin{lemma}
\label{lem:supp-loo-score-comparison}
Under {the conditions of} \Cref{lem:supp_loo_localization}, there exist nonnegative random variables \(C_n=O_p(1)\) such that, with probability tending to one, the following two bounds hold simultaneously over \(x,i\):
\begin{equation}\label{eq:supp-loo-score-column-bound}
\begin{aligned}
 \|nh^p d_{\star,i}(x)-\widehat\psi^{\mathrm A}_{\star,i}(x)\|
 &\le C_n\chi_{i,x}E_i
       \{\mathds{1}\{\star=\rbc\}r_{n,\nu}+E_i(nh^p)^{-1}\},\\
 nh^p\|d_{\star,i}(x)\|&\le C_n\chi_{i,x}E_i.
\end{aligned}
\end{equation}
\end{lemma}
\begin{proof}
\noindent{\textit{Step 1. Deletion of the local linear fit.}}
Work on the event of \Cref{lem:supp_loo_localization}. The score equation and
\eqref{eq:supp-exact-regression-coefficient-identity} give
\begin{equation}\label{eq:supp-loo-deleted-primary-score}
 P_n^{(-i)}[\widehat L_{1,h}^{(-i)}(X;x)\Log_{\widehat m_{h,x}}Y]
 =-n^{-1}K_{1,h}(X_i,x)\widehat\varepsilon_{1,i}(x)
 \left\langle(\widehat S_{1,h}^{(-i)}(x))^{-1}
 r_{1,x}(U_x(X_i)/h),e_{1,x}\right\rangle.
\end{equation}
By \eqrefs{eq:supp-loo-hessian-lower-bound,eq:supp-loo-design-path,eq:supp-loo-regression-coefficient-bounds,eq:supp-loo-weighted-derivative-moments},
\begin{equation}\label{eq:supp-loo-primary-deletion-bounds}
{\|d_{\us,i}(x)\|\le C_n\chi_{i,x}E_i(nh^p)^{-1},\quad \|\widehat A_{h,x}^{(-i)}(\widehat m_{h,x})
        -\widehat A_{h,x}(\widehat m_{h,x})\|_{\mathrm{op}}
  \le C_n\chi_{i,x}E_i(nh^p)^{-1}.}
\end{equation}
Taylor's formula and \eqref{eq:supp-loo-weighted-derivative-moments} yield
\begin{equation}\label{eq:supp-loo-primary-deletion-taylor}
\begin{aligned}
 &\left\|P_n^{(-i)}[\widehat L_{1,h}^{(-i)}(X;x)\Log_{\widehat m_{h,x}}Y]
       +\widehat A_{h,x}^{(-i)}(\widehat m_{h,x})d_{\us,i}(x)\right\|
 \le C_n\|d_{\us,i}(x)\|^2.
\end{aligned}
\end{equation}
Equations
\eqrefs{eq:supp-loo-deleted-primary-score,eq:supp-loo-primary-deletion-bounds,eq:supp-loo-primary-deletion-taylor,eq:supp-inverse-difference-identity}
therefore give
\begin{equation}\label{eq:supp-loo-primary-score-difference}
 \left\|d_{\us,i}(x)-n^{-1}\widehat A_{h,x}(\widehat m_{h,x})^{-1}
                  \widehat L_{1,h,i}(x)\widehat\varepsilon_{1,i}(x)\right\|
 \le C_n\chi_{i,x}E_i^2(nh^p)^{-2}.
\end{equation}
This and \eqref{eq:supp-auxiliary-influence} prove the US claim.

\noindent{\textit{Step 2. Deletion of the estimated leading smoothing displacement.}}
For RRBC, \eqref{eq:supp-exact-regression-coefficient-identity} at the fixed
center \(\widehat m_{h,x}\) gives
\begin{equation}\label{eq:supp-loo-fixed-center-pilot-deletion}
\begin{aligned}
 \widehat\Theta_{2,b,x}(\widehat m_{h,x})
       -\widehat\Theta_{2,b,x}^{(-i)}(\widehat m_{h,x})
 &=n^{-1}K_{2,b}(X_i,x)\widehat\varepsilon_{2,i}(x)
       \otimes\{(\widehat S_{2,b}^{(-i)}(x))^{-1}r_{2,x}(U_x(X_i)/b)\},\\
 \|\widehat M_{2,h}(x)-\widehat M_{2,h}^{(-i)}(x)\|
 &\le C_n\chi_{i,x}(nh^p)^{-1}.
\end{aligned}
\end{equation}
Together with \eqrefs{eq:supp-quadratic-block-bound,eq:supp-loo-regression-coefficient-bounds}, this implies
\begin{equation}\label{eq:supp-loo-deleted-quadratic-block}
 \sup_{\|M\|\le1}
 \|\widehat\Theta_{2,b,x}^{(-i)}(\widehat m_{h,x})(0,0,M)\|
 \le C_n\{r_{n,\nu}+\chi_{i,x}E_i(nh^p)^{-1}\}.
\end{equation}
{Holding \(\widehat m_{h,x}\) fixed, we obtain}
\begin{equation}\label{eq:supp-loo-fixed-center-bias-decomposition}
\begin{aligned}
 &P_n[\widehat L_{\bias,h,b}(X;x)\Log_{\widehat m_{h,x}}Y]
  -P_n^{(-i)}[\widehat L_{\bias,h,b}^{(-i)}(X;x)\Log_{\widehat m_{h,x}}Y]\\
 &\quad=\frac{h^2}{2b^2}
  \{\widehat\Theta_{2,b,x}(\widehat m_{h,x})
       -\widehat\Theta_{2,b,x}^{(-i)}(\widehat m_{h,x})\}
        (0,0,\widehat M_{2,h}(x))+\frac{h^2}{2b^2}\widehat\Theta_{2,b,x}^{(-i)}(\widehat m_{h,x})
       (0,0,\widehat M_{2,h}(x)-\widehat M_{2,h}^{(-i)}(x)).
\end{aligned}
\end{equation}
By \eqrefs{eq:supp-loo-design-path,eq:supp-loo-fixed-center-pilot-deletion,eq:supp-loo-deleted-quadratic-block,eq:supp-loo-fixed-center-bias-decomposition},
\begin{equation}\label{eq:supp-loo-fixed-center-bias-remainder}
\begin{aligned}
 &\left\|P_n[\widehat L_{\bias,h,b}(X;x)\Log_{\widehat m_{h,x}}Y]
 -P_n^{(-i)}[\widehat L_{\bias,h,b}^{(-i)}(X;x)\Log_{\widehat m_{h,x}}Y]
 -n^{-1}\widehat L_{\bias,h,b,i}(x)\widehat\varepsilon_{2,i}(x)\right\|\\
 &\quad\le C_n\chi_{i,x}E_i
       \{r_{n,\nu}(nh^p)^{-1}+E_i(nh^p)^{-2}\}.
\end{aligned}
\end{equation}

Equation \eqref{eq:supp-bias-weight-empirical-hessian-rate},
\Cref{lem:supp_local_empirical_frechet_linearization_mh}, and
\eqrefs{eq:supp-transported-hessian-lipschitz,eq:supp-loo-weighted-derivative-moments}
give
\[
 \sup_x\|P_n[\widehat L_{\bias,h,b}(X;x)\cH_{\widehat m_{h,x}}(Y)]\|_{\mathrm{op}}
 =O_p(r_{n,\nu}).
\]
{Apply} \eqref{eq:supp-exact-regression-coefficient-identity} {with response \(\cH_{\widehat m_{h,x}}(Y)\), holding \(\widehat m_{h,x}\) fixed. This identity, the bound for \(\widehat M_{2,h}-\widehat M_{2,h}^{(-i)}\) in} \eqref{eq:supp-loo-fixed-center-pilot-deletion}{, and Taylor's formula in the response center give}
\begin{equation}\label{eq:supp-loo-bias-center-change}
\begin{aligned}
 &\|P_n^{(-i)}[\widehat L_{\bias,h,b}^{(-i)}(X;x)
                      \cH_{\widehat m_{h,x}}(Y)]\|
 \le C_n\{r_{n,\nu}+\chi_{i,x}E_i(nh^p)^{-1}\},\\
 &\left\|\Pi_{\widehat m_{h,x}^{(-i)}\to\widehat m_{h,x}}
     P_n^{(-i)}[\widehat L_{\bias,h,b}^{(-i)}(X;x)
                      \Log_{\widehat m_{h,x}^{(-i)}}Y]
       -P_n^{(-i)}[\widehat L_{\bias,h,b}^{(-i)}(X;x)
                      \Log_{\widehat m_{h,x}}Y]\right\|\\
 &\quad\le C_n\{r_{n,\nu}+\chi_{i,x}E_i(nh^p)^{-1}\}\|d_{\us,i}(x)\|
        +C_n\|d_{\us,i}(x)\|^2.
\end{aligned}
\end{equation}
Equations
\eqrefs{eq:supp-transported-hessian-lipschitz,eq:supp-loo-primary-deletion-bounds,eq:supp-loo-deleted-quadratic-block,eq:supp-loo-bias-center-change}
also give
\begin{equation}\label{eq:supp-loo-transported-deletion-bounds}
\begin{aligned}
 &\left\|\Pi_{\widehat m_{h,x}^{(-i)}\to\widehat m_{h,x}}
      \widehat A_{h,x}^{(-i)}(\widehat m_{h,x}^{(-i)})
      \Pi_{\widehat m_{h,x}\to\widehat m_{h,x}^{(-i)}}
         -\widehat A_{h,x}(\widehat m_{h,x})\right\|_{\mathrm{op}}
  \le C_n\chi_{i,x}E_i(nh^p)^{-1},\\
 &\|\widehat B_{h,b,x}^{(-i)}(\widehat m_{h,x}^{(-i)})\|
 \le C_n\{r_{n,\nu}+\chi_{i,x}E_i(nh^p)^{-1}\}.
\end{aligned}
\end{equation}

The full-sample and LOO {estimates of the leading smoothing displacement} satisfy
\[
\begin{aligned}
 \widehat A_{h,x}(\widehat m_{h,x})\widehat B_{h,b}(x)
  &=P_n[\widehat L_{\bias,h,b}(X;x)\Log_{\widehat m_{h,x}}Y],\\
 \widehat A_{h,x}^{(-i)}(\widehat m_{h,x}^{(-i)})
       \widehat B_{h,b,x}^{(-i)}(\widehat m_{h,x}^{(-i)})
  &=P_n^{(-i)}[\widehat L_{\bias,h,b}^{(-i)}(X;x)
                       \Log_{\widehat m_{h,x}^{(-i)}}Y].
\end{aligned}
\]
Transport the second identity to \(T_{\widehat m_{h,x}}\cY\). Adding and subtracting {\(P_n^{(-i)}[\widehat L_{\bias,h,b}^{(-i)}(X;x)\Log_{\widehat m_{h,x}}Y]\)} gives
\[
\begin{aligned}
 &\widehat A_{h,x}(\widehat m_{h,x})
 \left\{\widehat B_{h,b}(x)
 -\Pi_{\widehat m_{h,x}^{(-i)}\to\widehat m_{h,x}}
       \widehat B_{h,b,x}^{(-i)}(\widehat m_{h,x}^{(-i)})\right\}\\
 &\quad=
 \Bigl\{P_n[\widehat L_{\bias,h,b}(X;x)\Log_{\widehat m_{h,x}}Y]
 -P_n^{(-i)}[\widehat L_{\bias,h,b}^{(-i)}(X;x)
                         \Log_{\widehat m_{h,x}}Y]\Bigr\}\\
 &\qquad+
 \Bigl\{P_n^{(-i)}[\widehat L_{\bias,h,b}^{(-i)}(X;x)
                         \Log_{\widehat m_{h,x}}Y]
 -\Pi_{\widehat m_{h,x}^{(-i)}\to\widehat m_{h,x}}
 P_n^{(-i)}[\widehat L_{\bias,h,b}^{(-i)}(X;x)
                         \Log_{\widehat m_{h,x}^{(-i)}}Y]\Bigr\}\\
 &\qquad+
 \Bigl\{\Pi_{\widehat m_{h,x}^{(-i)}\to\widehat m_{h,x}}
 \widehat A_{h,x}^{(-i)}(\widehat m_{h,x}^{(-i)})
 \Pi_{\widehat m_{h,x}\to\widehat m_{h,x}^{(-i)}}
 -\widehat A_{h,x}(\widehat m_{h,x})\Bigr\}
 \Pi_{\widehat m_{h,x}^{(-i)}\to\widehat m_{h,x}}
       \widehat B_{h,b,x}^{(-i)}(\widehat m_{h,x}^{(-i)}).
\end{aligned}
\]
Apply \eqref{eq:supp-loo-fixed-center-bias-remainder} to the first term,
\eqrefs{eq:supp-loo-primary-deletion-bounds,eq:supp-loo-bias-center-change}
to the second, and \eqref{eq:supp-loo-transported-deletion-bounds} to the third.
Since \(\sup_x\|\widehat A_{h,x}(\widehat m_{h,x})^{-1}\|_{\mathrm{op}}\le c^{-1}\)
by \eqref{eq:supp-loo-hessian-lower-bound}, these estimates yield
\begin{equation}\label{eq:supp-loo-correction-score-increment}
\begin{aligned}
  &\left\|\widehat B_{h,b}(x)
       -\Pi_{\widehat m_{h,x}^{(-i)}\to\widehat m_{h,x}}\widehat B_{h,b,x}^{(-i)}(\widehat m_{h,x}^{(-i)})
       -n^{-1}\widehat A_{h,x}(\widehat m_{h,x})^{-1}
                    \widehat L_{\bias,h,b,i}(x)\widehat\varepsilon_{2,i}(x)\right\|\\
  &\quad\le C_n\chi_{i,x}E_i
       \{r_{n,\nu}(nh^p)^{-1}+E_i(nh^p)^{-2}\}.
\end{aligned}
\end{equation}

\noindent{\textit{Step 3. }}{\textit{Effect of deleting an observation on the RRBC estimator.}}
For the RRBC deletion displacement \(d_{\rbc,i}(x)\), use
\(\widehat m_{\rbc,x}=\Exp_{\widehat m_{h,x}}[-\widehat B_{h,b}(x)]\)
and consider, for \(u,v\in T_{\widehat m_{h,x}}\cY\) near zero,
\[
 -\Pi_{\widehat m_{\rbc,x}\to\widehat m_{h,x}}
 \Log_{\widehat m_{\rbc,x}}
 \Exp_{\Exp_{\widehat m_{h,x}}[-u]}
 \left[-\Pi_{\widehat m_{h,x}\to\Exp_{\widehat m_{h,x}}[-u]}
             (\widehat B_{h,b}(x)-v)\right].
\]
At
\[
 (u,v)=\left(d_{\us,i}(x),\widehat B_{h,b}(x)
 -\Pi_{\widehat m_{h,x}^{(-i)}\to\widehat m_{h,x}}
       \widehat B_{h,b,x}^{(-i)}(\widehat m_{h,x}^{(-i)})\right),
\]
this expression equals
\(\Pi_{\widehat m_{\rbc,x}\to\widehat m_{h,x}}d_{\rbc,i}(x)\)
in \(T_{\widehat m_{h,x}}\cY\).
For every sufficiently small \(\widehat B_{h,b}(x)\), it vanishes at \(u=v=0\).
When \(\widehat B_{h,b}(x)=0\), its derivatives in \(u\) and \(v\) at the origin
are \(I\) and \(-I\), respectively.
Joint smoothness in \(\widehat B_{h,b}(x),u,v\) on fixed compact neighborhoods
contained in these local domains
bounds the change in this differential by \(C\|\widehat B_{h,b}(x)\|\)
and bounds the second derivatives in \((u,v)\) uniformly.
Taylor's formula and the isometry of parallel transport therefore give
\begin{equation}\label{eq:supp-loo-rbc-output-linearization}
\begin{aligned}
 &\left\|d_{\rbc,i}(x)-\Pi_{\widehat m_{h,x}\to\widehat m_{\rbc,x}}
       \left(d_{\us,i}(x)-\widehat B_{h,b}(x)
        +\Pi_{\widehat m_{h,x}^{(-i)}\to\widehat m_{h,x}}
                     \widehat B_{h,b,x}^{(-i)}(\widehat m_{h,x}^{(-i)})\right)\right\|\\
  &\quad\lesssim\|\widehat B_{h,b}(x)\|
   \left\{\|d_{\us,i}(x)\|
       +\|\widehat B_{h,b}(x)
          -\Pi_{\widehat m_{h,x}^{(-i)}\to\widehat m_{h,x}}
                       \widehat B_{h,b,x}^{(-i)}(\widehat m_{h,x}^{(-i)})\|\right\}\\
  &\qquad+\left\{\|d_{\us,i}(x)\|
       +\|\widehat B_{h,b}(x)
          -\Pi_{\widehat m_{h,x}^{(-i)}\to\widehat m_{h,x}}
                     \widehat B_{h,b,x}^{(-i)}(\widehat m_{h,x}^{(-i)})\|\right\}^{2}\\
 &\quad\le C_n\chi_{i,x}E_i
       \{r_{n,\nu}(nh^p)^{-1}+E_i(nh^p)^{-2}\}.
\end{aligned}
\end{equation}
The last line uses
\eqrefs{eq:supp_unscaled_feasible_bias_vector_size,eq:supp-loo-primary-score-difference,eq:supp-loo-correction-score-increment}.
Equations
\eqrefs{eq:supp-auxiliary-influence,eq:supp-loo-primary-score-difference,eq:supp-loo-correction-score-increment,eq:supp-loo-rbc-output-linearization} prove
the first assertion of \eqref{eq:supp-loo-score-column-bound}.
The envelope \(\|\widehat\psi^{\mathrm A}_{\star,i}(x)\|\le C_n\chi_{i,x}E_i\)
and \eqref{eq:supp-loo-local-first-and-max} give the second.
All deletion differences vanish outside the kernel supports used by the procedure.
\end{proof}

{Using the LOO contributions in \eqref{eq:supp-loo-contributions} and the frames in \eqref{eq:supp-oracle-cross-covariance}, define the transported LOO covariance estimator, the cross-covariance estimator, and the Gaussian multiplier process by}
\begin{equation}\label{eq:supp-loo-covariance-generator}
\begin{aligned}
 \widehat\Omega_\star^{\parallel}(x)
 &\coloneqq \Pi_{\widehat m_\star(x)\to m_x}^{\cY}\widehat\Omega_\star(x)
   \Pi_{m_x\to\widehat m_\star(x)}^{\cY},\\
 \widehat C^{\ell,\ell'}_{\star,n}(x,x')
 &\coloneqq nh^p\sum_i
  \bigl\{E_{x,\ell}^\ast\Pi_{\widehat m_\star(x)\to m_x}^{\cY}d_{\star,i}(x)\bigr\}
  \bigl\{E_{x',\ell'}^\ast\Pi_{\widehat m_\star(x')\to m_{x'}}^{\cY}
                  d_{\star,i}(x')\bigr\}^\top,\\
 \widetilde{\mathfrak Z}_{n,\star}^\ast(x)
 &\coloneqq \sqrt{nh^p}\sum_i\xi_i
   \Pi_{\widehat m_\star(x)\to m_x}^{\cY}d_{\star,i}(x).
\end{aligned}
\end{equation}
Use the same \(\xi_i\) at every location and zero extensions off the
common localization event. The process \(\mathfrak Z_{n,\star}^{\ast o}\) is unchanged.

\begin{proposition}
\label{prop:supp-loo-replacement}
Under the respective pointwise or uniform conditions of
\Cref{lem:supp_loo_localization},
\begin{equation}\label{eq:supp-loo-covariance-rate}
 \sup_x\|\widehat\Omega_\star^{\parallel}(x)-\Omega_{\star,n}(x)\|_{\mathrm{op}}
 =\begin{cases}o_p(1),&\nu=\mathrm{pt},\\
 O_p(r_{n,\mathrm{unif}}+c_n(h)),&\nu=\mathrm{unif}.
 \end{cases}
\end{equation}
Uniformly,
\begin{align}
 \max_{\ell,\ell'\le L}\sup_{\substack{x\in\cX_\ell\\x'\in\cX_{\ell'}}}\|\widehat C^{\ell,\ell'}_{\star,n}(x,x')-C^{\ell,\ell'}_{\star,n}(x,x')\|_{\mathrm{op}}
 &=o_p(\{\log(1/h)\}^{-1}),
 \label{eq:supp-loo-cross-covariance-rate}\\
 \sup_{x\in\cX_0}\|\widetilde{\mathfrak Z}_{n,\star}^\ast(x)
                    -\mathfrak Z_{n,\star}^{\ast o}(x)\|
 &=o_p^\ast(\{\log(1/h)\}^{-1/2}).
 \label{eq:supp-loo-unstandardized-multiplier-replacement}
\end{align}
\end{proposition}
\begin{proof}
\noindent{\textit{Step 1. Covariance comparison.}}
Use the fixed frames in \eqref{eq:supp-loo-covariance-generator}, and put
\begin{equation}\label{eq:supp-loo-column-error-definition}
 \Delta_{i,\ell}(x)\coloneqq E_{x,\ell}^\ast\Pi_{\widehat m_\star(x)\to m_x}^{\cY}
 \{nh^p d_{\star,i}(x)-\widehat\psi^{\mathrm A}_{\star,i}(x)\}.
\end{equation}
Uniform bounds below use the local truncation event of
\Cref{lem:supp_loo_localization}.
On the common localization event, the isometries of the frames and parallel transport,
\eqref{eq:supp-loo-score-column-bound}, and the triangle inequality in
\(\ell_2\) give
\begin{equation}\label{eq:supp-loo-column-l2-bound}
\begin{aligned}
 &\left\{\max_{\ell\le L}\sup_{x\in\cX_\ell}(nh^p)^{-1}\sum_i\|\Delta_{i,\ell}(x)\|^2\right\}^{1/2}=\left\{\sup_x(nh^p)^{-1}\sum_i
   \|nh^p d_{\star,i}(x)-\widehat\psi^{\mathrm A}_{\star,i}(x)\|^2
   \right\}^{1/2}\\
 &\quad \le C_n\mathds{1}\{\star=\rbc\}r_{n,\nu}
   \left\{\sup_x(nh^p)^{-1}\sum_i\chi_{i,x}E_i^2\right\}^{1/2}+C_n\sup_x(nh^p)^{-3/2}
   \left(\sum_i\chi_{i,x}E_i^4\right)^{1/2}\\
  &\quad \le C_n\mathds{1}\{\star=\rbc\}r_{n,\nu}
   \left\{\sup_x(nh^p)^{-1}\sum_i\chi_{i,x}E_i^2\right\}^{1/2}+C_n(nh^p)^{-1/2}
   \sup_x(nh^p)^{-1}\sum_i\chi_{i,x}E_i^2\\
  &\quad=O_p(r_{n,\nu}).
\end{aligned}
\end{equation}
The last inequality uses \((\sum_i\chi_{i,x}E_i^4)^{1/2}\le\sum_i\chi_{i,x}E_i^2\){. The} last bound follows from \eqref{eq:supp-loo-local-moments} and \((nh^p)^{-1/2}\le r_{n,\nu}\) for all sufficiently large \(n\). Equations \eqrefs{eq:supp-loo-local-moments,eq:supp-loo-score-column-bound} also give \(nh^p\sup_x\sum_i\|d_{\star,i}(x)\|^2=O_p(1)\). Equations \eqrefs{eq:supp-column-second-moment-transfer,eq:supp-auxiliary-column-second-moments,eq:supp-loo-column-l2-bound} yield
\begin{equation}\label{eq:supp-loo-auxiliary-covariance-transfer}
\begin{aligned}
 \sup_x\|\widehat\Omega_\star^{\parallel}(x)
            -\widehat\Omega_\star^{\mathrm A,\parallel}(x)\|_{\mathrm{op}}
 &=O_p(r_{n,\nu}),\\
 \max_{\ell,\ell'\le L}\sup_{\substack{x\in\cX_\ell\\x'\in\cX_{\ell'}}}\|\widehat C^{\ell,\ell'}_{\star,n}(x,x')
            -\widehat C^{\mathrm A,\ell,\ell'}_{\star,n}(x,x')\|_{\mathrm{op}}
 &=O_p(r_{n,\mathrm{unif}}),\qquad \nu=\mathrm{unif}.
\end{aligned}
\end{equation}
Equations
\eqrefs{eq:supp-uniform-rate-consequences,eq:supp-auxiliary-covariance-replacement,eq:supp-auxiliary-cross-covariance-replacement,eq:supp-loo-auxiliary-covariance-transfer}
prove \eqrefs{eq:supp-loo-covariance-rate,eq:supp-loo-cross-covariance-rate}.

\noindent{\textit{Step 2. Reduction of the multiplier bound to a finite net.}}
{To bound the difference between the LOO and residual-based multiplier processes uniformly over \(x\), define the set of scalar coefficient vectors by}
\[
 \mathcal V_n\coloneqq\{0\}\cup\bigcup_{\ell=1}^L
 \left\{\left((nh^p)^{-1/2}\langle u,\Delta_{i,\ell}(x)\rangle\right)_{i=1}^n:
 x\in\cX_\ell,\ \|u\|=1\right\}.
\]
Equation \eqref{eq:supp-loo-column-l2-bound} gives
\begin{equation}\label{eq:supp-loo-array-radius}
 \sup_{f\in\mathcal V_n}\|f\|_2
 \le\left\{\max_{\ell\le L}\sup_{x\in\cX_\ell}(nh^p)^{-1}\sum_i\|\Delta_{i,\ell}(x)\|^2\right\}^{1/2}
 =O_p(r_{n,\mathrm{unif}}).
\end{equation}
For some fixed \(K_0<\infty\), with probability tending to one,
there exists a finite set \(\mathcal F_n\) satisfying
\begin{equation}\label{eq:supp-loo-finite-net}
 \mathcal F_n\subset\mathcal V_n,\qquad
 1\le\#\mathcal F_n\le n^{K_0},\qquad
 \sup_{f\in\mathcal V_n}\min_{g\in\mathcal F_n}\|f-g\|_2
 \le4n^{-2}.
\end{equation}
Here \(\#A\) denotes the cardinality of a finite set \(A\).
We prove \eqref{eq:supp-loo-finite-net} below.
For each fixed data set admitting a set as in
\eqref{eq:supp-loo-finite-net}, take any such \(\mathcal F_n\).

{Write \(\xi\coloneqq(\xi_1,\ldots,\xi_n)^\top\).} Conditional on the data, \(\langle\xi,g\rangle\), \(g\in\mathcal F_n\), are centered Gaussian variables with variances \(\|g\|_2^2\). The finite Gaussian maximum bound, \({\E^\ast[\|\xi\|_2]}\le\sqrt n\), and \eqrefs{eq:supp-loo-array-radius,eq:supp-loo-finite-net} give
\begin{equation}\label{eq:supp-loo-gaussian-maximum}
\begin{aligned}
 {\E^\ast\Bigl[\sup_{f\in\mathcal V_n}|\langle\xi,f\rangle|\Bigr]}
 &\le {\E^\ast\Bigl[\max_{g\in\mathcal F_n}|\langle\xi,g\rangle|\Bigr]}
       +4n^{-2}{\E^\ast[\|\xi\|_2]}\\
 &\le\left(\sup_{f\in\mathcal V_n}\|f\|_2\right)
       \sqrt{2\log(2\#\mathcal F_n)}+4n^{-3/2}\\
 &=O_p(r_{n,\mathrm{unif}}\sqrt{\log n}).
\end{aligned}
\end{equation}
The last bound uses
\(r_{n,\mathrm{unif}}\ge n^{-1/2}\sqrt{\log(1/h)}\) and
\(\log(1/h)\asymp\log n\).
The representation of the norm as a supremum over unit directions, conditional Markov's inequality, and
\eqrefs{eq:supp-uniform-rate-consequences,eq:supp-loo-gaussian-maximum}
yield
\begin{equation}\label{eq:supp-loo-continuum-auxiliary-replacement}
 \sup_{x\in\cX_0}\|\widetilde{\mathfrak Z}_{n,\star}^\ast(x)
                    -\widetilde{\mathfrak Z}_{n,\star}^{\ast,\mathrm A}(x)\|
 =\sup_{f\in\mathcal V_n}|\langle\xi,f\rangle|=O_p^\ast(r_{n,\mathrm{unif}}\sqrt{\log n})=o_p^\ast(\{\log(1/h)\}^{-1/2}).
\end{equation}

\noindent{\textit{Step 3. Control of kernel oscillation within predictor cells.}}
To prove \eqref{eq:supp-loo-finite-net}, cover each \(\cX_\ell\) from \Cref{lem:supp_uniform_finite_shrunken_cover} separately by cells \(C\subset\cX_\ell\), {labeled} by \(\ell\), with \(c_C\in C\), satisfying
\begin{equation}\label{eq:supp-loo-cell-cover}
 \operatorname{diam}(C)\le chn^{-10},\qquad
 \#\{C\}\lesssim(hn^{-10})^{-p}\lesssim n^{1+10p}.
\end{equation}

Use \((j,a)=(1,h)\) for US and
\((j,a)\in\{(1,h),(2,b)\}\) for RRBC.
Since \(a\asymp h\), choose \(c\) in
\eqref{eq:supp-loo-cell-cover} sufficiently small that
\[
 a^{-1}\operatorname{diam}(C)\le n^{-10}.
\]
Apply \Cref{lem:supp_local_polynomial_kernel_factors_are_localized}
separately to the classes
\(\{K_j(d_{\cX}(\cdot,x)/a):x\in C\}\),
with {\((M_a,m,\delta)=(1,0,n^{-10})\)}.
Sum the resulting oscillation envelopes and set the sum to zero
outside \(\{z:d_{\cX}(z,c_C)\le(R_\ast+1)h\}\), where all kernel
differences vanish.
Since \(a\asymp h\), this gives a deterministic measurable
\(w_C\ge0\) satisfying
\begin{equation}
\begin{aligned}
 &\sup_{x\in C}\sum_{(j,a)}
 |K_j(d_{\cX}(z,x)/a)-K_j(d_{\cX}(z,c_C)/a)|
 \le w_C(z),\\
 &{\E}[w_C(X)]\lesssim h^pn^{-10},\qquad w_C(z)=0\quad\text{if }d_{\cX}(z,c_C)>(R_\ast+1)h.
\end{aligned}
\end{equation}
Put \(B_C\coloneqq\{i:w_C(X_i)>n^{-4}\}\).
Markov's inequality, independence of \(X_1,\ldots,X_n\), the bound based on binomial factorial moments, and \eqref{eq:supp-loo-cell-cover} give
\begin{equation}
\begin{aligned}
 \mathbb P(w_C(X)>n^{-4})
 &\le n^4{\E}[w_C(X)]\lesssim h^pn^{-6},\\
 \mathbb P({\#B_C}\ge2p+2)
 &\le{\E\left[\binom{\#B_C}{2p+2}\right]}
 \lesssim\{(nh^p)n^{-6}\}^{2p+2},\\
 \mathbb P(\max_C{\#B_C}\ge2p+2)
 &\lesssim n^{1+10p}\{(nh^p)n^{-6}\}^{2p+2}
 \lesssim n^{-9}\to0,
\end{aligned}
\end{equation}
where the last inequality uses \(nh^p\le n\).

\noindent{\textit{Step 4. Approximation by fits with finitely many free kernel values.}}
On \(\{\max_C{\#B_C}\le2p+1\}\), {let \(v=(v_{j,i})_{(j,a),\,i\in B_C}\) collect the assigned kernel values.} Interpret the design coordinates on the common local support in fixed smooth orthonormal predictor frames and replace
\[
\begin{gathered}
 r_{j,x}(U_x(X_i)/a)
 \longmapsto r_{j,c_C}(U_{c_C}(X_i)/a),\\
 K_j(d_{\cX}(X_i,x)/a)\longmapsto
 \begin{cases}
 K_j(d_{\cX}(X_i,c_C)/a),&i\notin B_C,\\
 v_{j,i},&i\in B_C.
 \end{cases}
\end{gathered}
\]
Retain the factors \(a^{-p}\). For \(x\in C\), set
\[
 v_C(x)\coloneqq
 \bigl(K_j(d_{\cX}(X_i,x)/a)\bigr)_{(j,a),\,i\in B_C}.
\]
{For comparison with the original fits at \(x\in C\), take \(v=v_C(x)\).} For the required pairs \((j,a)\),
\[
\begin{aligned}
 &|K_j(d_{\cX}(X_i,c_C)/a)-K_j(d_{\cX}(X_i,x)/a)|
 \le w_C(X_i)\le n^{-4},\qquad i\notin B_C,\\
 &v_{j,i}=K_j(d_{\cX}(X_i,x)/a),\qquad i\in B_C.
\end{aligned}
\]
Put
\(\chi_{C,i}=\mathds{1}\{d_{\cX}(X_i,c_C)\le(R_\ast+1)h\}\).
For all sufficiently large \(n\) and \(x\in C\),
\[
\begin{aligned}
 &K_j(d_{\cX}(X_i,x)/a)\ne0
 \ \Longrightarrow\ d_{\cX}(X_i,c_C)\le R_\ast h+\operatorname{diam}(C)\le(R_\ast+1)h,\\
 &i\in B_C\ \Longrightarrow\ w_C(X_i)>0\ \Longrightarrow\ \chi_{C,i}=1.
\end{aligned}
\]
Since \(c_C\in C\), both the original and frozen kernels vanish when
\(\chi_{C,i}=0\).
For \(\chi_{C,i}=1\), smoothness in the fixed predictor frames gives
\[
 \|r_{j,x}(U_x(X_i)/a)-r_{j,c_C}(U_{c_C}(X_i)/a)\|
 \le C\frac{d_{\cX}(x,c_C)}a
 \le C\frac{\operatorname{diam}(C)}a\le Cn^{-10}.
\]
For all sufficiently small \(h\),
\eqref{eq:supp-matched-predictor-radius} gives, uniformly over cells,
\[
 c_C\in\cX_\ell\quad\Longrightarrow\quad
 \overline B_{\cX}(c_C,(R_\ast+1)h)\subset\cV_\ell.
\]
The common response envelopes satisfy \(1+E_i^2+J_i\le3E_i^2\).
Thus \Cref{lem:supp_localized_support_count,lem:supp_loo_localization},
applied with the fixed support radius \(R_\ast+1\), give
\[
 \sup_C(nh^p)^{-1}\sum_i\chi_{C,i}(1+E_i^2+J_i)\le3\sup_{z\in\cX_0}(nh^p)^{-1}\sum_i
 \mathds{1}\{d_{\cX}(X_i,z)\le(R_\ast+1)h\}E_i^2=O_p(1).\]
Deleting an observation only decreases these nonnegative sums. All subsequent bounds are uniform over cells, {vectors \(v_C(x)\) with \(x\in C\)}, and full or deleted samples, with a common \(C_n=O_p(1)\).

Overbars on empirical quantities denote recomputation after the displayed substitutions, using the same full or retained sample. Responses and bandwidths are unchanged{. We suppress} \(C\), \(v=v_C(x)\), and deletion superscripts. Recompute \(\overline L_{1,h}\) from the design matrices and \(\overline M_{2,h}\) from the {frozen predictor coordinates} and these weights. With sums over the retained observations,
\[
 \overline Q_h(\omega)
 =\frac1{2n}\sum_k\overline L_{1,h,k}d_{\cY}^2(Y_k,\omega),\qquad
 \overline A_{h,x}(\omega)
 =\Hess_\omega^{\cY}\overline Q_h.
\]

These input errors preserve the design lower bounds in
\Cref{lem:supp_loo_localization}.
The inverse identity \eqref{eq:supp-inverse-difference-identity}
and the weight formula {in \eqref{eq:main-population-local-linear-weight}} give
\[
\begin{aligned}
 &\|\overline S_{j,a}-\widehat S_{j,a}\|_{\mathrm{op}}
 +\|\overline S_{j,a}^{-1}-\widehat S_{j,a}^{-1}\|_{\mathrm{op}}
 \le C_n n^{-4},\\
 &n^{-1}|\overline L_{1,h,i}-\widehat L_{1,h,i}|
 \le C_n n^{-4}\chi_{C,i}(nh^p)^{-1}.
\end{aligned}
\]
On \(\mathcal W_\ell^{\cY}\), {the envelopes for derivatives with respect to the response center \(\omega\)} and
local moments give
\[
 \sup_\omega\left\{
 \|\operatorname{grad}\overline Q_h(\omega)
       -\operatorname{grad}\widehat Q_{h,x}(\omega)\|
 +\|\overline A_{h,x}(\omega)-\widehat A_{h,x}(\omega)\|_{\mathrm{op}}
 \right\}\le C_n n^{-4}.
\]

Choose a common \(c>0\) such that \(\overline A_{h,x}(\omega)\succeq cI\)
on this domain with probability tending to one.
For each \(x\in C\), take \(z_0=\widehat m_{h,x}\) for the full sample and \(z_0=\widehat m_{h,x}^{(-i)}\) for deletion \(i\), using the corresponding frozen objective \(\overline Q_h\). Localization and
\eqref{eq:supp-matched-response-domains} give a fixed \(\rho>0\) with
\(\overline B_{\cY}(z_0,\rho)\subset\mathcal W_\ell^{\cY}\).
For \(d_{\cY}(z,z_0)=\rho\),
\[
 \overline Q_h(z)-\overline Q_h(z_0)
 \ge-C_n n^{-4}\rho+\tfrac c2\rho^2>0.
\]
Minimization on this compact ball gives an interior zero \(\overline m_h\) of the score, unique in \(\mathcal W_\ell^{\cY}\) by strong geodesic convexity{. The} score bound gives \(d_{\cY}(\overline m_h,z_0)\le C_n n^{-4}\).

In these {fixed} frames, the frozen objectives depend only on \(v\). {For two kernel-value vectors \(v=v_C(x)\) and \(v'=v_C(x')\), with \(x,x'\in C\), both frozen centers lie in the same strongly geodesically convex response domain.} The inverse identity and weight formula, followed by strong convexity, give
\[
 c\,d_{\cY}(\overline m_h(v),\overline m_h(v'))
 \le\|\operatorname{grad}\overline Q_{h,v}
                  (\overline m_h(v'))\|\le C_n\|v-v'\|.
\]
The coefficient formula \eqref{eq:supp-regression-coefficients} for the required pairs \((j,a)\), the weight formulas {in \eqrefs{eq:supp-population-bias-weight,eq:main-population-local-linear-weight,eq:supp-empirical-bias-weight}}, derivative envelopes, and the inverse identity extend the \(n^{-4}\) and \(\|v-v'\|\) bounds to \(A^{-1},\Theta_j,M_{2,h},h^pL_{1,h}\) and, for RRBC, \(h^pL_{\bias,h,b}\), after parallel transport where needed. The {formula for \(\widehat B_{h,b,x}\)} {in \eqref{eq:main-empirical-bias-displacement}}, \(h^2/b^2=O(1)\), and smoothness of \(\Exp\) give the corresponding {displacement} and RRBC-center bounds. Together with \eqref{eq:supp-loo-rbc-localization}, these estimates keep the corrected centers in the common logarithm domains.

Define \(\Delta_{C,i}(v,x)\) by \eqref{eq:supp-loo-column-error-definition}, recomputing the full-sample and LOO fits and using frozen predictor coordinates in the residuals, but retaining \(m_x,E_{x,\ell}\). If \(\chi_{C,i}=0\), the full and deleted frozen fits coincide and \(\Delta_{C,i}(v,x)=0\).

The number of free kernel coordinates satisfies
\[
 \dim v\le2{\#B_C}\le2(2p+1).
\]
{The sets}
\[
 {\mathcal A_C\coloneqq\{(v_C(x),x,u):x\in C\cap\cX_0,\ u\in\mathbb S^{q-1}\}}
\]
{are uniformly bounded in fixed Euclidean coordinates.}
In fixed local coordinates for \(x\), the preceding bounds and
smoothness of \(m_x,E_{x,\ell}\) give, for
\((v,x,u),(v',x',u')\in\mathcal A_C\),
\[
 |\langle u,\Delta_{C,i}(v,x)\rangle
       -\langle u',\Delta_{C,i}(v',x')\rangle|\le C_n\chi_{C,i}(nh^p+E_i)
                 \|(v,x,u)-(v',x',u')\|.\]
Since
\[
 \left\{(nh^p)^{-1}\sum_i
       \chi_{C,i}(nh^p+E_i)^2\right\}^{1/2}
 \le C_n(nh^p+1),
\]
there are fixed \(d_0,K<\infty\) such that, with probability tending
to one, simultaneously over cells,
\begin{equation}\label{eq:supp-loo-synthetic-lipschitz}
 \dim(v,x,u)\le d_0,\qquad
 \operatorname{Lip}_{\mathcal A_C}
 \left((nh^p)^{-1/2}
       \langle u,\Delta_{C,i}(v,x)\rangle\right)_{i=1}^n
 \le n^K.
\end{equation}
{Here \(\operatorname{Lip}_{\mathcal A_C}\) denotes the Lipschitz constant on \(\mathcal A_C\), using Euclidean norms.} Indeed, \(nh^p\le n\) permits \(K=2\).

The center and residual bounds above, applied to the frozen and original inputs
at the same \(x\), give
\begin{equation}
 \|\Delta_{i,\ell}(x)-\Delta_{C,i}(v_C(x),x)\|\le C_n n^{-4}
 \mathds{1}\{d_{\cX}(X_i,c_C)\le(R_\ast+1)h\}(nh^p+E_i).
\end{equation}
The enlarged-support count and local moment bounds imply
\begin{equation}\label{eq:supp-loo-array-approximation}
 \sup_C\sup_{x\in C}
 \left\{(nh^p)^{-1}\sum_i
 \|\Delta_{i,\ell}(x)-\Delta_{C,i}(v_C(x),x)\|^2\right\}^{1/2}=O_p((nh^p)n^{-4})=O_p(n^{-3}).
\end{equation}
For \(\|u\|=1\), \(|\langle u,z\rangle|\le\|z\|\), so the same
bound holds for the scalar arrays.

\noindent{\textit{Step 5. Construction of the finite net.}}
Fix data on the localization event, with
\(\max_C{\#B_C}\le2p+1\), satisfying
\eqref{eq:supp-loo-synthetic-lipschitz}, and for which the left-hand side
of \eqref{eq:supp-loo-array-approximation} is at most \(n^{-2}\).
The probability of the intersection tends to one.
Choose a Euclidean \(n^{-(K+2)}\)-net of each \(\mathcal A_C\), with centers
in \(\mathcal A_C\) and at most \(Cn^{(K+2)d_0}\) points.
By \eqrefs{eq:supp-loo-synthetic-lipschitz,eq:supp-loo-array-approximation},
the total approximation error is at most
\[
 n^{-2}+n^K n^{-(K+2)}=2n^{-2}.
\]
Thus the balls of radius \(2n^{-2}\) centered at zero and at the images
of the points in the parameter net cover \(\mathcal V_n\).
By \eqref{eq:supp-loo-cell-cover}, their number is at most
\[
 1+Cn^{1+10p+(K+2)d_0}\le n^{K_0}
\]
for some fixed \(K_0>1+10p+(K+2)d_0\) and all sufficiently large \(n\).
{Discard balls disjoint from \(\mathcal V_n\), and choose one representative in \(\mathcal V_n\) from each remaining ball. The triangle inequality doubles the covering radius to \(4n^{-2}\). Since \(0\in\mathcal V_n\), the representatives form a nonempty set \(\mathcal F_n\subset\mathcal V_n\) with \(\#\mathcal F_n\le n^{K_0}\), proving} \eqref{eq:supp-loo-finite-net}{.}
Equations
\eqrefs{eq:supp-unstandardized-feasible-multiplier-replacement,eq:supp-loo-continuum-auxiliary-replacement}
prove \eqref{eq:supp-loo-unstandardized-multiplier-replacement}.
\end{proof}

By \eqrefs{eq:supp-unstandardized-multiplier-size,eq:supp-loo-unstandardized-multiplier-replacement},
\begin{equation}\label{eq:supp-loo-multiplier-size}
 \sup_{x\in\cX_0}\{\|\widetilde{\mathfrak Z}_{n,\star}^\ast(x)\|
                    +\|\mathfrak Z_{n,\star}^{\ast o}(x)\|\}
 =O_p^\ast(\sqrt{\log(1/h)}).
\end{equation}

The following lemma shows that even a single Newton update yields LOO approximations that preserve covariance consistency and, under the uniform regime, the validity of Gaussian multiplier calibration.

\begin{lemma}
\label{lem:supp-loo-newton-approximation}
Under the conditions of \Cref{lem:supp_loo_localization},
let \(\widetilde m_h^{(-i)}(x)\) denote, in this lemma only, the first Newton iterate \(\widetilde m_{h,1}^{(-i)}(x)\) in \eqref{eq:main-loo-newton-update}.
Set \(\widetilde m_{\us}^{(-i)}(x)\coloneqq\widetilde m_h^{(-i)}(x)\) and construct \(\widetilde m_{\rbc}^{(-i)}(x)\) using \eqref{eq:main-loo-rbc-update}, with the exact full-sample fit \(\widehat m_h(x)\) and all empirical quantities evaluated exactly.
Set \(\widetilde m_\star^{(-i)}(x)=\widehat m_\star(x)\) when observation \(i\) lies outside all kernel supports used by \(\star\). Define \(\widetilde d_{\star,i}(x)\coloneqq -\Log_{\widehat m_\star(x)}^{\cY}\widetilde m_\star^{(-i)}(x)\). Then, with probability tending to one, simultaneously over \(x\in\mathcal I_\nu\) and \(1\le i\le n\),
\begin{equation}\label{eq:supp-loo-newton-column-error}
 \bigl\|\widetilde d_{\star,i}(x)-d_{\star,i}(x)\bigr\|_{\widehat m_\star(x)}
 \le C_n\chi_{i,x}E_i^2(nh^p)^{-2},\qquad C_n=O_p(1).
\end{equation}
{Consequently, these approximations preserve covariance consistency
and, under the uniform regime, the validity of Gaussian multiplier calibration.}
\end{lemma}
\begin{proof}
Work on the localization event of \Cref{lem:supp_loo_localization},
and enlarge \(C_n=O_p(1)\) as needed.
If \(\chi_{i,x}=0\), then
\[
 \widetilde m_\star^{(-i)}(x)=\widehat m_\star^{(-i)}(x)
 =\widehat m_\star(x),\qquad
 \widetilde d_{\star,i}(x)=d_{\star,i}(x)=0.
\]
By \eqrefs{eq:supp-loo-hessian-lower-bound,eq:supp-loo-deleted-primary-score}
and \eqref{eq:main-loo-newton-update},
\[
 d_{\cY}(\widetilde m_h^{(-i)}(x),\widehat m_h(x))
 \le\left\|
 \widehat A_{h,x}^{(-i)}(\widehat m_h(x))^{-1}
 P_n^{(-i)}[\widehat L_{1,h}^{(-i)}\Log_{\widehat m_h(x)}Y]
 \right\|\le C_n\chi_{i,x}E_i(nh^p)^{-1}.
\]
By \Cref{prop:supp_pointwise_empirical_smoothed_target_consistency} and
\eqref{eq:supp-loo-local-first-and-max},
\[
\begin{aligned}
 \sup_{x,i}d_{\cY}(\widetilde m_h^{(-i)}(x),m_x)
 &\le\sup_x d_{\cY}(\widehat m_h(x),m_x)
 +C_n\sup_{x,i}\frac{\chi_{i,x}E_i}{nh^p}=o_p(1).
\end{aligned}
\]
Pointwise shrink \(\rho\) in \eqref{eq:supp-localized-center-containment},
if necessary, so that
\(\overline B_{\cY}(m_x,2\rho)\subset\cW_x\).
Uniformly use \eqref{eq:supp-matched-response-domains}.
Consequently, \(\widehat m_h(x)\), \(\widehat m_h^{(-i)}(x)\), and
\(\widetilde m_h^{(-i)}(x)\) belong to the inner balls in
\eqref{eq:supp-localized-center-containment} simultaneously over \(x,i\),
with probability tending to one.

Using \eqref{eq:supp-loo-local-first-and-max} and the normal domains in
{\Cref{lem:supp_fixed_local_log_domain} \crefplainname{and} \Cref{lem:supp_uniform_finite_shrunken_cover}~\textup{(ii)}}
in \eqref{eq:main-loo-newton-update} gives
\[
 \widetilde d_{\us,i}(x)
 =-\widehat A_{h,x}^{(-i)}(\widehat m_h(x))^{-1}
 P_n^{(-i)}[\widehat L_{1,h}^{(-i)}\Log_{\widehat m_h(x)}Y].
\]
Hence \eqrefs{eq:supp-loo-primary-deletion-bounds,eq:supp-loo-primary-deletion-taylor}
imply
\[
\begin{aligned}
 \|\widetilde d_{\us,i}(x)-d_{\us,i}(x)\|
 &\le c^{-1}\left\|
 P_n^{(-i)}[\widehat L_{1,h}^{(-i)}\Log_{\widehat m_h(x)}Y]
 +\widehat A_{h,x}^{(-i)}(\widehat m_h(x))d_{\us,i}(x)
 \right\|\\
 &\le C_n\|d_{\us,i}(x)\|^2
 \le C_n\chi_{i,x}E_i^2(nh^p)^{-2}.
\end{aligned}
\]
By \eqrefs{eq:supp-loo-local-first-and-max,eq:supp-loo-primary-deletion-bounds},
\({\sup_{x,i}\{\|d_{\us,i}(x)\|+\|\widetilde d_{\us,i}(x)\|\}}=o_p(1)\).
Smoothness of \(\Exp\) on the closed unit disk bundles over
\(\overline{\cW_x}\) pointwise and over
\(\bigcup_{\ell=1}^L\overline{\cW_\ell^{\cY}}\) uniformly therefore gives
\[
 d_{\cY}(\widetilde m_h^{(-i)}(x),\widehat m_h^{(-i)}(x))
 \le C\|\widetilde d_{\us,i}(x)-d_{\us,i}(x)\|
 \le C_n\chi_{i,x}E_i^2(nh^p)^{-2}.
\]

For RRBC, take \(\omega\in\cW_x\) pointwise. Uniformly take \(x\in\cX_\ell\), \(\omega\in\cW_\ell^{\cY}\), and include the supremum over \(\ell=1,\ldots,L\). The {estimated leading smoothing displacement} and its covariant derivative satisfy, for \(v\in T_\omega\cY\),
\[
\begin{aligned}
 \widehat B_{h,b,x}^{(-i)}(\omega)
 &=\widehat A_{h,x}^{(-i)}(\omega)^{-1}
 P_n^{(-i)}[\widehat L_{\bias,h,b}^{(-i)}\Log_\omega Y],\\
 {\nabla_v^{\cY}\widehat B_{h,b,x}^{(-i)}(\omega)}
 &=-\widehat A_{h,x}^{(-i)}(\omega)^{-1}
 ({\nabla_v^{\cY}\widehat A_{h,x}^{(-i)}(\omega)})
 \widehat B_{h,b,x}^{(-i)}(\omega)-\widehat A_{h,x}^{(-i)}(\omega)^{-1}
 P_n^{(-i)}[\widehat L_{\bias,h,b}^{(-i)}\cH_\omega(Y)v].
\end{aligned}
\]
The weights \(\widehat L_{1,h}^{(-i)}\) and
\(\widehat L_{\bias,h,b}^{(-i)}\) do not depend on \(\omega\), and
\({\nabla_v^{\cY}\Log_\omega Y}=-\cH_\omega(Y)v\).
The Hessian lower bound in
\eqref{eq:supp-loo-hessian-lower-bound},
the weighted moment bound in
\eqref{eq:supp-loo-weighted-derivative-moments},
and the derivative bounds in
\Cref{lem:supp_fixed_local_log_domain}~\textup{(ii)} pointwise and
{\Cref{lem:supp_uniform_finite_shrunken_cover}~\textup{(iv)}} uniformly give
\[
 \sup_{x,i,\omega}\left\{
 \|\widehat B_{h,b,x}^{(-i)}(\omega)\|
 +{\sup_{v\in T_\omega\cY,\,\|v\|_\omega\le1}\|\nabla_v^{\cY}\widehat B_{h,b,x}^{(-i)}(\omega)\|_\omega}
 \right\}\le C_n.
\]
Integrating along the geodesic from \(\widehat m_h^{(-i)}(x)\) to
\(\widetilde m_h^{(-i)}(x)\), contained in \(\cW_x\) pointwise and in
\(\cW_\ell^{\cY}\) for \(x\in\cX_\ell\) uniformly, gives
\[
 \left\|
 \widehat B_{h,b,x}^{(-i)}(\widetilde m_h^{(-i)}(x))
 -\Pi_{\widehat m_h^{(-i)}(x)\to\widetilde m_h^{(-i)}(x)}
 \widehat B_{h,b,x}^{(-i)}(\widehat m_h^{(-i)}(x))
 \right\|\le C_n
 d_{\cY}(\widetilde m_h^{(-i)}(x),\widehat m_h^{(-i)}(x)).
\]
Together with \eqref{eq:supp-loo-rbc-localization}, this yields
\[
 \sup_{x,i}\|\widehat B_{h,b,x}^{(-i)}(\widetilde m_h^{(-i)}(x))\|\le\sup_{x,i}\|\widehat B_{h,b,x}^{(-i)}(\widehat m_h^{(-i)}(x))\|
 +C_n\sup_{x,i}
 d_{\cY}(\widetilde m_h^{(-i)}(x),\widehat m_h^{(-i)}(x))=o_p(1).
\]

The points \(\widehat m_h^{(-i)}(x)\) and \(\widetilde m_h^{(-i)}(x)\) lie in \(\overline{\cW_x}\) pointwise and in \(\bigcup_{\ell=1}^L\overline{\cW_\ell^{\cY}}\) uniformly. {The norms of both} {displacement estimates} {are \(o_p(1)\), uniformly over \(x,i\),} so {both estimates} lie in the closed unit disk bundles over these compact sets with probability tending to one. Smoothness of \(\Exp\) on these compact disk bundles, \eqref{eq:main-loo-rbc-update}, and the bound for {the difference between the transported displacement estimates} give
\[
 d_{\cY}(\widetilde m_{\rbc}^{(-i)}(x),\widehat m_{\rbc}^{(-i)}(x))
 \le C_n
 d_{\cY}(\widetilde m_h^{(-i)}(x),\widehat m_h^{(-i)}(x))
 \le C_n\chi_{i,x}E_i^2(nh^p)^{-2}.
\]
Equation \eqref{eq:supp-loo-rbc-localization} and
\(\sup_x d_{\cY}(\widehat m_{\rbc}(x),m_x)=o_p(1)\), established in the
proof of \Cref{lem:supp_loo_localization}, together with the approximation bound
for \(\widetilde m_{\rbc}^{(-i)}(x)\), imply that
\(\widehat m_{\rbc}(x)\), \(\widehat m_{\rbc}^{(-i)}(x)\), and
\(\widetilde m_{\rbc}^{(-i)}(x)\) belong to the inner balls in
\eqref{eq:supp-localized-center-containment} with probability tending to one.
The derivative bound for
\(\omega\mapsto\Log_{\widehat m_{\rbc}(x)}\omega\) in
\Cref{lem:supp_fixed_local_log_domain}~\textup{(iii)} and
{\Cref{lem:supp_uniform_finite_shrunken_cover}~\textup{(iv)}} therefore gives
\[
\begin{aligned}
 \|\widetilde d_{\rbc,i}(x)-d_{\rbc,i}(x)\|
 &=\left\|
 \Log_{\widehat m_{\rbc}(x)}\widehat m_{\rbc}^{(-i)}(x)
 -\Log_{\widehat m_{\rbc}(x)}\widetilde m_{\rbc}^{(-i)}(x)
 \right\|\\
 &\le C\,d_{\cY}
 (\widetilde m_{\rbc}^{(-i)}(x),\widehat m_{\rbc}^{(-i)}(x))
 \le C_n\chi_{i,x}E_i^2(nh^p)^{-2}.
\end{aligned}
\]
{Thus \eqref{eq:supp-loo-newton-column-error} holds for both
procedures. Together with \eqref{eq:supp-loo-local-moments} and
\((\sum_i\chi_{i,x}E_i^4)^{1/2}\le\sum_i\chi_{i,x}E_i^2\), it gives
\[
 \left\{nh^p\sup_x\sum_i
   \|\widetilde d_{\star,i}(x)-d_{\star,i}(x)\|^2\right\}^{1/2}
 =O_p((nh^p)^{-1/2}).
\]
Since \(nh^p\sup_x\sum_i\|d_{\star,i}(x)\|^2=O_p(1)\) by
\eqrefs{eq:supp-loo-local-moments,eq:supp-loo-score-column-bound}, \eqref{eq:supp-column-second-moment-transfer} implies
\[
 \sup_x\left\|nh^p\sum_i
   \{\widetilde d_{\star,i}(x)^{\otimes2}
      -d_{\star,i}(x)^{\otimes2}\}\right\|_{\mathrm{op}}
 =O_p((nh^p)^{-1/2}).
\]
Hence covariance consistency is preserved.
Under the uniform regime,
\[
 \sup_x\left\|\sqrt{nh^p}\sum_i\xi_i
   \{\widetilde d_{\star,i}(x)-d_{\star,i}(x)\}\right\|
 \le C_n\max_i|\xi_i|(nh^p)^{-1/2}
 =O_p^\ast\!\left(\sqrt{\frac{\log n}{nh^p}}\right).
\]
By \eqref{eq:supp-uniform-bandwidth-rate}, the preceding covariance and multiplier error bounds,
together with \Cref{prop:supp-loo-replacement}, verify
\eqrefs{eq:supp-generic-uniform-covariance,eq:supp-generic-generator-replacement};
\Cref{thm:supp-modular-inference-transfer} therefore gives uniform calibration.}
\end{proof}

Under their respective conditions,
\Cref{prop:supp-score-replacement,prop:supp-loo-replacement} give
\eqref{eq:supp-generic-pointwise-covariance} pointwise and, by
\eqref{eq:supp-uniform-rate-consequences},
\eqrefs{eq:supp-generic-uniform-covariance,eq:supp-generic-generator-replacement}
uniformly. For fixed \(\star\) and bandwidths, either the residual-based or the LOO covariance estimator may
be paired with either of the Gaussian multiplier processes defined in \eqrefs{eq:supp-auxiliary-influence,eq:supp-loo-covariance-generator}.

\section{Proofs of the main results}

\begin{proof}[Proof of \Cref{thm:local_linear_expansion_bias}]
For part~\textup{(i)}, \Cref{prop:supp_intrinsic_oracle_representations} with
\(\nu=\mathrm{pt}\), \(\star=\us\), and \(B_{\us,n}(x)=B_h(x)\), together with
\eqref{eq:supp-rbc-rate-consequences}, gives
\[
 \sqrt{nh^p}\{\Log_{m(x)}^{\cY}\widehat m_h(x)-B_h(x)\}
 =\mathfrak Z_{n,\us}(x)+o_p(1).
\]
\Cref{lem:supp_oracle_residualized_score_covariance_consistency} and
Slutsky's theorem give the limit. Part~\textup{(ii)} is
\Cref{prop:supp_smoothing_bias_expansion}.
The final rate follows from these conclusions and the covariance upper bound in
\Cref{lem:supp_oracle_residualized_score_covariance_consistency}.
\end{proof}

\begin{proof}[Proof of \Cref{cor:undersmoothing-limit}]
\Cref{thm:local_linear_expansion_bias}, \(\sqrt{nh^p}h^2\to0\), and Slutsky's theorem give
the result.
\end{proof}

\begin{proof}[Proof of \Cref{thm:rbc-limit}]
{Equation} \eqref{eq:supp_pointwise_intrinsic_oracle_representation}, \Cref{lem:supp_oracle_residualized_score_covariance_consistency}, and Slutsky's theorem give the result.
\end{proof}

\begin{proof}[Proof of \Cref{thm:pointwise-confidence-validity}]
By \eqref{eq:supp-loo-covariance-rate}, the covariance consistency
condition in \Cref{prop:supp-generic-pointwise-wald} holds for
\(\widehat\Omega_\star\). Hence
\[
 W_{n,\star}(x)\coloneqq nh^p
 \langle\Log_{\widehat m_\star(x)}m_x,
 \widehat\Omega_\star(x)^{-1}\Log_{\widehat m_\star(x)}m_x\rangle
 \dto\chi_q^2.
\]
Since the cdf \(F_q\) of \(\chi_q^2\) is continuous,
P\'olya's theorem
\citep[Lemma~2.11]{vanderVaart1998} gives
\[
 \sup_t|\mathbb P(W_{n,\star}(x)\le t)-F_q(t)|\to0.
\]

{To obtain coverage of the confidence sets, choose} \(M_n\to\infty\) {with} \(M_n/(nh^p)\to0\), and put \(u_n\coloneqq F_q(M_n)\to1\). The squared radii of the tangent Wald ellipsoids with {critical values} at most \(M_n\) are bounded by
\[
 \lambda_{\max}\{\widehat\Omega_\star(x)\}\,M_n/(nh^p)=o_p(1).
\]
Localization and geodesic reversal therefore give, with probability
tending to one, simultaneously for \(0<u\le u_n\),
\[
 m_x\in\widehat{\cC}_n^\star(x;F_q^{-1}(u))
 \quad\Longleftrightarrow\quad
 W_{n,\star}(x)\le F_q^{-1}(u).
\]
Hence
\[
 \sup_{0<u\le u_n}
 |\mathbb P(m_x\in\widehat{\cC}_n^\star(x;F_q^{-1}(u)))-u|\to0.
\]
Nestedness gives
\[
 \sup_{u_n<u<1}
 |\mathbb P(m_x\in\widehat{\cC}_n^\star(x;F_q^{-1}(u)))-u|
 \le1-u_n+
 |\mathbb P(m_x\in\widehat{\cC}_n^\star(x;M_n))-u_n|\to0.
\]
Set \(u=1-\alpha\).
\end{proof}

\begin{proof}[Proof of \Cref{thm:uniform-multiplier-validity}]
By \Cref{prop:supp-loo-replacement} and
\eqref{eq:supp-uniform-rate-consequences}, the LOO covariance
\(\widehat\Omega_\star\) and the multiplier process in
\eqref{eq:supp-loo-covariance-generator} satisfy
\eqrefs{eq:supp-generic-uniform-covariance,eq:supp-generic-generator-replacement}.
The {distributional approximations for the squared statistics} in \Cref{thm:supp-modular-inference-transfer}
therefore give
\begin{align}
 \sup_t\left|\mathbb P(T_{n,\star}(\cG_n)\le t)
 -\mathbb P(\{S_{n,\star}^G(\cG_n)\}^2\le t)\right|
 &\to0,\label{eq:supp-feasible-sampling-gaussian-approximation}\\
 \sup_t\left|\mathbb P^\ast(\widehat T_{n,\star}^\ast(\cG_n)\le t)
 -\mathbb P(\{S_{n,\star}^G(\cG_n)\}^2\le t)\right|
 &=o_p(1).\label{eq:supp-feasible-multiplier-gaussian-approximation}
\end{align}
The triangle inequality proves the first assertion.

For the coverage assertion, suppose \(B_n\to\infty\).
\Cref{thm:supp-modular-inference-transfer} gives
\[
 \sup_{0<\alpha<1}\left|
 \mathbb P(T_{n,\star}(\cG_n)\le
       \widehat c_{1-\alpha,\star}^{\ast,B_n}(\cG_n))-(1-\alpha)\right|
 \le o(1)+(B_n+1)^{-1}\to0.
\]

Take \(M_n\coloneqq\sqrt{nh^p\log(1/h)}\), so \(M_n/\log(1/h)\to\infty\) and \(M_n/(nh^p)\to0\). {Equations} \eqrefs{eq:supp-oracle-supremum-size,eq:supp-feasible-multiplier-gaussian-approximation} give \(\eta_n\coloneqq \mathbb P^\ast (\widehat T_{n,\star}^\ast (\cG_n)>M_n)=o_p(1)\). Since \(0\le\eta_n\le1\), \({\E}[\eta_n]\to0\). Choose deterministic \(\epsilon_n\downarrow0\) with \({\E}[\eta_n]/\epsilon_n\to0\), and put \(u_n\coloneqq B_n(1-\epsilon_n)/(B_n+1)\to1\). {Let \(\widehat T_{n,\star}^{\ast(j)}(\cG_n)\), \(j=1,\ldots,B_n\), {denote the realizations of} \(\widehat T_{n,\star}^\ast(\cG_n)\) {obtained from conditionally independent multiplier draws.}} Conditional independence and Markov's inequality give
\begin{equation}\label{eq:supp-uniform-cutoff-localization}
 \mathbb P\left(\sup_{0<u\le u_n}
       \widehat c_{u,\star}^{\ast,B_n}(\cG_n)>M_n\right)\le\mathbb P\left(\frac1{B_n}\sum_{j=1}^{B_n}
       \mathds{1}\{\widehat T_{n,\star}^{\ast(j)}(\cG_n)>M_n\}>
       \epsilon_n\right)
 \le\frac{{\E}[\eta_n]}{\epsilon_n}\to0.
\end{equation}

By \eqrefs{eq:supp_uniform_intrinsic_oracle_representation,eq:supp-oracle-supremum-size,eq:supp-generic-uniform-covariance},
\begin{equation}
\label{eq:supp-uniform-tube-localization}
\begin{aligned}
M_n(nh^p)^{-1}
\sup_{x\in\cG_n}
\lambda_{\max}(\widehat\Omega_\star^{\parallel}(x))
&=O_p(M_n/(nh^p))=o_p(1),\\
\sup_{x\in\cG_n}
\|\Log_{\widehat m_{\star,x}}^{\cY}m_x\|_{\widehat m_{\star,x}}
&=O_p\left(\sqrt{\frac{\log(1/h)}{nh^p}}\right)=o_p(1).
\end{aligned}
\end{equation}
By \eqrefs{eq:supp-uniform-cutoff-localization,eq:supp-uniform-tube-localization}
and localization in the common response neighborhoods, the ellipsoids and target logarithms lie
in common injectivity balls with probability tending to one,
uniformly for \(0<u\le u_n\). On this event,
\begin{equation}
\label{eq:supp-uniform-tube-wald-equivalence}
m_x\in\widehat{\cC}_n^\star
\bigl(x;\widehat c_{u,\star}^{\ast ,B_n}(\cG_n)\bigr)
\quad\text{for every }x\in\cG_n\quad\Longleftrightarrow\quad
T_{n,\star}(\cG_n)
\le\widehat c_{u,\star}^{\ast ,B_n}(\cG_n).
\end{equation}
The rank calibration and \eqref{eq:supp-uniform-tube-wald-equivalence}
give uniform coverage on this range. Nestedness and \(u_n\to1\) extend it to \(0<u<1\).
\end{proof}

\begin{corollary}
\label{cor:supp-grid-calibrated-continuum-tube}
Fix \(\star\in\{\us,\rbc\}\). Suppose the corresponding uniform
regime holds. Let \(\cG_n\subseteq\cX_0\) be nonempty,
deterministic, and finite, and suppose
\begin{equation}
\label{eq:supp-grid-fill-distance}
\rho_n\coloneqq
\sup_{x\in\cX_0}\min_{z\in\cG_n}d_{\cX}(x,z),
\qquad
{\rho_n=o\!\left(h/(\log n)^2\right)}.
\end{equation}
Use any covariance satisfying \eqref{eq:supp-generic-uniform-covariance}
over \(\cX_0\), and any conditionally centered Gaussian process satisfying
\eqref{eq:supp-generic-generator-replacement} on \(\cG_n\).
The Wald statistics and sets use this covariance.
Then
\begin{equation}
\label{eq:supp-grid-continuum-cdf}
\sup_{t\in\mathbb R}\left|
\mathbb P(T_{n,\star}(\cX_0)\le t)
-\mathbb P^\ast(\widehat T_{n,\star}^\ast(\cG_n)\le t)
\right|\pto0.
\end{equation}
If \(B_n\to\infty\), the {critical value} obtained from the {rank rule} in \Cref{subsec:uniform-intrinsic-confidence-tubes} satisfies
\begin{equation}
\label{eq:supp-grid-continuum-coverage}
\sup_{0<\alpha<1}\left|
\mathbb P\left(
m_x\in\widehat{\cC}_n^\star
\bigl(x;\widehat c_{1-\alpha,\star}^{\ast ,B_n}(\cG_n)\bigr)
\text{ for every }x\in\cX_0
\right)-(1-\alpha)
\right|\to0.
\end{equation}
\end{corollary}

\begin{proof}
\noindent{\textit{Step 1. Entropy and score increments.}}
{Using the class inclusion in}
\eqref{eq:supp-oracle-score-kernel-class-selection}{, apply}
\Cref{lem:supp_local_polynomial_kernel_factors_are_localized}
{with \((a,M_a,H,m)=(h,h^{-p/2},\cE_R,1)\), for which \(\sigma_h=1\).
The bracketing bound and finite-union conclusion give}
\begin{equation}\label{eq:supp_oracle_score_class_bounds}
{\log\cN(\cF_{\star,n},e_P,r)
\le\log N_{[]}(r/2,\cF_{\star,n},L_2(P))
\lesssim\log\frac{C}{hr},\qquad 0<r\le1.}
\end{equation}
{Since \(\log(1/h)\asymp\log n\), the right-hand side is
\(\lesssim\log n+\log(C/r)\).}

{Fix \(x,x'\in\cX_\ell\) with \(d_{\cX}(x,x')\le h\), and put
\(\delta\coloneqq h^{-1}d_{\cX}(x,x')\).
Express all differences in the chosen smooth predictor and response frame
fields. Use the zero extension in}
\eqref{eq:supp-oracle-score-class-definition}
{for all score functions in this proof. Define}
\[
{\begin{aligned}
 s_{\star,n,\ell,x}
 &\coloneqq h^{-p/2}E_{x,\ell}^\ast
 \Omega_{\star,n}(x)^{-1/2}A_x^{-1}
 \varphi_{\star,n}(\cdot;x).
\end{aligned}}
\]
{Then \(f_{\star,n,\ell,x,u}=u^\top s_{\star,n,\ell,x}\) and
\(P[s_{\star,n,\ell,x}s_{\star,n,\ell,x}^\top]=I_q\).
Thus, for every \(u\in\mathbb S^{q-1}\),}
\[
{e_P(f_{\star,n,\ell,x,u},f_{\star,n,\ell,x',u})
=\|u^\top(s_{\star,n,\ell,x}-s_{\star,n,\ell,x'})\|_{P,2}
\le\|s_{\star,n,\ell,x}-s_{\star,n,\ell,x'}\|_{P,2}.}
\]
{Assume \(\delta>0\). For RRBC, first suppose that
\(d_{\cX}(x,x')\le\min\{h,b\}\).
For \((j,a)=(1,h)\) and, for RRBC, \((2,b)\), compare
\(C_{j,a,x}(m_x)\) and \(C_{j,a,x'}(m_x)\) with the response center and
response frame fixed at \(m_x\). Apply the first-moment oscillation bound in}
\cref{lem:supp_local_polynomial_kernel_factors_are_localized}
{coordinatewise to the integrand in}
\eqref{eq:supp-regression-coefficients}{, with
\((M_a,H,m)=(a^{-p},\cE_R,1)\) and parameter-cell diameter
\(a^{-1}d_{\cX}(x,x')\le1\).
The framed logarithm bounds in}
\cref{lem:supp_uniform_finite_shrunken_cover}~\textup{(i), (iii), and (iv)}
{and the conditional moment bound}
\eqref{eq:supp-response-envelope-moment}
{control the remaining increment from \(C_{j,a,x'}(m_x)\) to
\(C_{j,a,x'}(m_{x'})\) by \(Cd_{\cX}(x,x')\).
Equations}
\eqrefs{eq:supp-regression-coefficients,eq:supp-population-design-bounds,eq:supp-population-regression-coefficient-bounds,eq:supp-population-design-inverse-modulus}
{then give}
\[
{\|C_{j,a,x}(m_x)-C_{j,a,x'}(m_{x'})\|_{\mathrm{op}}
 +\|\Theta_{j,a,x}(m_x)-\Theta_{j,a,x'}(m_{x'})\|_{\mathrm{op}}
 \lesssim(a^{-1}+1)d_{\cX}(x,x')\lesssim\delta.}
\]

{Use \(\zeta_{\star,n,\ell}\) from}
\eqref{eq:supp-oracle-cross-covariance}
{. For \(z\in\{x,x'\}\), define}
\[
 \Sigma_z
 \coloneqq E_{z,\ell}^\ast\Omega_{\star,n}(z)E_{z,\ell}=P[\zeta_{\star,n,\ell}(\cdot;z)
       \zeta_{\star,n,\ell}(\cdot;z)^\top].
\]
{The covariance bounds in}
\eqrefs{eq:supp-oracle-covariance-upper-bound,eq:supp-oracle-covariance-lower-bound}
{give}
\[
{cI_q\preceq\Sigma_z\preceq CI_q,
\qquad
\|\zeta_{\star,n,\ell}(\cdot;z)\|_{P,2}^2
=\operatorname{tr}(\Sigma_z)\lesssim1.}
\]
{For \(W=(v,y)\) with \(y\in\cK_\ell\) and
\(\min\{d_{\cX}(v,x),d_{\cX}(v,x')\}\le R_ja\), the uniform coefficient
bound in}
\eqref{eq:supp-population-regression-coefficient-bounds}{,
the increment bound for \(\Theta_{j,a,z}(m_z)\),}
\eqrefs{eq:supp-regression-residual,eq:supp-population-hessian-modulus},
and \cref{lem:supp_uniform_finite_shrunken_cover}
{give}
\[
{\begin{aligned}
 &\|E_{z,\ell}^{\ast}A_z^{-1}\varepsilon_{j,a,z}(m_z;W)\|
 \lesssim\cE_R(y),\qquad z\in\{x,x'\},\\
 &\Bigl\|E_{x,\ell}^{\ast}A_x^{-1}\varepsilon_{j,a,x}(m_x;W)
 -E_{x',\ell}^{\ast}A_{x'}^{-1}\varepsilon_{j,a,x'}(m_{x'};W)\Bigr\|
 \lesssim\cE_R(y)\delta.
\end{aligned}}
\]
{Choose \(w_1\) and, for RRBC, \(w_b\) for the pair \((x,x')\) as in}
\eqrefs{eq:supp_population_weight_basepoint_modulus,eq:supp_population_bias_weight_basepoint_modulus}{.
The weighted second-moment estimate following}
\eqref{eq:supp_population_bias_weight_basepoint_modulus}{,
the score formula}
\eqref{eq:supp_generic_oracle_score}{,}
\eqref{eq:supp-local-polynomial-weight-envelopes}, and
\eqref{eq:supp-response-envelope-moment}
{yield, omitting the \(w_b\) term for US,}
\[
{\begin{aligned}
 \|\zeta_{\star,n,\ell}(\cdot;x)-\zeta_{\star,n,\ell}(\cdot;x')\|_{P,2}^2
 &\lesssim h^{-p}P\Bigl[\cE_R(Y)^2\Bigl\{
 w_1(X)^2+\mathds{1}\{\star=\rbc\}w_b(X)^2
 +\delta^2\mathds{1}\{d_{\cX}(X,x')\le Ch\}\Bigr\}\Bigr]\\
 &\lesssim h^{-p}\{h^p\delta+h^p\delta^2\}\lesssim\delta.
\end{aligned}}
\]
{Cauchy--Schwarz gives}
\begin{equation}\label{eq:supp-oracle-covariance-basepoint-modulus}
{\begin{aligned}
 \|\Sigma_x-\Sigma_{x'}\|_{\mathrm{op}}
 &=\bigl\|P[
 \zeta_{\star,n,\ell}(\cdot;x)\zeta_{\star,n,\ell}(\cdot;x)^\top
 -\zeta_{\star,n,\ell}(\cdot;x')\zeta_{\star,n,\ell}(\cdot;x')^\top]
 \bigr\|_{\mathrm{op}}\\
 &=\Bigl\|P\Bigl[
 \{\zeta_{\star,n,\ell}(\cdot;x)-\zeta_{\star,n,\ell}(\cdot;x')\}
 \zeta_{\star,n,\ell}(\cdot;x)^\top+\zeta_{\star,n,\ell}(\cdot;x')
 \{\zeta_{\star,n,\ell}(\cdot;x)-\zeta_{\star,n,\ell}(\cdot;x')\}^\top
 \Bigr]\Bigr\|_{\mathrm{op}}\\
 &\le\bigl(
 \|\zeta_{\star,n,\ell}(\cdot;x)\|_{P,2}
 +\|\zeta_{\star,n,\ell}(\cdot;x')\|_{P,2}\bigr)
 \|\zeta_{\star,n,\ell}(\cdot;x)-\zeta_{\star,n,\ell}(\cdot;x')\|_{P,2}\\
 &\lesssim\delta^{1/2}.
\end{aligned}}
\end{equation}
{Orthonormality gives
\(s_{\star,n,\ell,z}=\Sigma_z^{-1/2}\zeta_{\star,n,\ell}(\cdot;z)\).
Hence, by}
\eqref{eq:supp-oracle-covariance-basepoint-modulus} and
\cref{lem:supp_covariance_transfer_spectral_stability}{,}
\[
{\begin{aligned}
 \|s_{\star,n,\ell,x}-s_{\star,n,\ell,x'}\|_{P,2}
 &=\Bigl\|\Sigma_x^{-1/2}
 \{\zeta_{\star,n,\ell}(\cdot;x)-\zeta_{\star,n,\ell}(\cdot;x')\}
 +(\Sigma_x^{-1/2}-\Sigma_{x'}^{-1/2})
 \zeta_{\star,n,\ell}(\cdot;x')\Bigr\|_{P,2}\\
 &\le\|\Sigma_x^{-1/2}\|_{\mathrm{op}}
 \|\zeta_{\star,n,\ell}(\cdot;x)-\zeta_{\star,n,\ell}(\cdot;x')\|_{P,2}
 +\|\Sigma_x^{-1/2}-\Sigma_{x'}^{-1/2}\|_{\mathrm{op}}
 \|\zeta_{\star,n,\ell}(\cdot;x')\|_{P,2}\\
 &\lesssim\delta^{1/2}+\delta^{1/2}\lesssim\delta^{1/2}.
\end{aligned}}
\]
{When \(\delta=0\), \(x=x'\), so the score difference is zero.
For RRBC pairs with \(b<d_{\cX}(x,x')\le h\), use
\(\delta>b/h\ge\underline t>0\) and
\(\|s_{\star,n,\ell,z}\|_{P,2}=\sqrt q\) to obtain}
\[
{\|s_{\star,n,\ell,x}-s_{\star,n,\ell,x'}\|_{P,2}
 \le2\sqrt q\le2\sqrt{q/\underline t}\,\delta^{1/2}.}
\]
{Thus, uniformly in \(u\in\mathbb S^{q-1}\),}
\begin{equation}\label{eq:supp-oracle-score-l2-modulus}
{e_P(f_{\star,n,\ell,x,u},f_{\star,n,\ell,x',u})
 \lesssim\{h^{-1}d_{\cX}(x,x')\}^{1/2}.}
\end{equation}

\noindent{\textit{Step 2. Discretization of the Gaussian supremum.}}
If \(\rho_n=0\), then \(\cG_n=\cX_0\), and the assertions reduce
to \Cref{thm:supp-modular-inference-transfer} and
\eqref{eq:supp-uniform-tube-wald-equivalence}. For \(n\) with \(\rho_n>0\), put
\(\varepsilon_n\coloneqq\rho_n/h>0\).

{Use the bounded, uniformly \(e_P\)-continuous Gaussian process
\(G_{\star,n}\) on \(\cF_{\star,n}\) constructed in}
\Cref{thm:supp_oracle_cck_gaussian_approximation_uniform_wald}{.
Its restrictions define \(S_{n,\star}^G(\cX_0)\) and
\(S_{n,\star}^G(\cG_n)\).}

On overlapping neighborhoods, an orthonormal frame change reparametrizes
\(u\in\mathbb S^{q-1}\). The chosen uniformly \(e_P\)-continuous
Gaussian version assigns equal values to indices at zero
\(e_P\)-distance, so the directional supremum is unchanged almost surely.
A Lebesgue number for the open cover \((\cU_\ell)_{\ell\le L}\) in
\Cref{lem:supp_uniform_finite_shrunken_cover} places each \(x\in\cX_0\)
and a closest grid point in a common neighborhood
for all sufficiently large \(n\). {With the same \(u\) at \(x\) and its closest grid point
in the common local frame,}
\eqref{eq:supp-oracle-score-l2-modulus} {gives}
\[
0\le S_{n,\star}^{G}(\cX_0)-S_{n,\star}^{G}(\cG_n)
\le
\sup_{\substack{f,g\in\cF_{\star,n}\\
                 e_P(f,g)\le C\sqrt{\varepsilon_n}}}
|G_{\star,n}f-G_{\star,n}g|.
\]

Dudley's entropy bound
\citep[Corollary~2.2.9]{VanDerVaartandWellner2023},
together with \eqref{eq:supp_oracle_score_class_bounds}
and the comparison between covering and packing numbers, yields
\[
{\begin{aligned}
\E[S_{n,\star}^{G}(\cX_0)-S_{n,\star}^{G}(\cG_n)]
&\lesssim \int_0^{C\sqrt{\varepsilon_n}}
       \sqrt{{\log n}+\log(C/r)}\,dr\\
&\lesssim \sqrt{\varepsilon_n}
       \sqrt{{\log n}+\log(e/\varepsilon_n)}
=o(\{{\log n}\}^{-1/2}).
\end{aligned}}
\]

{By} \eqref{eq:supp-grid-fill-distance}{,
\(\varepsilon_n\{{\log n}\}^2\to0\).} Since
\(\varepsilon_n\log(e/\varepsilon_n)\lesssim \sqrt{\varepsilon_n}\),
\[
\varepsilon_n{\log n}\{{\log n}+\log(e/\varepsilon_n)\}
\lesssim\varepsilon_n\{{\log n}\}^2
+\sqrt{\varepsilon_n}{\log n}\to0.
\]

Markov's inequality {and} \eqrefs{eq:supp_oracle_gaussian_anticoncentration,eq:supp-cdf-perturbation-bound} show that {the distribution functions of the two Gaussian suprema differ by \(o(1)\), uniformly in their argument}. The same holds after squaring, since they are nonnegative. The sampling part of \Cref{thm:supp-modular-inference-transfer} applies over \(\cX_0\), and its multiplier part over \(\cG_n\). Combining the cdf comparison for the squared Gaussian suprema over \(\cX_0\) and \(\cG_n\) with \eqref{eq:supp-feasible-sampling-gaussian-approximation} on \(\cX_0\) and \eqref{eq:supp-feasible-multiplier-gaussian-approximation} on \(\cG_n\) proves \eqref{eq:supp-grid-continuum-cdf}.

\noindent{\textit{Step 3. Quantile calibration and coverage.}}
Apply \Cref{lem:supp_uniform_random_quantile_calibration} with
statistic \(T_{n,\star}(\cX_0)\), deterministic cdf
\(F_n(t)\coloneqq \mathbb P((S_{n,\star}^{G}(\cX_0))^2\le t)\), and
\(\widehat F_n(t)\coloneqq \mathbb P^\ast(\widehat T_{n,\star}^\ast(\cG_n)\le t)\).
The two cdf approximation conditions follow from
\eqrefs{eq:supp-feasible-sampling-gaussian-approximation,eq:supp-grid-continuum-cdf}
and the triangle inequality.
By \eqref{eq:supp_oracle_gaussian_anticoncentration}, \(F_n\) is
continuous. The lemma's rank bound therefore gives
\[
\sup_{0<\alpha<1}\left|
\mathbb P(T_{n,\star}(\cX_0)
\le\widehat c_{1-\alpha,\star}^{\ast ,B_n}(\cG_n))-(1-\alpha)
\right|
\le o(1)+\frac1{B_n+1}.
\]

{For the grid draws in this corollary, choose a deterministic sequence
\(u_n\to1\) as in the argument proving}
\eqref{eq:supp-uniform-cutoff-localization}{.
Apply the localization argument for}
\eqref{eq:supp-uniform-tube-localization}
{over \(\cX_0\), using}
\eqref{eq:supp-generic-uniform-covariance}{. As in}
\eqref{eq:supp-uniform-tube-wald-equivalence}{, on an event whose
probability tends to one, the event of simultaneous coverage over
\(\cX_0\) is equivalent to
\(\{T_{n,\star}(\cX_0)\le\widehat c_{1-\alpha,\star}^{\ast,B_n}(\cG_n)\}\),
simultaneously for \(0<1-\alpha\le u_n\).
Nestedness and \(u_n\to1\) extend the resulting coverage bound to all
\(0<\alpha<1\), proving}
\eqref{eq:supp-grid-continuum-coverage}{.}
\end{proof}

\section{Extensions and auxiliary tools}

\subsection{Pushforward response distributions and moment regularity}
\label{subsec:supp-smooth-pushforward-laws}

{The following conditions for conditional moment regularity allow response distributions supported on lower-dimensional sets, including the angular fibers in the ocean analysis.} Let $\nu$ be a $\sigma$-finite measure on a standard Borel space $A$. On a predictor chart $N$, fix the conditional distribution
\[
 \mathbb P_{Y\mid X=z}
 =g(z,\cdot)_\#\{\rho(\cdot\mid z)\nu\},\qquad
 \rho\ge0,\qquad \int_A\rho(u\mid z)\,\nu(du)=1.
\]
{Here \(g:N\times A\to\cY\), and \(g(z,\cdot)_\#\) denotes pushforward under \(u\mapsto g(z,u)\).}

Let \(g,\rho,T\) be jointly measurable, with \(\theta\) in an open finite-dimensional coordinate domain and \(T\) valued in a fixed finite-dimensional Hilbert space. In fixed frames, {define the integrand \(f_T\) and the conditional moment \(M_T\) by}
\[ f_T(z,\theta,u)\coloneqq T(z,\theta,g(z,u))\rho(u\mid z),\qquad
 M_T(z,\theta)\coloneqq\int_A f_T(z,\theta,u)\,\nu(du).
\]
{Here \(\alpha,\beta\) are multi-indices, and \(D_z,D_\theta\) denote ordinary derivatives in the fixed coordinates, with all other arguments held fixed.} For $0<\gamma\le1$ and compact coordinate products
$K=K_z\times K_\theta$, write
\[
 \|f\|_{0;K}\coloneqq\sup_K\|f\|,\quad \|f\|_{2;K}\coloneqq\sum_{|\alpha|+|\beta|\le2}\sup_K\|D_z^\alpha D_\theta^\beta f\|,\quad \|f\|_{2,\gamma;K}\coloneqq\|f\|_{2;K}+\sup_{\theta\in K_\theta}[D_z^2f(\cdot,\theta)]_{\gamma;K_z}.
\]

{The H\"older seminorm in \(\|f\|_{2,\gamma;K}\) concerns only the predictor coordinates, uniformly over \(\theta\in K_\theta\).}
\begin{proposition}
\label{prop:supp-smooth-pushforward-moment-regularity}
Require each envelope condition below for every compact coordinate
product $K$.

\noindent\textup{(i) Continuity.}
If $\int_A\|f_T(\cdot,\cdot,u)\|_{0;K}\,\nu(du)<\infty$ and
$f_T(\cdot,\cdot,u)$ is continuous for $\nu$-almost every $u$, then $M_T$ is continuous.

\noindent\textup{(ii) Differentiation.}
If $f_T(\cdot,\cdot,u)$ is $C^2$ for $\nu$-almost every $u$, with jointly
measurable derivatives and
$\int_A\|f_T(\cdot,\cdot,u)\|_{2;K}\,\nu(du)<\infty$, then $M_T$ is $C^2$ and
\begin{equation}\label{eq:supp-pushforward-moment-derivatives}
 D_z^\alpha D_\theta^\beta M_T(z,\theta)
 =\int_A D_z^\alpha D_\theta^\beta f_T(z,\theta,u)\,\nu(du),
 \qquad |\alpha|+|\beta|\le2.
\end{equation}

\noindent\textup{(iii) H\"older regularity.}
Under \textup{(ii)}, if
$\int_A\|f_T(\cdot,\cdot,u)\|_{2,\gamma;K}\,\nu(du)<\infty$, then
\[
 \|M_T\|_{2,\gamma;K}
 \le\int_A\|f_T(\cdot,\cdot,u)\|_{2,\gamma;K}\,\nu(du).
\]
\end{proposition}
\begin{proof}
Dominated convergence proves \textup{(i)}. On compact coordinate
neighborhoods, the derivative envelopes dominate difference quotients,
giving \eqref{eq:supp-pushforward-moment-derivatives} and continuity of
the derivatives. Integrating the H\"older bound proves \textup{(iii)}.
\end{proof}

For compact metric $A$ and finite $\nu$, suppose $g$ and its predictor derivatives through order three are jointly continuous. For the {continuity, differentiation, and H\"older envelope conditions}, respectively, it suffices that $T$ is continuous, $C^2$, or $C^3$ near the compact parameter--response sets, and $\rho(u\mid\cdot)$ is $C^2$ for $\nu$-almost every $u$, with
\begin{equation}
\label{eq:supp-pushforward-density-bounds}
\sup_z\sum_{j=0}^2\|D_z^j\rho(u\mid z)\|
+[D_z^2\rho(u\mid\cdot)]_\gamma\le e(u),
\qquad e\in L^1(\nu),
\end{equation}
locally in $z$. The chain and product rules give the stated envelopes.

For inference, require
\Cref{prop:supp-smooth-pushforward-moment-regularity}~\textup{(i)} for squared
loss on every compact center set. On compact center subcharts in the
smooth logarithm domain, apply \textup{(ii)} pointwise and
\textup{(iii)} uniformly to
\[
 \Log_\omega^{\cY}y,\qquad \cH_\omega(y),\qquad
 \nabla^{\cY}\cH_\omega(y),\qquad
 D_s^j\mathcal L_x(s,y),\quad 0\le j\le2,
\]
and their pairwise tensor products. Include {all response center arguments} in $\theta$, varying them independently and holding them fixed under predictor differentiation. Use the local compact set containing the conditional response supports under \textup{(C)} and $\cY$ under \textup{(H)}.

\begin{corollary}
\label{cor:supp-smooth-pushforward-main-results}
{Replace the existence and regularity requirements for the
conditional response density by the conditions of}
\Cref{prop:supp-smooth-pushforward-moment-regularity}:
\textup{(i)} for squared loss, and \textup{(ii)} pointwise or
\textup{(iii)} uniformly for the listed geometric functions and their
pairwise tensor products. For uniform results, impose these conditions
on a finite predictor cover with a common $\gamma$.
Under all remaining assumptions of each result, {including those
on \(p_X\),} the US and RRBC expansions and the results for the
auxiliary scores, LOO covariances, and multiplier processes remain valid. These moment conditions do not imply
covariance nondegeneracy. Studentization in the full tangent space or after projection requires
the corresponding covariance lower bounds.
\end{corollary}
\begin{proof}
For the listed geometric functions,
\Cref{prop:supp-smooth-pushforward-moment-regularity}
gives the regularity of their conditional moments stated in
\Cref{lem:supp_conditional_moment_regularity}.

For local uniform integrability of the squared distances, fix an admissible center \(m\) and a compact predictor set \(K\). {For fixed \(u\in A\), set}
\[
\begin{gathered}
 r_z\coloneqq\|\Log_m g(z,u)\|,\qquad
 a_z\coloneqq\rho(u\mid z)r_z^2,\qquad
 b_z\coloneqq\rho(u\mid z)\Log_m g(z,u),\\
 a_\cdot,b_\cdot\in C(K)\quad(\nu\text{-a.e. }u),\qquad
 \int_A\sup_{z\in K}a_z\,\nu(du)<\infty.
\end{gathered}
\]
If \(\sup_{z\in K}a_z\mathds{1}\{r_z>T\}\not\to0\), compactness gives
\[
\begin{gathered}
 z_k\to z\in K,\qquad r_{z_k}\to\infty,\qquad
 0<\epsilon\le a_{z_k}\le\sup_Ka_z<\infty,\\
 \|b_{z_k}\|=a_{z_k}/r_{z_k}\to0
 \ \Longrightarrow\ b_z=0\ \Longrightarrow\ a_z=0,
 \qquad a_z=\lim_k a_{z_k}\ge\epsilon,
\end{gathered}
\]
a contradiction. Since
\(a_z\mathds{1}\{r_z>T\}=a_z\mathds{1}\{a_z>T\|b_z\|\}\), lower
semicontinuity gives a measurable countable supremum. Dominated
convergence yields
\begin{equation}\label{eq:supp-pushforward-local-ui}
 \sup_{z\in K}\E[\{1+d_{\cY}(m,Y)^2\}\mathds{1}\{d_{\cY}(m,Y)>T\}\mid X=z]\le2\int_A\sup_{z\in K}\{a_z\mathds{1}\{r_z>T\}\}\,\nu(du)\to0,\qquad T\ge1.
\end{equation}

The envelope for {\(\nabla^{\cY}\cH_\omega\)} gives the local bounds on first moments. These bounds, \eqref{eq:supp-pushforward-local-ui}, and the unchanged geometric assumptions suffice for the moment and replacement results in \crefrange{sec:supp_consistency_localization}{sec:supp-covariance-constructions}. The uniform approximation bounds for \(\widehat Q_{h,x}^{\mathrm{ext}}\) and the covariance bounds use \(c_n(h)\). Under uniform \textup{(H)}, the retained $2r_0$-moment and bandwidth conditions give the required bounds for the {function classes} with $m\le2$ in \Cref{lem:supp_local_polynomial_kernel_factors_are_localized,lem:supp_localized_finite_moment_maximal_inequality} and for the moments of the uncentered scores in \Cref{prop:cck-multiplier-bounded-envelope}.
\end{proof}

\subsection{Inference for smooth transformations and comparisons}

For compatible linear maps, we use the identity
\begin{equation}\label{eq:supp-covariance-product-replacement}
 \widehat B\widehat V\widehat B^\ast-BVB^\ast
 =(\widehat B-B)\widehat V\widehat B^\ast
  +B(\widehat V-V)\widehat B^\ast+BV(\widehat B-B)^\ast.
\end{equation}

\begin{theorem}
\label{thm:smooth-functional-inference}
Under the corresponding pointwise US or RRBC regime, let
\(H:\cY\to\cZ\) be Borel and \(C^1\) near \(m_x\), where \(\cZ\)
is a complete Riemannian manifold of dimension \(d\ge1\). Set
\(\theta_x\coloneqq H(m_x)\), \(\widehat\theta_{\star,x}\coloneqq H(\widehat m_\star(x))\), and
\(\Xi_{\star,n}(x)\coloneqq (dH)_{m_x}\Omega_{\star,n}(x)(dH)_{m_x}^\ast\).
If \(\liminf_n\lambda_{\min}(\Xi_{\star,n}(x))>0\), then
\[
 \Xi_{\star,n}(x)^{-1/2}\sqrt{nh^p}
 \Log_{\theta_x}^{\cZ}\widehat\theta_{\star,x}
 \dto N(0,I_{T_{\theta_x}\cZ}).
\]
Let \(\widehat V_\star\) satisfy
\eqref{eq:supp-generic-pointwise-covariance}. Define
\(\widehat\Xi_\star(x)\coloneqq (dH)_{\widehat m_\star(x)}
\widehat V_\star(x)(dH)_{\widehat m_\star(x)}^\ast\) and
\[
 \widehat{\cC}_{n,1-\alpha}^{H,\star}(x)
 \coloneqq \Exp_{\widehat\theta_{\star,x}}^{\cZ}
 \left[\{u:nh^p\langle u,\widehat\Xi_\star(x)^{-1}u\rangle
 \le\chi^2_{d,1-\alpha}\}\right].
\]
{The transformed covariance estimator \(\widehat\Xi_\star(x)\)} is positive definite with probability
tending to one, and
\[
 \sup_{0<\alpha<1}
 \left|\mathbb P(\theta_x\in\widehat{\cC}_{n,1-\alpha}^{H,\star}(x))
 -(1-\alpha)\right|\to0.
\]
\end{theorem}
\begin{proof}
The map
\(\Log_{\theta_x}^{\cZ}\circ H\circ\Exp_{m_x}^{\cY}\)
is \(C^1\) near zero, with value zero and derivative \((dH)_{m_x}\) at zero.
Consistency of \(\widehat m_\star(x)\) and the unstandardized representation in
\eqref{eq:supp_pointwise_intrinsic_oracle_representation} give
\[
 \sqrt{nh^p}\Log_{\theta_x}^{\cZ}\widehat\theta_{\star,x}
 =(dH)_{m_x}\mathfrak Z_{n,\star}(x)+o_p(1).
\]
The projected CLT \eqref{eq:supp-projected-standardized-clt}
proves the standardized limit.

Continuity of \(dH\),
\eqrefs{eq:supp-generic-pointwise-covariance,eq:supp-covariance-product-replacement},
and boundedness of \(\Omega_{\star,n}\) give
\[
 \Pi^{\cZ}_{\widehat\theta_{\star,x}\to\theta_x}
 (dH)_{\widehat m_\star(x)}\Pi^{\cY}_{m_x\to\widehat m_\star(x)}
 \pto (dH)_{m_x},\qquad
 \|\widehat\Xi_\star^{\parallel}-\Xi_{\star,n}\|=o_p(1),
\]
where the parallel superscript denotes transport to \(T_{\theta_x}\cZ\).
\Cref{lem:supp_covariance_transfer_spectral_stability} and geodesic
reversal give the \(\chi_d^2\) Wald limit. The argument using a ball
within the injectivity radius and nestedness in the proof of
\Cref{thm:pointwise-confidence-validity} gives coverage uniformly over the confidence level.
\end{proof}
{For simultaneous comparisons, we allow each contrast to depend on finitely many regression values.} Let $\cM$ be a $q$-dimensional Riemannian manifold, let $S_0$ be a compact metric space, and let $M:S_0\to\cM$ be Borel measurable. Let $\widehat M_n(s)$ be jointly Borel in the data and $s$, and choose measurable orthonormal frames $E_s:\mathbb R^q\to T_{M(s)}\cM$. Let $h\to0$, $nh^p\to\infty$, and let $\zeta_n(W;s)\in\mathbb R^q$ be centered and jointly Borel. Set $Z_n(s)\coloneqq \mathbb G_n\zeta_n(\cdot;s)$ and $e_n\coloneqq \sup_{s\in S_0}d_{\cM}(\widehat M_n(s),M(s))$.

{Fix \(r,d\ge1\), where \(r\) is the number of regression values entering each comparison and \(d\) is its output dimension. Let \(\mathcal T_n\) be a nonempty deterministic standard Borel index space, and let the Borel maps \(s_{n,\ell}:\mathcal T_n\to S_0\), \(1\le\ell\le r\), select the indices of these regression values. Define their estimation errors in tangent coordinates by}
\[
v_n(t)\coloneqq \bigoplus_{\ell=1}^r E_{s_{n,\ell}(t)}^\ast
\Log_{M(s_{n,\ell}(t))}\widehat M_n(s_{n,\ell}(t)).
\]
Let the deterministic maps $\Phi_{n,t}:\mathbb R^{qr}\to\mathbb R^d$
be jointly Borel in $(t,v)$ and $C^2$ in $v$ near a fixed closed ball
$\overline B_\delta(0)$. Let $\widehat\Delta_n$ be jointly Borel and suppose,
with probability tending to one,
$\widehat\Delta_n(t)=\Phi_{n,t}(v_n(t))-\Phi_{n,t}(0)$ for every $t$.

Set \(L_{n,t}\coloneqq D\Phi_{n,t}(0)\). {Define the transformed oracle score, its covariance, and the corresponding processes by}
\[
{\eta_{n,t}\coloneqq L_{n,t}\bigoplus_{\ell=1}^r\zeta_n(\cdot;s_{n,\ell}(t)),\quad
\Xi_n(t)\coloneqq {\E[\eta_{n,t}(W)^{\otimes2}]},\quad
U_n(t)\coloneqq \mathbb G_n\eta_{n,t},\quad U_n^{\ast o}(t)\coloneqq \mathbb G_n^\ast\eta_{n,t}.}
\]

Let $\widehat\Xi_n(t)$ be a jointly Borel positive semidefinite covariance estimator, and let $\widehat U_n^\ast(t)$ be a jointly Borel, conditionally centered Gaussian process. {For observation-level contributions \(\widehat\eta_{n,i}(t)\in\R^d\) that are measurable with respect to the data, one construction sets}
\begin{equation}\label{eq:supp-sf-observation-level-construction}
\widehat\Xi_n(t)\coloneqq \frac1n\sum_i
\widehat\eta_{n,i}(t)^{\otimes2},\qquad
\widehat U_n^\ast(t)\coloneqq \frac1{\sqrt n}\sum_i\xi_i
\widehat\eta_{n,i}(t).
\end{equation}

Define
\begin{equation}\label{eq:supp-sf-maximum-wald-statistics}
{\begin{gathered}
S_n^o\coloneqq \sup_t\|\Xi_n(t)^{-1/2}U_n(t)\|,
\qquad S_n^{\ast o}\coloneqq \sup_t\|\Xi_n(t)^{-1/2}U_n^{\ast o}(t)\|,\\
T_n\coloneqq \sup_t nh^p
\left\langle\widehat\Delta_n(t),
\widehat\Xi_n(t)^{-1}\widehat\Delta_n(t)\right\rangle,\quad
\widehat T_n^\ast\coloneqq \sup_t
\left\langle\widehat U_n^\ast(t),
\widehat\Xi_n(t)^{-1}\widehat U_n^\ast(t)\right\rangle.
\end{gathered}}
\end{equation}
The norm of a bilinear map is its operator norm on two unit balls.

\begin{theorem}
\label{thm:uniform-smooth-functional-transfer}
Assume:

{\textup{(i)} The base estimators satisfy \(\sqrt{nh^p\log(1/h)}e_n^2=o_p(1)\). Their linearization error and the derivatives of \(\Phi_{n,t}\) satisfy, respectively,}
\begin{align}
{\sqrt{\log(1/h)}\sup_{s\in S_0}
\|\sqrt{nh^p}E_s^\ast \Log_{M(s)}\widehat M_n(s)-Z_n(s)\|}
&{=o_p(1),}\label{eq:supp-sf-base-linearization}\\
{\sup_{n,t,\,\|v\|\le\delta}
\{\|D\Phi_{n,t}(v)\|_{\mathrm{op}}+
\|D^2\Phi_{n,t}(v)\|_{\mathrm{op}}\}}
&{<\infty.}\label{eq:supp-sf-uniform-c2}
\end{align}

\textup{(ii)} For fixed $0<c_\Xi\le C_\Xi<\infty$ and all sufficiently
large $n$,
\begin{equation}
\label{eq:supp-sf-projected-nondegeneracy}
c_\Xi I_d\preceq\Xi_n(t)\preceq C_\Xi I_d,
\qquad t\in\mathcal T_n.
\end{equation}
There is a nonnegative Gaussian supremum
$S_n^G=O_p(\sqrt{\log(1/h)})$ such that
\[
\sup_z|\mathbb P(S_n^o\le z)-\mathbb P(S_n^G\le z)|\to0,
\qquad
\sup_z|\mathbb P^\ast (S_n^{\ast o}\le z)-\mathbb P(S_n^G\le z)|\pto 0,
\]
and $\sup_z\mathbb P(|S_n^G-z|\le\varepsilon)
\lesssim \varepsilon\sqrt{\log(1/h)}$ for every $\varepsilon>0$ and all
sufficiently large $n$.

\textup{(iii)} The covariance and multiplier replacements satisfy
\begin{align}
{\sup_t\|\widehat\Xi_n(t)-\Xi_n(t)\|_{\mathrm{op}}}
 &{=\allowbreak o_p(1/\log(1/h)),}\label{eq:supp-sf-covariance-rate}\\
{\sup_t\|\widehat U_n^\ast (t)-U_n^{\ast o}(t)\|}
 &{=o_p^\ast (1/\sqrt{\log(1/h)}).}\label{eq:supp-sf-multiplier-rate}
\end{align}
Then $\inf_t\lambda_{\min}(\widehat\Xi_n(t))\ge c_\Xi/2$ with
probability tending to one. If $\widehat c^\ast _{n,1-\alpha}$ is the
conditional generalized quantile of $\widehat T_n^\ast $, then
\[
\sup_z|\mathbb P(T_n\le z)-\mathbb P^\ast (\widehat T_n^\ast \le z)|\pto 0,
\qquad
\sup_{0<\alpha<1}|\mathbb P(T_n\le\widehat c^\ast _{n,1-\alpha})-(1-\alpha)|\to0.
\]
\end{theorem}
\begin{proof}
By part~\textup{(i)}, \(\sup_t\|v_n(t)\|\le\sqrt r\,e_n=o_p(1)\).
Taylor's theorem gives
\begin{equation}\label{eq:supp-sf-summary-linearization}
\begin{aligned}
 \sup_t\|\sqrt{nh^p}\widehat\Delta_n(t)-U_n(t)\|
 &\lesssim\sup_s\|\sqrt{nh^p}E_s^\ast\Log_{M(s)}\widehat M_n(s)-Z_n(s)\|
 +\sqrt{nh^p}e_n^2\\
 &=o_p(\{\log(1/h)\}^{-1/2}).
\end{aligned}
\end{equation}

Using \eqref{eq:supp-sf-summary-linearization}, apply
\Cref{thm:supp-modular-inference-transfer} on \(\mathcal T_n\)
in \(\mathbb R^d\), with center zero, estimator \(\widehat\Delta_n\),
and
\[
 \mathfrak Z_{n,\star}=U_n,\qquad
 \Omega_{\star,n}=\Xi_n,\qquad
 \widehat V_\star=\widehat\Xi_n,\qquad
 \mathfrak Z_{n,\star}^{\ast o}=U_n^{\ast o},\qquad
 \widetilde{\mathfrak Z}_{n,\star}^\ast=\widehat U_n^\ast.
\]
{Condition \textup{(ii)} of \Cref{thm:uniform-smooth-functional-transfer} supplies the spectral bounds, the two distributional approximations, the order bound for \(S_n^G\), and anti-concentration.}
Equations \eqrefs{eq:supp-sf-covariance-rate,eq:supp-sf-multiplier-rate}
verify the covariance and multiplier assumptions. Thus \Cref{thm:supp-modular-inference-transfer}
gives positive definiteness, {the two distributional approximations for the squared statistics}, and quantile
calibration. The triangle inequality proves the stated comparison of
sampling and conditional multiplier distributions.
\end{proof}

For the next lemma, use the setup of
\Cref{thm:uniform-smooth-functional-transfer}. Let
\(\widehat L_{n,t}\) and \(\widehat\zeta_{n,i}(s)\) be jointly Borel
in the data and their index arguments.
Use the observation-level construction {in \eqref{eq:supp-sf-observation-level-construction}} for \(\widehat\Xi_n(t)\) and \(\widehat U_n^\ast(t)\), with contributions
\[
 \widehat\eta_{n,i}(t)\coloneqq \widehat L_{n,t}
 \bigoplus_{\ell=1}^r\widehat\zeta_{n,i}(s_{n,\ell}(t)).
\]

\begin{lemma}
\label{lem:supp-sf-finite-arity-projected-replacement}
Assume \eqref{eq:supp-sf-uniform-c2} and
\begin{equation}\label{eq:supp-sf-jacobian-replacement}
 \sup_t\|\widehat L_{n,t}-L_{n,t}\|_{\mathrm{op}}
 =o_p(\{\log(1/h)\}^{-1}).
\end{equation}
Then:
\begin{enumerate}[(i)]
\item If, with
\(C_n(s,s')\coloneqq {\E[\zeta_n(W;s)\zeta_n(W;s')^\top]}\) and
\(\widehat C_n(s,s')\coloneqq n^{-1}\sum_i
\widehat\zeta_{n,i}(s)\widehat\zeta_{n,i}(s')^\top\),
\[
 \sup_{s,s'}\|\widehat C_n(s,s')-C_n(s,s')\|_{\mathrm{op}}
 =o_p(\{\log(1/h)\}^{-1}),\qquad
 \sup_{s,s'}\|C_n(s,s')\|_{\mathrm{op}}=O(1),
\]
then \eqref{eq:supp-sf-covariance-rate} holds.
\item If, with
\(Z_n^{\ast o}(s)\coloneqq\mathbb G_n^\ast\zeta_n(\cdot;s)\) and
\(\widehat Z_n^\ast(s)\coloneqq n^{-1/2}\sum_i\xi_i\widehat\zeta_{n,i}(s)\),
\[
 \sup_s\|\widehat Z_n^\ast(s)-Z_n^{\ast o}(s)\|
 =o_p^\ast(\{\log(1/h)\}^{-1/2}),\qquad
 \sup_s\|\widehat Z_n^\ast(s)\|
 =O_p^\ast(\sqrt{\log(1/h)}),
\]
then \eqref{eq:supp-sf-multiplier-rate} holds.
\end{enumerate}
\end{lemma}

\begin{proof}
For part~\textup{(i)}, since \(r\) is fixed,
\eqrefs{eq:supp-covariance-product-replacement,eq:supp-sf-uniform-c2,eq:supp-sf-jacobian-replacement}
give, with probability tending to one,
\[
 \sup_t\|\widehat\Xi_n(t)-\Xi_n(t)\|_{\mathrm{op}}
 \lesssim
 \sup_{s,s'}\|\widehat C_n(s,s')-C_n(s,s')\|_{\mathrm{op}}
 +\sup_t\|\widehat L_{n,t}-L_{n,t}\|_{\mathrm{op}}
 =o_p(\{\log(1/h)\}^{-1}).
\]

For part~{\textup{(ii)}},
\[
\sup_t\|\widehat U_n^\ast (t)-U_n^{\ast o}(t)\|
\le\sqrt r\Big\{
\sup_t
\|\widehat L_{n,t}-L_{n,t}\|_{\mathrm{op}}
\sup_s\|\widehat Z_n^\ast (s)\|+
\sup_t\|L_{n,t}\|_{\mathrm{op}}
\sup_s
\|\widehat Z_n^\ast (s)-Z_n^{\ast o}(s)\|
\Big\}.
\]
The assumed rates and the uniform Jacobian bound imply \eqref{eq:supp-sf-multiplier-rate}.
\end{proof}

For the plug-in choice \(\widehat L_{n,t}=D\Phi_{n,t}(v_n(t))\), suppose \(e_n\log(1/h)=o_p(1)\). With probability tending to one, \(\sup_t\|v_n(t)\|\le\sqrt r\,e_n\le\delta\). On this event, \eqref{eq:supp-sf-uniform-c2} and the mean value theorem give
\[
\sup_t\|D\Phi_{n,t}(v_n(t))-L_{n,t}\|_{\mathrm{op}}\lesssim\sqrt r\,e_n=o_p\left(\frac1{\log(1/h)}\right),
\]
which verifies \eqref{eq:supp-sf-jacobian-replacement}.

For smooth transformations of a fixed number of US or RRBC estimates in
\Cref{cor:supp-projected-rbc-uniform-inference}, assume
\eqref{eq:supp-sf-uniform-c2} and
\[
 \sup_t\|\widehat L_{n,t}-L_{n,t}\|_{\mathrm{op}}
 =O_p(r_{n,\mathrm{unif}}).
\]
Under the corresponding uniform regime,
\Cref{prop:supp-score-replacement,prop:supp-loo-replacement},
\eqref{eq:supp-uniform-rate-consequences}, and
\Cref{lem:supp-sf-finite-arity-projected-replacement}
then give \eqrefs{eq:supp-sf-covariance-rate,eq:supp-sf-multiplier-rate}.

\begin{corollary}
\label{cor:supp-projected-rbc-uniform-inference}
Fix $\star\in\{\us,\rbc\}$.
Use the setup of \Cref{thm:uniform-smooth-functional-transfer} with $\cM=\cY$,
$S_0=\cX_0$, $M(s)=m_s$, $\widehat M_n(s)=\widehat m_{\star,s}$, and
\[
 \zeta_n(W;s)=h^{-p/2}E_s^\ast A_s^{-1}\varphi_{\star,n}(W;s).
\]
Assume the corresponding uniform regime and
\eqrefs{eq:supp-sf-uniform-c2,eq:supp-sf-projected-nondegeneracy}.
Let the covariance estimator and conditionally centered Gaussian
process satisfy
\eqrefs{eq:supp-sf-covariance-rate,eq:supp-sf-multiplier-rate}.
Then \Cref{thm:uniform-smooth-functional-transfer} applies.
\end{corollary}
\begin{proof}
{Equation} \eqref{eq:supp_uniform_intrinsic_oracle_representation} gives \eqref{eq:supp-sf-base-linearization}. The unstandardized oracle maximal bound in \Cref{rem:supp-unstandardized-oracle-score-class} and the same linear representation give \(e_n=O_p(\delta_{n,\mathrm{unif}}(h))\). Equation \eqref{eq:supp-rbc-rate-consequences} verifies the quadratic remainder condition in \Cref{thm:uniform-smooth-functional-transfer}~\textup{(i)}.

{To verify the oracle Gaussian approximation in part~\textup{(ii)} of \Cref{thm:uniform-smooth-functional-transfer}, define the class of standardized scalar projections by}
\[
\mathcal H_n\coloneqq \left\{
 w\mapsto u^\top\Xi_n(t)^{-1/2}\eta_{n,t}(w):
 t\in\mathcal T_n,\ u\in\mathbb S^{d-1}\right\}.
\]
Difference quotients make \(t\mapsto L_{n,t}\) Borel.
Composition and parameter integration make
\((t,w)\mapsto\eta_{n,t}(w)\) and \(t\mapsto\Xi_n(t)\) Borel.
By continuity of the map taking a positive definite matrix to its inverse square root
and \eqref{eq:supp-sf-projected-nondegeneracy},
\(\mathcal H_n\) is jointly Borel parametrized on the standard Borel space
\(\mathcal T_n\times\mathbb S^{d-1}\) for all sufficiently large \(n\).

Orthogonal changes to the local frames preserve the replacement bounds.
The rows of $\Xi_n(t)^{-1/2}L_{n,t}$ are uniformly bounded by
\eqrefs{eq:supp-sf-uniform-c2,eq:supp-sf-projected-nondegeneracy}.
Thus every element of $\mathcal H_n$ is a sum of $r$ scalar projections of the unstandardized scores from \Cref{rem:supp-unstandardized-oracle-score-class}
with bounded coefficients, absorbing the orthogonal frame changes.

{The preceding finite-sum representation satisfies the {conditions for the function classes}
with \((a,M_a,H,m)=(h,h^{-p/2},\cE_R,1)\). Hence}
\Cref{lem:supp_local_polynomial_kernel_factors_are_localized}
{and its conclusion for fixed finite unions give the centered bracketing
bounds for \(\mathcal H_n\).}

By construction,
\[
 Pf=0,\qquad Pf^2=1\qquad(f\in\mathcal H_n).
\]
Hence
\Cref{prop:cck-multiplier-bounded-envelope} applies to \(\mathcal H_n\). The representation of the norm as a supremum over unit directions identifies its empirical and
multiplier suprema with $S_n^o$ and $S_n^{\ast o}$. Symmetry of
$\mathcal H_n$ makes them nonnegative. This verifies part \textup{(ii)}
of \Cref{thm:uniform-smooth-functional-transfer}.

The assumed covariance and multiplier replacement conditions verify
part \textup{(iii)} of \Cref{thm:uniform-smooth-functional-transfer}.
\end{proof}

\subsection{Random critical values and perturbation bounds}

For real-valued random variables \(U,V\) on a common probability
space, any distribution function \(F\), and \(\varepsilon>0\),
\begin{equation}
\label{eq:supp-cdf-perturbation-bound}
\sup_{t\in\mathbb R}|\mathbb P(U\le t)-F(t)|
\le\mathbb P(|U-V|>\varepsilon)
+\sup_{t\in\mathbb R}|\mathbb P(V\le t)-F(t)|+\sup_{t\in\mathbb R}
\{F(t+\varepsilon)-F(t-\varepsilon)\}.
\end{equation}
Indeed, outside \(\{|U-V|>\varepsilon\}\),
\(\{V\le t-\varepsilon\}\subseteq\{U\le t\}
\subseteq\{V\le t+\varepsilon\}\).
Taking probabilities and comparing each endpoint with \(F\)
gives \eqref{eq:supp-cdf-perturbation-bound}.
The conditional bound holds almost surely for any reference cdf
measurable with respect to the conditioning sigma-field.

The generalized inverse of a data-dependent cdf, defined through an infimum over rational arguments, is measurable on the completed probability space.
\begin{lemma}
\label{lem:supp_uniform_random_quantile_calibration}
Let $T_n$ be real-valued, $F_n$ a deterministic cdf,
and $\widehat F_n$ a random cdf satisfying
\[
\sup_t|\mathbb P(T_n\le t)-F_n(t)|\to0,\qquad
\|\widehat F_n-F_n\|_\infty\pto0,\qquad
\sup_t\{F_n(t)-F_n(t-)\}\to0.
\]
For $0<u<1$, set
$\widehat c_n(u)\coloneqq \inf\{t\in\mathbb Q:\widehat F_n(t)\ge u\}$.
Then
\[
\sup_{0<u<1}|\mathbb P(T_n\le\widehat c_n(u))-u|\to0.
\]

On an extension of the probability space, conditionally on
$(T_n,\widehat F_n)$, let $(T_{n,j}^\ast)_{j\ge1}$
be i.i.d. with cdf $\widehat F_n$, and set
$\widehat p_{n,B}^+\coloneqq (1+\sum_{j=1}^B\mathds{1}\{T_{n,j}^\ast \ge T_n\})/(B+1)$
for $B\ge1$. Then
\[
\sup_{\substack{B\ge1\\0<\alpha<1}}
\left|
\mathbb P(\widehat p_{n,B}^+\le\alpha)
-\frac{\lfloor(B+1)\alpha\rfloor}{B+1}
\right|\to0.
\]
In particular, $\widehat p_{n,B}^+$ is asymptotically
super-uniform, uniformly over $B\ge1$.
\end{lemma}
\begin{proof}
Write
$J_n\coloneqq \sup_t\{F_n(t)-F_n(t-)\}$,
$V_n\coloneqq F_n(T_n-)$, and $U_n\coloneqq \widehat F_n(T_n-)$.
For $0<u<1$, put $r_n(u)\coloneqq \inf\{t:F_n(t)>u\}$.
Then $\{V_n\le u\}=\{T_n\le r_n(u)\}$ and
$u\le F_n(r_n(u))\le u+J_n$.
Choose $\varepsilon_n\downarrow0$ with
$\mathbb P(\|\widehat F_n-F_n\|_\infty>\varepsilon_n)\to0$.
Since $|U_n-V_n|\le\|\widehat F_n-F_n\|_\infty$, including endpoints by
limits gives
\[
\begin{aligned}
\kappa_n\coloneqq \sup_{0\le u\le1}|\mathbb P(U_n\le u)-u|
\le\sup_t|\mathbb P(T_n\le t)-F_n(t)|+J_n+\varepsilon_n+\mathbb P(\|\widehat F_n-F_n\|_\infty>\varepsilon_n)\to0.
\end{aligned}
\]
The same bound holds with $U_n<u$ for $0\le u\le1$. The inclusions
\[
\{U_n<u\}\subset\{T_n\le\widehat c_n(u)\}\subset\{U_n\le u\}
\]
therefore prove the quantile conclusion.

Conditionally on $(T_n,\widehat F_n)$,
$L_{n,B}\coloneqq \sum_{j=1}^B\mathds{1}\{T_{n,j}^\ast <T_n\}$ is
$\operatorname{Bin}(B,U_n)$.
{For $1\le k\le B$, the function $g_{B,k}(u)\coloneqq \mathbb P(\operatorname{Bin}(B,u)\le k-1)$ decreases from one to zero and has integral $k/(B+1)$.}
Integration by parts gives
\[
\left|{\E}[g_{B,k}(U_n)]-\frac{k}{B+1}\right|
\le\kappa_n.
\]
Since $\widehat p_{n,B}^+=1-L_{n,B}/(B+1)$, its rejection event is empty
when $\lfloor(B+1)\alpha\rfloor=0$ and otherwise equals
$\{L_{n,B}\ge B+1-\lfloor(B+1)\alpha\rfloor\}$.
The preceding bound therefore yields
\[
\left|\mathbb P(\widehat p_{n,B}^+\le\alpha)
-\frac{\lfloor(B+1)\alpha\rfloor}{B+1}\right|\le\kappa_n,
\qquad B\ge1,\quad0<\alpha<1.
\]
Since the fraction is at most $\alpha$, the last conclusion follows.
\end{proof}

\begin{lemma}
\label{lem:supp-regression-calculus}
Let \(P,Q\) be finite measures, \(k\) a scalar kernel, \(r\) a finite-dimensional design {vector}, and \(f\) a vector- or tensor-valued response. For \(R\in\{P,Q\}\), define \(S_R\coloneqq R[kr^{\otimes2}]\) and \(\Theta_R\coloneqq R[kf\otimes r]S_R^{-1}\), assuming the integrands are \(R\)-integrable and \(S_R\) is invertible. With \(\varepsilon_P\coloneqq f-\Theta_Pr\),
\begin{equation}\label{eq:supp-exact-regression-coefficient-identity}
 \Theta_Q-\Theta_P=(Q-P)[k\varepsilon_P\otimes r]S_Q^{-1}.
\end{equation}
\end{lemma}
\begin{proof}
The residual satisfies the normal equation
\[
 P[k\varepsilon_P\otimes r]
 =P[kf\otimes r]-\Theta_PS_P=0.
\]
Consequently,
\[
 (Q-P)[k\varepsilon_P\otimes r]
 =Q[kf\otimes r]-\Theta_PS_Q
 =(\Theta_Q-\Theta_P)S_Q.
\]
Right multiplication by \(S_Q^{-1}\) proves
\eqref{eq:supp-exact-regression-coefficient-identity}.
\end{proof}

\begin{lemma}
\label{lem:supp_covariance_transfer_spectral_stability}
Let $A,B$ be self-adjoint operators on the same finite-dimensional
Hilbert space, and let $c>0$. If $A,B\succeq cI$, then
\[
\|A^{-1}\|_{\mathrm{op}},\|B^{-1}\|_{\mathrm{op}}\le c^{-1},
\qquad
\|B^{-1}-A^{-1}\|_{\mathrm{op}}\le c^{-2}\|B-A\|_{\mathrm{op}},
\]
and $\|B^{-1/2}-A^{-1/2}\|_{\mathrm{op}}
\le(2c^{3/2})^{-1}\|B-A\|_{\mathrm{op}}$.
If $A\succeq2cI$ and $\|B-A\|_{\mathrm{op}}\le c$, then
$B\succeq cI$, so the same bounds apply.
\end{lemma}
\begin{proof}
For invertible \(A,B\),
\begin{equation}\label{eq:supp-inverse-difference-identity}
 B^{-1}-A^{-1}=B^{-1}(A-B)A^{-1}.
\end{equation}
For \(A,B\succeq cI\),
\[
B^{-1/2}-A^{-1/2}
=\frac1\pi\int_0^\infty t^{-1/2}
\{(B+tI)^{-1}-(A+tI)^{-1}\}\,dt.
\]
Equation \eqref{eq:supp-inverse-difference-identity} applied to \(A+tI,B+tI\) and
$\int_0^\infty t^{-1/2}(c+t)^{-2}\,dt=\pi/(2c^{3/2})$
give the bounds. Weyl's inequality gives the last assertion.
\end{proof}

\begin{lemma}
\label{lem:supp_local_logarithmic_additivity}
Let \(K_0\subset\cY\) be compact with
\(K_0\times K_0\subset\cD_{\log}^{\cY}\). There exists \(\delta>0\) such that, for all
\(a,z\in K_0\) and \(v\in T_z\cY\) with \(\|v\|_z\le\delta\),
\(
(a,\Exp_z^{\cY}\left[v\right])\in\cD_{\log}^{\cY}
\)
and
\[
\left\|
\Log_a^{\cY}(\Exp_z^{\cY}\left[v\right])
-\Log_a^{\cY}z
-\Pi_{z\to a}^{\cY}v
\right\|_a
\lesssim
\left\{
d_{\cY}(a,z)\|v\|_z+\|v\|_z^2
\right\}.
\]
\end{lemma}

\begin{proof}
Set
\[
 R(a,z,v)\coloneqq\Log_a^{\cY}(\Exp_z^{\cY}\left[v\right])
 -\Log_a^{\cY}z-\Pi_{z\to a}^{\cY}v.
\]
Compactness and openness give a common \(\delta>0\) such that
\((a,\Exp_z^{\cY}\left[v\right])\in\cD_{\log}^{\cY}\) for the stated \(a,z,v\).
{Here \(D_v\) differentiates the vector variable in \(T_z\cY\), with \(a,z\) held fixed.} The map \(R\) is smooth on a neighborhood of this compact disk bundle,
so \(D_v^2R\) is uniformly bounded on the bundle. Since
\((d\Exp_z^{\cY})_0=I\),
\[
 {(D_vR)(a,z,0)}=(d\Log_a^{\cY})_z-\Pi_{z\to a}^{\cY},
 \qquad {(D_vR)(a,a,0)}=0.
\]
Smoothness near the compact diagonal gives a common
\(0<r<\inf_{a\in K_0}\inj_{\cY}(a)\) such that
\[
 u\longmapsto
 {(D_vR)(a,\Exp_a^{\cY}\left[u\right],0)}\circ
 \Pi_{a\to\Exp_a^{\cY}\left[u\right]}^{\cY}
 \in\operatorname{End}(T_a\cY)
\]
is smooth with uniformly bounded first \(u\)-derivative for
\(a\in K_0\), \(\|u\|_a\le r\). For \(d_{\cY}(a,z)<r\), put
\(u=\Log_a^{\cY}z\). Parallel transport is isometric, so the mean
value theorem in \(T_a\cY\) gives
\[
 \|{(D_vR)(a,z,0)}\|_{\mathrm{op}}
 =\|{(D_vR)(a,\Exp_a^{\cY}\left[u\right],0)}\circ
       \Pi_{a\to\Exp_a^{\cY}\left[u\right]}^{\cY}\|_{\mathrm{op}}
 \lesssim\|u\|_a=d_{\cY}(a,z).
\]

For \(d_{\cY}(a,z)\ge r\), compactness of \(K_0\times K_0\)
and the fixed choice \(r>0\) give
\[
 \|{(D_vR)(a,z,0)}\|_{\mathrm{op}}
 \lesssim d_{\cY}(a,z).
\]
As \(R(a,z,0)=0\), integral Taylor expansion yields
\[
 \|R(a,z,v)\|_a
 =\left\|{(D_vR)(a,z,0)}[v]
   +\int_0^1(1-t){(D_v^2R)(a,z,tv)}[v,v]\,dt\right\|_a
 \lesssim d_{\cY}(a,z)\|v\|_z+\|v\|_z^2.
\]
\end{proof}

\subsection{{Empirical process bounds and Gaussian approximation}}

\begin{lemma}
\label{lem:supp_localized_bracketing_maximal_inequality}
{Let \(W_1,\ldots,W_n\) be i.i.d. with distribution \(P\).
Let \(a=a_n\in(0,1)\) and \(M_a>0\) be deterministic, with \(a\to0\),
and let \(\mathcal F_a\) be a nonempty deterministic class of measurable real-valued functions.
{Suppose that the following three bounds hold for a constant \(C\ge2\) independent of \(n\) and \(a\):}}
\begin{equation}\label{eq:supp-bounded-class-bounds}
{\begin{gathered}
 \sup_{f\in\mathcal F_a}\|f\|_\infty\le CM_a,\qquad
 \sup_{f\in\mathcal F_a}Pf^2\le CM_a^2a^p,\\
 \log N_{[]}(CM_aa^{p/2}\varepsilon,\mathcal F_a,L_2(P))
 \le C\log(C/(a\varepsilon)),\qquad 0<\varepsilon\le1.
\end{gathered}}
\end{equation}
{Assume that \(\sup_{f\in\mathcal F_a}|(P_n-P)f|\) is measurable
on the completed probability space.
If \(\log(1/a)/(na^p)\to0\), then}
\[
\sup_{f\in\mathcal F_a}|(P_n-P)f|
=O_p\left(
M_a\sqrt{a^p\log(1/a)/n}\right).
\]
\end{lemma}
\begin{proof}
{Choose a fixed \(c_0>\sqrt C\) and set
\(\sigma_a\coloneqq c_0M_aa^{p/2}\).
Then the second-moment bound in}
\eqref{eq:supp-bounded-class-bounds}
{gives \(\sup_{f\in\mathcal F_a}Pf^2<\sigma_a^2\).
The bracketing bound in} \eqref{eq:supp-bounded-class-bounds}
{and monotonicity of bracketing numbers give}
\[
\int_0^{\sigma_a}
\sqrt{1+\log N_{[]}(u,\mathcal F_a,L_2(P))}\,du
\lesssim M_aa^{p/2}\sqrt{\log(1/a)}.
\]
{Applying the bracketing maximal inequality}
\citep[Theorem~$2.14.17'$]{VanDerVaartandWellner2023}{,
using the envelope bound in} \eqref{eq:supp-bounded-class-bounds}
{and the assumed measurability, yields}
\[
{\E}
\left[\sup_{f\in\mathcal F_a}|(P_n-P)f|\right]
\lesssim M_a\left\{
\sqrt{a^p\log(1/a)/n}+\log(1/a)/n
\right\}.
\]
Since \(\log(1/a)/(na^p)\to0\), the second term is negligible.
 Markov's inequality completes the proof.
\end{proof}

{We establish uniform bounds for the empirical and multiplier processes indexed by localized classes of kernel-weighted functions. We then derive Gaussian approximations for their suprema over standardized oracle score classes.} {The classes allow kernels of bounded variation and factors controlled by a finite-moment response envelope.} {For the {function classes} considered below, let \(\cX_0\subset\cX\) be compact. Fix a finite local frame cover with enlarged normal coordinate neighborhoods \(V_1,\ldots,V_L\), and set \(V\coloneqq\bigcup_{\ell=1}^{L}V_\ell\). Assume that the predictor density \(p_X\) is bounded on \(V\).} Fix \(r_0>1\) and a measurable response envelope \(H:\cY\to[1,\infty)\), independent of \(n\), \(a\), and \(\vartheta\), such that
\begin{equation}\label{eq:supp-response-envelope-moment}
 \sup_{z\in V}\E[H(Y)^{2r_0}\mid X=z]<\infty.
\end{equation}
{For {these classes}, take \(H\equiv1\) in regime \textup{(C)}.}

{The choices in this paragraph do not depend on \(n\) or \(a\). Let \(J\ge1\) be the number of summands, and let \(s_\nu\in\{1,2\}\) be the number of kernel factors in summand \(\nu\), \(1\le\nu\le J\). Fix an integer \(D\ge0\), a compact set \(E\subset\R^D\), and \(0<\underline t\le\overline t<\infty\). For \(1\le\nu\le J\) and \(1\le k\le s_\nu\), fix a bounded kernel \(K_{\nu k}\) of bounded variation, supported on \([0,R_{\nu k}]\). Fix Borel functions \(\chi_\nu:\cX\times\cY\to[0,1]\) that vanish outside the corresponding response domains. Fix a nonempty compact set}
\[
 {\Theta\subset
 \left[\prod_{\nu=1}^{J}
 \left(\cX_0^{s_\nu}\times[\underline t,\overline t]^{s_\nu}\right)\right]\times E,}
\]
{and write its elements as \(\vartheta=((x_{\nu k},t_{\nu k})_{\nu,k},\eta)\).}
{Within each class, each center \(x_{\nu k}\) varies in a fixed coordinate neighborhood with a fixed local frame. Different assignments of these neighborhoods and frames are treated as separate classes.}

{Choose \(C_0>1+\overline t\max_{\nu,k}R_{\nu k}\), and then choose \(a_0\in(0,1)\) sufficiently small that, for every \(0<a\le a_0\), the enlarged balls used below lie in \(V\) and in the chosen normal coordinate domains.}

{Define}
\[
{\begin{aligned}
 d_a(\vartheta,\vartheta')\coloneqq{}&
 \sum_{\nu=1}^{J}\sum_{k=1}^{s_\nu}
 \left\{a^{-1}d_{\cX}(x_{\nu k},x'_{\nu k})
       +|t_{\nu k}-t'_{\nu k}|\right\}+\|\eta-\eta'\|.
\end{aligned}}
\]

{Fix \(m\in\{0,1,2\}\). For each \(0<a\le a_0\), let \(Q_{\nu,a}(z,y;\vartheta)\) be deterministic factors, jointly Borel measurable in \((z,y,\vartheta)\). Scalar coefficients are included in these factors. To select the subclass of interest, let \(\mathcal T_a\) be a nonempty standard Borel space and let \(\psi_a:\mathcal T_a\to\Theta\) be a deterministic Borel map. Both the indexing space and the map may depend on \(n\) and \(a\); no continuity of \(\psi_a\) is required. For deterministic \(M_a>0\), define}
\begin{equation}\label{eq:supp-kernel-factor-class}
{\begin{aligned}
 f_{a,\vartheta}(z,y)
 &\coloneqq M_a\sum_{\nu=1}^{J}
 \chi_\nu(z,y)Q_{\nu,a}(z,y;\vartheta)
 \prod_{k=1}^{s_\nu}
 K_{\nu k}\!\left(\frac{d_{\cX}(z,x_{\nu k})}{a t_{\nu k}}\right),\\
 \mathcal F_a
 &\coloneqq\left\{f_{a,\psi_a(\tau)}:\tau\in\mathcal T_a\right\}.
\end{aligned}}
\end{equation}
{Each summand is evaluated only on its response domain and kernel support; elsewhere it is defined to be zero without evaluating its factors.}

{For every \(1\le\nu\le J\), \(0<a\le a_0\), and \(\vartheta\in\Theta\), define the enlarged evaluation domain}
\[
{\mathcal D_{\nu,a}(\vartheta)\coloneqq
 \left\{(z,y):\chi_\nu(z,y)\ne0,\quad
 d_{\cX}(z,x_{\nu k})\le C_0a
 \text{ for all }1\le k\le s_\nu\right\}.}
\]
{Assume that the following two conditions hold with a constant \(C>0\), uniform in \(a\) and all admissible arguments.}

\begin{enumerate}[label={\textup{(\roman*)}}]
\item {For every \(\nu\), every \(\vartheta\in\Theta\), and every \((z,y)\in\mathcal D_{\nu,a}(\vartheta)\),}
\begin{equation}\label{eq:supp-kernel-factor-bounds}
 {|Q_{\nu,a}(z,y;\vartheta)|\le C H(y)^m.}
\end{equation}
\item {For every \(\nu\), every \(\vartheta,\vartheta'\in\Theta\) with \(d_a(\vartheta,\vartheta')\le1\), and every \((z,y)\in\mathcal D_{\nu,a}(\vartheta)\cap\mathcal D_{\nu,a}(\vartheta')\),}
\begin{equation}\label{eq:supp-kernel-factor-increment-bound}
 {|Q_{\nu,a}(z,y;\vartheta)-Q_{\nu,a}(z,y;\vartheta')|
 \le C H(y)^m d_a(\vartheta,\vartheta').}
\end{equation}
\end{enumerate}
{All constants in the results below are independent of \(\psi_a\).}

{For the uniform applications under \textup{(H)} in}
\crefrange{sec:supp_consistency_localization}{sec:supp-covariance-constructions}{, take \(H=\cE_R\). Condition}
\eqref{eq:supp-response-envelope-moment}
{then follows from} \Cref{ass:uniform-inference-conditions}{.}

{For \(T\ge1\), set}
\[
 {v_T\coloneqq T^{(4-2r_0)_+},
 \qquad u_+\coloneqq\max\{u,0\}.}
\]
{When \(m=2\), define the truncated class}
\[
 {\mathcal F_a^T\coloneqq
 \left\{f\mathds{1}\{H\le T\}:f\in\mathcal F_a\right\}.}
\]

\begin{lemma}
\label{lem:supp_local_polynomial_kernel_factors_are_localized}
{Under the preceding setup, let \(T\ge1\) and set}
\[
{(\mathcal G_a,\sigma_a)\coloneqq
 \begin{cases}
  (\mathcal F_a,\;M_aa^{p/2}),&m=0,1,\\[2pt]
  (\mathcal F_a^T,\;M_aa^{p/2}\sqrt{v_T}),&m=2.
 \end{cases}}
\]
{For \(\mathcal C\subset\Theta\), let \(\mathcal G_a(\mathcal C)\) denote the subclass indexed by \(\psi_a^{-1}(\mathcal C)\). All constants below are independent of \(n,a,M_a,T,\delta,\varepsilon\) and \(\psi_a\).}

{For every nonempty parameter cell \(\mathcal C\subset\Theta\) and every \(0<\delta\le1\) satisfying \(\operatorname{diam}_{d_a}(\mathcal C)\le\delta\), there exists a nonnegative measurable function \(w_{\mathcal C}\) such that, for all \(f,g\in\mathcal G_a(\mathcal C)\),}
\[
 {|f-g|\le w_{\mathcal C},\qquad
 Pw_{\mathcal C}\lesssim M_aa^p\delta,\qquad
 Pw_{\mathcal C}^2\lesssim\sigma_a^2\delta.}
\]

{Moreover, uniformly for \(0<\varepsilon\le1\),}
\begin{equation}\label{eq:supp-truncated-product-brackets}
{\sup_{f\in\mathcal G_a}Pf^2\lesssim\sigma_a^2,\qquad
 \log N_{[]}\bigl(C\sigma_a\varepsilon,\mathcal G_a,L_2(P)\bigr)
 \lesssim\log(C/(a\varepsilon)).}
\end{equation}
{The second-moment and bracketing bounds also hold for the centered class \(\{f-Pf:f\in\mathcal G_a\}\) and for fixed finite unions of such classes. If \(H\equiv1\), all these classes satisfy the bounds in}
\eqref{eq:supp-bounded-class-bounds}{.}
\end{lemma}
\begin{proof}
{Define}
\[
{I_{m,T}(y)\coloneqq
 \begin{cases}
  1,&m=0,1,\\
  \mathds{1}\{H(y)\le T\},&m=2,
 \end{cases}
 \qquad g_{a,\vartheta}\coloneqq I_{m,T}f_{a,\vartheta}.}
\]
{Then \(\mathcal G_a=\{g_{a,\psi_a(\tau)}:\tau\in\mathcal T_a\}\).
Write \(g_{a,\vartheta}=\sum_{\nu=1}^{J}g^{(\nu)}_{a,\vartheta}\), where
\(g^{(\nu)}_{a,\vartheta}\) is its \(\nu\)-th summand.
To build one bracket for each parameter cell, we first control individual
kernel factors and then the oscillation of the full function.
Fix a nonempty parameter cell \(\mathcal C\subset\Theta\) with
\(\operatorname{diam}_{d_a}(\mathcal C)\le\delta\), where \(0<\delta\le1\).
If \(\mathcal G_a(\mathcal C)=\varnothing\), take \(w_{\mathcal C}=0\).
Otherwise, fix an arbitrary representative \(\vartheta^\circ\in\mathcal C\).}

{Uniform volume and density bounds give, uniformly in \(x\in\cX_0\),}
\begin{align}
 {\mathbb P\bigl(d_{\cX}(X,x)\le C_0a\bigr)}
 &{\lesssim a^p,}
 \label{eq:supp-kernel-ball-probability-bound}\\
 {\mathbb P\bigl(d_{\cX}(X,x)\le C_0a,\,
       |d_{\cX}(X,x)-r|\le Ca\delta\bigr)}
 &{\lesssim a^p\delta,\qquad 0\le r\le Ca.}
 \label{eq:supp-kernel-shell-probability-bound}
\end{align}
{The choice of \(a_0\) ensures that these balls lie in \(V\).}

{Extend each \(K_{\nu k}\) by zero to \(\mathbb R\), and let
\(\mu_{\nu k}\) be the finite positive Stieltjes measure induced by the
right-continuous modification of its total variation function}
\citep[Theorem~1.16{{\normalcolor{} and Lemma~3.26}}]{Folland1999}{. Then}
\[
 {|K_{\nu k}(s)-K_{\nu k}(r)|\le\mu_{\nu k}([r,s]),
 \qquad 0\le r<s.}
\]
{For a sufficiently large fixed \(A\), define}
\[
{\begin{aligned}
 u_{\nu k}^\circ(z)&\coloneqq
 \frac{d_{\cX}(z,x_{\nu k}^\circ)}{at_{\nu k}^\circ},\quad
 w_{\nu k,\mathcal C}(z)\coloneqq
 \mathds{1}\{d_{\cX}(z,x_{\nu k}^\circ)\le C_0a\}\mu_{\nu k}\!\left(
 [u_{\nu k}^\circ(z)-A\delta,\,u_{\nu k}^\circ(z)+A\delta]
 \cap[0,\infty)\right).
\end{aligned}}
\]
{A kernel factor for a parameter in \(\mathcal C\) can be nonzero
only on this ball, since its center is within \(a\delta\) of the
representative center and \(C_0>1+\overline t\max_{\nu,k}R_{\nu k}\).
On the ball, its argument differs from \(u_{\nu k}^\circ(z)\) by at most
\(A\delta\). Thus, uniformly for \(\vartheta,\vartheta'\in\mathcal C\),}
\[
{\left|
 K_{\nu k}\!\left(d_{\cX}(z,x_{\nu k})/(at_{\nu k})\right)
 -K_{\nu k}\!\left(d_{\cX}(z,x'_{\nu k})/(at'_{\nu k})\right)
 \right|\le w_{\nu k,\mathcal C}(z).}
\]
{These envelopes are Borel by integration of Borel indicators against the
fixed measures \(\mu_{\nu k}\). An atom at zero is covered by a ball with
probability \(\lesssim a^p\delta^p\le a^p\delta\).
Tonelli's theorem and}
\eqref{eq:supp-kernel-shell-probability-bound}
{give}
\[
{\begin{aligned}
 \E[w_{\nu k,\mathcal C}(X)]
 &=\int_{[0,\infty)}
 \mathbb P\bigl(d_{\cX}(X,x_{\nu k}^\circ)\le C_0a,\,
       |u_{\nu k}^\circ(X)-u|\le A\delta\bigr)\,\mu_{\nu k}(du)
 \lesssim a^p\delta,\\
 \E[w_{\nu k,\mathcal C}(X)^2]
 &\le\mu_{\nu k}([0,\infty))\E[w_{\nu k,\mathcal C}(X)]
 \lesssim a^p\delta.
\end{aligned}}
\]

{For \(\vartheta,\vartheta'\in\mathcal C\),
\(d_a(\vartheta,\vartheta')\le\delta\le1\).
Where \(\chi_\nu(z,y)\ne0\) and either of the two \(\nu\)-th kernel
products is nonzero, the choice of \(C_0\) places \((z,y)\) in
\(\mathcal D_{\nu,a}(\vartheta)\cap\mathcal D_{\nu,a}(\vartheta')\).
On this common domain, split each summand difference into the change in
\(Q_{\nu,a}\) and the change in the kernel product. Applying}
\eqrefs{eq:supp-kernel-factor-bounds,eq:supp-kernel-factor-increment-bound}{,
expanding each kernel product difference one factor at a time, and using
boundedness of the remaining kernels and the one-factor bounds above gives}
\[
{\begin{aligned}
 |g_{a,\vartheta}(z,y)-g_{a,\vartheta'}(z,y)|
 &\leq\sum_{\nu=1}^{J}
 \left|g^{(\nu)}_{a,\vartheta}(z,y)
       -g^{(\nu)}_{a,\vartheta'}(z,y)\right|\\
 &\leq C_*M_aI_{m,T}(y)H(y)^m
 \sum_{\nu=1}^{J}\sum_{k=1}^{s_\nu}
 \left[
 w_{\nu k,\mathcal C}(z)
 +\delta\,\mathds{1}\{d_{\cX}(z,x^\circ_{\nu k})\le C_0a\}
 \right].
\end{aligned}}
\]
{Here \(C_*>0\) is independent of
\(a,\delta,\mathcal C,\vartheta,\vartheta'\).
The \(\nu\)-th summand difference vanishes when \(\chi_\nu(z,y)=0\)
or both kernel products vanish, so the bound holds for all \((z,y)\).
Take the final right-hand side as \(w_{\mathcal C}(z,y)\).
Thus \(|f-g|\le w_{\mathcal C}\) for all
\(f,g\in\mathcal G_a(\mathcal C)\).}

{Since \(H\ge1\), \(m\le2\), and \(I_{m,T}^2=I_{m,T}\),}
\eqref{eq:supp-response-envelope-moment}
{and the definitions of \(v_T\) and \(\sigma_a\) imply}
\[
{\sup_{z\in V}\E[I_{m,T}(Y)H(Y)^m\mid X=z]\lesssim1,\quad
 \sup_{z\in V}\E[I_{m,T}(Y)H(Y)^{2m}\mid X=z]
 \lesssim\frac{\sigma_a^2}{M_a^2a^p}.}
\]
{Every predictor envelope above is supported in \(V\).
Since \(\sum_{\nu=1}^{J}s_\nu\le2J\) is fixed, conditioning on \(X\)
and using the kernel-envelope estimates and}
\eqref{eq:supp-kernel-ball-probability-bound}
{gives}
\[
{\begin{aligned}
 Pw_{\mathcal C}
 &\lesssim M_a\sum_{\nu=1}^{J}\sum_{k=1}^{s_\nu}
 \left[
 \E[w_{\nu k,\mathcal C}(X)]
 +\delta\,\mathbb P(d_{\cX}(X,x^\circ_{\nu k})\le C_0a)
 \right]
 \lesssim M_aa^p\delta,\\
 Pw_{\mathcal C}^2
 &\lesssim \frac{\sigma_a^2}{a^p}
 \sum_{\nu=1}^{J}\sum_{k=1}^{s_\nu}
 \left[
 \E[w_{\nu k,\mathcal C}(X)^2]
 +\delta^2\mathbb P(d_{\cX}(X,x^\circ_{\nu k})\le C_0a)
 \right]\lesssim\sigma_a^2(\delta+\delta^2)
 \lesssim\sigma_a^2\delta.
\end{aligned}}
\]
{The size bound}
\eqref{eq:supp-kernel-factor-bounds}{, bounded kernels, and the same
conditional moments and support probabilities also give
\(\sup_{g\in\mathcal G_a}Pg^2\lesssim\sigma_a^2\).}

{We construct one bracket for each parameter cell.
For a cell \(\mathcal C\subset\Theta\) with
\(\operatorname{diam}_{d_a}(\mathcal C)\le\delta\), where \(0<\delta\le1\),
and \(\mathcal G_a(\mathcal C)\ne\varnothing\), choose
\(g_{\mathcal C}\in\mathcal G_a(\mathcal C)\) and set
\(l_{\mathcal C}\coloneqq g_{\mathcal C}-w_{\mathcal C}\) and
\(u_{\mathcal C}\coloneqq g_{\mathcal C}+w_{\mathcal C}\).
The preceding oscillation and moment bounds give}
\[
{\begin{aligned}
 l_{\mathcal C}
 \le g\le u_{\mathcal C},\qquad g\in\mathcal G_a(\mathcal C),\qquad
 \|u_{\mathcal C}-l_{\mathcal C}\|_{P,2}
 =2\sqrt{Pw_{\mathcal C}^2}
 \lesssim\sigma_a\sqrt{\delta}.
\end{aligned}}
\]
{The endpoints are measurable and belong to \(L_2(P)\).}

{Now fix \(0<\varepsilon\le1\) and take \(\delta=\varepsilon^2\).
For sufficiently small fixed \(c>0\), cover \(\Theta\) by finitely many
Borel cells of \(d_a\)-diameter at most \(\delta\), using meshes \(ca\delta\)
for the centers and \(c\delta\) for the remaining parameters.
Let \(\mathfrak C\) denote the cells in this cover for which
\(\mathcal G_a(\mathcal C)\ne\varnothing\).
For each \(\mathcal C\in\mathfrak C\), use the bracket constructed above.
Every \(\psi_a(\tau)\), \(\tau\in\mathcal T_a\), belongs to a retained cell, so}
\[
{\mathcal G_a
 =\bigcup_{\mathcal C\in\mathfrak C}\mathcal G_a(\mathcal C)
 \subseteq\bigcup_{\mathcal C\in\mathfrak C}[l_{\mathcal C},u_{\mathcal C}].}
\]
{Since \(\delta=\varepsilon^2\), these brackets satisfy
\(\sup_{\mathcal C\in\mathfrak C}
\|u_{\mathcal C}-l_{\mathcal C}\|_{P,2}\lesssim\sigma_a\varepsilon\).
Hence the definition of the bracketing number and {the number of cells in the
parameter grid, whose dimension is fixed,} give}
\[
{N_{[]}\bigl(C\sigma_a\varepsilon,\mathcal G_a,L_2(P)\bigr)
 \le\#\mathfrak C
 \le\left(C/(a\delta)\right)^{C_1}\left(C/\delta\right)^{C_2},}
\]
{where \(C_1,C_2\) are fixed.
Taking logarithms and using \(\delta=\varepsilon^2\) yields}
\[
{\log N_{[]}\bigl(C\sigma_a\varepsilon,\mathcal G_a,L_2(P)\bigr)
 \lesssim\log\!\left(C/(a\varepsilon^2)\right)
 +\log\!\left(C/\varepsilon^2\right)
 \lesssim\log\!\left(C/(a\varepsilon)\right).}
\]
{This proves}
\eqref{eq:supp-truncated-product-brackets}{.
The constants are independent of \(\psi_a\), since the cells are
constructed in the fixed parameter space \(\Theta\).}

{For a bracket \([l,u]\), the centered bracket is
\([l-Pu,u-Pl]\), whose \(L_2(P)\)-width is at most twice that of
\([l,u]\). Also, \(P(g-Pg)^2\le Pg^2\).
For a fixed finite union, take the union of the brackets of its component
classes. Finally, if \(H\equiv1\), then \(I_{m,T}=1\) and truncation has
no effect. The same construction gives the moment and bracketing bounds
with scale \(M_aa^{p/2}\), and
\(\sup_{g\in\mathcal G_a}\|g\|_\infty\lesssim M_a\).
These bounds are preserved under centering and fixed finite unions,
which verifies} \eqref{eq:supp-bounded-class-bounds} {when \(H\equiv1\).}
\end{proof}

For the finite-moment bounds below, let \(a=a_n\to0\) satisfy \eqref{eq:supp-uniform-bandwidth-rate}. The truncation level \(t_n\) from \Cref{sec:supp-notation} then satisfies
\begin{equation}\label{eq:supp-truncation-rate-consequences}
\begin{aligned}
\frac{t_n(\log n)^{3/2}}{\sqrt{na^p}}&\to0,\quad \frac{t_n}{na^p}=o((na^p)^{-1/2}),\quad \frac{t_n}{na^p}=o((\log n)^{-1}),\\
\sqrt{\frac{\log n}{na^p}}&=o((\log n)^{-1/2}),\quad \sqrt{na^p\log n}\,t_n^{1-2r_0}+t_n^{1-r_0}\log n\to0\quad\text{under \textup{(H)}}.
\end{aligned}
\end{equation}
Under \textup{(H)}, the last limit follows from \(na^p\le n\) and \(\sqrt n\,t_n^{1-2r_0}=t_n^{1-r_0}=n^{1/(2r_0)-1/2}\).

\begin{lemma}
\label{lem:supp_localized_finite_moment_maximal_inequality}
{Consider the classes \(\mathcal F_a\) in}
\eqref{eq:supp-kernel-factor-class}
{under the preceding setup.}
{Use \(t_n\) and \(c_n(a)\) from} \Cref{sec:supp-notation}
{, with \(r_0\) as in}
\eqref{eq:supp-response-envelope-moment}{.}
{Then}
\begin{equation}
\label{eq:supp-localized-finite-moment-rate}
\sup_{f\in\mathcal F_a}|(P_n-P)f|
=\begin{cases}
 O_p(M_aa^p\sqrt{\log n/(na^p)}),&m=0,1,\\
 O_p(M_aa^pc_n(a)),&m=2.
 \end{cases}
\end{equation}
For {\((M_a,m)=(a^{-p/2},1)\)},
\[
\sup_f|\mathbb G_nf|=O_p(\sqrt{\log n}),\qquad
\sup_f|\mathbb G_n^\ast f|=O_p^\ast(\sqrt{\log n}).
\]
\end{lemma}
\begin{proof}
{We first derive an empirical process bound for a bounded class, to be
applied after response truncation. Brackets reduce the supremum to finite
maxima over representative functions and bracket widths.
Let \(\mathcal A\) be a nonempty class of measurable functions on
\(\mathcal X\times\mathcal Y\) such that}
\[
{
 \sup_{f\in\mathcal A}\|f\|_\infty\le B,\qquad
 \sup_{f\in\mathcal A}Pf^2\le\sigma^2,\qquad
 N_{[]}(\epsilon,\mathcal A,L_2(P))<\infty.
}
\]
{Choose such a bracket cover. After clipping the endpoints to \([-B,B]\) and discarding brackets that
do not meet \(\mathcal A\), write the remaining brackets as
\([l_j,u_j]\), \(1\le j\le N\).
Choose \(f_j\in\mathcal A\cap[l_j,u_j]\) and set
\(w_j\coloneqq u_j-l_j\). Then}
\[
{|f-f_j|\le w_j\quad(f\in\mathcal A\cap[l_j,u_j]),\quad
 0\le w_j\le2B,\quad Pw_j\le\epsilon,\quad Pw_j^2\le\epsilon^2.}
\]
{Consequently,}
\[
{\begin{aligned}
 \sup_{f\in\mathcal A}|\mathbb G_nf|
 &\le\max_{j\le N}
 \left\{|\mathbb G_nf_j|+\sqrt n(P_n+P)w_j\right\}\\
 &=\max_{j\le N}
 \left\{|\mathbb G_nf_j|+\mathbb G_nw_j+2\sqrt n\,Pw_j\right\}\\
 &\le\max_{j\le N}|\mathbb G_nf_j|
 +\max_{j\le N}|\mathbb G_nw_j|+2\sqrt n\,\epsilon.
\end{aligned}}
\]
{Applying Bernstein's inequality}
\citep[Lemma~2.2.10 and Example~2.2.11]{VanDerVaartandWellner2023}
{to the \(f_j\) and \(w_j\), followed by a union bound, gives,
for \(u\ge1\), with probability at least \(1-4e^{-u}\),}
\begin{equation}\label{eq:supp-finite-bracket-bernstein}
{\sup_{f\in\mathcal A}|\mathbb G_nf|
 \lesssim(\sigma+\epsilon)\sqrt{\log(2N)+u}
 +\frac{B\{\log(2N)+u\}}{\sqrt n}+\sqrt n\,\epsilon.}
\end{equation}

{We now apply} \eqref{eq:supp-finite-bracket-bernstein}
{to a bounded class obtained by truncating the response envelope.
Recall that \(V=\bigcup_{\ell=1}^{L}V_\ell\) is the fixed predictor
neighborhood from the setup. By the choice of \(a_0\), every
\(f\in\mathcal F_a\) satisfies \(f(z,y)=0\) whenever \(z\notin V\).
Set}
\[
{f^T\coloneqq f\mathds{1}\{H\le t_n\},\qquad
 A_n\coloneqq\{H(Y_i)\le t_n\text{ whenever }X_i\in V,\ 1\le i\le n\}.}
\]
{Under \textup{(C)}, \(f^T=f\) and \(A_n=\Omega\).
Under \textup{(H)}, \(t_n^{2r_0}=n\) and the local integrability in}
\eqref{eq:supp-response-envelope-moment} {give}
\begin{equation}\label{eq:supp-localized-truncation-probability}
{\mathbb P(A_n^c)\le n\mathbb P(X\in V,H>t_n)
 \le\E[H^{2r_0}\mathds{1}\{X\in V,H>t_n\}]\to0.}
\end{equation}
{On \(A_n\), all functions satisfy \(P_nf=P_nf^T\), so}
\begin{equation}\label{eq:supp-localized-truncation-identities}
{(P_n-P)f=(P_n-P)f^T-P(f-f^T),\qquad f\in\mathcal F_a.}
\end{equation}
{Consequently, on \(A_n\),}
\[
{\sup_{f\in\mathcal F_a}|(P_n-P)f|
 \le\sup_{f\in\mathcal F_a}|(P_n-P)f^T|
 +\sup_{f\in\mathcal F_a}P|f-f^T|.}
\]
{Under \textup{(H)}, the conditional moment bound gives}
\[
{\sup_{z\in V}\E[H^m\mathds{1}\{H>t_n\}\mid X=z]
 \lesssim t_n^{m-2r_0}.}
\]
{Combining this with the factor bound}
\eqref{eq:supp-kernel-factor-bounds} {and the support probability bound}
\eqref{eq:supp-kernel-ball-probability-bound} {yields}
\[
{\sup_{f\in\mathcal F_a}P|f-f^T|
 \lesssim\mathds{1}_{\mathrm{(H)}}M_aa^pt_n^{m-2r_0}.}
\]

{It remains to bound the empirical process of the truncated class.
Define}
\[
{\mathcal A\coloneqq
 \{f^T/(M_aa^{p/2}):f\in\mathcal F_a\},\qquad
 v_{m,n}\coloneqq
 \begin{cases}
  1,&m=0,1,\\
  t_n^{(4-2r_0)_+},&m=2.
 \end{cases}}
\]
{The factor and second-moment bounds in}
\eqrefs{eq:supp-kernel-factor-bounds,eq:supp-truncated-product-brackets}
{give}
\[
{\sup_{g\in\mathcal A}\|g\|_\infty
 \lesssim a^{-p/2}t_n^m,\qquad
 \sup_{g\in\mathcal A}Pg^2\lesssim v_{m,n}.}
\]
{For \(m=0,1\), multiply the untruncated brackets by
\(\mathds{1}\{H\le t_n\}\); for \(m=2\), use the directly constructed
truncated brackets. With absolute width \(\epsilon_n\coloneqq n^{-4}\),}
\eqref{eq:supp-truncated-product-brackets} {gives}
\[
{\log N_{[]}(\epsilon_n,\mathcal A,L_2(P))
 \lesssim\log\!\left(C\sqrt{v_{m,n}}/(a\epsilon_n)\right)
 \lesssim\log n,}
\]
{since \(a^{-1}\) and \(t_n\) are polynomially bounded in \(n\).
Thus apply} \eqref{eq:supp-finite-bracket-bernstein}
{with \(B\lesssim a^{-p/2}t_n^m\),
\(\sigma^2\lesssim v_{m,n}\), \(\epsilon=\epsilon_n\),
\(N\le n^C\), and \(u=C\log n\).
After rescaling, the bracket-width remainder
\(O(M_aa^{p/2}n^{-4})\) is negligible relative to the first term below, and}
\begin{equation}
{\sup_{f\in\mathcal F_a}|(P_n-P)f^T|
 =O_p\!\left(M_a\left\{
 \sqrt{\frac{a^pv_{m,n}\log n}{n}}
 +\frac{t_n^m\log n}{n}\right\}\right).}
\label{eq:supp-truncated-empirical-rate}
\end{equation}
{The reduction above and \(\mathbb P(A_n)\to1\) therefore yield}
\[
{\sup_{f\in\mathcal F_a}|(P_n-P)f|
 =O_p\!\left(M_aa^p\left\{
 \sqrt{\frac{v_{m,n}\log n}{na^p}}
 +\frac{t_n^m\log n}{na^p}
 +\mathds{1}_{\mathrm{(H)}}t_n^{m-2r_0}\right\}\right).}
\]
{Moreover,}
\[
{\sqrt{\frac{v_{m,n}\log n}{na^p}}
 +\frac{t_n^m\log n}{na^p}
 +\mathds{1}_{\mathrm{(H)}}t_n^{m-2r_0}
 \lesssim
 \begin{cases}
  \displaystyle\sqrt{\frac{\log n}{na^p}},&m=0,1,\\
  c_n(a),&m=2.
 \end{cases}}
\]
{For \(m=0,1\), \(v_{m,n}=1\), and}
\eqref{eq:supp-truncation-rate-consequences}
{shows that the last two terms on the left are
\(o(\sqrt{\log n/(na^p)})\).
For \(m=2\), use \(v_{2,n}\le1+t_n^4/n\) and
\(t_n^{2-2r_0}=t_n^2/n\) under \textup{(H)};
under \textup{(C)}, \(v_{2,n}=t_n=1\) and the tail term vanishes.
This proves}
\eqref{eq:supp-localized-finite-moment-rate}{.}

{{The empirical process bound follows from}}
\eqref{eq:supp-localized-finite-moment-rate}
{after multiplying by \(\sqrt n\) and setting {\((M_a,m)=(a^{-p/2},1)\)}.}

{For the multiplier bound, reuse the finite-bracket construction
for the bounded class \(\mathcal A\) used to derive}
\eqref{eq:supp-finite-bracket-bernstein}{.}
{Write \(\xi\sim N(0,1)\) for a multiplier independent of \(W\).}
{The same representatives \(f_j\) and widths \(w_j\) give}
\[
{\begin{aligned}
 \sup_{f\in\mathcal A}\left|\sum_i\xi_i f(W_i)\right|
 &\le\max_{j\le N}\left\{
 \left|\sum_i\xi_i f_j(W_i)\right|
 +\sum_i|\xi_i|w_j(W_i)\right\}\\
 &\le\max_{j\le N}\left|\sum_i\xi_i f_j(W_i)\right|+\max_{j\le N}\left|\sum_i
 \bigl\{|\xi_i|w_j(W_i)-\E[|\xi|]Pw_j\bigr\}\right|
 +\E[|\xi|]n\epsilon.
\end{aligned}}
\]
{{Both finite families are centered under the joint distribution of the data and multipliers.}
The second moments satisfy}
\[
{\E[\xi^2f_j(W)^2]=Pf_j^2\le\sigma^2,\qquad
 \operatorname{Var}(|\xi|w_j(W))\le Pw_j^2\le\epsilon^2.}
\]
{{The moments of the Gaussian multipliers}, together with \(|f_j|\le B\) and
\(0\le w_j\le2B\), give, for every integer \(k\ge3\),}
\[
{\begin{aligned}
 \E[|\xi f_j(W)|^k]
 &\le\frac{k!}{2}C\sigma^2(CB)^{k-2},\\
 \E\!\left[\bigl||\xi|w_j(W)-\E[|\xi|]Pw_j\bigr|^k\right]
 &\le\frac{k!}{2}C\epsilon^2(CB)^{k-2}.
\end{aligned}}
\]
{Under the joint distribution of the data and multipliers,
applying the moment form of Bernstein's inequality}
\citep[Lemma~2.2.10]{VanDerVaartandWellner2023}
{to these two finite families, followed by a union bound and division by \(\sqrt n\), gives,
for \(u\ge1\), with probability at least \(1-4e^{-u}\),}
\[
{\sup_{f\in\mathcal A}\left|n^{-1/2}\sum_i\xi_i f(W_i)\right|
 \lesssim(\sigma+\epsilon)\sqrt{\log(2N)+u}
 +\frac{B\{\log(2N)+u\}}{\sqrt n}+\sqrt n\,\epsilon.}
\]

{Now take {\((M_a,m)=(a^{-p/2},1)\)}.
Since \(M_aa^{p/2}=1\), \(\mathcal A=\{f^T:f\in\mathcal F_a\}\), with}
\[
{B\lesssim a^{-p/2}t_n,\qquad\sigma^2\lesssim1,\qquad
 \epsilon=n^{-4},\qquad\log N\lesssim\log n.}
\]
{Taking \(u=C\log n\) and using}
\eqref{eq:supp-truncation-rate-consequences} {yields}
\[
{\begin{aligned}
 \sup_{f\in\mathcal F_a}\left|n^{-1/2}\sum_i\xi_i f^T(W_i)\right|
 &=O_p\!\left(\sqrt{\log n}
 +\frac{t_n\log n}{\sqrt{na^p}}+n^{-7/2}\right)=O_p(\sqrt{\log n}).
\end{aligned}}
\]

{By Cauchy--Schwarz and the second-moment bound in}
\eqref{eq:supp-truncated-product-brackets}{,
\(\sup_{f\in\mathcal F_a}P|f^T|\lesssim1\).
The empirical bound in} \eqref{eq:supp-truncated-empirical-rate}{,
with {\((M_a,m)=(a^{-p/2},1)\)}, gives
\(\sup_{f\in\mathcal F_a}|(P_n-P)f^T|=O_p(\sqrt{\log n/n})\).
Together with the preceding multiplier bound and
\(n^{-1/2}\sum_i\xi_i=O_p(1)\), these estimates yield}
\[
{\begin{aligned}
\sup_{f\in\mathcal F_a}|\mathbb G_n^\ast f^T|
&=n^{-1/2}\sup_f\left|\sum_i\xi_i f^T(W_i)
-\left(\sum_i\xi_i\right)P_nf^T\right|\\
&\leq n^{-1/2}\sup_f\left\{
\left|\sum_i\xi_i f^T(W_i)\right|
+\left|\sum_i\xi_i\right||P_nf^T|\right\}\\
&\leq n^{-1/2}{\sup_f\left\{
\left|\sum_i\xi_i f^T(W_i)\right|
+\left|\sum_i\xi_i\right|
\left\{P|f^T|+|(P_n-P)f^T|\right\}\right\}}
=O_p(\sqrt{\log n}).
\end{aligned}}
\]
{On \(A_n\), \((f-f^T)(W_i)=0\) for every \(i\) and
\(f\in\mathcal F_a\), so \(\mathbb G_n^\ast f=\mathbb G_n^\ast f^T\).
Since} \eqref{eq:supp-localized-truncation-probability}
{gives \(\mathbb P(A_n)\to1\), the same bound holds for
\(\sup_{f\in\mathcal F_a}|\mathbb G_n^\ast f|\) under the joint distribution.
Finally, for every \(M,\eta>0\), Markov's inequality gives}
\[
{\begin{aligned}
 \mathbb P\!\left(
 \mathbb P^\ast\!\left(\sup_{f\in\mathcal F_a}|\mathbb G_n^\ast f|
 >M\sqrt{\log n}\right)>\eta\right)\le\eta^{-1}\mathbb P\!\left(
 \sup_{f\in\mathcal F_a}|\mathbb G_n^\ast f|>M\sqrt{\log n}\right).
\end{aligned}}
\]
{Taking \(\lim_{M\to\infty}\limsup_{n\to\infty}\) proves the stated
conditional order.}
\end{proof}

Under the uniform bandwidth restrictions for bias control,
$n^{-1/p}<h\le Cn^{-1/(p+4+2\gamma)}$ eventually, which implies
$\log(1/h)\asymp\log n$. For RRBC, $b\asymp h$.
Under the alternative conditions on the population displacements in
\Cref{cor:supp-bias-free-score-representations},
{the condition \(\log(1/h)\asymp\log n\)} is retained explicitly.

\begin{proposition}
\label{prop:cck-multiplier-bounded-envelope}
Assume the conditions of
\Cref{lem:supp_localized_finite_moment_maximal_inequality}
with {\((M_a,m)=(a^{-p/2},1)\)}.
{Let \(g\) range over a class of the form}
\eqref{eq:supp-kernel-factor-class}{.}
Suppose $\mathcal F_n\coloneqq\{g-Pg\}$ is nonempty, symmetric,
and deterministic, with $Pf^2=1$ for every $f\in\mathcal F_n$.
There is a separable centered Gaussian process $G_{P,n}$ on $\mathcal F_n$
with covariance $P(f_1f_2)$ and bounded, uniformly $e_P$-continuous paths. Put
\[
 S_n^G\coloneqq\sup_{f\in\mathcal F_n}G_{P,n}f.
\]
{The Gaussian supremum satisfies}
\begin{equation}\label{eq:supp-localized-gaussian-bounds}
 S_n^G=O_p(\sqrt{\log n}),\qquad
 \sup_z\mathbb P(|S_n^G-z|\le\epsilon)
 \lesssim\epsilon\sqrt{\log n},\qquad\epsilon>0.
\end{equation}
{Moreover,}
\begin{equation}\label{eq:supp-localized-gaussian-cdf}
 \sup_z\left|\mathbb P\bigl({\sup_{f\in\mathcal F_n}}\mathbb G_nf\le z\bigr)-\mathbb P(S_n^G\le z)\right|
 \to0,\quad
 \sup_z\left|\mathbb P^\ast\bigl({\sup_{f\in\mathcal F_n}}\mathbb G_n^\ast f\le z\bigr)-\mathbb P(S_n^G\le z)\right|
 \pto0.
\end{equation}
The cdf assertions also hold after squaring the nonnegative suprema.
\end{proposition}
\begin{proof}
The centered brackets in
\Cref{lem:supp_local_polynomial_kernel_factors_are_localized} give
\begin{equation}\label{eq:supp-localized-gaussian-entropy}
 \log\cN(\mathcal F_n,e_P,\epsilon)
 \leq\log N_{[]}(\epsilon,\mathcal F_n,L_2(P))
 \lesssim\log(C/(a\epsilon)),\quad 0<\epsilon\le1,
 \qquad\operatorname{diam}_{e_P}(\mathcal F_n)\le2.
\end{equation}
Thus the restriction of an isonormal process to $\mathcal F_n$ has a
bounded, uniformly $e_P$-continuous separable version, which we take as
$G_{P,n}$. Dudley's inequality
\citep[Corollary~2.2.9]{VanDerVaartandWellner2023} and anti-concentration for suprema of Gaussian variables with unit variance \citep[Theorem~2.1]{Chernozhukovetal2014b} yield
\[
 \E[S_n^G]\lesssim\sqrt{\log n},\qquad
 \sup_z\mathbb P(|S_n^G-z|\le\epsilon)
 \le4\epsilon(\E[S_n^G]+1),
\]
proving \eqref{eq:supp-localized-gaussian-bounds}.

{It remains to prove the cdf approximations in}
\eqref{eq:supp-localized-gaussian-cdf}{.}

{Define the truncated functions and their centered versions by}
\[
 g^T\coloneqq g\mathds{1}\{H\le t_n\},\qquad
 f^T\coloneqq g^T-Pg^T,
 \qquad g^T=g\quad\text{under \textup{(C)}}.
\]

{Choose deterministic brackets of \(L_2(P)\) width \(n^{-1}\) for the class of uncentered functions \(g\).
From each bracket containing a class member, select a representative \(g_j\) belonging to the class and the bracket.
The bracketing bound in}
\Cref{lem:supp_local_polynomial_kernel_factors_are_localized}
{and \(\log(1/a)\asymp\log n\) give \(J_n\le n^C\).
Duplicate representatives until \(J_n\ge3\). Then}
\begin{equation}\label{eq:supp-localized-finite-representatives}
 3\le J_n\le n^C,\qquad
 {\sup_g\min_{j\le J_n}\|g-g_j\|_{P,2}\le n^{-1},}\qquad
 f_j\coloneqq g_j-Pg_j.
\end{equation}

{Fix \(\varepsilon>0\). For \(n\ge3\),}
\[
{\sup_{z\in\mathbb R}\left|
\mathbb P\!\left(\sup_{f\in\mathcal F_n}\mathbb G_nf\le z\right)
-\mathbb P(S_n^G\le z)\right|
\le\mathrm I_n(\varepsilon)+\mathrm{II}_n(\varepsilon)
+\mathrm{III}_n+\mathrm{IV}_n(\varepsilon),}
\]
{where}
{\small
\[
{\begin{gathered}
\mathrm I_n(\varepsilon)\coloneqq
\mathbb P\!\left(\left|\sup_{f\in\mathcal F_n}\mathbb G_nf
-\max_{j\le J_n}\mathbb G_nf_j^T\right|>
\frac{\varepsilon}{\sqrt{\log n}}\right),\quad
\mathrm{II}_n(\varepsilon)\coloneqq
\mathbb P\!\left(\left|S_n^G-\max_{j\le J_n}G_{P,n}f_j\right|>
\frac{\varepsilon}{\sqrt{\log n}}\right),\\
\mathrm{III}_n\coloneqq\sup_{z\in\mathbb R}\left|
\mathbb P\!\left(\max_{j\le J_n}\mathbb G_nf_j^T\le z\right)
-\mathbb P\!\left(\max_{j\le J_n}G_{P,n}f_j\le z\right)\right|,\quad
\mathrm{IV}_n(\varepsilon)\coloneqq\sup_{z\in\mathbb R}
\mathbb P\!\left(|S_n^G-z|\le\frac{2\varepsilon}{\sqrt{\log n}}\right).
\end{gathered}}
\]
}
{For the conditional multiplier distribution,}
\[
{\sup_{z\in\mathbb R}\left|
\mathbb P^\ast\!\left(\sup_{f\in\mathcal F_n}\mathbb G_n^\ast f\le z\right)
-\mathbb P(S_n^G\le z)\right|
\le\mathrm I_n^\ast(\varepsilon)+\mathrm{II}_n(\varepsilon)
+\mathrm{III}_n^\ast+\mathrm{IV}_n(\varepsilon),}
\]
{where}
\[
\resizebox{\linewidth}{!}{$\displaystyle
{\mathrm I_n^\ast(\varepsilon)\coloneqq
\mathbb P^\ast\!\left(\left|\sup_{f\in\mathcal F_n}\mathbb G_n^\ast f
-\max_{j\le J_n}\mathbb G_n^\ast f_j^T\right|>
\frac{\varepsilon}{\sqrt{\log n}}\right),\quad
\mathrm{III}_n^\ast\coloneqq\sup_{z\in\mathbb R}\left|
\mathbb P^\ast\!\left(\max_{j\le J_n}\mathbb G_n^\ast f_j^T\le z\right)
-\mathbb P\!\left(\max_{j\le J_n}G_{P,n}f_j\le z\right)\right|.}
$}
\]
{The two cdf bounds follow from the event inclusions used to derive}
\eqref{eq:supp-cdf-perturbation-bound}{.
The two replacements each use tolerance \(\varepsilon/\sqrt{\log n}\), giving a total threshold shift of \(2\varepsilon/\sqrt{\log n}\).
For the multiplier bound, probabilities involving multiplier statistics are conditional on the data; the Gaussian probabilities defining \(\mathrm{II}_n(\varepsilon)\) and \(\mathrm{IV}_n(\varepsilon)\) are unconditional.
By} \eqref{eq:supp-localized-gaussian-bounds}{,}
\[
{\mathrm{IV}_n(\varepsilon)\le C\varepsilon,}
\]
{with \(C\) independent of \(n\) and \(\varepsilon\).
We prove \(\mathrm I_n(\varepsilon)\to0\), \(\mathrm{II}_n(\varepsilon)\to0\), and \(\mathrm{III}_n\to0\), together with
\(\mathrm I_n^\ast(\varepsilon)\pto0\) and \(\mathrm{III}_n^\ast\pto0\).}

{We first bound the mean and covariance errors caused by truncation; the mean error enters \(\mathrm I_n(\varepsilon)\), and the covariance error enters \(\mathrm{III}_n\) and \(\mathrm{III}_n^\ast\).}

{For arbitrary class members \(g,g'\), write \(f=g-Pg\) and \(f'=g'-Pg'\).
Truncating both functions on \(\{H\le t_n\}\) gives}
\[
{P(ff')-P(f^T(f')^T)
 =P[gg'\mathds{1}\{H>t_n\}]-Pg\,P(g'-(g')^T)-P(g-g^T)\,P[(g')^T].}
\]
{Under \textup{(H)},} \eqref{eq:supp-response-envelope-moment} {gives}
\[
 \sup_{z\in V}\E[H^{{k}}\mathds{1}\{H>t_n\}\mid X=z]\lesssim t_n^{{k}-2r_0},\qquad {k}=1,2.
\]
{Conditioning on \(X\) and integrating over the kernel supports gives
\({\sup_g\{P|g|+P|g^T|\}}\lesssim a^{p/2}\),
\(\sup_gP|g-g^T|\lesssim a^{p/2}t_n^{1-2r_0}\), and
\(P(|gg'|\mathds{1}\{H>t_n\})\lesssim t_n^{2-2r_0}\).
The covariance identity for \(f,f'\) bounds the covariance truncation error by \(Ct_n^{2-2r_0}\), since
\(a^pt_n^{1-2r_0}\le t_n^{2-2r_0}\).
Under \textup{(C)}, \(g^T=g\), so the mean and covariance truncation errors are zero. Hence}
\begin{equation}\label{eq:supp-localized-gaussian-truncation}
{\begin{gathered}
{\sup_g\{P|g|+P|g^T|\}}\lesssim a^{p/2},\quad
\sup_gP|g-g^T|\lesssim\mathds{1}_{\mathrm{(H)}}a^{p/2}t_n^{1-2r_0},\\
\sup_{f_1,f_2\in\mathcal F_n}
|P(f_1f_2)-P(f_1^Tf_2^T)|
\lesssim\mathds{1}_{\mathrm{(H)}}t_n^{2-2r_0}.
\end{gathered}}
\end{equation}

{To bound \(\mathrm I_n(\varepsilon)\) and \(\mathrm I_n^\ast(\varepsilon)\), control the bracket replacement errors after truncation.}

{Truncate the bracket endpoints by the common response cutoff and clip
them to \([-Ca^{-p/2}t_n,Ca^{-p/2}t_n]\). Write \(w_j\) for the resulting
widths. For \(g\) in bracket \(j\),}
\[
{|g^T-g_j^T|\le w_j,\qquad Pw_j^2\le n^{-2},\qquad
\|w_j\|_\infty\lesssim a^{-p/2}t_n.}
\]
{The Bernstein argument in the proof of}
\Cref{lem:supp_localized_finite_moment_maximal_inequality}
{applied to \(w_j\) and \(|\xi|w_j\), followed by a union bound over
\(J_n\le n^C\), gives}
\[
{\max_{j\le J_n}\left\{
\sqrt n(P_n+P)w_j+n^{-1/2}\sum_i|\xi_i|w_j(W_i)\right\}
=O_p\!\left(n^{-1/2}+\frac{t_n\log n}{\sqrt{na^p}}\right).}
\]
{The quantities \(\sqrt n(P_n+P)w_j\) bound the empirical replacement errors within the truncated brackets.
The sums \(n^{-1/2}\sum_i|\xi_i|w_j(W_i)\) bound the uncentered multiplier replacement errors for \(g^T-g_j^T\).
Centering by \(P_n\) adds at most
\(\left|n^{-1/2}\sum_i\xi_i\right|\max_jP_nw_j
=O_p(n^{-1}+t_n\log n/(na^{p/2}))\).
On the event
\(A_n\) in}
\eqref{eq:supp-localized-truncation-probability}{, truncation leaves
the multiplier process unchanged, while its empirical effect is at most
\(C\mathds{1}_{\mathrm{(H)}}\sqrt{na^p}t_n^{1-2r_0}\), by}
\eqrefs{eq:supp-localized-truncation-identities,eq:supp-localized-gaussian-truncation}{.
Combining} \eqrefs{eq:supp-truncation-rate-consequences,eq:supp-localized-truncation-probability}
{with the bracket replacement bounds gives}
\begin{equation}\label{eq:supp-localized-finite-max-reduction}
 \left|{\sup_{f\in\mathcal F_n}}\mathbb G_nf-\max_{j\le J_n}\mathbb G_nf_j^T\right|
 =o_p((\log n)^{-1/2}),\qquad
 \left|{\sup_{f\in\mathcal F_n}}\mathbb G_n^\ast f-\max_{j\le J_n}\mathbb G_n^\ast f_j^T\right|
 =o_p^\ast((\log n)^{-1/2}).
\end{equation}
{For every fixed \(\varepsilon>0\), the empirical replacement bound in}
\eqref{eq:supp-localized-finite-max-reduction} {gives \(\mathrm I_n(\varepsilon)\to0\).
The joint multiplier replacement bound gives \(\E[\mathrm I_n^\ast(\varepsilon)]\to0\), so the Markov argument in the proof of}
\Cref{lem:supp_localized_finite_moment_maximal_inequality}
{gives \(\mathrm I_n^\ast(\varepsilon)\pto0\).}

{Since centering contracts \(L_2(P)\) distances,}
\eqref{eq:supp-localized-finite-representatives} {gives}
\begin{equation}\label{eq:supp-localized-gaussian-net-radius}
 \sup_g\min_{j\le J_n}\|f-f_j\|_{P,2}
 {\le\sup_g\min_{j\le J_n}\|g-g_j\|_{P,2}\le n^{-1}.}
\end{equation}
{Covering the two arguments of \(f-f_j\) separately gives the entropy bound}
\eqref{eq:supp-localized-gaussian-entropy}
{for the class of differences within the \(n^{-1}\) brackets.
Dudley's inequality}
\citep[Corollary~2.2.9]{VanDerVaartandWellner2023}{, with}
\eqrefs{eq:supp-localized-gaussian-entropy,eq:supp-localized-gaussian-net-radius}{, gives}
\begin{equation}\label{eq:supp-localized-gaussian-net-reduction}
 {\E\left[\left|S_n^G-\max_{j\le J_n}G_{P,n}f_j\right|\right]}
 {\lesssim n^{-1}\sqrt{\log n}}
 =o((\log n)^{-1/2}).
\end{equation}
{Markov's inequality and} \eqref{eq:supp-localized-gaussian-net-reduction} {give
\(\mathrm{II}_n(\varepsilon)\lesssim(\log n)/(n\varepsilon)\to0\) for every fixed \(\varepsilon>0\).}

{To bound \(\mathrm{III}_n\) and \(\mathrm{III}_n^\ast\), we record the covariance, fourth-moment, and envelope bounds.
The vectors \((f_j^T(W_i))_{j\le J_n}\) are i.i.d. and centered, and \((G_{P,n}f_j)_{j\le J_n}\) has covariance \((P(f_jf_k))_{j,k\le J_n}\).
By} \eqref{eq:supp-response-envelope-moment}{,}
\[
\begin{gathered}
 \sup_{z\in V}\E[H(Y)^2\mid X=z]\lesssim1,\quad
 \sup_{z\in V}
 \E[H(Y)^4\mathds{1}\{H(Y)\le t_n\}\mid X=z]
 \lesssim v_{t_n}=t_n^{(4-2r_0)_+}.
\end{gathered}
\]
Localization and centering then give
\begin{equation}\label{eq:supp-localized-cckk-moments}
\begin{gathered}
 \max_jP(f_j^T)^2\le\max_jP(g_j^T)^2\lesssim1,\quad
 \max_jP|f_j^T|^4\le16\max_jP|g_j^T|^4\lesssim a^{-p}v_{t_n}.
\end{gathered}
\end{equation}
Using \(Pf_j^2=1\), the covariance bound in
\eqref{eq:supp-localized-gaussian-truncation},
{the moment bounds in} \eqref{eq:supp-localized-cckk-moments}{, and truncation at \(t_n\), we obtain}
\begin{equation}\label{eq:supp-localized-cckk-inputs}
{\begin{gathered}
\min_{j\le J_n}P f_j^2=1,\quad
\Delta_0\coloneqq\max_{j,k\le J_n}
|P(f_j^Tf_k^T)-P(f_jf_k)|
\lesssim\mathds{1}_{\mathrm{(H)}}t_n^{2-2r_0},\\
\Delta_1\coloneqq n^{-1}\max_{j\le J_n}P|f_j^T|^4
\lesssim v_{t_n}/(na^p),\quad
\kappa_n\coloneqq\max_{j\le J_n}\|f_j^T\|_\infty
\lesssim a^{-p/2}t_n.
\end{gathered}}
\end{equation}

{For \(\mathrm{III}_n\), the high-dimensional Gaussian approximation in}
\citet[Supplementary Material, Lemma~A.1, equation~(A.2)]{Chernozhukovetal2022Improved}
{with the quantities in} \eqref{eq:supp-localized-cckk-inputs}
{and \(\log J_n\lesssim\log n\) from} \eqref{eq:supp-localized-finite-representatives} {yields}
\begin{equation}\label{eq:supp-localized-cckk-comparison-rate}
\begin{aligned}
 {\mathrm{III}_n}
 &\lesssim\sqrt{\Delta_0}\log J_n
 +(\Delta_1\log^5J_n)^{1/4}
 +\frac{\kappa_n(\log J_n)^{3/2}}{\sqrt n}\\
 &\lesssim\mathds{1}_{\mathrm{(H)}}t_n^{1-r_0}\log n
 +\left\{\frac{v_{t_n}(\log n)^5}{na^p}\right\}^{1/4}
 +\frac{t_n(\log n)^{3/2}}{\sqrt{na^p}}\to0.
\end{aligned}
\end{equation}

{The convergence in} \eqref{eq:supp-localized-cckk-comparison-rate} {follows from}
\eqrefs{eq:supp-uniform-bandwidth-rate,eq:supp-truncation-rate-consequences}{, using \(v_{t_n}=1\) under \textup{(C)} and
\(v_{t_n}t_n^{-2}(\log n)^2
=n^{-\min(r_0-1,1)/r_0}(\log n)^2\to0\)
under \textup{(H)}.}

{For \(\mathrm{III}_n^\ast\), condition on the data.
The vector \((\mathbb G_n^\ast f_j^T)_{j\le J_n}\) is centered Gaussian with covariance matrix \((P_n(f_j^Tf_k^T)-P_nf_j^TP_nf_k^T)_{j,k\le J_n}\).
The vector \((G_{P,n}f_j)_{j\le J_n}\) has covariance \((P(f_jf_k))_{j,k\le J_n}\) and coordinate variances \(Pf_j^2=1\).
Decompose the covariance difference as}
\[
 P_n(f_j^Tf_k^T)-P_nf_j^TP_nf_k^T-P(f_jf_k)=(P_n-P)(f_j^Tf_k^T)-P_nf_j^TP_nf_k^T+\{P(f_j^Tf_k^T)-P(f_jf_k)\}.
\]
By \eqrefs{eq:supp-localized-cckk-moments,eq:supp-localized-cckk-inputs},
\[
 \max_{j,k}P[(f_j^Tf_k^T)^2]\lesssim a^{-p}v_{t_n},\qquad
 \max_{j,k}\|f_j^Tf_k^T\|_\infty\lesssim a^{-p}t_n^2.
\]
Also, \(Pf_j^T=0\) and \(\max_jP[(f_j^T)^2]\lesssim1\).
{Apply} \eqref{eq:supp-finite-bracket-bernstein}
{to \(\{f_j^Tf_k^T:j,k\le J_n\}\) and \(\{f_j^T:j\le J_n\}\), using singleton brackets and \(J_n\le n^C\), to obtain}
\begin{equation}\label{eq:supp-localized-cckk-bernstein}
\begin{aligned}
 \max_{j,k}|(P_n-P)(f_j^Tf_k^T)|
 &=O_p\!\left(\sqrt{\frac{v_{t_n}\log n}{na^p}}
             +\frac{t_n^2\log n}{na^p}\right),\\
 \max_j|P_nf_j^T|
 &=O_p\!\left(\sqrt{\frac{\log n}{n}}
             +\frac{t_n\log n}{na^{p/2}}\right).
\end{aligned}
\end{equation}
Since $a\le1\le t_n$, \eqref{eq:supp-localized-cckk-bernstein} gives
$\max_j|P_nf_j^T|^2=O_p(t_n^2\log n/(na^p))$.
{Adding the bounds for the empirical product, empirical centering, and truncation in}
\eqrefs{eq:supp-localized-gaussian-truncation,eq:supp-localized-cckk-bernstein} {gives}
\begin{equation}\label{eq:supp-localized-cckk-covariance-comparison}
 \max_{j,k\le J_n}
 \left|P_n(f_j^Tf_k^T)-P_nf_j^T P_nf_k^T-P(f_jf_k)\right|=O_p\!\left(
 \mathds{1}_{\mathrm{(H)}}t_n^{2-2r_0}
 +\sqrt{\frac{v_{t_n}\log n}{na^p}}
 +\frac{t_n^2\log n}{na^p}\right).
\end{equation}

{The Gaussian covariance comparison in} \citet[Proposition~2.1]{Chernozhukovetal2022Improved}
{applied conditionally on the data, with \(Pf_j^2=1\) and}
\eqrefs{eq:supp-localized-finite-representatives,eq:supp-localized-cckk-covariance-comparison}{, gives}
\begin{equation}\label{eq:supp-localized-cckk-multiplier-comparison}
\begin{aligned}
 {\mathrm{III}_n^\ast}
 &\lesssim\log J_n
 \left[
 \max_{j,k\le J_n}
 \left|P_n(f_j^Tf_k^T)-P_nf_j^T P_nf_k^T-P(f_jf_k)\right|
 \right]^{1/2}\\
 &=O_p\Biggl(
 \mathds{1}_{\mathrm{(H)}}t_n^{1-r_0}\log n
  +\left\{\frac{v_{t_n}(\log n)^5}{na^p}\right\}^{1/4}
  +\frac{t_n(\log n)^{3/2}}{\sqrt{na^p}}
 \Biggr)=o_p(1).
\end{aligned}
\end{equation}

{By} \eqref{eq:supp-localized-cckk-multiplier-comparison}{, \(\mathrm{III}_n^\ast\pto0\).
The Gaussian approximation for \(\mathrm{III}_n\) and the Gaussian covariance comparison for \(\mathrm{III}_n^\ast\) require the target coordinate variances \(Pf_j^2=1\) to be bounded below; the covariance matrices need not be nonsingular.}

{For every fixed \(\varepsilon>0\), {the unconditional cdf distance} has limit superior at most \(C\varepsilon\).
Letting \(\varepsilon\downarrow0\) proves {the unconditional cdf approximation} in}
\eqref{eq:supp-localized-gaussian-cdf}{.
{For the conditional cdf approximation}, fix \(\delta>0\) and choose \(0<\varepsilon<\delta/(2C)\).
The probability that the conditional cdf distance exceeds \(\delta\) is at most
\({\mathbb P\bigl(\mathrm I_n^\ast(\varepsilon)+\mathrm{II}_n(\varepsilon)+\mathrm{III}_n^\ast>\delta/2\bigr)}\to0\).
This proves the conditional cdf approximation in}
\eqref{eq:supp-localized-gaussian-cdf}{.}
{Symmetry of \(\mathcal F_n\) makes \(\sup_{f\in\mathcal F_n}\mathbb G_nf\), \(\sup_{f\in\mathcal F_n}\mathbb G_n^\ast f\), and \(S_n^G\) nonnegative.
Squaring therefore preserves each cdf distance in}
\eqref{eq:supp-localized-gaussian-cdf}{.}
\end{proof}

\section{{Details of the simulation studies}}
\label{sec:supp-simulation-details}

\subsection{Simulation designs with spherical responses}
\label{sec:supp-sphere-simulation}

The designs in \Cref{subsec:sim-sphere-smooth-short} use the orthonormal
frame \(E(p)\coloneqq(e_E(p),e_N(p))\), where, for
\(p=(p_1,p_2,p_3)^\top\in\mathbb S^2\) with \(p_1^2+p_2^2>0\),
\[
 e_E(p)\coloneqq\frac{(-p_2,p_1,0)^\top}{\sqrt{p_1^2+p_2^2}},
 \qquad e_N(p)\coloneqq p\times e_E(p).
\]
Central symmetry and \(\|\varepsilon\|\le5\pi/32<\pi/2\) identify
\(m_0(x)\) as the unique conditional Fr\'echet mean
\citep[Theorem~2.1]{Afsari2011}.

For the uniform disk of radius \(0.2\),
\(\Cov(\varepsilon\mid X=x)=0.01I_2\).
For the beta-disk distribution,
\[
    \Cov(\varepsilon\mid X=x)
    =
    \frac34 R_{\pi x/2}
    \diag(s_1(x)^2,s_2(x)^2)
    R_{\pi x/2}^\top.
\]

For \Cref{fig:sim-sphere-dgp-bias}, we compute the {localized Fr\'echet mean} \(m_h(x)\) under both designs with the uniform kernel and fixed
\(h=(1-2^{-1/2})1000^{-1/5}\), and evaluate
\(B_h(x)=\Log_{m_0(x)}m_h(x)\).

\subsection{Regression designs with SPD predictors and responses}
\label{sec:supp-spd-target}
Write \(\Sym^2\coloneqq\{S\in\R^{2\times2}:S=S^\top\}\) and
\(\Sym_{++}^2\coloneqq\{P\in\Sym^2:P\succ0\}\), with
\(T_P\Sym_{++}^2=\Sym^2\). Both predictor and response manifolds use
the AIRM:
\[
 g_P^{\mathrm{AI}}(U,V)\coloneqq \tr(P^{-1}UP^{-1}V),\qquad
 d_{\mathrm{AI}}(P,Q)\coloneqq \|\log(P^{-1/2}QP^{-1/2})\|_{\mathrm F}.
\]
Thus \(d_{\cX}=d_{\cY}=d_{\mathrm{AI}}\) in this design, and
\[
 \Exp_P^{\mathrm{AI}}\left[V\right]
 =P^{1/2}\exp(P^{-1/2}VP^{-1/2})P^{1/2},\qquad
 \Log_P^{\mathrm{AI}}(Q)
 =P^{1/2}\log(P^{-1/2}QP^{-1/2})P^{1/2}.
\]
Lowercase \(\exp\) and \(\log\) denote matrix functions.

Let
\[
    E_1\coloneqq \frac{1}{\sqrt2}
    \begin{pmatrix}1&0\\0&-1\end{pmatrix},\qquad
    E_2\coloneqq \frac{1}{\sqrt2}
    \begin{pmatrix}0&1\\1&0\end{pmatrix},\qquad
    E_3\coloneqq \frac{1}{\sqrt2}
    \begin{pmatrix}1&0\\0&1\end{pmatrix}.
\]
The uniform ball of radius \(\sqrt5\,\sigma_\varepsilon\) and its image
under \(D\) have covariances \(\sigma_\varepsilon^2I_3\) and
\(\sigma_\varepsilon^2D^2\), matching the Gaussian pairs in
\Cref{subsec:sim-spd2spd}; \(\tr(D^2)=3\) fixes total variance in the tangent space.
Central symmetry and uniqueness of the Fr\'echet mean on the Hadamard manifold
identify \(m_0(x)\) as the conditional Fr\'echet mean at \(X=x\).

\makeatletter
\setlength{\@fptop}{0pt}
\setlength{\@fpsep}{8pt plus 2pt minus 2pt}
\setlength{\@fpbot}{0pt plus 1fil}
\makeatother
\subsection{Bandwidth selection and simulation summaries}
\label{subsec:supp-simulation-bandwidths}

At each evaluation point \(x\), we compute a local linear Fr\'echet
pilot estimate at bandwidth \(c_0>0\), fixing this center and the
empirical local linear Hessian evaluated at that center throughout the
plug-in calculation. We use a degree-six reference regression of the
transformed tangent scores to select bandwidths for cubic and quartic
pilot fits. The resulting fits determine the quadratic pilot bandwidth
\(g_2\). The quadratic fit and local linear weights at \(c_0\) give an
estimate \(\widehat B_{\mathrm{pil}}(x)\) of the leading local linear
bias displacement at \(c_0\). Using HC3-adjusted residuals from the
quadratic fit and holding the pilot design fixed, we obtain covariance
approximations \(\widehat V_{B,\mathrm{pil}}(x)\) and
\(\widehat V_{\mathrm{LL,pil}}(x)\) for this bias estimate and the
uncorrected local linear estimator, respectively. Define the scaled
local linear covariance estimate by
\[
\widehat\Omega_{\mathrm{LL,pil}}(x)
\coloneqq nc_0^p\widehat V_{\mathrm{LL,pil}}(x).
\]
Holding the pilot bandwidths and estimates fixed as the candidate
bandwidth \(a>0\) varies, we use the regularized plug-in AMSE criterion
\[
\widehat{\operatorname{AMSE}}_\lambda(a;x)=\left(\frac{a}{c_0}\right)^4\left[\|\widehat B_{\mathrm{pil}}(x)\|^2+3\lambda\operatorname{tr}\bigl(\widehat V_{B,\mathrm{pil}}(x)\bigr)\right]+\frac{\operatorname{tr}\bigl(\widehat\Omega_{\mathrm{LL,pil}}(x)\bigr)}{na^p},\qquad\lambda=1/2.
\]
If \(\operatorname{tr}\bigl(\widehat\Omega_{\mathrm{LL,pil}}(x)\bigr)>0\) and
\(\|\widehat B_{\mathrm{pil}}(x)\|^2
+3\lambda\operatorname{tr}\bigl(\widehat V_{B,\mathrm{pil}}(x)\bigr)>0\),
its unique unconstrained minimizer is
\[
h_{\mathrm{raw}}(x)
=\left[
\frac{p\,c_0^4\operatorname{tr}\bigl(\widehat\Omega_{\mathrm{LL,pil}}(x)\bigr)}
{4n\left[\|\widehat B_{\mathrm{pil}}(x)\|^2
+3\lambda\operatorname{tr}\bigl(\widehat V_{B,\mathrm{pil}}(x)\bigr)\right]}
\right]^{1/(p+4)}.
\]
We use this minimizer as the initial bandwidth \(h_0\) in the reported simulations.

For each fixed \(b/h\), we enlarge the bandwidth pair proportionally so
that the local linear (\(d=1\)) and local quadratic (\(d=2\)) fits each have at
least \(\lceil1.5\binom{p+d}{d}\rceil\) observations with positive kernel
weight in the full sample.
At each candidate pair, let \(v_{\min}\) be a unit eigenvector corresponding to the smallest eigenvalue of the \textsc{RRBC LOO} covariance at \(x\). The directional effective sample size (ESS) along \(v_{\min}\), computed from the LOO contributions in \eqref{eq:supp-loo-contributions}, is
\[
 \frac{\left\{\sum_i
 \left\langle v_{\min},d_{\rbc,i}(x)\right\rangle^2\right\}^2}
 {\sum_i\left\langle v_{\min},d_{\rbc,i}(x)\right\rangle^4}.
\]
Starting from the bandwidth pair satisfying the support requirements, we examine common enlargement
factors \(1,1.1,\ldots,2\) in increasing order. We adopt the first candidate
with a defined \textsc{RRBC LOO} covariance and directional ESS of at least
20 for \(n=250\), or 30 for \(n=500,1000,2000\).
If no candidate meets the target, we retain the largest pair and record
the shortfall. The adopted \(h,b\) are held fixed during deletion.
\textsc{US} rescaling follows
\Cref{subsec:simulation-implementation}.
The score norm in the Newton stopping rule is computed with deleted local linear weights normalized to sum to one. Finite fits that reach the update cap without meeting the score tolerance are retained and flagged.

In the main simulations using the uniform kernel with \(b=1.5h\), enlargement to meet the ESS requirement was needed in only 0.37\% of point--replication pairs at
\(n=1000\) and in none at \(n=2000\).

All reported confidence sets and volume summaries were computed in each of the 1,000 replications. {Critical values} use the rank rule in
\Cref{subsec:uniform-intrinsic-confidence-tubes}.

{We report the median intrinsic volume across replications at each evaluation point.}
Simultaneous summaries and transformed volume ratios follow
\Cref{subsec:sim-us-volume}.

\subsection{Additional comparisons and sensitivity analyses}
\label{subsec:supp-simulation-sensitivity}

Unless varied below, comparisons use the settings of \Cref{sec:simulation}
and the uniform kernel.

\subsubsection{Covariance estimators for \textsc{RRBC}}
\label{subsubsec:supp-covariance-comparison}

At \(b/h=1.5\), we compare the {unresidualized, residual-based, and LOO covariance estimators} for the same \textsc{RRBC} center.
Unresidualized uses the covariance and Gaussian multiplier process in
\eqref{eq:supp-auxiliary-influence}, with
\(\widehat\varepsilon_{j,i}(x)\), \(j=1,2\), replaced by
\(\Log_{\widehat m_{h,x}}^{\cY}Y_i\).
Explicitly,
\begin{equation+}
 \widehat\Omega^{\mathrm U}_{\rbc}(x)
 \coloneqq \frac{h^p}{n}\sum_{i=1}^n
 \{\widehat L_{1,h,i}(x)-\widehat L_{\bias,h,b,i}(x)\}^2
 \left\{\Pi_{\widehat m_{h,x}\to\widehat m_{\rbc,x}}^{\cY}
 \widehat A_{h,x}(\widehat m_{h,x})^{-1}
 \Log_{\widehat m_{h,x}}^{\cY}Y_i\right\}^{\otimes2}.
\end{equation+}
Residual, labeled \textsc{RRBC Residual} in the main figures, uses
\(\widehat\Omega_{\rbc}^{\mathrm A}(x)\) in
\eqref{eq:supp-auxiliary-influence}.
LOO uses \(\widehat\Omega_{\rbc}(x)\) in
\eqref{eq:main-loo-covariance}.
Each covariance estimator is paired with its corresponding Gaussian
multiplier process. For the uniform kernel and {both Wald and ball sets}, the empirical simultaneous coverage of LOO lies between those of Residual and Unresidualized in all 24 settings (\Cref{fig:supp-covariance-comparison,fig:supp-shape-sphere,fig:supp-shape-spd-isotropic,fig:supp-shape-spd-anisotropic}).

With the uniform kernel, Residual undercovers in all 24 Wald settings,
while Unresidualized exceeds 95\% coverage in 23; the remaining value is
94.6\%. At \(n=250\), Unresidualized is closest to nominal coverage among
the three covariance choices in all six designs. {Its higher coverage is accompanied by larger} {medians of the grid-averaged intrinsic volumes}{.} Its coverage remains 97.6--99.0\% at \(n=2000\).

The overcoverage of Unresidualized in these simulations is consistent with
retaining variation in the conditional tangent mean across the local window.
At a fixed population base point, the conditional second moment of the response logarithm equals its conditional covariance plus the outer product of its
conditional mean. Residual subtracts fitted local linear and quadratic
tangent means, whereas Unresidualized retains these components. This
interpretation is analogous to the distinction between score second moments
and conditional score variances under misspecification in
\citet[Section~2 and Theorem~2]{Abadieetal2014}. It provides a mechanism for
the observed overcoverage without implying a finite-sample ordering of the
fitted covariance estimators or conservative coverage in general.

\begin{figure}[!htbp]
 \centering
 \includegraphics[width=\textwidth]{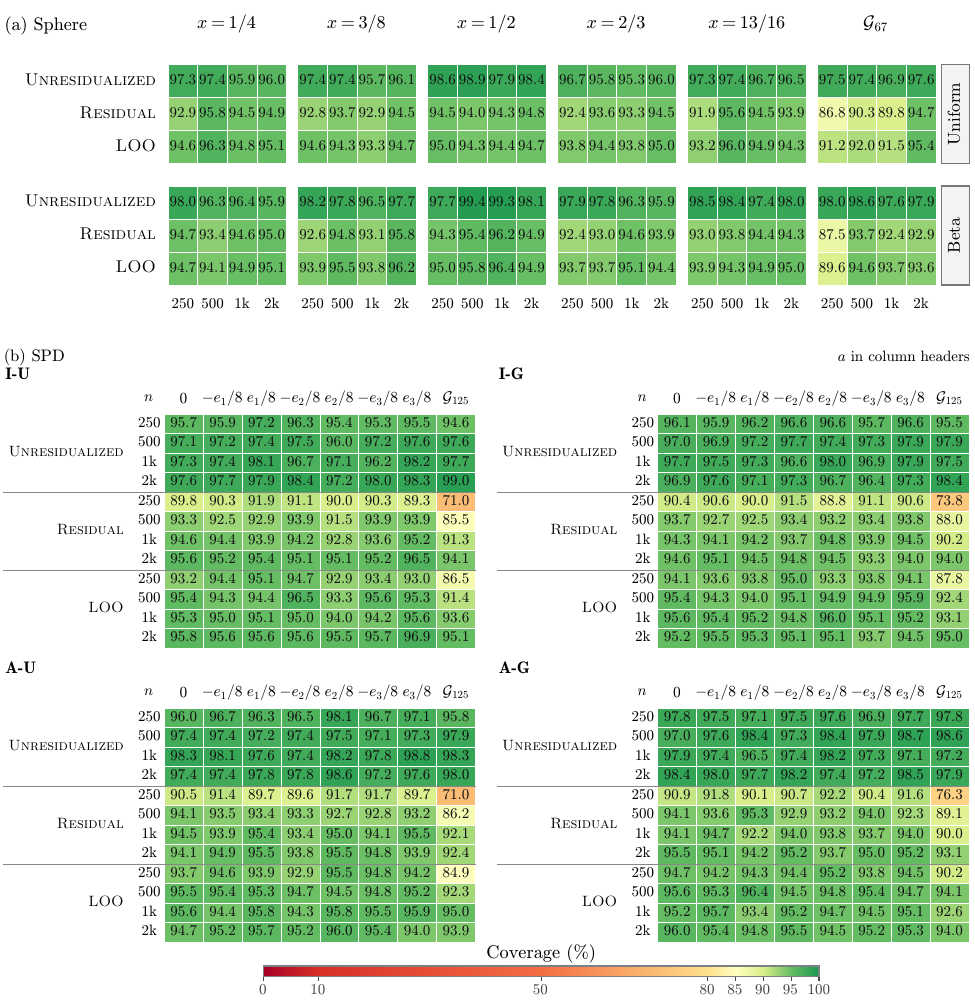}
 \caption{\textsc{RRBC} coverage (\%) by covariance estimator:
 (a) sphere; (b) SPD.}
 \label{fig:supp-covariance-comparison}
 \label{fig:supp-cov-sphere}\label{fig:supp-cov-spd}
\end{figure}

\subsubsection{Intrinsic volumes of the four procedures}
\label{subsubsec:supp-size-comparison}

{We compare the intrinsic volumes of \textsc{Naive}, \textsc{BC + Naive Cov}, \textsc{US}, and \textsc{RRBC LOO}.} In these geometries, an isometry inducing the parallel transport maps the \textsc{Naive} set to the \textsc{BC + Naive Cov} set, so their intrinsic volumes agree. {\textsc{RRBC LOO}} has larger median intrinsic volumes than \textsc{Naive} and \textsc{BC + Naive Cov} (\Cref{fig:supp-procedure-sizes}). \textsc{BC + Naive Cov} has empirical simultaneous coverage below 95\% in all 24 settings (\Cref{fig:sim-sphere-results,fig:sim-spd2spd-results}).

\begin{table}[!htbp]
\centering
\caption{\(q\)th roots of the \textsc{RRBC LOO}/\textsc{US}
ratios of {the medians of the grid-averaged intrinsic volumes} for simultaneous
Wald sets, with \(q=2\) for the sphere and \(q=3\) for SPD.
Results use the uniform kernel and \(b=1.5h\) for RRBC.}
\label{tab:rrbc-us-volume-root-ratios}
\begin{tabular}{lrrrr}
\toprule
Design & \(n=250\) & \(n=500\) & \(n=1000\) & \(n=2000\) \\
\midrule
Sphere: uniform disk & 1.079 & 1.063 & 1.047 & 1.030 \\
Sphere: beta disk & 1.082 & 1.070 & 1.051 & 1.034 \\
\addlinespace
SPD: I-U & 1.158 & 1.151 & 1.136 & 1.115 \\
SPD: I-G & 1.168 & 1.158 & 1.141 & 1.117 \\
SPD: A-U & 1.158 & 1.152 & 1.137 & 1.114 \\
SPD: A-G & 1.174 & 1.157 & 1.137 & 1.116 \\
\bottomrule
\end{tabular}
\end{table}

\begin{figure}[!htbp]
 \centering
 \includegraphics[width=\textwidth]{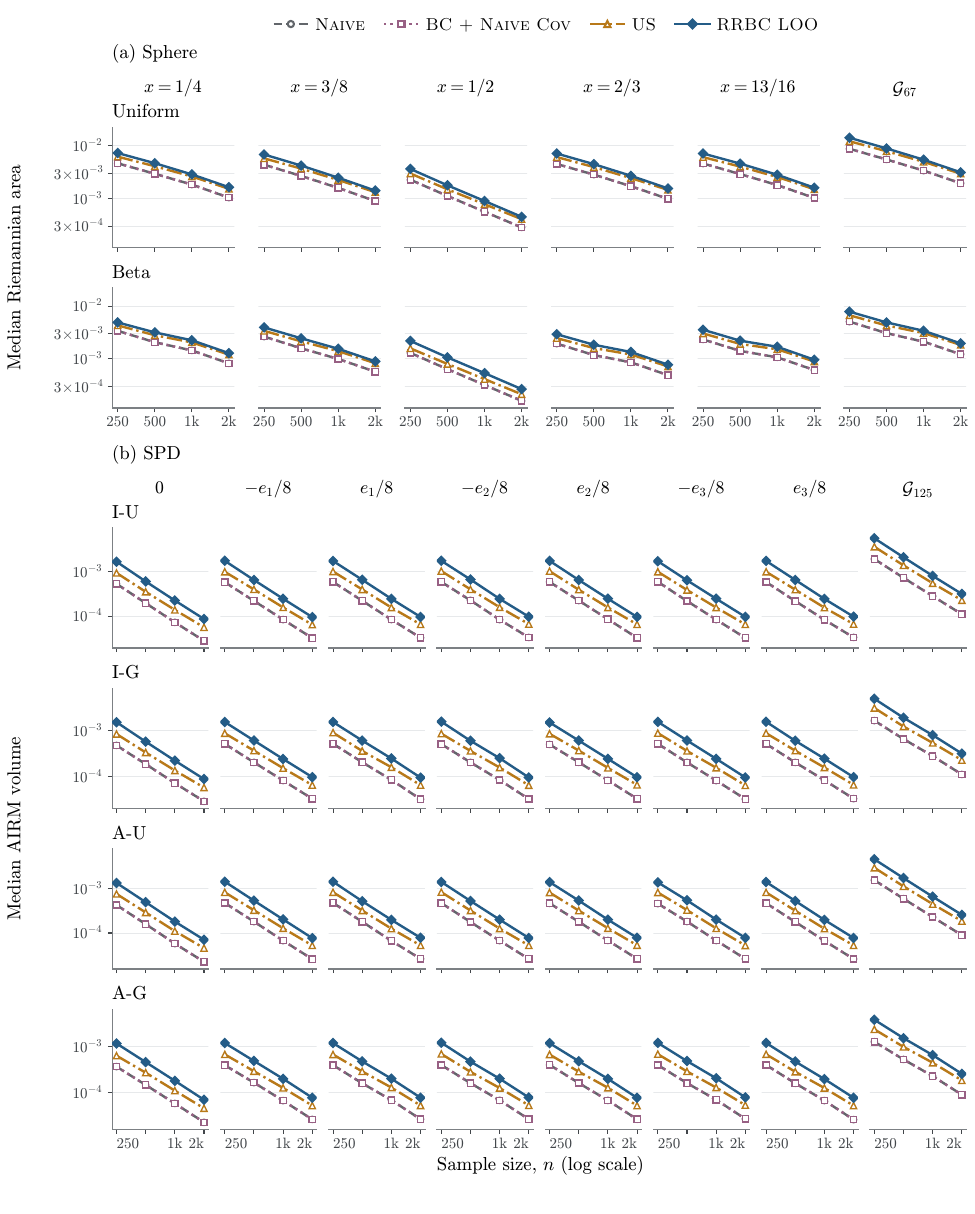}
 \caption{Median intrinsic volume for the four procedures:
 (a) Riemannian area; (b) AIRM volume.}
 \label{fig:supp-procedure-sizes}
 \label{fig:supp-size-sphere}\label{fig:supp-size-spd}
\end{figure}

\subsubsection{\textsc{RRBC} bandwidth ratio sensitivity}
\label{subsubsec:supp-ratio-comparison}

We compare \(b/h\in\{1,1.25,1.5,2\}\) for \textsc{RRBC}
(\Cref{fig:supp-ratio-coverage,fig:supp-ratio-sizes}).
The ESS rule selects the local linear bandwidth \(h\) separately
for each ratio, so this compares the implemented bandwidth selection
procedures rather than varying \(b\) at a fixed selected \(h\).
For example, under sphere beta-disk noise at \(n=500,b/h=1\), the ESS
target is recorded as unmet at the selected bandwidths in 34.32\% of
point--replication pairs. The corresponding fraction at \(b/h=1.5\)
is 0.22\%.

\begin{figure}[!htbp]
 \centering
 \includegraphics[width=\textwidth]{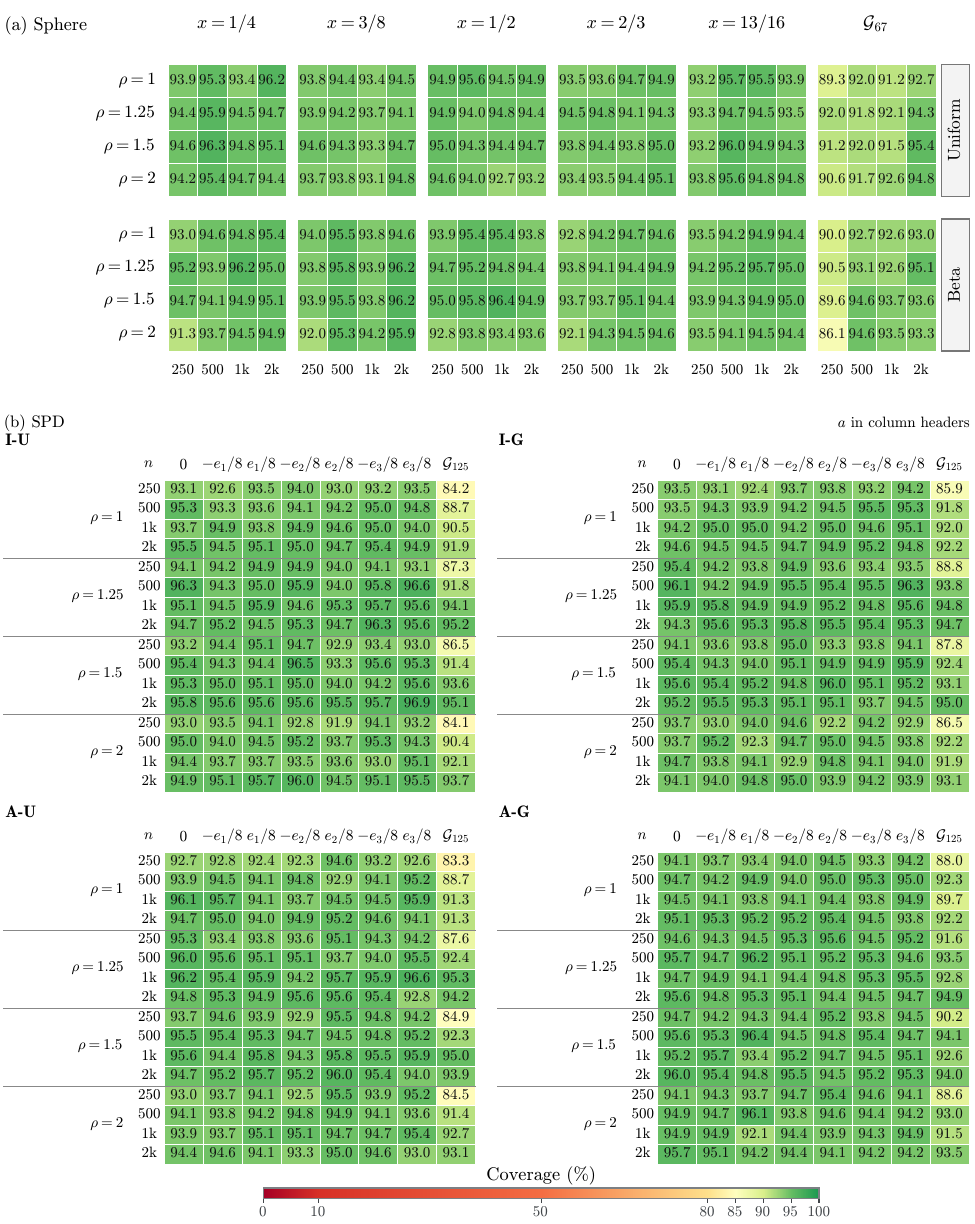}
 \caption{\textsc{RRBC} coverage (\%) across bandwidth ratios {($\rho=b/h$)}: (a) sphere; (b) SPD.}
 \label{fig:supp-ratio-coverage}
 \label{fig:supp-ratio-coverage-sphere}\label{fig:supp-ratio-coverage-spd}
\end{figure}
\begin{figure}[!htbp]
 \centering
 \includegraphics[width=\textwidth]{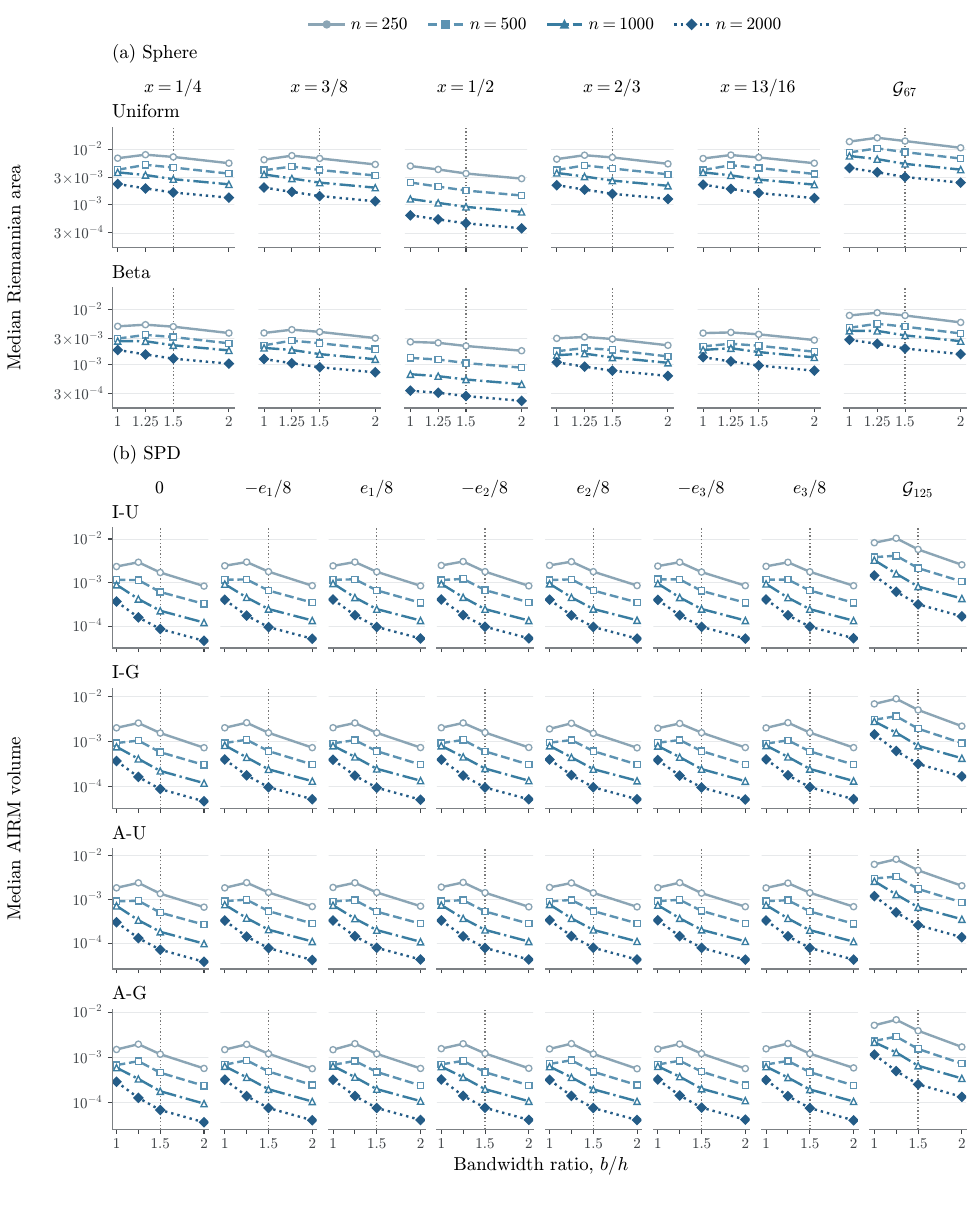}
 \caption{Median \textsc{RRBC} intrinsic volume across bandwidth ratios:
 (a) Riemannian area; (b) AIRM volume.}
 \label{fig:supp-ratio-sizes}
 \label{fig:supp-ratio-volume-sphere}\label{fig:supp-ratio-volume-spd}
\end{figure}

\subsubsection{Ball and Wald sets for \textsc{RRBC}}
\label{subsubsec:supp-shape-comparison}

For each covariance estimator, we compare Wald sets with ball sets
obtained by replacing \(\widehat\Omega_{\rbc}\) with
\(\tr(\widehat\Omega_{\rbc})I_q/q\) in the Studentization of both
the observed statistic and the multiplier statistic defined in
\eqref{eq:feasible-multiplier-supremum-statistic}.
We retain the same observation-level Gaussian contributions and
recalibrate the pointwise and simultaneous critical values using
the rank rule in
\Cref{subsec:uniform-intrinsic-confidence-tubes}.
With LOO covariance at \(b/h=1.5\), ball sets have higher simultaneous
coverage than Wald sets in all 24 settings.
\Cref{fig:supp-shape-sphere,fig:supp-shape-spd-isotropic,fig:supp-shape-spd-anisotropic}
report coverage. \Cref{tab:rrbc-ball-wald-size-ratios} reports
ratios of median intrinsic volume (Ball/Wald). Pointwise
volumes use the central target \(x=1/2\) on the sphere and \(a=0\) for
SPD; simultaneous summaries follow \Cref{subsec:supp-simulation-bandwidths}.
Each ratio is formed after taking the two medians across replications.

At \(n=2000\), the simultaneous Ball/Wald ratios of
{the medians of the grid-averaged intrinsic volumes} are \(1.010\) for sphere
uniform-disk noise and \(1.015\)--\(1.020\) for the isotropic
SPD designs, compared with \(1.735\) for sphere beta-disk
noise and \(1.791\)--\(1.798\) for the anisotropic SPD designs.

\begin{table}[!htbp]
\centering
\caption{Ratios of median intrinsic volumes {(Ball/Wald)} for \textsc{RRBC} with LOO covariance.}
\label{tab:rrbc-ball-wald-size-ratios}
\begingroup
\setlength{\tabcolsep}{4pt}
\begin{tabular}{lrrrrrrrr}
\toprule
& \multicolumn{4}{c}{Pointwise} & \multicolumn{4}{c}{Simultaneous} \\
\cmidrule(lr){2-5}\cmidrule(lr){6-9}
Design & 250 & 500 & 1,000 & 2,000 & 250 & 500 & 1,000 & 2,000 \\
\midrule
Sphere: uniform disk & 1.012 & 1.007 & 1.005 & 1.000 & 1.041 & 1.024 & 1.014 & 1.010 \\
Sphere: beta disk & 1.443 & 1.458 & 1.470 & 1.471 & 1.694 & 1.715 & 1.725 & 1.735 \\
\addlinespace
SPD: I-U & 1.054 & 1.031 & 1.014 & 1.009 & 1.138 & 1.064 & 1.034 & 1.015 \\
SPD: I-G & 1.087 & 1.051 & 1.022 & 1.012 & 1.160 & 1.081 & 1.041 & 1.020 \\
SPD: A-U & 1.499 & 1.450 & 1.424 & 1.415 & 1.886 & 1.826 & 1.799 & 1.798 \\
SPD: A-G & 1.482 & 1.455 & 1.438 & 1.417 & 1.931 & 1.850 & 1.807 & 1.791 \\
\bottomrule
\end{tabular}
\endgroup
\end{table}

\begin{figure}[!htbp]
 \centering
 \includegraphics[width=\textwidth]{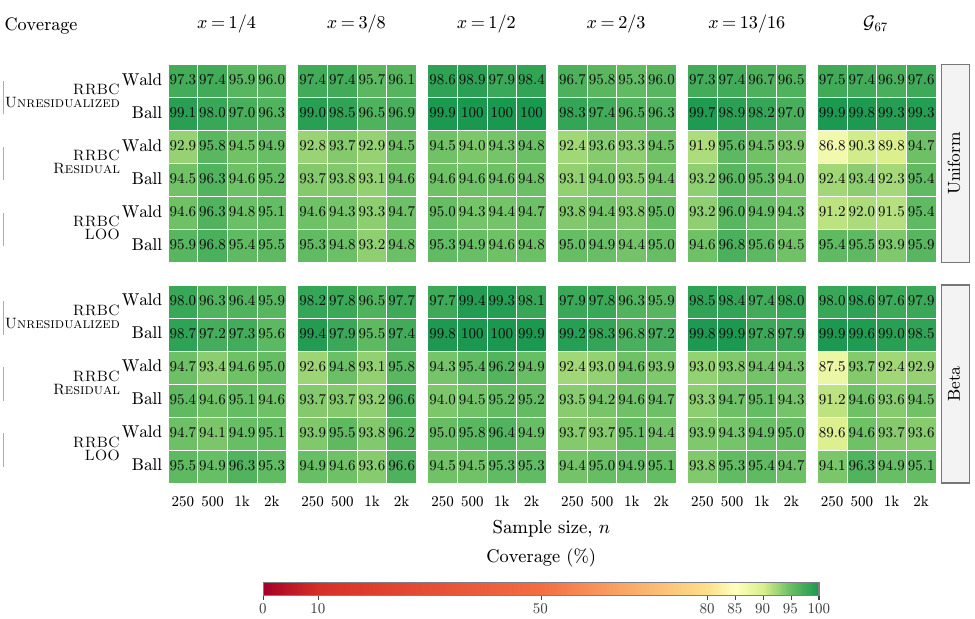}
 \caption{Sphere \textsc{RRBC} coverage (\%) for Wald and ball sets.}
 \label{fig:supp-shape-sphere}
\end{figure}
\begin{figure}[!htbp]
 \centering
 \includegraphics[width=\textwidth]{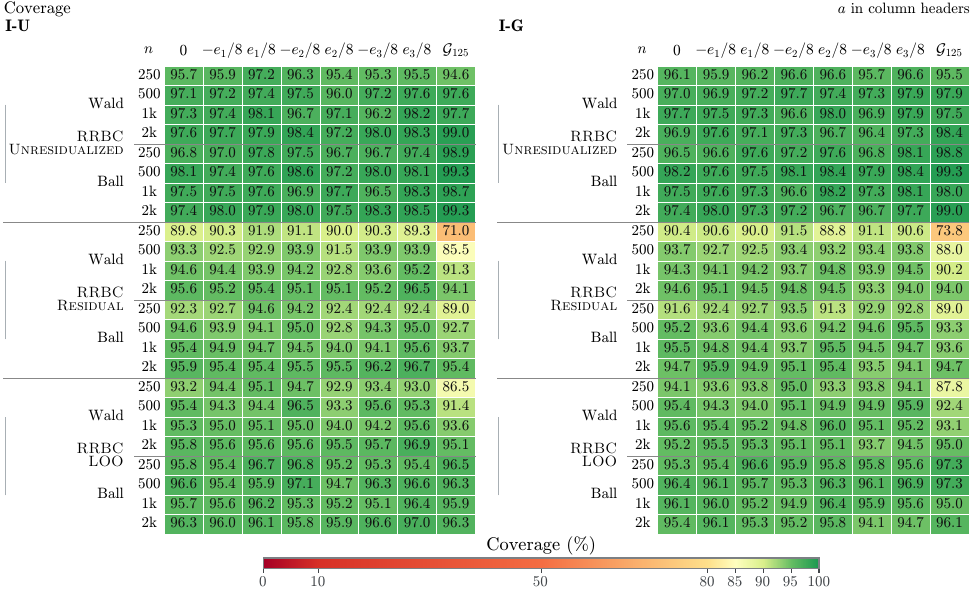}
 \caption{Isotropic SPD \textsc{RRBC} coverage (\%) for Wald and ball sets.}
 \label{fig:supp-shape-spd-isotropic}
\end{figure}
\begin{figure}[!htbp]
 \centering
 \includegraphics[width=\textwidth]{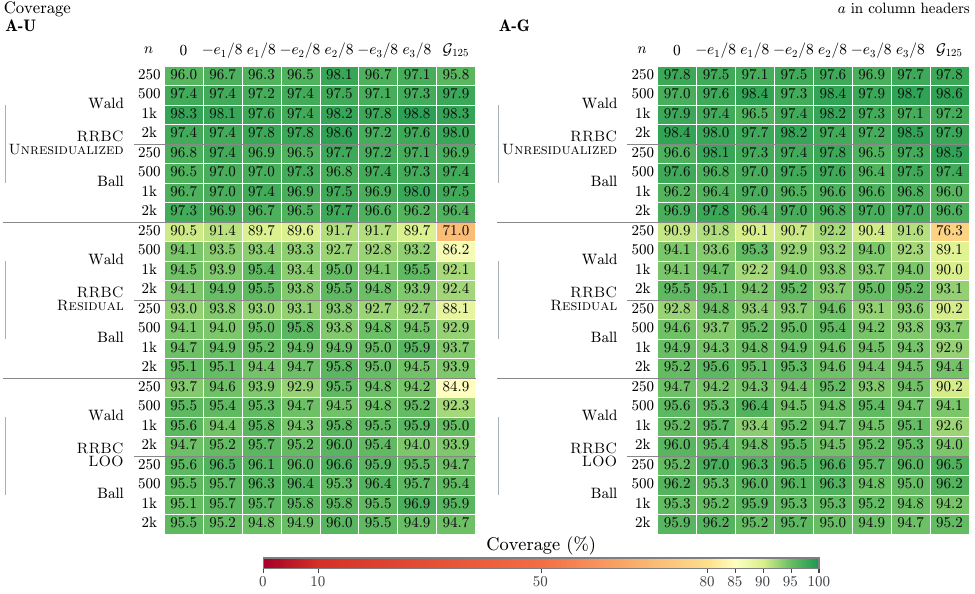}
 \caption{Anisotropic SPD \textsc{RRBC} coverage (\%) for Wald and ball sets.}
 \label{fig:supp-shape-spd-anisotropic}
\end{figure}

\subsubsection{Uniform and Epanechnikov kernels}
\label{subsubsec:supp-epan-kernel}

\Cref{fig:supp-epan-sphere-wald,fig:supp-epan-spd-wald}
show the results for Wald coverage with the Epanechnikov kernel for the
comparison in \Cref{subsec:sim-kernel-sensitivity}.

\Cref{tab:epan-uniform-size-ratios} reports Epanechnikov/uniform
ratios of {the median intrinsic volumes of the Wald sets} for \textsc{RRBC LOO},
using the central pointwise targets in
\Cref{subsubsec:supp-shape-comparison} and the volume summaries in
\Cref{subsec:supp-simulation-bandwidths}.

\begin{figure}[!htbp]
 \centering
 \includegraphics[width=\textwidth]{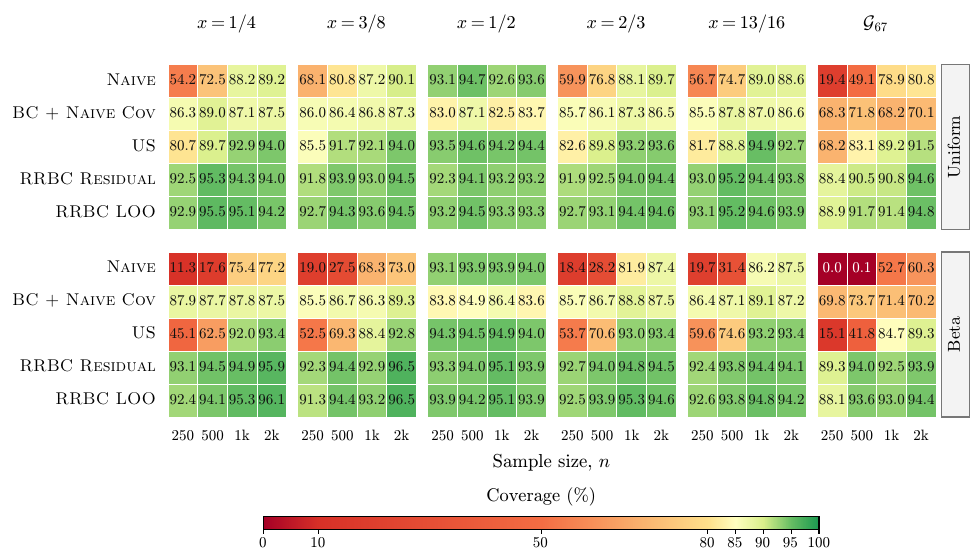}
 \caption{Wald coverage with the Epanechnikov kernel (\%) for the sphere designs.}
 \label{fig:supp-epan-sphere-wald}
\end{figure}
\begin{figure}[!htbp]
 \centering
 \includegraphics[width=\textwidth]{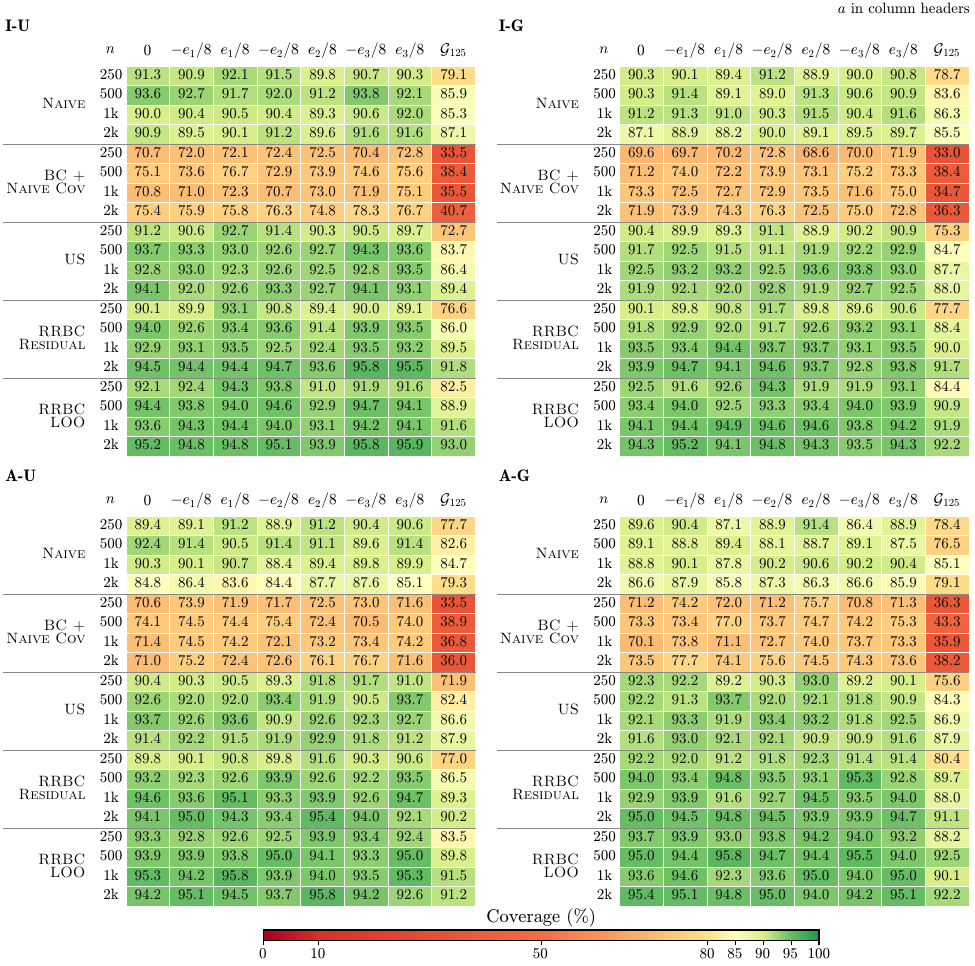}
 \caption{Wald coverage with the Epanechnikov kernel (\%) for the SPD designs.}
 \label{fig:supp-epan-spd-wald}
\end{figure}

\begin{table}[!htbp]
\centering
\begingroup
\singlespacing
\small
\setlength{\tabcolsep}{4pt}
\caption{Epanechnikov/uniform-kernel ratios of {the median intrinsic volumes of the Wald sets} for \textsc{RRBC LOO}. Values below one indicate smaller Epanechnikov volumes.}
\label{tab:epan-uniform-size-ratios}
\begin{tabular}{lrrrrrrrr}
\toprule
& \multicolumn{4}{c}{Pointwise} & \multicolumn{4}{c}{Simultaneous} \\
\cmidrule(lr){2-5}\cmidrule(lr){6-9}
Design & 250 & 500 & 1,000 & 2,000 & 250 & 500 & 1,000 & 2,000 \\
\midrule
Sphere: uniform disk & 0.855 & 0.865 & 0.854 & 0.860 & 0.652 & 0.685 & 0.795 & 0.804 \\
Sphere: beta disk & 0.823 & 0.826 & 0.828 & 0.829 & 0.636 & 0.644 & 0.773 & 0.804 \\
\addlinespace
SPD: I-U & 0.576 & 0.745 & 0.847 & 0.839 & 0.476 & 0.590 & 0.784 & 0.762 \\
SPD: I-G & 0.535 & 0.658 & 0.848 & 0.831 & 0.459 & 0.536 & 0.757 & 0.758 \\
SPD: A-U & 0.574 & 0.724 & 0.863 & 0.827 & 0.457 & 0.576 & 0.780 & 0.761 \\
SPD: A-G & 0.546 & 0.634 & 0.841 & 0.835 & 0.449 & 0.510 & 0.749 & 0.759 \\
\bottomrule
\end{tabular}
\endgroup
\end{table}

\section{{Details of the real-data analyses}}
\label{sec:supp-empirical-details}

At each grid point, we adjust the bandwidth pair {to meet the support requirements for the local polynomial fits} and enlarge it proportionally, by at most a factor of two, toward an ESS target of 20. We use the minimum ESS over covariance eigendirections in the cardiovascular analysis and the heading ESS in the ocean analysis. We select the first numerically adequate pair meeting the target, otherwise the last numerically adequate pair, and record any shortfall.

Within each stratum or phase, we use the contributions in
\eqref{eq:supp-loo-contributions}, updating all RRBC components after
deletion while holding the selected bandwidths fixed.
The empirical LOO fits use a score tolerance of \(10^{-12}\), at most ten Newton updates, and an acceptance threshold of \(10^{-12}\) for the deleted-score norm. Finite fits reaching the update cap without meeting the score tolerance are retained and flagged.

\subsection{Cardiovascular analysis}
\label{sec:supp-cardiovascular}

\subsubsection{Data and covariance construction}

We analyzed the first 480 seconds of recordings from eligible
subjects in version 1.0.0 of the publicly available Autonomic Aging database
\citep{SchumannandBar2021}, available through PhysioNet
\citep{Pollardetal2026}.
Electrocardiogram (ECG) recordings were band-pass filtered at
0.5--40 Hz, as in \citet{Parlatoetal2023}, and noninvasive blood
pressure (NIBP) recordings were low-pass filtered at 20 Hz,
as in \citet{Balletal2024}. We used forward--backward Butterworth
filtering with prototype order four, applied separately to
finite waveform runs without imputing missing waveform values.

R peaks were detected by a Pan--Tompkins-type procedure
\citep{PanandTompkins1985}.
Let \(t_0<t_1<\cdots\) denote the detected times of the R peaks and define
\[
\mathrm{RR}_j\coloneqq t_{j+1}-t_j,\qquad
\mathrm{SBP}_j\coloneqq \max_{t\in[t_j,t_{j+1})}\mathrm{NIBP}_{\mathrm{filt}}(t).
\]
RR intervals outside 0.3--2.0 seconds or differing by more than 20\%
from a local median computed from up to 11 usable neighboring intervals
were declared invalid. Runs of one or two invalid intervals bracketed
by valid intervals were interpolated on the logarithmic RR scale. We retained
\(\mathrm{SBP}_j\) only when the corresponding RR interval had valid NIBP
throughout and \(\mathrm{SBP}_j\in[40,260]\) mmHg.

For each valid lag-1 pair, we formed a bivariate observation
\[
\begin{pmatrix}
100\log \mathrm{SBP}_j\\
100\log \mathrm{RR}_{j+1}^{\mathrm{clean}}
\end{pmatrix}
\]
and linearly interpolated the resulting series to a 4-Hz regular
grid within the observed time span. Cleaned RR intervals include the
permitted interpolations. Eligibility required a resampled time span of
at least 420 seconds, no gap exceeding 5 seconds between consecutive
valid pair times, at least 180 cleaned RR intervals, and at least
160 valid lag-1 pairs. The number of cleaned RR intervals had to be at least 70\%
of the total number of detected RR intervals, and the number of valid pairs
at least 70\% of the number of cleaned RR intervals.

The 4-Hz series was centered by subtracting its mean and tapered using a
split-cosine-bell taper over 10\% of each end. We constructed a matrix
periodogram and applied two-pass modified-Daniell smoothing with spans
\((5,5)\) in Fourier bins. We evaluated the response integral in
\Cref{sec:cardiovascular} from the resulting spectral matrices.
Further preprocessing details are provided in the accompanying
reproducibility materials.

\subsubsection{Additional diagnostics}
\label{subsec:supp-cardiovascular-joint-calibration}

For test~\textup{(iii)} in \Cref{subsec:supp-age-contrasts}, all 28
estimator-scale covariance estimates for
\(\widehat\Delta_n^{\mathrm{(iii)}}(j)\) are positive definite at the
adopted bandwidths, with
\(\min_j\lambda_{\min}(\widehat\Xi_n(j)/(nh^p))=3.98\times10^{-5}\).

\begin{figure}[H]
\centering
\captionsetup{skip=3pt}
\includegraphics[width=0.98\textwidth]{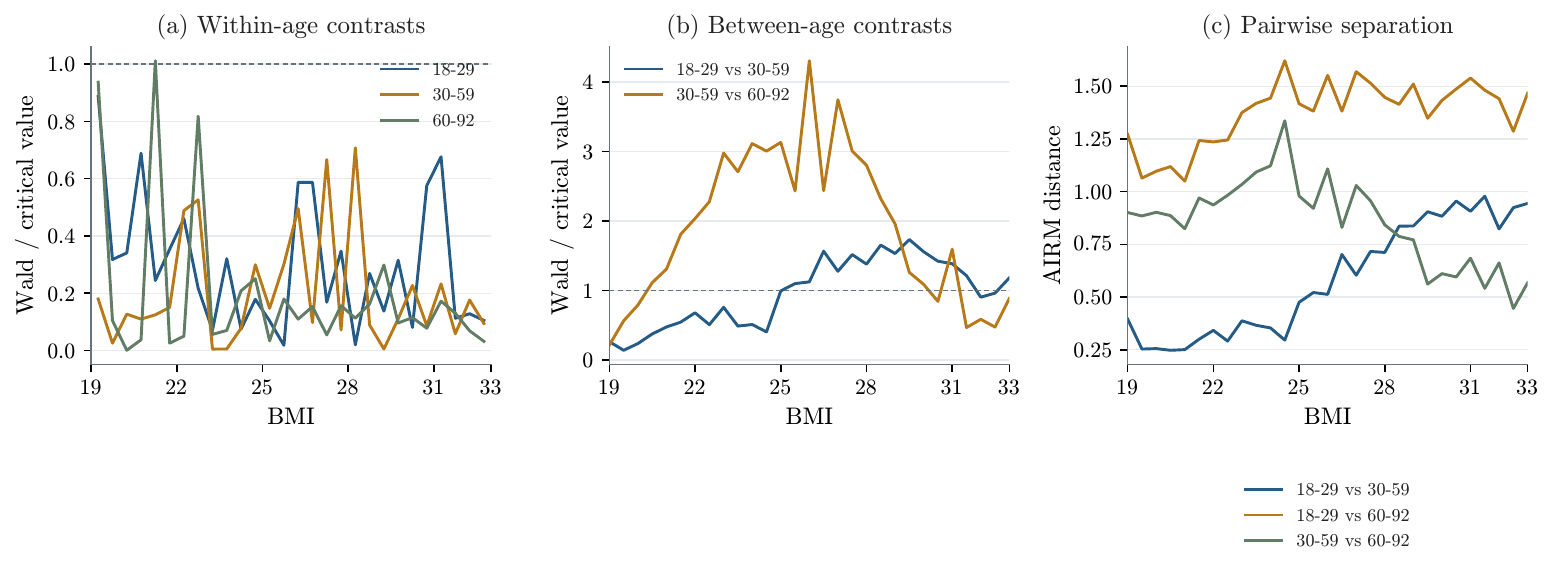}
\caption{Contrasts as functions of BMI and separation between age strata. Panels (a,b) show
Wald statistics divided by the corresponding family's 95\% {critical value}.
Panel (c) shows pairwise AIRM distances.}
\end{figure}

\subsection{Tropical Pacific surface current analysis}
\label{sec:supp-ocean}

\subsubsection{Sample construction and response geometry}
\label{subsec:supp-ocean-acquisition}

We used six-hourly interpolated data from the Global Drifter
Program \citep{LumpkinandCenturioni2019}, restricted to
1990--2024, \(20^\circ\mathrm{S}\)--\(20^\circ\mathrm{N}\), and
\(120^\circ\mathrm{E}\)--\(290^\circ\mathrm{E}\).
For each deployment, we retained at most one observation:
the drogue-attached observation closest to day seven within a
\(\pm3\)-hour window.
We required available position and velocity data and positive
horizontal speed.

Each retained observation was assigned the ERSSTv6 RONI phase
of the three-month season centered on its UTC observation month
\citep{NOAACPC2026}.
An El Ni\~no (La Ni\~na) episode was defined by at least five
consecutive overlapping seasons with RONI at least
\(0.5^\circ\mathrm{C}\) (at most \(-0.5^\circ\mathrm{C}\)).
Observations outside qualifying El Niño or La Niña episodes were
excluded. The retained El Ni\~no and La Ni\~na samples span
eleven and twelve episodes, respectively. The first retained observation
is May 18, 1991, and the
last is December 13, 2024.

\label{subsec:supp-ocean-sasaki}
The three-dimensional unit tangent bundle is
\(\mathrm{UT}\mathbb S^2\coloneqq \{(s,u)\in\mathbb S^2\times\R^3:
s^\top u=0,\ \|u\|=1\}\). The fiber at each fixed position is one-dimensional.
For \(s=(\cos\phi\cos\lambda,\cos\phi\sin\lambda,\sin\phi)^\top\)
with longitude \(\lambda\) and latitude \(|\phi|<\pi/2\) in radians,
the frame in \Cref{sec:supp-sphere-simulation} has coordinate expressions
\[
    e_E(s)=(-\sin\lambda,\cos\lambda,0)^\top,\qquad
    e_N(s)=(-\sin\phi\cos\lambda,-\sin\phi\sin\lambda,\cos\phi)^\top.
\]
For a differentiable curve \((s(t),u(t))\), its tangent vector
\((\xi,\eta)=(s'(0),u'(0))\) satisfies
\(s^\top\xi=u^\top\eta=0\) and \(u^\top\xi+s^\top\eta=0\).
Let \(\nu\coloneqq s\times u\). Every such tangent vector has unique
coordinates \((a,b,c)\) through
\(\xi=au+b\nu\) and \(\eta=-as+c\nu\); these coordinates are orthonormal for
the standard Sasaki metric.  Under this normalization, the map
\(\mathcal R:\mathrm{UT}\mathbb S^2\to SO(3)\),
\(\mathcal R(s,u)\coloneqq [s,u,s\times u]\), is an isometry onto \(SO(3)\) equipped
with the bi-invariant metric
\(\langle RA,RB\rangle_R\coloneqq \tfrac12\tr(A^\top B)\), where
\(A,B\in\mathfrak{so}(3)\coloneqq\{A\in\R^{3\times3}:A^\top=-A\}\).
Identifying
\((s_j,u_j)\) with \(r_j\coloneqq \mathcal R(s_j,u_j)\), the implementation uses
\(d_S(r_0,r_1)=\arccos([\tr(r_0^\top r_1)-1]/2)\),
\(\Log_{r_0}(r_1)=r_0\log(r_0^\top r_1)\), and
\(\Exp_{r_0}\left[r_0A\right]=r_0\exp(A)\).
The logarithm formula applies for \(d_S(r_0,r_1)<\pi\). Along \(r_0\exp(tA)\),
left-trivialized \(B\) is transported
as \(\exp(-tA/2)B\exp(tA/2)\).  This normalization has curvature
\(1/4\) and injectivity radius \(\pi\)
\citep{KlingenbergandSasaki1975}.

\subsubsection{Heading contrasts and simultaneous calibration}
\label{subsec:supp-ocean-heading-inference}

For \(z=(s,u)\), write
\(c_E(z)\coloneqq \langle u,e_E(s)\rangle\) and
\(c_N(z)\coloneqq \langle u,e_N(s)\rangle\).
Suppressing the phase index, set \(\theta(x)\coloneqq H(m_S(x))\)
and \(\widehat\theta(x)\coloneqq H(\widehat m_{\rbc}(x))\),
where \(H\) is the heading map defined in \Cref{sec:ocean-currents}.
For \(\theta\in\mathbb S^1\), set \(e(\theta)\coloneqq (-\theta_2,\theta_1)\)
and identify \(T_\theta\mathbb S^1\) with \(\mathbb R\) through
\(v\mapsto\langle e(\theta),v\rangle\).
In Sasaki coordinates, \((dH)_z\) has row representation
\((-c_E(z)\tan\phi,\ c_N(z)\tan\phi,\ 1)\).

Let \(G\) be the grid in \Cref{sec:ocean-currents}.
Within this subsection, let \(G_i\in\{\mathrm{EN},\mathrm{LN}\}\)
denote the ENSO phase of deployment \(i\). For each phase \(g\), define
\begin{equation}\label{eq:supp-heading-projected-contributions}
 \widehat a_{g,i}(x)\coloneqq
 \left\langle e(\widehat\theta_g(x)),
 (dH)_{\widehat m_{\rbc,g}(x)}d_{\rbc,g,i}(x)\right\rangle,
 \qquad
 \widehat V_g(x,y)\coloneqq \sum_{i:G_i=g}
 \widehat a_{g,i}(x)\widehat a_{g,i}(y),
\end{equation}
where \(d_{\rbc,g,i}(x)\coloneqq -\Log_{\widehat m_{\rbc,g}(x)}
\widehat m_{\rbc,g}^{(-i)}(x)\) is the exact LOO displacement.
The computations substitute the iterative LOO fits described in \Cref{subsec:linear-representation-covariance}.

For \(x\in G\) with
\(\widehat\theta_{\mathrm{EN}}(x)\ne-\widehat\theta_{\mathrm{LN}}(x)\),
define the signed phase contrast
\begin{equation}\label{eq:supp-heading-phase-contrast}
 \widehat\delta(x)\coloneqq
 \left\langle
 e(\widehat\theta_{\mathrm{LN}}(x)),
 \Log_{\widehat\theta_{\mathrm{LN}}(x)}^{\mathbb S^1}
 \widehat\theta_{\mathrm{EN}}(x)
 \right\rangle\in(-\pi,\pi).
\end{equation}
For the contrast in \eqref{eq:supp-heading-phase-contrast}, set
\[
 \widehat V_\Delta(x)\coloneqq
 \widehat V_{\mathrm{EN}}(x,x)+\widehat V_{\mathrm{LN}}(x,x),
 \qquad
 T_{\mathrm{phase}}\coloneqq \max_{x\in G}
 \widehat\delta(x)^2/\widehat V_\Delta(x).
\]
For draw \(k\), use
\begin{equation}\label{eq:supp-heading-phase-multiplier}
 Z_{g,k}^\ast(x)\coloneqq \sum_{i:G_i=g}\xi_{g,i,k}\widehat a_{g,i}(x),
 \qquad
 T_{\mathrm{phase},k}^\ast\coloneqq
 \max_{x\in G}\{Z_{\mathrm{EN},k}^\ast(x)-Z_{\mathrm{LN},k}^\ast(x)\}^2/\widehat V_\Delta(x).
\end{equation}
Each deployment's Gaussian multiplier is shared across every location;
the two phase multiplier vectors are independent.

For mirror symmetry, identify points of \(G\) with their
longitude--latitude coordinates \((\lambda,\phi)\). Within each phase, test
\[
 \theta_g(\lambda,\phi)=\operatorname{diag}(1,-1)\theta_g(\lambda,-\phi),
 \qquad (\lambda,\phi)\in G,\quad \phi\ge0.
\]
The signed logarithm from the reflected southern heading to the northern
heading is the wrapped sum of their angles. Its scalar deletion
contribution is
\[
 \widehat a_{g,i}(\lambda,\phi)+\widehat a_{g,i}(\lambda,-\phi).
\]
At the equator this is \(2\widehat a_{g,i}(\lambda,0)\).
Squaring and summing these combined contributions
retains the covariance between latitudes. Each symmetry statistic {is the maximum of
the} squared studentized contrasts over the 51 pairs of longitude and absolute latitude.
Its multipliers use
\(Z_{g,k}^\ast(\lambda,\phi)+Z_{g,k}^\ast(\lambda,-\phi)\) from \eqref{eq:supp-heading-phase-multiplier}.

We compute add-one multiplier \(p\)-values for the three
maximum Wald statistics and apply Holm's method
\citep{Holm1979}.

{For the phase-specific simultaneous confidence arcs, we maximize}
\(\lvert Z_{g,k}^\ast(x)\rvert/\sqrt{\widehat V_g(x,x)}\)
over \(x\in G\).
% The 95\% simultaneous arcs use the 19,000th order statistic
% of the 19,999 maxima.

\subsection{Theoretical justification for the empirical analyses}
\label{subsec:supp-empirical-technical}

\subsubsection{Age-stratified contrasts}
\label{subsec:supp-age-contrasts}

{We apply the uniform inference results for smooth transformations and comparisons to the three cardiovascular tests, accounting for random stratum sizes and possible degeneracy of the squared-distance Wald statistic under equality across strata.} For LOO calibration, we combine the fits for the three age strata, using one multiplier per subject. Assume the uniform RRBC conditions in every
stratum. Write \(W_i\coloneqq (X_i,Y_i,G_i)\), where
\(G_i\in\{1,2,3\}\) indexes the age strata in \Cref{sec:cardiovascular}, and let \(\widehat\pi_a\coloneqq n_a/n\).
Under the i.i.d. subject model, let \(\pi_a\coloneqq \mathbb P(G=a)>0\). Then
\begin{equation}
\label{eq:supp-group-proportion-replacement-rate}
\max_{1\le a\le3}
\left(
|\widehat\pi_a-\pi_a|
+|\widehat\pi_a^{-1}-\pi_a^{-1}|
+|\widehat\pi_a^{-1/2}-\pi_a^{-1/2}|
\right)=O_p(n^{-1/2}),
\end{equation}
on the event that all strata are nonempty, whose probability tends to one.
The rate follows from the indicator variances and the mean value theorem.
Conditional on the group indicators, the samples are independent and
i.i.d. within strata. The uniform RRBC bounds within each group apply uniformly
when \(n_a/n\) stays away from zero.
\Cref{rem:supp-unstandardized-oracle-score-class}, the localized maximal
inequalities, and \eqref{eq:supp-loo-multiplier-size} give
oracle and multiplier sizes of order \(\sqrt{\log(1/h)}\).
By \eqref{eq:supp-group-proportion-replacement-rate}, {the proportion replacement bounds follow from}
\[
{\begin{aligned}
 O_p(n^{-1/2})O_p(\sqrt{\log(1/h)})
 &=O_p(n^{-1/2}\sqrt{\log(1/h)}),\\
 O_p(n^{-1/2})O_p^\ast(\sqrt{\log(1/h)})
 &=O_p^\ast(n^{-1/2}\sqrt{\log(1/h)}).
\end{aligned}}
\]
Since
\(\log(1/h)=O(\log n)\),
\[
 \frac{n^{-1/2}\sqrt{\log(1/h)}}{\{\log(1/h)\}^{-1/2}}
 =n^{-1/2}\log(1/h)
 =O\!\left(\frac{\log n}{\sqrt n}\right)\to0.
\]
For any
stratum-\(a\) empirical summand \(f_i\),
\(n^{-1}\sum_i\widehat\pi_a^{-1}\mathds{1}\{G_i=a\}f_i
=n_a^{-1}\sum_{i:G_i=a}f_i\).

For \(G_i=a\), let \(d_{\mathrm{rbc},i}^{(a)}(x)\) denote the
contribution for that stratum in \eqref{eq:supp-loo-contributions}.
The normalized contribution for the full sample is
\[
 \widehat b_{(a,x),i}
 \coloneqq nh^{p/2}\mathds{1}\{G_i=a\}d_{\mathrm{rbc},i}^{(a)}(x),
\]
with value zero for \(G_i\ne a\), using \(n_a/\widehat\pi_a=n\).
Since the indicators have disjoint support,
\[
 \frac1n\sum_i\left(\bigoplus_{a=1}^3\widehat b_{(a,x),i}\right)^{\otimes2}
 =\bigoplus_{a=1}^3\widehat\pi_a^{-1}
        \widehat\Omega_{\mathrm{rbc}}^{(a)}(x).
\]
The estimated covariance within each stratum before normalization is
\(\widehat\Omega_{\mathrm{rbc}}^{(a)}(x)/(n_ah^p)\).
The centered oracle coordinates have covariance blocks
\(\pi_a^{-1}\Omega_{\mathrm{rbc},n}^{(a)}(x)\), where
\(\Omega_{\mathrm{rbc},n}^{(a)}\) is the oracle quantity for the conditional distribution within that stratum in
\eqref{eq:rbc-asymptotic-covariance}.

Use the grid points \(x_1,\ldots,x_{29}\) from
\Cref{sec:cardiovascular}. Let \(\widehat S(x)\) be \(S(x)\) with each
\(m_a(x)\) replaced by \(\widehat m_{\mathrm{rbc},a}(x)\), \(a=1,2,3\).
The estimated contrast families are
\[
\begin{aligned}
 \widehat\Delta_n^{\mathrm{(i)}}(a,j)
 &\coloneqq\Log_{\widehat m_{\mathrm{rbc},a}(x_j)}\widehat m_{\mathrm{rbc},a}(x_{j+1}),
 &&a=1,2,3,\quad j=1,\ldots,28,\\
 \widehat\Delta_n^{\mathrm{(ii)}}(a,j)
 &\coloneqq\Log_{\widehat m_{\mathrm{rbc},a}(x_j)}\widehat m_{\mathrm{rbc},a+1}(x_j),
 &&a=1,2,\quad j=1,\ldots,29,\\
 \widehat\Delta_n^{\mathrm{(iii)}}(j)
 &\coloneqq\widehat S(x_{j+1})-\widehat S(x_j),
 &&j=1,\ldots,28.
\end{aligned}
\]
For tests~\textup{(i)}--\textup{(iii)}, \((r,d)\) equals
\((2,3),(2,3),(6,3)\), respectively.

For \Cref{thm:uniform-smooth-functional-transfer}, take
\(\cM=\Sym_{++}^2\), \(q=3\), \(p=1\),
\(S_0=\{1,2,3\}\times\cX_0\), \(M(a,x)=m_a(x)\), and
\(\widehat M_n(a,x)=\widehat m_{\mathrm{rbc},a}(x)\).
For (i), the two selected indices are \((a,x_j),(a,x_{j+1})\);
for (ii), they are \((a,x_j),(a+1,x_j)\); for (iii), the six are
\((1,x_j),(2,x_j),(3,x_j),(1,x_{j+1}),(2,x_{j+1}),(3,x_{j+1})\).
These lists define \(s_{n,\ell}\); \(\mathcal T_n\) is the corresponding
finite set of \((a,j)\) or \(j\) indices.

For \(P\in\Sym_{++}^2\), use the frame
\(E_P:\R^3\to T_P\Sym_{++}^2\) given by
\[
 E_Pu\coloneqq P^{1/2}\Bigl(\sum_{j=1}^3u_jE_j\Bigr)P^{1/2},\qquad
 g_P^{\mathrm{AI}}(E_Pu,E_Pv)=u^\top v.
\]
Here \(u,v\in\R^3\), and \(E_1,E_2,E_3\) are the basis in
\Cref{sec:supp-spd-target}.
For selected indices \(s,t\), write \(P\coloneqq M(s)\), \(Q\coloneqq M(t)\), and define
the logarithm map in fixed tangent coordinates
\[
 \Phi_{s,t}(u,v)
 \coloneqq E_P^\ast \Pi_{\Exp_P\left[E_Pu\right]\to P}^{\mathrm{AI}}
 \Log_{\Exp_P\left[E_Pu\right]}^{\mathrm{AI}}(\Exp_Q\left[E_Qv\right]).
\]
For (i) and (ii), use
\(\Phi_{n,t}\coloneqq \Phi_{s_{n,1}(t),s_{n,2}(t)}\).
Under each null hypothesis for the logarithmic contrast, \(P=Q\), so
\(\Phi_{s,t}(0,0)=0\) and
\({D\Phi_{s,t}(0,0)[(u,v)]}=v-u\).
The theorem's \(\widehat\Delta_n\) is
\(E_P^\ast \Pi_{\widehat M_n(s)\to P}^{\mathrm{AI}}
\widehat\Delta_n^{\mathrm{(k)}}\), \(k=\mathrm{i},\mathrm{ii}\).
This isometric change of coordinates leaves the Wald statistic unchanged.
For (iii), let \(\mathcal S(P_1,P_2,P_3)\) be the vector of the three
squared distances defining \(S(x)\). With
\(P_\ell(v_\ell)\coloneqq \Exp_{M(s_{n,\ell})}\left[E_{M(s_{n,\ell})}v_\ell\right]\), take
\[
 \Phi_{n,j}(v)\coloneqq \mathcal S(P_4(v_4),P_5(v_5),P_6(v_6))
 -\mathcal S(P_1(v_1),P_2(v_2),P_3(v_3)).
\]
Its derivative uses
\[
 (d(d_{\mathrm{AI}}^2))_{(P,Q)}(U,V)
 =-2\langle\Log_P(Q),U\rangle_P
  -2\langle\Log_Q(P),V\rangle_Q.
\]

The base oracle score in the theorem is
\[
 \zeta_n(W;(a,x))\coloneqq \pi_a^{-1}\mathds{1}\{G=a\}
 h^{-p/2}E_{m_a(x)}^\ast(A_x^{(a)})^{-1}
 \varphi_{\rbc,n}^{(a)}(W;x),
\]
where superscript \((a)\) denotes conditioning on \(G=a\).
For (i) and (ii), LOO calibration uses the null differential:
\[
 \widehat\eta_{n,i}(s,t)
 \coloneqq \Pi_{\widehat M_n(t)\to\widehat M_n(s)}^{\mathrm{AI}}
 \widehat b_{t,i}-\widehat b_{s,i}.
\]
The LOO contribution expressed in the theorem's coordinates is
\(E_P^\ast \Pi_{\widehat M_n(s)\to P}^{\mathrm{AI}}
\widehat\eta_{n,i}(s,t)\).
For (iii), write \(\widehat{\boldsymbol M}(x)
\coloneqq (\widehat M_n(1,x),\widehat M_n(2,x),\widehat M_n(3,x))\).
LOO calibration uses the differential of the squared-distance transformation evaluated at the fitted means:
\[
 \widehat c_i(x)
 \coloneqq (d\mathcal S)_{\widehat{\boldsymbol M}(x)}
 (\widehat b_{(1,x),i},\widehat b_{(2,x),i},\widehat b_{(3,x),i}),\qquad
 \widehat\eta_{n,i}^{\mathrm{(iii)}}(j)
 \coloneqq \widehat c_i(x_{j+1})-\widehat c_i(x_j).
\]
Use the observation-level construction {in \eqref{eq:supp-sf-observation-level-construction}} and the maximum Wald statistics {in \eqref{eq:supp-sf-maximum-wald-statistics} for} \Cref{thm:uniform-smooth-functional-transfer}, with the
uncentered feasible scores above in the theorem's coordinates and the
same subject multipliers across all selected indices.
The estimated covariance before normalization is \(\widehat\Xi_n(t)/(nh^p)\).
Smoothness on the compact center region gives \eqref{eq:supp-sf-uniform-c2}.
Let \(\widehat L_{n,t}\) denote the representation in fixed coordinates of the
displayed null differential for {\textup{(i)} {\normalcolor and} \textup{(ii)}} and of
the fitted \(d\mathcal S\) maps for \textup{(iii)}.
Each input is mapped to the fitted tangent space by
\(\Pi_{M(s_{n,\ell}(t))\to\widehat M_n(s_{n,\ell}(t))}^{\mathrm{AI}}
E_{M(s_{n,\ell}(t))}\); for {\textup{(i)} {\normalcolor and} \textup{(ii)}}, the output map is
\(E_P^\ast\Pi_{\widehat M_n(s)\to P}^{\mathrm{AI}}\),
with \(s=s_{n,1}(t)\) and \(P=M(s)\).
Under each stated null, smoothness and the groupwise center rates give
\(\sup_t\|\widehat L_{n,t}-D\Phi_{n,t}(v_n(t))\|_{\mathrm{op}}
=O_p(r_{n,\mathrm{unif}})\) and
\(\sup_t\|v_n(t)\|\le\sqrt r\,e_n=O_p(r_{n,\mathrm{unif}})\).
On \(\{\sup_t\|v_n(t)\|\le\delta\}\), whose probability tends to one,
\eqref{eq:supp-sf-uniform-c2} and \(L_{n,t}=D\Phi_{n,t}(0)\) yield
\[
 \sup_t\|\widehat L_{n,t}-L_{n,t}\|_{\mathrm{op}}
 \le\sup_t\|\widehat L_{n,t}-D\Phi_{n,t}(v_n(t))\|_{\mathrm{op}}
       +C\sup_t\|v_n(t)\|=O_p(r_{n,\mathrm{unif}})
   =o_p(\{\log(1/h)\}^{-1}),
\]
where the last step follows from \eqref{eq:supp-uniform-rate-consequences}.
Thus \eqref{eq:supp-sf-jacobian-replacement} holds.
Equations \eqrefs{eq:supp-loo-cross-covariance-rate,eq:supp-loo-unstandardized-multiplier-replacement,eq:supp-group-proportion-replacement-rate}
therefore give the replacements in
\Cref{lem:supp-sf-finite-arity-projected-replacement}.

Under their respective nulls, assume
\eqref{eq:supp-sf-projected-nondegeneracy} for families
\textup{(i)} and \textup{(ii)}. For \textup{(iii)}, assume either
\(H_0^{\mathrm{(ii)}}\) or
\eqref{eq:supp-sf-projected-nondegeneracy} for family \textup{(iii)}.
Let \(p_1,p_2,p_3\) denote the \(p\)-values.
Set \(p_3=1\) if its Wald statistic is undefined.
Since \(H_0^{\mathrm{(ii)}}\subset H_0^{\mathrm{(iii)}}\), define
\(
\widetilde p_3\coloneqq\max(p_2,p_3).
\)
For the oracle score \(\zeta_n\),
\Cref{cor:supp-projected-rbc-uniform-inference,lem:supp_uniform_random_quantile_calibration}
give asymptotic super-uniformity of \(p_1\) and \(p_2\) under
their respective nulls and of \(p_3\) on the nondegenerate part of
\(H_0^{\mathrm{(iii)}}\setminus H_0^{\mathrm{(ii)}}\).
For a fixed distribution in the stated class under \(H_0^{\mathrm{(iii)}}\),
for every \(0<u<1\),
\[
\mathbb P(\widetilde p_3\le u)
\le
\begin{cases}
\mathbb P(p_2\le u),&\text{if }H_0^{\mathrm{(ii)}}\text{ holds},\\
\mathbb P(p_3\le u),&\text{otherwise},
\end{cases}
\le u+o(1).
\]
Thus Holm's method \citep[Theorem~1]{Holm1979}, applied to
\((p_1,p_2,\widetilde p_3)\), controls asymptotic familywise error
at the nominal level for every fixed distribution satisfying the stated
conditions, including equality of the means in all three age strata.

\subsubsection{Response geometry and heading inference}
\label{subsec:supp-heading-summaries}

The geometry and heading differential are given in
\Cref{subsec:supp-ocean-sasaki,subsec:supp-ocean-heading-inference}. {Because the conditional response distribution is supported on a one-dimensional fiber, we impose conditions on its angular density rather than requiring a density with respect to Riemannian volume on \(\mathrm{UT}\mathbb S^2\).}

\begin{condition}[Conditional distributions on angular fibers]
\label{cond:supp-smooth-pushforward}
Let $U_0$ be a relatively compact open neighborhood of $\cX_0$ in
$\{x\in\mathbb S^2:|x_3|<1\}$.
With the frame from \Cref{sec:supp-sphere-simulation}, set
$\iota(x,v)\coloneqq (x,v_1e_E(x)+v_2e_N(x))$ for $v\in\mathbb S^1$.
Let $\lambda$ be normalized arclength and $A_{\mathrm{ang}}\subset\mathbb S^1$
be compact. Assume
\[
\mathbb P_{Y\mid X=x}
=\iota(x,\cdot)_\#(f_{\mathrm{ang}}(\cdot\mid x)
\lambda|_{A_{\mathrm{ang}}}),
\qquad
\int_{A_{\mathrm{ang}}}f_{\mathrm{ang}}(v\mid x)\,d\lambda(v)=1.
\]
Here $f_{\mathrm{ang}}\ge0$ is jointly Borel, is $C^2$ in $x$, and
satisfies \eqref{eq:supp-pushforward-density-bounds} locally on $U_0$ with a common
$\gamma$ and a bounded envelope. The conditional Fr\'echet mean
$m_S(x)$ under the Sasaki metric is unique on $U_0$, and $m_S(U_0)$ has a compact neighborhood
$C\Subset\{(s,u)\in \mathrm{UT}\mathbb S^2:|s_3|<1\}$ such that, for some
$\eta\in(0,\pi)$,
\begin{equation}
\label{eq:supp-fiber-cut-locus-margin}
\sup_{x\in U_0,\,\omega\in C,\,v\in A_{\mathrm{ang}}}
 d_S(\omega,\iota(x,v))\le\pi-\eta.
\end{equation}
\end{condition}
On relatively compact predictor charts, $\iota$ is smooth with compact
image.
The Hessian formula for a manifold of constant sectional curvature
\citep[Proposition~11.3]{Lee2018} gives eigenvalues $1$ and
$(r/2)\cot(r/2)$ for $r=d_S(\omega,y)<\pi$, with value $1$ at $r=0$. Thus
\eqref{eq:supp-fiber-cut-locus-margin} gives
\[
 \cH_\omega(y)\succeq
 \left\{\frac{\pi-\eta}{2}\cot\!\left(\frac{\pi-\eta}{2}\right)\right\}I
 \succ0.
\]
The cut-locus margin in \eqref{eq:supp-fiber-cut-locus-margin} and this Hessian
bound verify
\Cref{ass:local_support_log_geometry}~\textup{(C)}.

{The conditional Fr\'echet mean lies on the fiber
\(\{\iota(x,v):v\in\mathbb S^1\}\), which contains the support of the conditional response distribution.}
After a left translation in the \(SO(3)\) representation of
\Cref{subsec:supp-ocean-sasaki}, the supporting fiber has unit quaternion
representatives \(\{(r_0,r_1,0,0):r_0^2+r_1^2=1\}\).
Write a unit quaternion as \(q=(q_0,q_1,q_2,q_3)\), with
\(\rho\coloneqq (q_0^2+q_1^2)^{1/2}\).
For \(\rho>0\), its normalized projection
\(q^\parallel\coloneqq \rho^{-1}(q_0,q_1,0,0)\) belongs to the fiber and satisfies
\[
 |\langle q^\parallel,r\rangle|
 =\rho^{-1}|\langle q,r\rangle|
 \ge |\langle q,r\rangle|
\]
for every unit quaternion fiber representative \(r\), so the Sasaki
distance \(2\arccos|\langle q,r\rangle|\) weakly decreases.
For \(\rho=0\), all distances equal \(\pi\); any fiber point weakly
decreases them. Projection onto the fiber cannot increase the criterion,
so uniqueness forces the conditional mean to lie on the fiber.
\(\Log_{m_S(x)}^{\mathrm{UT}\mathbb S^2}Y\) is vertical almost surely given
\(X=x\). Under \Cref{cond:supp-smooth-pushforward}, {the Riemannian Hessian of the conditional Fr\'echet function at \(m_S(x)\) acts as the identity on this vertical tangent line.} The leading covariance obtained by
holding the conditional distribution fixed at \(x\) therefore has rank at most one;
the finite-bandwidth oracle covariance also includes nearby predictor
fibers and need not have rank one.
 Since \((dH)_{m_S(x)}\) maps the vertical unit tangent to a unit tangent
of \(\mathbb S^1\), the projected variance is positive whenever the
conditional angular residual variance is positive.

Under \Cref{cond:supp-smooth-pushforward} and the predictor, kernel, and
bandwidth conditions of the uniform RRBC regime,
\Cref{cor:supp-smooth-pushforward-main-results} extends
\Cref{prop:supp_smoothing_bias_expansion,lem:supp_local_empirical_frechet_linearization_mh,lem:supp_feasible_to_oracle_rbc_replacement} to these response distributions. Hence
\begin{equation}
\label{eq:supp-heading-uniform-center-rate}
\sup_{x\in\cX_0}d_S(\widehat m_{\rbc}(x),m_S(x))
=O_p(r_{n,\mathrm{unif}})=o_p(1).
\end{equation}
\Cref{cor:supp-smooth-pushforward-main-results} also gives the unstandardized RRBC expansion.
\Cref{prop:supp-loo-replacement}
gives LOO covariance consistency and multiplier replacement.

For the phase-specific heading arcs, use a finite local frame cover and
the setup for uniform inference on smooth transformations with \(r=d=1\),
\[
 S_0=\bigsqcup_{\ell=1}^L(\cX_\ell\times\{\ell\}),\qquad \mathcal T_n=\{(x,\ell)\in S_0:x\in G\},\qquad s_{n,1}(x,\ell)=(x,\ell),
\]
\(M(x,\ell)=m_S(x)\), \(\widehat M_n(x,\ell)=\widehat m_{\rbc}(x)\),
\(E_{(x,\ell)}=E_{x,\ell}\), and
\(\zeta_n(W;(x,\ell))=\zeta_{\rbc,n,\ell}(W;x)\). Set
\[
\Phi_{x,\ell}(v)
\coloneqq
\left\langle e(\theta(x)),
\Log_{\theta(x)}^{\mathbb S^1}
H\left(
\Exp_{m_S(x)}^{\mathrm{UT}\mathbb S^2}\left[E_{x,\ell}v\right]
\right)\right\rangle.
\]
Then \(\Phi_{x,\ell}(0)=0\) and
\(D\Phi_{x,\ell}(0)[v]=\langle e(\theta(x)),(dH)_{m_S(x)}E_{x,\ell}v\rangle\).
Using the same LOO fits as in
\eqref{eq:supp-heading-projected-contributions}, we obtain normalized feasible scalar
contributions \(nh\,\widehat a_{g,i}(x)\), which do not depend on \(\ell\).
Here \(n\) is the phase sample size, \(h\) is the realized local
bandwidth at \(x\), and \(p=2\).
Define \(\widehat L_{x,\ell}:\R^q\to\R\) by
\[
 \widehat L_{x,\ell}[v]\coloneqq
 \left\langle e(\widehat\theta(x)),
 (dH)_{\widehat m_{\rbc}(x)}
 \Pi_{m_S(x)\to\widehat m_{\rbc}(x)}E_{x,\ell}v\right\rangle.
\]
On the compact logarithm region, the first two derivatives of the heading
and coordinate maps are bounded uniformly on the finite grid \(G\), giving
\eqref{eq:supp-sf-uniform-c2}. Smoothness and
\eqref{eq:supp-heading-uniform-center-rate} give
\[
 \max_{\ell\le L}\sup_{x\in G\cap\cX_\ell}
 \left\|\widehat L_{x,\ell}-D\Phi_{x,\ell}(0)\right\|_{\mathrm{op}}
 \lesssim \sup_{x\in G}
 d_S(\widehat m_{\rbc}(x),m_S(x))
 =O_p(r_{n,\mathrm{unif}}).
\]
The derivative error and
\eqref{eq:supp-uniform-rate-consequences} imply
\eqref{eq:supp-sf-jacobian-replacement}, so
\Cref{lem:supp-sf-finite-arity-projected-replacement} applies.
If
\[
\liminf_n\inf_{x\in G}
\lambda_{\min}
((dH)_{m_S(x)}\Omega_{\rbc,n}(x)(dH)_{m_S(x)}^\ast )>0,
\]
then \Cref{thm:smooth-functional-inference} and
\Cref{cor:supp-projected-rbc-uniform-inference} give pointwise and simultaneous heading inference for deterministic bandwidth
sequences satisfying the stated uniform RRBC regime and the covariance and multiplier
approximation conditions.

\renewcommand{\refname}{References for the Supplementary Material}
\putbib[ref]
\end{arxivsupplementformat}
\end{bibunit}

\end{document}